\documentclass[aps,prc,10pt,twocolumn,amsmath,amssymb,superscriptaddress]{revtex4-2}
\usepackage{graphicx}
\usepackage{color}
\usepackage{lineno}
\newcommand{\be}{\begin{equation}}
\newcommand{\ee}{\end{equation}}
\newcommand{\beq}{\begin{eqnarray}}
\newcommand{\eeq}{\end{eqnarray}}
\newcommand{\ba}{\begin{array}}
\newcommand{\ea}{\end{array}}

\begin{document}

\title{Extended SAID Partial-Wave Analysis of Pion Photoproduction}

\vspace{3mm}
\date{\today}

%--------------------- AUTHOR LIST ---------------------------
\author{\mbox{William~J.~Briscoe}}
\affiliation{Institute for Nuclear Studies, Department of Physics, The
        George Washington University, Washington, DC 20052, USA}

\author{\mbox{Axel~Schmidt}}
\affiliation{Institute for Nuclear Studies, Department of Physics, 
    The George Washington University, Washington, DC 20052, USA}

\author{\mbox{Igor~Strakovsky}}
\altaffiliation{Corresponding author: \texttt{igor@gwu.edu}}
\affiliation{Institute for Nuclear Studies, Department of Physics, 
    The George Washington University, Washington, DC 20052, USA}

\author{\mbox{Ron~L.~Workman}}
\affiliation{Institute for Nuclear Studies, Department of Physics, 
    The George Washington University, Washington, DC 20052, USA}

\collaboration{SAID Group}

\author{\mbox{Alfred \v{S}varc}}
\affiliation{Rudjer Bo\v{s}kovi\'c Institute, 10000 Zagreb, Croatia}
\affiliation{Tesla Biotech d.o.o., 7 Mandlova, 10000 Zagreb, Croatia}

\noaffiliation

%---------------------- ABSTRACT -----------------------------
\begin{abstract}
A unified Chew-Mandelstam description of single-pion photoproduction data, together with pion- and eta-hadroproduction data, has been extended to include measurements carried out over the last decade. We consider photo-decay amplitudes evaluated at the pole with particular emphasis on $n\gamma$ couplings and the influence of weighting on our fits. Both energy-dependent and single-energy analysis (energy-binned data) are considered.
\end{abstract}

\maketitle

%------------------------------------------------------------
\clearpage
\section{Introduction}
\label{Sec:Intro}

Our knowledge of the baryon spectrum, as determined from analyses of experimental data, has advanced rapidly~\cite{ParticleDataGroup:2022pth} over the past decade. The progress has been most significant for non-strange baryons, due largely to the wealth of new and more precise measurements made at electron accelerators worldwide. The majority of these new measurements have been performed at Jefferson Lab, USA (using the CLAS and Hall~A detectors), with the MAMI accelerator in Mainz, Germany (the Crystal Ball/TAPS detector being particularly well suited for the measurement of neutral final states), and with the Crystal Barrel detector at ELSA in Bonn, Germany. While most of the early progress~\cite{Hoehler:1983g, 
Cutkosky:1979zv, Arndt:2006bf, Shrestha:2012ep} in baryon spectroscopy was based on the analysis of meson-nucleon scattering data, particularly pion-nucleon scattering ($\pi N\to \pi N$, $\pi N\to \pi \pi N$), photon-nucleon interactions offer the possibility of detecting unstable intermediate states with small branchings to the $\pi N$ channel. Many groups have performed either single-channel or multi-channel analyses of these photon-induced reactions. In the more recent single-channel analyses, fits have typically used isobar models~\cite{Drechsel:2007if, Aznauryan:2002gd} with unitarity constraints at the lower energies, $K$-matrix-based formalisms, having built-in cuts associated with inelastic channels~\cite{Workman:2012jf}, and dispersion-relation constraints~\cite{Aznauryan:2002gd, Hanstein:1997tp}. Multi-channel fits have analyzed data (or, in some cases, amplitudes) from
hadronic scattering experiments together with the photon-induced channels. These approaches have utilized unitarity more directly. Among others, analyses have been carried out by MAID~\cite{Drechsel:2007if}, the Bonn-Gatchina~\cite{Anisovich:2010mks},  ANL-Osaka~\cite{Kamano:2016bgm}, Kent State~\cite{Hunt:2018wqz}, and JPAC~\cite{Nys:2016uel} groups, SAID~\cite{Workman:2012jf} (Scattering Analysis Interactive Database) and 
J\"ulich-Bonn~\cite{Ronchen:2018ury}. Here we should also briefly mention the possibility of extracting reaction amplitudes directly  from scattering data with minimal model input. Examples of this approach are described in the analyses of kaon photoproduction data by the Jefferson Lab~\cite{Sandorfi:2010uv} and Bonn-Gatchina~\cite{Anisovich:2014yza} groups. The measurements required for an amplitude extraction with minimal model bias differ depending on whether the goal is to obtain helicity amplitudes (the usual {\it complete experiment} case~\cite{Chiang:1996em}) or partial-wave amplitudes~\cite{Workman:2016irf}. A number of recent studies have shown the limits to model independence~\cite{Workman:2011hi} and the convergence~\cite{Anisovich:2016vzt} of independent fits with the availability of more observables measured with high precision. The above studies have also recently been extended to pseudo-scalar-meson electroproduction~\cite{Tiator:2017cde}.

An objective of this program is the determination of all relevant characteristics of these resonances, \textit{i.e.}, pole positions, widths, principal decay channels, and branching ratios. In order to compare directly with QCD-inspired models and Lattice QCD predictions, there has also been a considerable effort to find ``hidden'' or ``missing'' resonances~\cite{Koniuk:1979vw}, predicted by quark models~\cite{Santopinto:2014opa} and LQCD~\cite{Edwards:2012fx} but not yet confirmed.  Actually, PDG~\cite{ParticleDataGroup:2022pth} reports a third of predicted states by QMs and LQCD. 

Knowledge of the $N$ and $\Delta$ resonance photodecay amplitudes has largely been restricted to the charged states. Apart from lower-energy inverse reaction $\pi^-p\to\gamma n$ measurements, the extraction of the two-body 
$\gamma n\to\pi^-p$ and $\gamma n\to\pi^{0}n$ observables requires the use of a model-dependent nuclear correction, which mainly comes from final state interaction (FSI) effects within the target deuteron~\cite{Migdal:1955ab, Watson:1952ji, Briscoe:2021siu}. As a result, the observables for proton-target experiments are most thoroughly explored and, among neutron-target (deuteron) measurements, the $\pi^0 n$ charge channel is least explored. This problem
is less severe if isospin relations are used to express the four charge-channel amplitudes in terms of 
three isospin amplitudes~\cite{Moorhouse:1973}. Then, in principle, the $\pi^0 n$ production channel can be predicted in terms of the $\pi^0 p$, $\pi^+ n$ and, $\pi^- p$ production channel amplitudes. This approach has been tested~\cite{A2atMAMI:2021iuz} with the improved availability of $\pi^0 n$ data; we will consider this again in the fits to data that follow.

%The importance of improving the $\gamma n$ database relative to the $\gamma p$ one is directly related to the fact that the
%electromagnetic interaction does not conserve isospin symmetry.  The amplitude for the reactions $\gamma N\to\pi X$
%distinct $I = 1/2$ and $I = 3/2$ isospin components, 
%\begin{equation}
%	A_{\gamma\pi^\pm} = \sqrt{2}~(A_{p/n}^{I = 1/2} \mp A_{}^{I=3/2}) \>.
%    \label{eq:eq1}
%\end{equation}
%This expression indicates that the excitation of the $I = 3/2$ $\Delta^\ast$ states can be entirely determined from proton target data. However, %measurements from datasets with both neutron and proton targets are required to determine the isospin $I = 1/2$ amplitudes and to separate the $\gamma %pN^\ast$ and $\gamma nN^\ast$ photocouplings. Only with good data on both proton and neutron targets, one can hope to disentangle isoscalar and isovector %electromagnetic couplings~\cite{Watson:1954uc,Walker:1968xu}.

The GW SAID pion photoproduction analyses have been updated periodically since 1990~\cite{Arndt:1989ww,Arndt:1990ej}, with more frequent updates published through our GW website~\cite{Briscoe:2020zzz}. Often, we present our results with CLAS and A2 Collaborations including determination of the resonance parameters (see, for instance, Refs.~\cite{Chen:2012yv, CLAS:2013pcs, CLAS:2017dco, CLAS:2017kua, A2:2019yud}) while our full analysis was reported 10 years ago~\cite{Workman:2012jf, Svarc:2014sqa}. The present work updates our SAID partial-wave analysis (PWA) results and reports a new determination of photodecay amplitudes and pole positions in the complex energy plane.

High activity of worldwide electromagnetic facilities (JLab, MAMI, CBELSA, MAX-lab, SPring-8, and ELPH) increased the body of the SAID database by a significant amount (see Table~\ref{tab:tbl1}). 60\% of these are $\gamma p\to\pi^{0}p$ data. 
%One can see that the ``neutron'' database grows rapidly which is important for the determination of the neutral couplings. 
A review of the last two decades of using photon beams to measure the production of mesons, and in particular the information that can be obtained on the spectrum of light, non-strange baryons is given in Ref.~\cite{Ireland:2019uwn}. A wealth of $\gamma N\to\pi N$ data, for single- and double-polarization observables, have been anticipated over the past ten years. These data are pivotal in determining the underlying amplitudes in nearly complete experiments, and in discerning between various microscopic models of multichannel reaction theory.

The amplitudes from these analyses can be utilized, in particular, in evaluating contributions to the Gerasimov-Drell-Hearn (GDH) sum rule and related integrals, as was reported recently~\cite{Strakovsky:2022tvu}. 

In the following section (Sec.~\ref{Sec:DB}), we summarize changes to the SAID database since 2012. The changes reflected in our multipoles are displayed in Section~\ref{Sec:ampl}. A comparison of past and recent photo-decay amplitudes, for resonances giving a significant contribution to pion photoproduction, is made in Section~\ref{Sec:res}. Finally, in Section~\ref{Sec:sum}, we summarize our results and comment on possible changes due to further measurements and changes in our parametrization form.

%------------------------------------------------------------
%\clearpage
\section{Extended SAID Database}
\label{Sec:DB}

At present, the SAID database~\cite{Briscoe:2020zzz} has 
35,898 $\gamma p\to\pi^{0}p$,
12,494 $\gamma p\to\pi^{+}n$,
13,473 $\gamma n\to\pi^{-}p$, and
 2,515  $\gamma n\to\pi^{0}n$ data below $E_\gamma = 2700~\mathrm{MeV}$. 
 %While SAID fits are using 15,946, 6,232, 6,670, and 1,205 data, respectively.

Table~\ref{tab:tbl1} accumulates 
21,190 $\gamma p\to\pi^{0}p$,
 1,502  $\gamma p\to\pi^{+}n$,
10,923 $\gamma n\to\pi^{-}p$, and 
 1,763  $\gamma n\to\pi^{0}n$ data published since 2012~\protect\cite{Briscoe:2020zzz}. New measurements mostly cover the $\pi^0p$ sector. Then there are a lot of single ($\Sigma$, $\mathbb P$, and $\mathbb T$,) and double ($\mathbb E$, $\mathbb G$, $\mathbb F$, and $\mathbb H$) polarized data which came recently. It is an essential input for the amplitude reconstruction of the pion photoproduction and determination photocouplings. One can see that the ``neutron'' database grows rapidly which is important for the determination of the neutral photocouplings. 

A full $\chi^2/\mathrm{data}$ contribution for each pion photoproduction reaction vs different PWAs reports in Table~\ref{tab:tbl2}. It presents a partial $\chi^2/\mathrm{data}$ contribution of data from Table~\ref{tab:tbl3} vs different PWAs. 

%---------------------------------------------------------
\begin{table*}[htb!]

\centering \protect\caption{Published data for $\gamma N\to\pi N$ reactions since 2012 as given in the SAID database~\protect\cite{Briscoe:2020zzz}:
    1st column is the reaction,
    2nd column is the observable, 
    3rd column is the number of energy bins,
    4th column is the number of data points.}
    
\vspace{2mm}
{%
\begin{tabular}{|c|c|cccccc|c|c|}
\hline
 Reaction          &Observable    & Nexp & Ndata & E$_\gamma$(min) & E$_\gamma$(max) & $\theta$(min) & $\theta$(max)& Laboratory/ & Ref \tabularnewline
                   &              &      &       &    (MeV)        &      (MeV)      &  (deg)
                   & (deg)     & Collaboration   & \tabularnewline
\hline
$\gamma p\to\pi^0p$& $d\sigma/d\Omega$&  30 &  600 & 147   &  218 & 18  & 162 & MAMI/A2   & 
\cite{A2:2012lnr}\tabularnewline
                   &                  & 269 & 7978 & 218   & 1573 & 15  & 165 & MAMI/A2   & 
\cite{A2:2015mhs}\tabularnewline
                   &                  & 41  &  560 &  862  & 2475 & 15  & 165 & CBELSA/CBELSA/TAPS &  
\cite{CBELSATAPS:2011nwh}\tabularnewline 
                   &                  & 80  & 2030 & 1275  & 5425 & 27  & 140 & JLab/CLAS &  
\cite{CLAS:2017kyf}\tabularnewline 
                   &                  & 22  & 350  & 1325  & 2375 & 47  & 162 & SPring-8/LEPS2\&BGOegg &  
\cite{LEPS2:2019bek}\tabularnewline 
                   &   $\Sigma$       & 26  &  220 & 147   & 206  & 25  & 155 & MAMI/A2   & 
\cite{A2:2012lnr}\tabularnewline
                   &                  & 78  & 1403 & 319   & 649  & 31  & 158 & MAMI/A2   & 
\cite{MAINZ-A2:2016iua}\tabularnewline
                   &                  & 39  & 700  & 1102  & 1862 & 32  & 148 & JLab/CLAS & 
\cite{CLAS:2013pcs}\tabularnewline
                   &                  & 16  & 252  & 1325  & 2350 & 57  & 162 & SPring-8/LEPS2\&BGOegg &  
\cite{LEPS2:2019bek}\tabularnewline 
                   &  $\mathbb P$     &  8  &  152 &  683  &  917 & 51  & 163 & CBELSA/CBELSA/TAPS &  
\cite{CBELSATAPS:2015rtp}\tabularnewline 
                   &                  & 11  & 11   & 1845  & 5631 & 79  & 143 & JLab/GEp-III & \tabularnewline
                   &                  &     &      &       &      &     &     & \& GEp2gamma &
\cite{GEp-III:2011kcr}\tabularnewline
                   &   $\mathbb T$    & 245 & 4343 & 151   &  419 & 5   & 175 & MAMI/A2   & 
\cite{MAINZ-A2:2015yzu}\tabularnewline
                   &                  & 34  & 397  & 440   & 1430 & 30  & 162 & MAMI/A2   & 
\cite{A2:2016tkj}\tabularnewline
                   &                  & 29  & 601  & 683   & 2805 & 29  & 163 & CBELSA/CBELSA/TAPS & 
\cite{CBELSATAPS:2015rtp}\tabularnewline
                   &   $\mathbb E$    & 33  & 456  & 615   & 2250 & 22  & 158 & CBELSA/CBELSA/TAPS & 
\cite{CBELSATAPS:2013btn}\tabularnewline 
                   &   $\mathbb G$    & 22  & 197  & 632   & 2187 & 37  & 144 & JLab/CLAS   & 
\cite{CLAS:2021udy}\tabularnewline 
                   &                  & 19  & 318  & 633   & 1300 & 23  & 156 & CBELSA/CBELSA/TAPS & 
\cite{Thiel:2012yj}\tabularnewline
                   &   $\mathbb F$    & 34  & 397  & 440   & 1430 & 30  & 162 & MAMI/A2   & 
\cite{A2:2016tkj}\tabularnewline
                   &  $\mathbb H$     &  8  &  154 &  683  &  917 & 51  & 163 & CBELSA/CBELSA/TAPS &  
\cite{CBELSATAPS:2015rtp}\tabularnewline 
                   & $\mathbb C_{x'}$ & 45  & 45   & 462   & 1337 & 75  & 140 & MAMI/A2   & 
\cite{Sikora:2013vfa}\tabularnewline
                   &                  & 13  & 13   & 1845  & 5643 & 82  & 143 & JLab/GEp-III & \tabularnewline
                   &                  &     &      &       &      &     &     & \& GEp2gamma &
\cite{GEp-III:2011kcr}\tabularnewline
                   & $\mathbb C_{z'}$ & 13  & 13   & 1845  & 5643 & 80  & 143 & JLab/GEp-III & \tabularnewline
                   &                  &     &      &       &      &     &     & \& GEp2gamma &
\cite{GEp-III:2011kcr}\tabularnewline
\hline
$\gamma p\to\pi^+n$& $\Sigma$         & 39  & 386  & 1102  & 1862 & 32  & 148 & JLab/CLAS & 
\cite{CLAS:2013pcs}\tabularnewline
                   & $\mathbb E$      & 35  & 900  & 363   & 2181 & 20  & 146 & JLab/CLAS &
\cite{CLAS:2015ykk}\tabularnewline
                   & $\mathbb G$      & 22  & 216  & 632   & 2229 & 29  & 142 & MAMI/A2   &
\cite{CLAS:2021udy}\tabularnewline 
\hline
$\gamma n\to\pi^-p$& $\sigma_{tot}$  & 6    & 6    & 150   & 162  &     &     & MAX-lab/PIONS$@$MAX-lab & \cite{Briscoe:2020qat}\tabularnewline
                   &$d\sigma/d\Omega$& 14   & 104  & 301   & 455  & 58  & 133  & MAMI/A2   & 
\cite{Briscoe:2012ni}\tabularnewline
                   &                 & 156  & 8428 & 445   & 2510 & 26  & 128  & JLab/CLAS &
\cite{CLAS:2017dco}\tabularnewline 
                   &                 & 68   & 816  & 1050  & 3500 & 32  & 157  & JLab/CLAS &  
\cite{Chen:2012yv}\tabularnewline
                   & $\Sigma$        & 93   & 1293 & 947   & 2498 & 24  & 145  & JLab/CLAS & 
\cite{Sokhan:2008ts}\tabularnewline
                   &$\mathbb E$      & 21   & 266  & 727   & 2345 & 26  & 154  & JLab/CLAS & 
\cite{CLAS:2017kua}\tabularnewline
\hline
$\gamma n\to\pi^0n$&$d\sigma/d\Omega$& 27   & 492  & 290   & 813  & 32   & 139 & MAMI/A2    &  
\cite{A2:2019yud}\tabularnewline
                   &                 & 49   & 931  & 446   & 1427 & 32   & 162 & MAMI/A2 &  
\cite{A2:2018jcd}\tabularnewline
                   & $\Sigma$        & 12   & 189  & 390   & 610  & 49   & 148 & MAMI/A2    & 
\cite{A2atMAMI:2021iuz}\tabularnewline
                   & $\mathbb E$     & 17   & 151  & 446   & 1427 & 46   & 154 & MAMI/A2    & 
\cite{Dieterle:2017myg}\tabularnewline
\hline
\end{tabular}} \label{tab:tbl1}
\end{table*}
%---------------------------------------------------------
%---------------------------------------------------------
\begin{table*}[htb!]

\centering \protect\caption{Comparison of $\chi^{2}$ per datum values for all charged and neutral channels covering fit energy range. 
%There are several new SAID fits: SM22, SM44, %NM22, and WM22 which are valid up to %E$_\gamma = 2700~\mathrm{MeV}$. 
The previous SAID fit, CM12, was published in Ref.~\cite{Workman:2012jf} (and is valid up to E$_\gamma = 2700~\mathrm{MeV}$). CM12 is compared to both the current database and data before 2012. All data are available in the SAID database (DB)~\protect\cite{Briscoe:2020zzz}. For the SM44 fit, $\pi^0n$ data were weighted by an arbitrary factor of 4. For the WM22 fit, all data with large $\chi^2/\mathrm{data}$ for the SM22 solution  (data are listed in Table~\ref{tab:tbl3}) were weighted by an arbitrary factor of 4. The NM22 solution represents a fit without the inclusion of $\pi^0n$ data.
The previous MAID2007 solution is valid up to E$_\gamma = 1680~\mathrm{MeV}$ ($W = 2~\mathrm{GeV}$)~\cite{Drechsel:2007if}.
}
\vspace{2mm}
{%
\begin{tabular}{|c|c|c|c|c|c|}
\hline
Solution &Observable& $\chi^2/(\pi^0p$ data) & $\chi^2/(\pi^+n$ data) &  $\chi^2/(\pi^-p$ data) & $\chi^2/(\pi^0n$ data) \tabularnewline
\hline
SM22     &Total     &30399/15901=  1.92&13945/6194=  2.25&12267/6662=  1.84&4190/1205=  3.48 \tabularnewline
         &UnPol     & 9842/5730=   1.72& 4984/2603=  1.91& 7497/4706=  1.59&1995/649=   3.07 \tabularnewline
         &SinglePol &16036/8249=   1.94& 6078/2483=  2.45& 4014/1684=  2.38&1258/405=   3.11 \tabularnewline
         &DoublePol & 4521/1922=   2.35& 2883/1108=  2.60&  765/275=   2.78& 937/151=   6.21 \tabularnewline
\hline
SM44     &Total     &30870/15901=  1.94&14293/6194=  2.31&12358/6662=  1.86&3361/1205=  2.79 \tabularnewline
         &UnPol     & 9880/5730=   1.72& 5154/2603=  1.98& 7832/4706=  1.66&1648/649=   2.54 \tabularnewline
         &SinglePol &16405/8249=   1.99& 6229/2483=  2.51& 3830/1684=  2.27& 823/405=   2.03 \tabularnewline
         &DoublePol & 4585/1922=   2.39& 2910/1108=  2.63&  696/275=   2.53& 890/151=   5.89 \tabularnewline
\hline
NM22     &Total     &29998/15901=  1.89&13592/6194= 2.19&11992/6662=   1.80&8531/1205= 7.08 \tabularnewline
         &UnPol     & 9887/5730=   1.73& 4757/2603= 1.83& 7262/4706=   1.54&2322/649=  3.58 \tabularnewline
         &SinglePol &15662/8240=   1.90& 5915/2483= 2.38& 3746/1684=   2.22&4570/405= 11.28 \tabularnewline
         &DoublePol & 4449/1922=   2.31& 2920/1108= 2.64&  984/275=    3.58&1639/151= 10.85 \tabularnewline
\hline
WM22     &Total     &31315/15901=  1.97&14038/6194=  2.27&12819/6662=  1.92&3853/1205= 3.20 \tabularnewline
         &UnPol     & 9816/5730=   1.71& 4659/2603=  1.79& 7735/4706=  1.64&2113/649=  3.26 \tabularnewline
         &SinglePol &16922/8249=   2.05& 6537/2483=  2.63& 4258/1684=  2.53& 885/405=  2.19 \tabularnewline
         &DoublePol & 4577/1922=   2.38& 2.842/1108= 2.57&  826/275=   3.00& 855/151=  5.66 \tabularnewline
\hline
\hline
CM12     &Total     &78254/15901=  4.92&27933/6194=  4.51&222454/6662=33.39&7024/1205=  5.89 \tabularnewline
(current &UnPol     &18074/5730=   3.15& 4565/2603=  1.75&65514/4706= 13.92&4063/649=   6.26 \tabularnewline
 DB)     &SinglePol &50016/8249=   6.06&12221/2483=  4.92&154303/1684=91.62& 976/405=   2.41 \tabularnewline
         &DoublePol &10164/1922=   5.26&11147/1108= 10.06&2637/275=    9.59&1985/151=  13.15 \tabularnewline
\hline
CM12     &Total     &10544/4507=   2.34&10444/4916=  2.12& 2486/1509=  1.65&  987/373=  2.65 \tabularnewline
(old DB) &UnPol     & 2682/1094=   2.45& 4247/2459=  1.73& 1769/1118=  1.58&  475/157=  3.03 \tabularnewline
         &SinglePol & 5846/2723=   2.15& 3312/1523=  2.18&  564/304=   1.86&  512/216=  2.37 \tabularnewline
         &DoublePol & 2016/690=    2.92& 2885/934=   3.09&  153/87=    0.82&                 \tabularnewline
\hline
MAID2007 &Total     &170832/14454=11.82&128063/5396=23.73&102968/5520=18.65&29390/1205=24.39 \tabularnewline
(current &UnPol     & 74153/5188= 14.29& 24533/2210=11.10& 40840/4166= 9.80& 2812/649=  4.33 \tabularnewline
 DB)     &SinglePol & 84286/7578= 11.12& 96337/2168=44.44& 59097/1182=50.00&22087/405= 54.54 \tabularnewline
         &DoublePol & 12393/1688=  7.34&  7193/1018= 7.07& 3031/172=  17.62& 4494/151= 29.76 \tabularnewline
\hline
\end{tabular}} \label{tab:tbl2}
\end{table*}
%---------------------------------------------------------
%---------------------------------------------------------
\begin{table*}[htb!]

\centering \protect\caption{List of data with large $\chi^2/\mathrm{data}$ for the SM22 and associated fits. Notation for solutions is given in the caption of Table~\ref{tab:tbl2}.}

\vspace{2mm}
{%
\begin{tabular}{|c|c|c|c|c|c|c|c|c|c|c|c|}
\hline
 Reaction          &       Obs        & E$_\gamma$   &Data& MAID2007& CM12& SM22& SM44& WM22& NM22& Ref. \tabularnewline
                   &                  & (MeV)        &              & $\chi^2/\mathrm{data}$  &$\chi^2/\mathrm{data}$
                   &                             $\chi^2/\mathrm{data}$  &$\chi^2/\mathrm{data}$
                   &                             $\chi^2/\mathrm{data}$  &$\chi^2/\mathrm{data}$ & \tabularnewline
\hline
$\gamma p\to\pi^0p$& $d\sigma/d\Omega$&675$-$2875    &620&40.56& 2.38& 3.28& 3.09& 2.18& 3.34& \cite{Dugger:2007bt} \tabularnewline
                   & $\mathbb P$      &1845$-$2776   &3&     & 242.& 107.& 83.1&26.13&89.01& \cite{GEp-III:2011kcr} \tabularnewline 
                   &                  &773$-$2472    &29& 8.47& 5.45&12.83&12.93& 8.69&13.10& \cite{Wijesooriya:2002uc} \tabularnewline
                   & $\mathbb G$      &632$-$2187    &197&11.45&46.34& 4.23& 4.43& 4.02& 3.87& \cite{CLAS:2021udy} \tabularnewline
                   & ~~$\mathbb C_{x'}$ &1845$-$2776 &3&     & 985.& 8.75& 5.18& 9.39& 7.53& \cite{GEp-III:2011kcr} \tabularnewline
                   &                  &773$-$2472    &28&28.25& 9.96& 7.64& 7.82& 4.89& 8.39& \cite{Wijesooriya:2002uc} \tabularnewline 
                   & ~~$\mathbb C_{z'}$ &1845$-$2776 &3&     &1370.& 8.68&14.40& 2.46& 7.87& \cite{GEp-III:2011kcr} \tabularnewline 
                   &                  &773$-$2472&25 &35.44&12.80&12.00& 8.44& 9.16&13.28& \cite{Wijesooriya:2002uc} \tabularnewline
\hline
$\gamma p\to\pi^+n$& $d\sigma/d\Omega$&725$-$2875&618&65.71& 2.08& 2.75& 2.83& 1.82& 2.44& \cite{CLAS:2009tyz} \tabularnewline
                   & $\mathbb G$      &632$-$2229&216&21.09&25.33& 4.42& 4.66& 3.57& 4.49& \cite{CLAS:2021udy} \tabularnewline
\hline
$\gamma n\to\pi^0n$& $\Sigma$         &703$-$1475&216&100.1& 2.37& 4.72& 2.81& 2.93&19.26& \cite{DiSalvo:2009zz} \tabularnewline
                   & $\mathbb E$      &446$-$1427&151&29.75&13.14& 6.21& 5.89& 5.66&10.85& \cite{Dieterle:2017myg} \tabularnewline
\hline
\end{tabular}} \label{tab:tbl3}
\end{table*}

%----------------------------------------------------------------------
\begin{figure*}[hbt!]
%\vspace{0.4cm}
\centering
{
    \includegraphics[width=0.32\textwidth,angle=90,keepaspectratio]{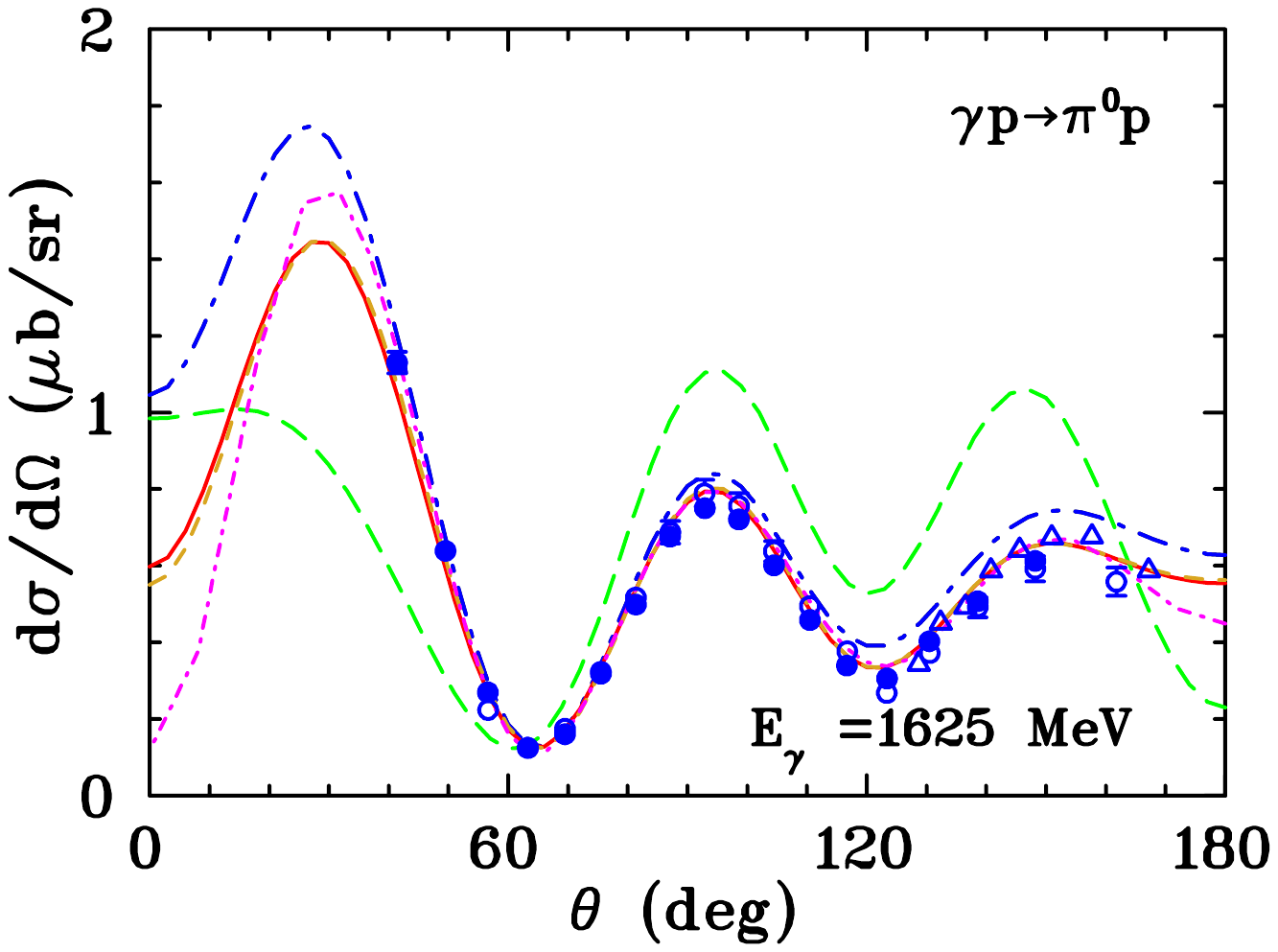}
    \includegraphics[width=0.32\textwidth,angle=90,keepaspectratio]{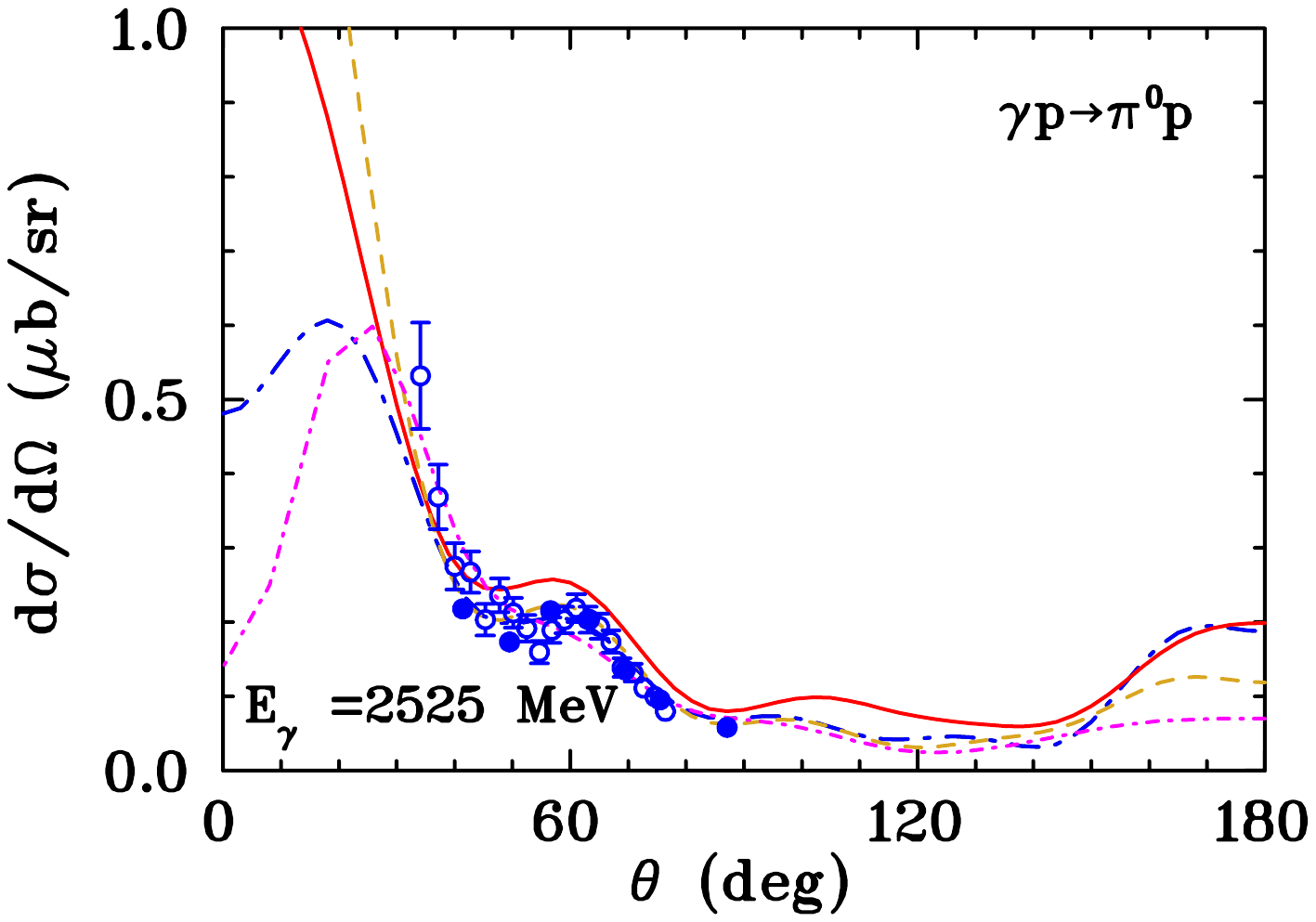}
}
\centering
{
    \includegraphics[width=0.32\textwidth,angle=90,keepaspectratio]{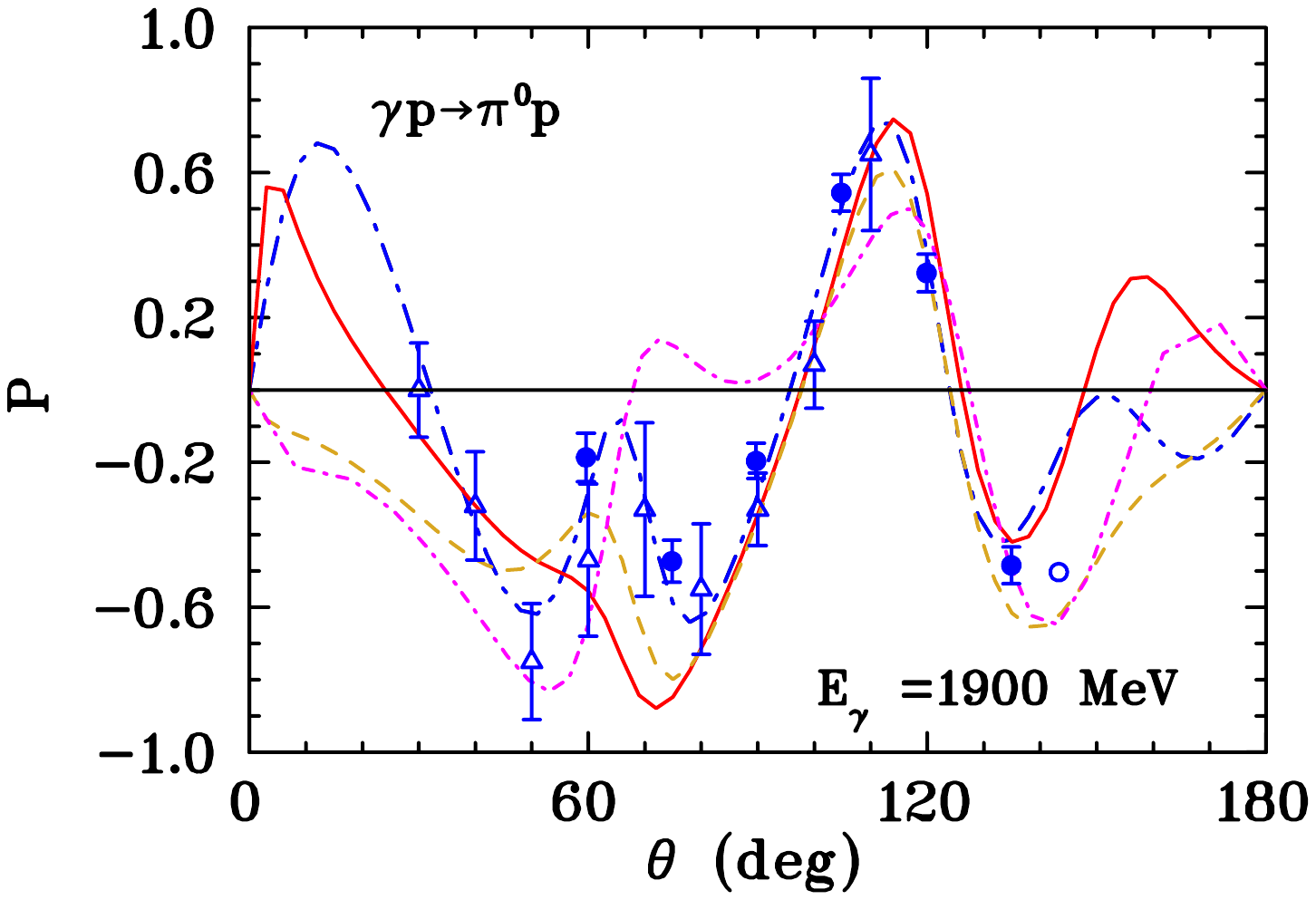}
    \includegraphics[width=0.32\textwidth,angle=90,keepaspectratio]{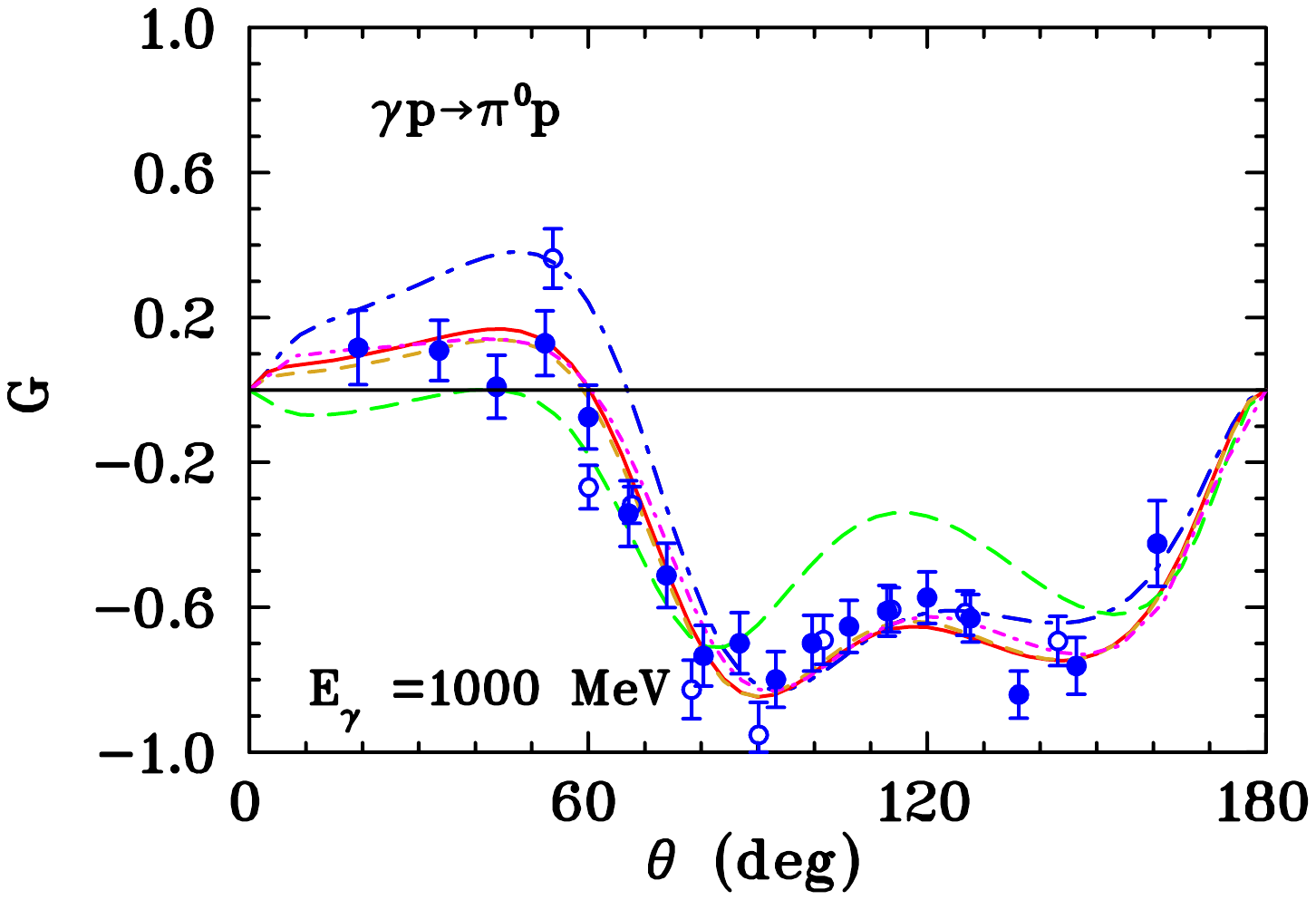}
}
\centering
{
    \includegraphics[width=0.32\textwidth,angle=90,keepaspectratio]{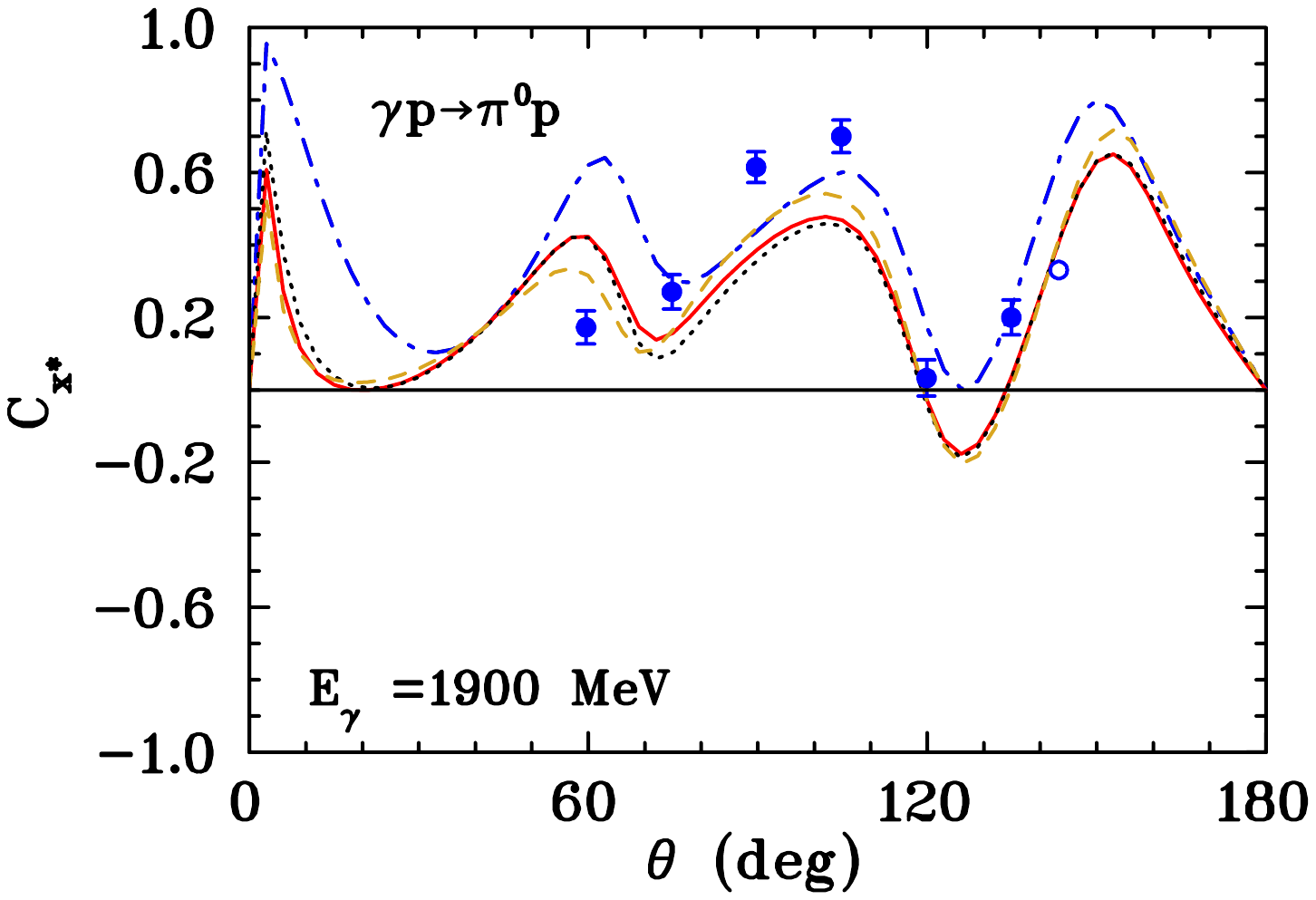}
    \includegraphics[width=0.32\textwidth,angle=90,keepaspectratio]{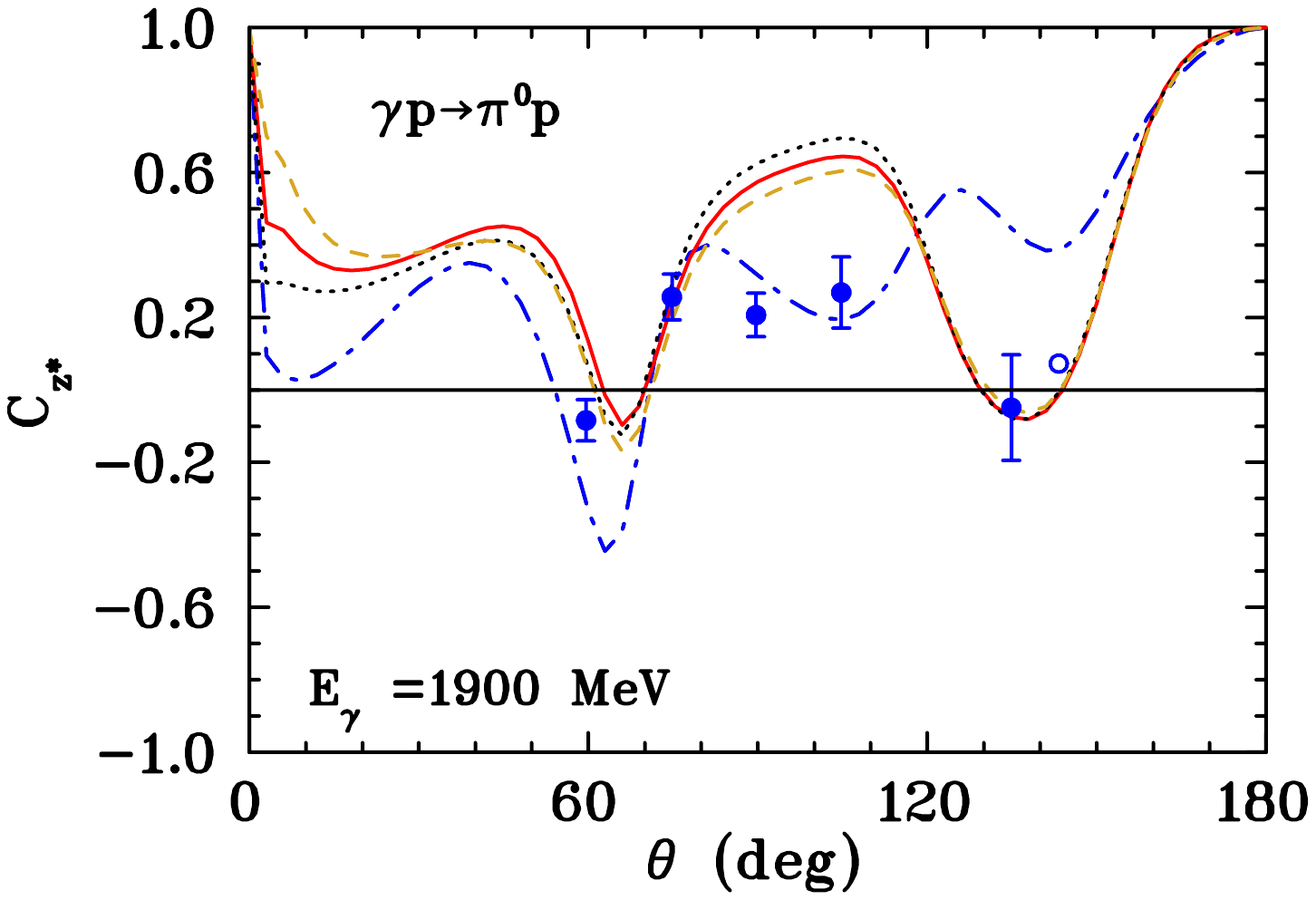}
}
\centering
{
    \includegraphics[width=0.32\textwidth,angle=90,keepaspectratio]{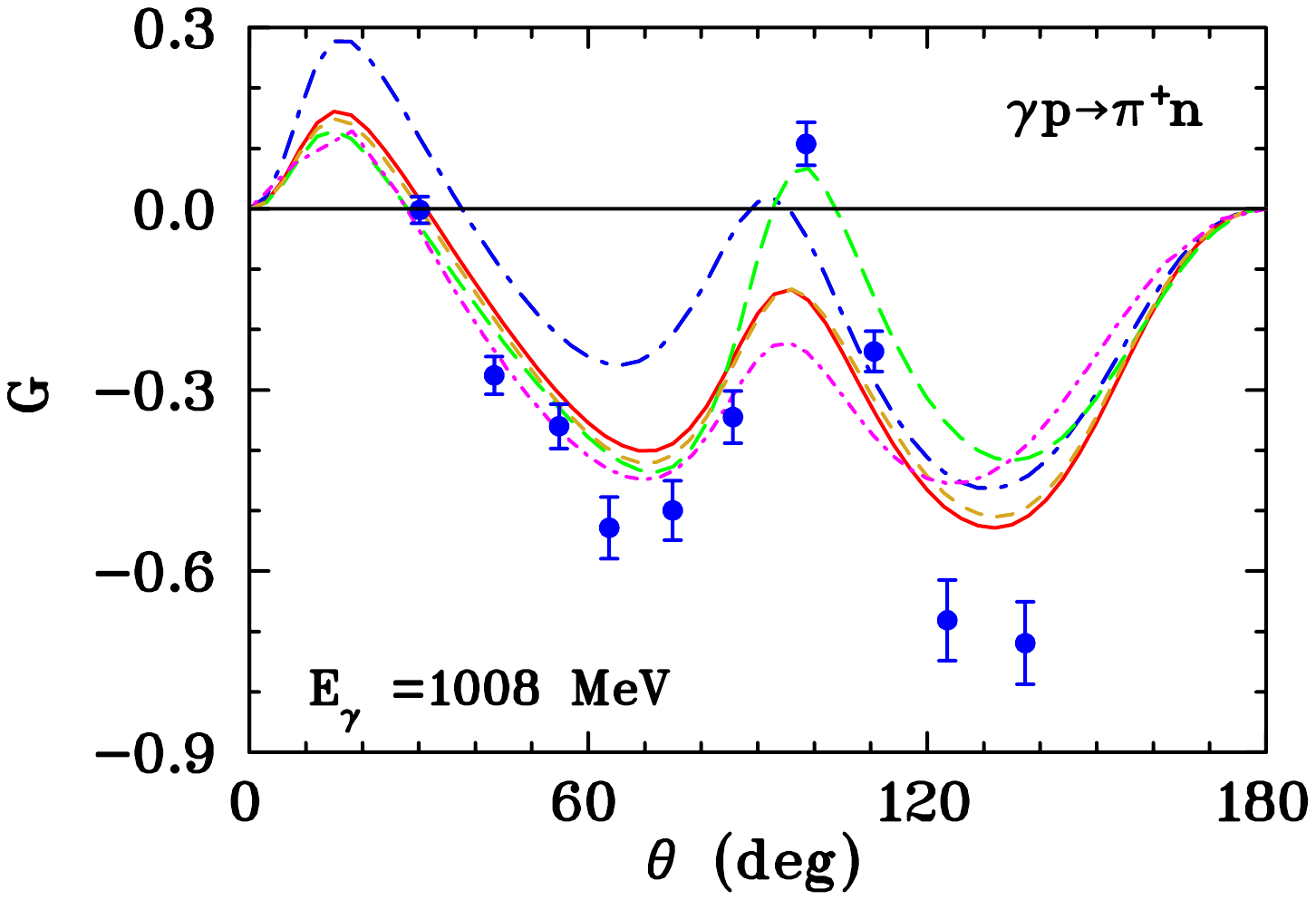}
}

\caption{Samples of pion photoproduction off the proton.  Data for $\gamma p\to\pi^0p$ are from Refs.~\cite{Dugger:2007bt, CLAS:2017kyf, Wijesooriya:2002uc, GEp-III:2011kcr, Bussey:1979wt, Thiel:2012yj, CLAS:2021udy, Sikora:2013vfa} and for $\gamma p\to\pi^+n$ are from Ref.~\cite{CLAS:2021udy}. Notation for solutions is given in the caption of Table~\ref{tab:tbl2}. The SAID SM22 (WM22) fit is shown as a red solid (yellow dashed) curve. SAID CM12~\cite{Workman:2012jf} (MAID2007~\cite{Drechsel:2007if}) predictions shown as blue dash-dotted (green dashed) curves. BG2019~\cite{CBELSATAPS:2014wvh} predictions are shown as magenta short dash-dotted curves.
}
\label{fig:obs1}
\end{figure*}

%----------------------------------------------------------------------
\begin{figure*}[hbt!]
%\vspace{0.4cm}
\centering
{
    \includegraphics[width=0.32\textwidth,angle=90,keepaspectratio]{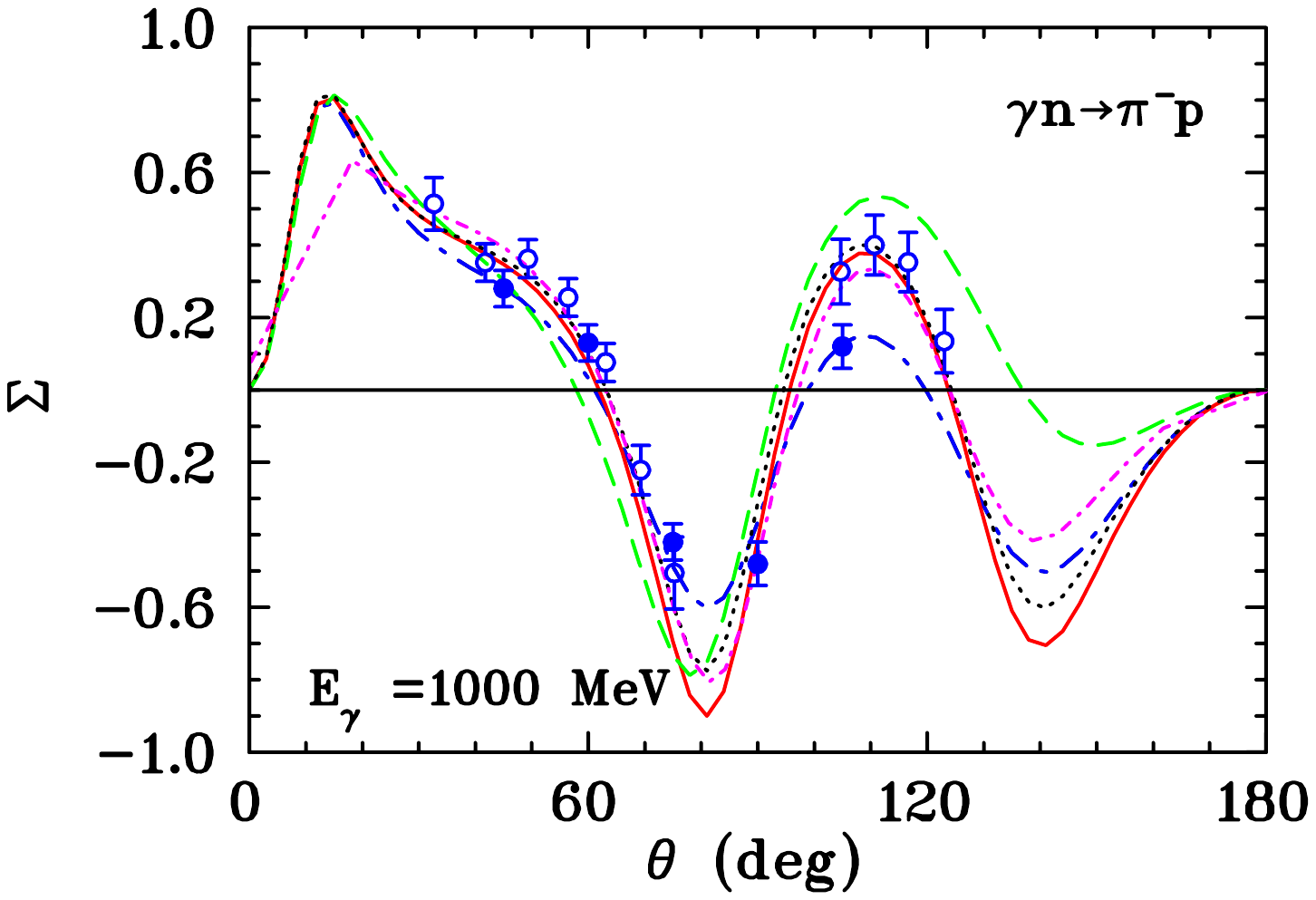}
    \includegraphics[width=0.32\textwidth,angle=90,keepaspectratio]{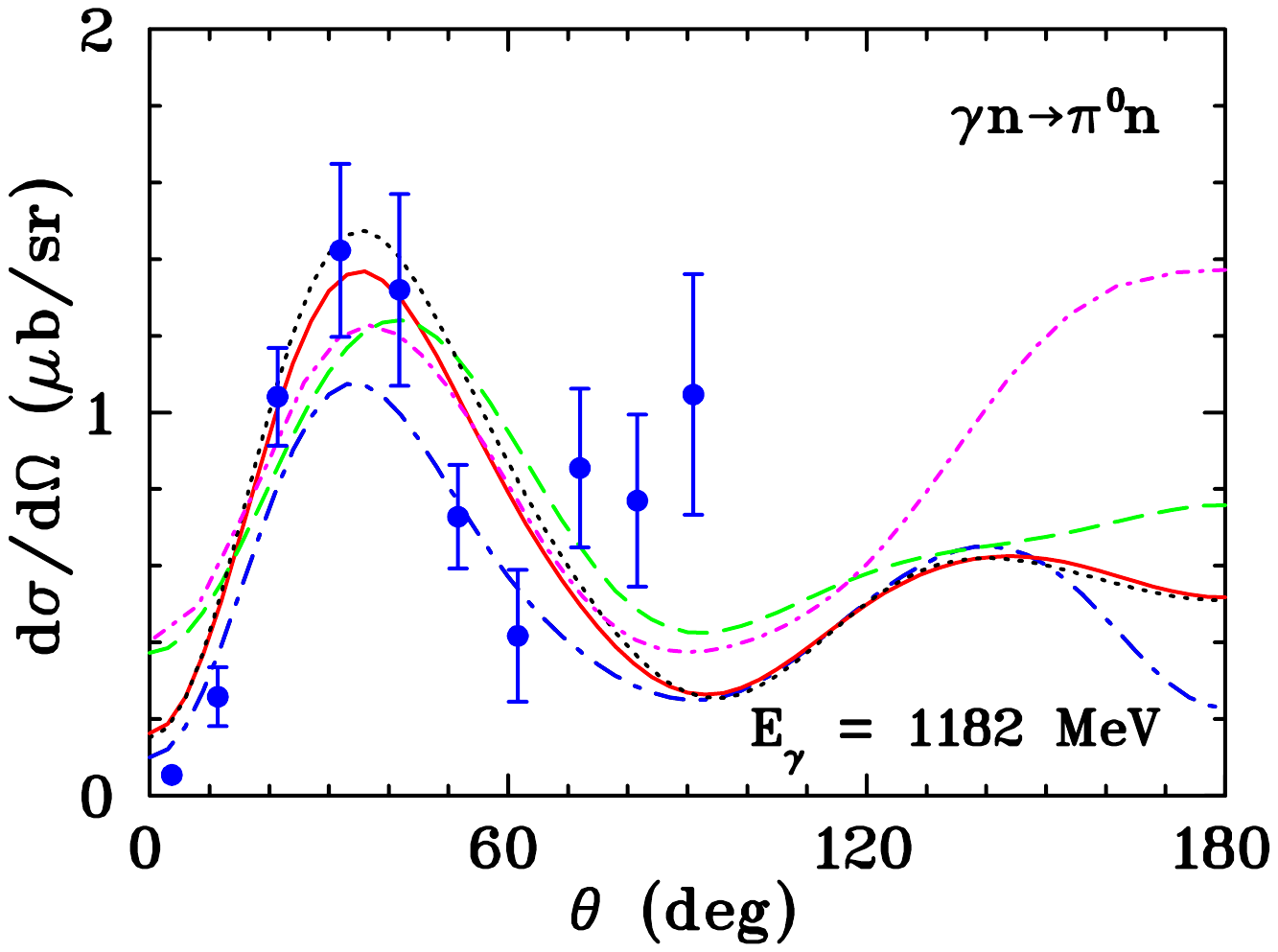}
}
\centering
{
    \includegraphics[width=0.32\textwidth,angle=90,keepaspectratio]{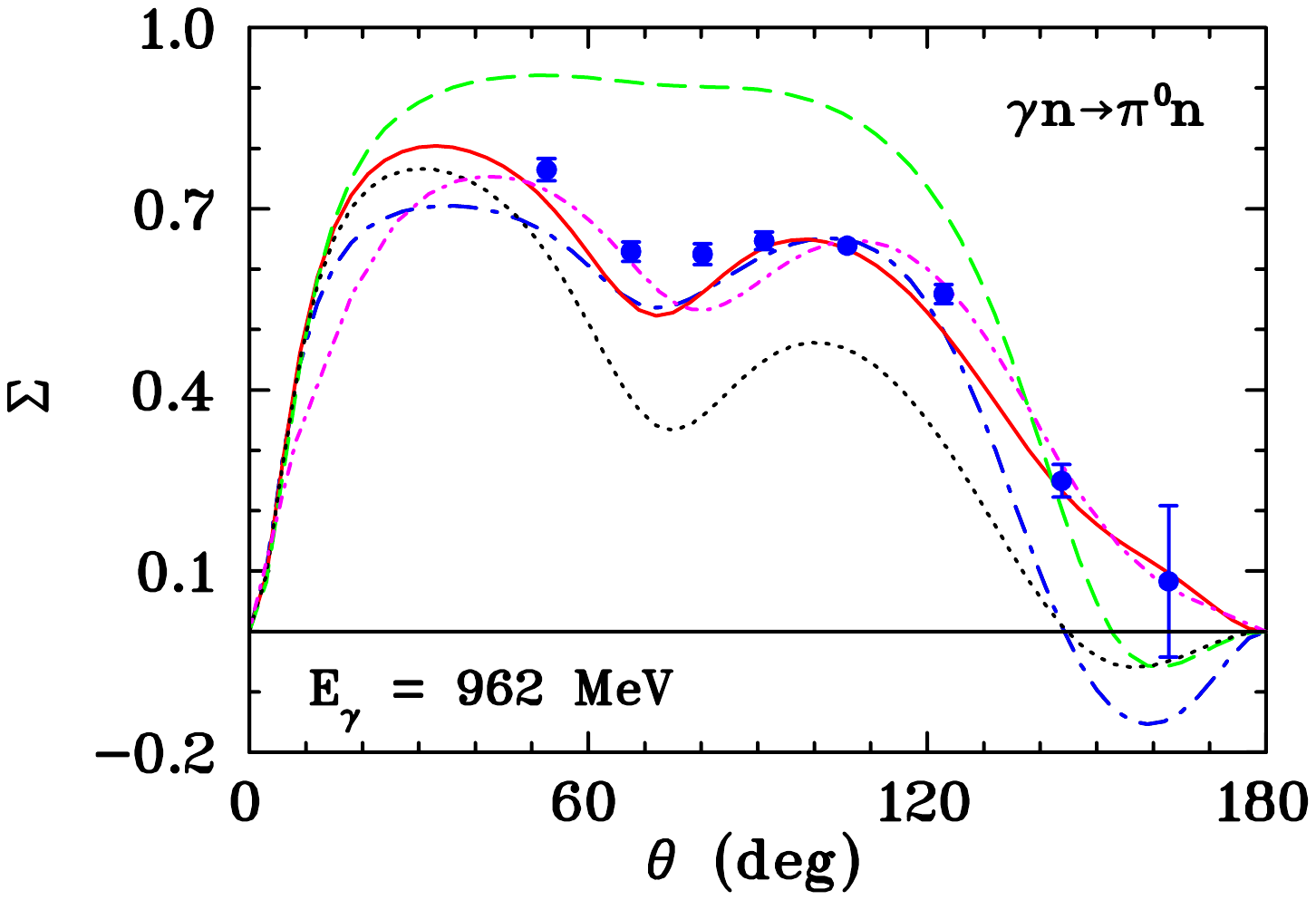}
    \includegraphics[width=0.32\textwidth,angle=90,keepaspectratio]{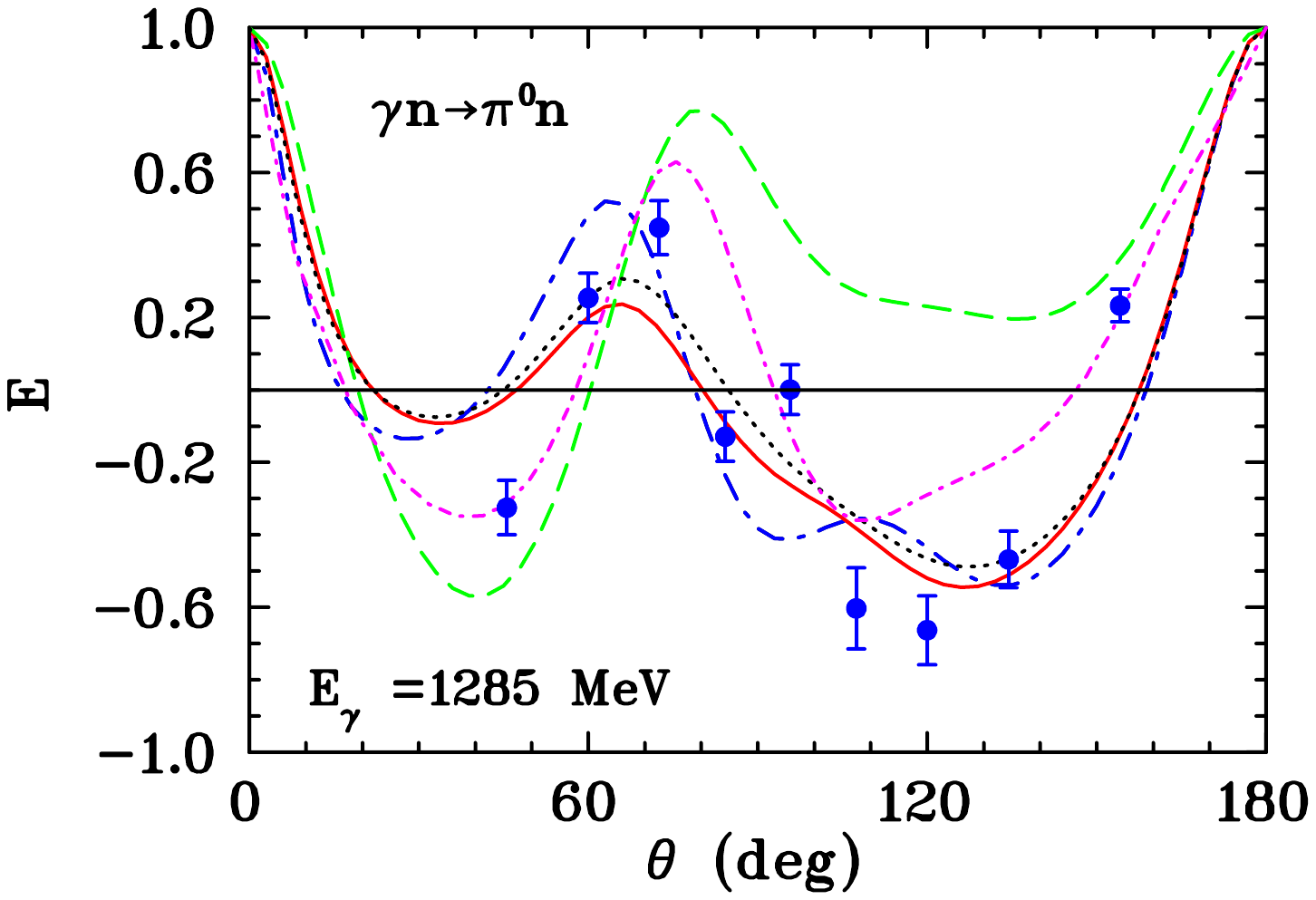}
}

\caption{Samples of pion photoproduction off the neutron.  Data for $\gamma n\to\pi^-p$ are from Refs.~\cite{Sokhan:2008ts, Adamian:1989kg} and for $\gamma n\to\pi^0n$ are from Refs.~\cite{Clinesmith:1967zn, DiSalvo:2009zz, Dieterle:2017myg}. Notation for solutions is given in the caption of Table~\ref{tab:tbl2}.  The SAID SM22 (NM22) fit is shown as a red solid (black dotted) curve. SAID CM12~\cite{Workman:2012jf} (MAID2007~\cite{Drechsel:2007if}) predictions are shown as blue dash-dotted (green dashed) curves.  BG2019~\cite{CBELSATAPS:2014wvh} predictions are shown as magenta short dash-dotted curves.
}
\label{fig:obs1a}
\end{figure*}

%---------------------------------------------------------
\begin{table*}[htb!]

\centering \protect\caption{Comparison $\chi^2/\mathrm{data}$ for published data since 2012 as given in Table~\ref{tab:tbl1} and available in the SAID database~\protect\cite{Briscoe:2020zzz}. Notation for solutions is given in the caption of Table~\ref{tab:tbl2}. Data, which are partially (completely) excluded in the SAID fits, denoted by $^\ddagger$ ($^\dagger$).}

\vspace{2mm}
{%
\begin{tabular}{|c|c|c|c|c|c|c|c|c|}
\hline
 Reaction          &       Obs    & MAID2007 & CM12 & SM22 & SM44 & WM22 & NM22 & Ref. \tabularnewline
                   &              &$\chi^2/\mathrm{data}$& $\chi^2/\mathrm{data}$
                   &$\chi^2/\mathrm{data}$
                   &$\chi^2/\mathrm{data}$&$\chi^2/\mathrm{data}$ 
                   &$\chi^2/\mathrm{data}$ & \tabularnewline
\hline
$\gamma p\to\pi^0p$& $d\sigma/d\Omega$&10.44& 7.08& 1.32& 1.36& 1.32& 1.33 & \cite{A2:2012lnr}$^\ddagger$ \tabularnewline
                   &                  &12.50& 3.01& 1.40& 1.44& 1.51& 1.40 & \cite{A2:2015mhs}$^\ddagger$ \tabularnewline
                   &                  & 4.44& 2.33& 3.46& 3.41& 3.22& 3.49 & \cite{CBELSATAPS:2011nwh}$^\dagger$ \tabularnewline 
                   &                  &18.28& 2.34& 2.69& 2.50& 2.37& 2.77 & \cite{CLAS:2017kyf}$^\dagger$ \tabularnewline 
                   &                  &16.15& 3.63& 2.39& 2.31& 2.74& 2.45 & \cite{LEPS2:2019bek} \tabularnewline 
                   &  $\Sigma$        &41.69& 0.99& 1.40& 1.39& 1.33& 1.39 & \cite{A2:2012lnr} \tabularnewline
                   &                  & 2.25& 1.42& 1.16& 1.12& 1.22& 1.17 & \cite{MAINZ-A2:2016iua}$^\ddagger$ \tabularnewline
                   &                  &72.13&43.81& 3.62& 3.87& 4.04& 3.47 & \cite{CLAS:2013pcs} \tabularnewline
                   &                  & 4.93&11.21& 1.95& 1.96& 2.46& 1.81 & \cite{LEPS2:2019bek} \tabularnewline 
                   &  $\mathbb P$     & 2.13& 1.50& 1.04& 1.09& 1.17& 1.05 & \cite{CBELSATAPS:2015rtp} \tabularnewline 
                   &                  &     &241.0& 6.47&82.62& 26.1&89.01 & \cite{GEp-III:2011kcr} \tabularnewline
                   &  $\mathbb T$     & 1.30& 1.41& 1.06& 1.07& 1.09& 1.04 & \cite{MAINZ-A2:2015yzu}$^\ddagger$ \tabularnewline
                   &                  & 9.15& 5.80& 3.09& 3.25& 3.28& 2.94 & \cite{A2:2016tkj} \tabularnewline
                   &                  &12.25& 4.14& 2.17& 2.24& 2.43& 2.05 & \cite{CBELSATAPS:2015rtp} \tabularnewline
                   &  $\mathbb E$     &15.14& 4.22& 2.11& 2.20& 2.62& 2.03 & \cite{CBELSATAPS:2013btn} \tabularnewline 
                   &  $\mathbb G$     &11.45& 6.38& 4.23& 4.43& 4.02& 4.20 & \cite{CLAS:2021udy} \tabularnewline 
                   &                  & 3.42& 3.90& 1.26& 1.26& 1.21& 1.20 & \cite{Thiel:2012yj} \tabularnewline
                   &  $\mathbb F$     & 3.48& 3.34& 2.33& 2.34& 2.26& 2.28 & \cite{A2:2016tkj} \tabularnewline
                   &  $\mathbb H$     & 4.38& 6.25& 1.70& 1.96& 1.89& 1.44 & \cite{CBELSATAPS:2015rtp} \tabularnewline 
                   & $\mathbb C_{x'}$ & 2.07& 2.36& 1.71& 1.71& 1.76& 1.73 & \cite{Sikora:2013vfa} \tabularnewline
                   &                  &     &984.0& 8.90& 5.28& 9.53& 7.53 & \cite{GEp-III:2011kcr} \tabularnewline
                   & $\mathbb C_{z'}$ &     &1370.& 8.74&14.49& 2.48& 7.87 & \cite{GEp-III:2011kcr} \tabularnewline
\hline
$\gamma p\to\pi^+n$& $\Sigma$         &285.1&18.37& 3.00& 3.14& 3.81& 2.97 & \cite{CLAS:2013pcs} \tabularnewline
                   & $\mathbb E$      & 5.09& 9.82& 1.96& 1.86& 2.21& 2.03 & \cite{CLAS:2015ykk}$^\ddagger$ \tabularnewline
                   & $\mathbb G$      &21.09&25.33& 4.42& 6.64& 3.57& 4.49 & \cite{CLAS:2021udy} \tabularnewline
\hline
$\gamma n\to\pi^-p$& $\sigma_{tot}$   & 0.33& 0.05& 0.06& 0.20& 0.10& 0.90 & \cite{Briscoe:2020qat} \tabularnewline
                   &$d\sigma/d\Omega$ & 5.99& 4.61& 3.27& 3.96& 2.78& 3.22 & \cite{Briscoe:2012ni} \tabularnewline
                   &                  &14.88&20.39& 1.28& 1.30& 1.33& 1.25 & \cite{CLAS:2017dco}$^\ddagger$
\tabularnewline 
                   &                  &30.39&76.83& 3.97& 3.97& 3.77& 4.17 & \cite{Chen:2012yv}$^\dagger$ \tabularnewline
                   & $\Sigma$         & 7.21&118.8& 2.38& 2.27& 2.57& 2.24 & \cite{Sokhan:2008ts} \tabularnewline
                   &$\mathbb E$       &18.25&17.43& 2.84& 2.62& 3.11& 3.68 & \cite{CLAS:2017kua} \tabularnewline
\hline
$\gamma n\to\pi^0n$&$d\sigma/d\Omega$ & 3.77& 7.29& 2.88& 2.43& 3.14& 3.89 & \cite{A2:2019yud} \tabularnewline
                   &                  &20.32&18.72&11.22& 9.52& 9.97&15.73 & \cite{A2:2018jcd}$^\dagger$ \tabularnewline
                   & $\Sigma$         & 2.44& 2.46& 1.25& 1.15& 1.33& 2.17 & \cite{A2atMAMI:2021iuz} \tabularnewline
                   & $\mathbb E$      &29.75&13.11& 6.21& 5.89& 5.66&10.85 & \cite{Dieterle:2017myg} \tabularnewline
\hline
\end{tabular}} \label{tab:tbl4}
\end{table*}
%---------------------------------------------------------

%------------------------------------------------------------
%\clearpage
\section{SAID Multipole Amplitudes}
\label{Sec:ampl}

The SAID parametrization of the transition amplitude $T_{\alpha\beta}$
used in the hadronic fits to the $\pi N$ scattering data is given as
\begin{equation}
	 T_{\alpha\beta} = \sum_\sigma [1 - \overline{K}C]_{\alpha\sigma}^{-1} \overline{K_{\sigma\beta}} \>,
    \label{eq:eq2}
\end{equation}
where $\alpha$, $\beta$, and $\sigma$ are channel indices for the $\pi N$, $\pi\Delta$, $\rho N$, and $\eta N$ channels. Here $\overline{K_{\sigma\beta}}$ are the Chew-Mandelstam $K$-matrices, which are parameterized as polynomials in the scattering energy. $C_\alpha$ is the Chew-Mandelstam function, an element of a diagonal matrix $C$ in channel space, which is expressed as a dispersion integral with an imaginary part equal to the 
two-body phase space~\cite{Arndt:1985vj}.

In Ref.~\cite{Workman:2012jf}, it was shown that this form could be extended to $T_{\alpha\gamma}$ to include the electromagnetic 
channel as
\begin{equation}
	 T_{\alpha\gamma} = \sum_\sigma [1 - \overline{K}C]_{\alpha\sigma}^{-1} \overline{K_{\sigma\gamma}} \>.
    \label{eq:eq3}
\end{equation}
Here, the Chew-Mandelstam K-matrix elements associated with the hadronic channels are kept fixed from the previous SAID solution SP06~\cite{Arndt:2006bf}, and only the electromagnetic elements are varied. The resonance pole and cut structures are also fixed from hadronic scattering. This provides a minimal
description of the photoproduction process, where only the $N^\ast$ and $\Delta^\ast$ states present in the SAID $\pi$N scattering amplitudes are included in this multipole analysis.

For each angular distribution, a normalization constant ($X$) and its uncertainty ($\epsilon_X$) were assigned. The quantity $\epsilon_X$ is generally associated with the normalization uncertainty (if known). The modified $\chi^2$ function to be minimized is given by
\begin{equation}
	 \chi^2 = \sum_i \bigg(\frac{X\theta_i - \theta_i^{exp}}{\epsilon_i} \bigg)^2 + \bigg(\frac{X - 1}{\epsilon_X} \bigg)^2 \>,
    \label{eq:eq4}
\end{equation}
where the subscript $i$ labels the data points within the distribution, $\theta_i^{exp}$ is an individual measurement, $\theta_i$ is the corresponding calculated value, and $\epsilon_i$ represents the total angle-dependent uncertainty. The total $\chi^2$ is then found by summing over all measurements. This re-normalization freedom is essential for obtaining the best SAID fit results. For other data analyzed in the fit, such as the total cross sections and excitation data, the statistical and systematic uncertainties were combined in quadrature and no re-normalization was allowed.

In the previous fits to differential cross sections, the unrestricted best fit gave re-normalization constants $X$ significantly different from unity. As can be seen from Eq.~(\ref{eq:eq4}), if an angular distribution contains many measurements with small statistical uncertainties, a change in the re-normalization may improve the fit with only a modest $\chi^2$ penalty. Here, however, the weight of the second term in Eq.~(\ref{eq:eq4}) has been adjusted by the fit for each dataset to keep the re-normalization constants approximately within $X$ of unity.

With the new quality datasets (Table~\ref{tab:tbl1}), a new SAID multipole analysis has been completed. This new global energy-dependent solution has been labeled as SM22. The overall fit quality of the present SM22 and previous SAID CM12 solutions are compared in Tables~\ref{tab:tbl3} and ~\ref{tab:tbl4}. There are many cases where the CM12 fit produces a $\chi^2$ per datum, for new measurements, which is significantly 
than greater than unity.  The new best fit, SM22, includes these new measurements, reducing the $\chi^2/\mathrm{data}$ to more acceptable values.

Both energy-dependent (ED) and single-energy (SE) solutions were obtained from fits to the combined proton and neutron  target database, extending from threshold to $E_\gamma = 2.7~\mathrm{GeV}$ for the ED fit and to $E_\gamma = 2.2~\mathrm{GeV}$ for SE fits.

%\textcolor{red}{In Table~\ref{tab:tbl4}, we compare the energy-dependent and single-energy results over the energy bins used in these single-energy analyses. Also listed are the number of parameters varied in each single-energy solution. The extended database allowed an increase in the number of SES versus our previous CM12 result~\cite{Workman:2012jf} over the same energy range to $2.7~\mathrm{GeV}$. The SES are also shown with uncertainties coming from the error matrix. In the SES fits, initial values for the partial-wave amplitudes and their (fixed) energy derivatives were obtained from the energy dependent solution. A comparison of global and single energy solutions then serves as a check for structures that could have been ``smoothed over'' in the energy-dependent analysis. Partial waves with $l<3$ are displayed, whereas the analysis fitted waves up to $J=5$. }  

Apart from the main ED result (SM22) several supplemental fits were done in order to gauge the importance of including $\pi^0 n$ data (which can, in principle, be at least qualitatively predicted from the remaining
more fully populated charge channels). Here fits were done with increased weight for the $\pi^0 n$ data and conversely the removal of all such data. In addition, a fit was done more heavily weighting all data poorly fitted by SM22. Figures~\ref{fig:obs1} and \ref{fig:obs1a} plot representative comparisons of SAID fits to data. In addition,
older MAID and more recent Bonn-Gatchina results are plotted for comparison. Numerical
comparisons of the various SAID fits are given in Tables~\ref{tab:tbl2} to \ref{tab:tbl4}. 
%\textcolor{red}{Samples of pion photoproduction off proton (neutron) shown on Fig.~\ref{fig:obs1} (Fig.~\ref{fig:obs1a}) is a plot of the ED fits SM22 and CM12 over the full-energy region.} \\

\begin{figure*}[hbt!]
%\vspace{0.4cm}
\centering
{
    \includegraphics[width=0.32\textwidth,angle=90,keepaspectratio]{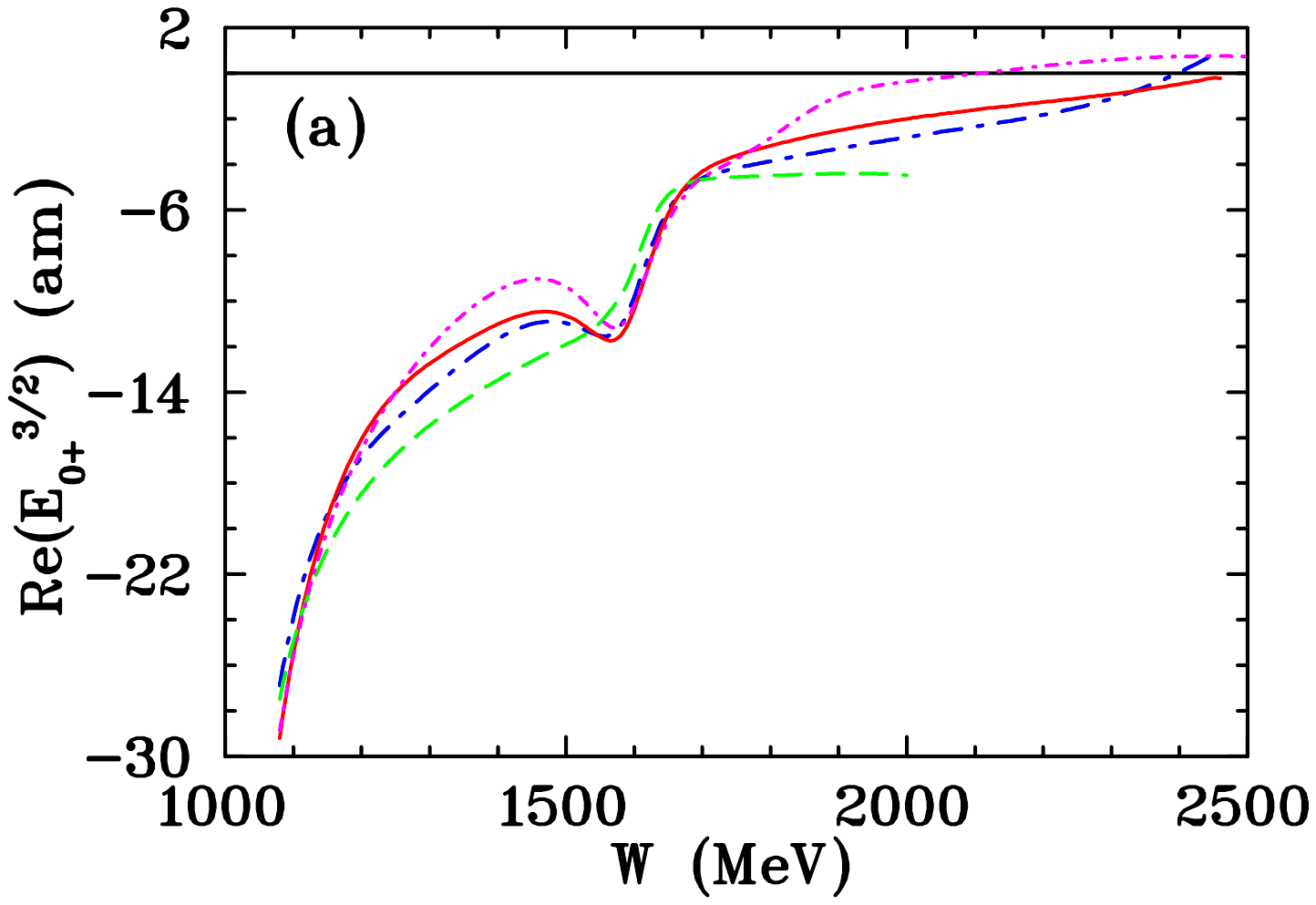}
    \includegraphics[width=0.32\textwidth,angle=90,keepaspectratio]{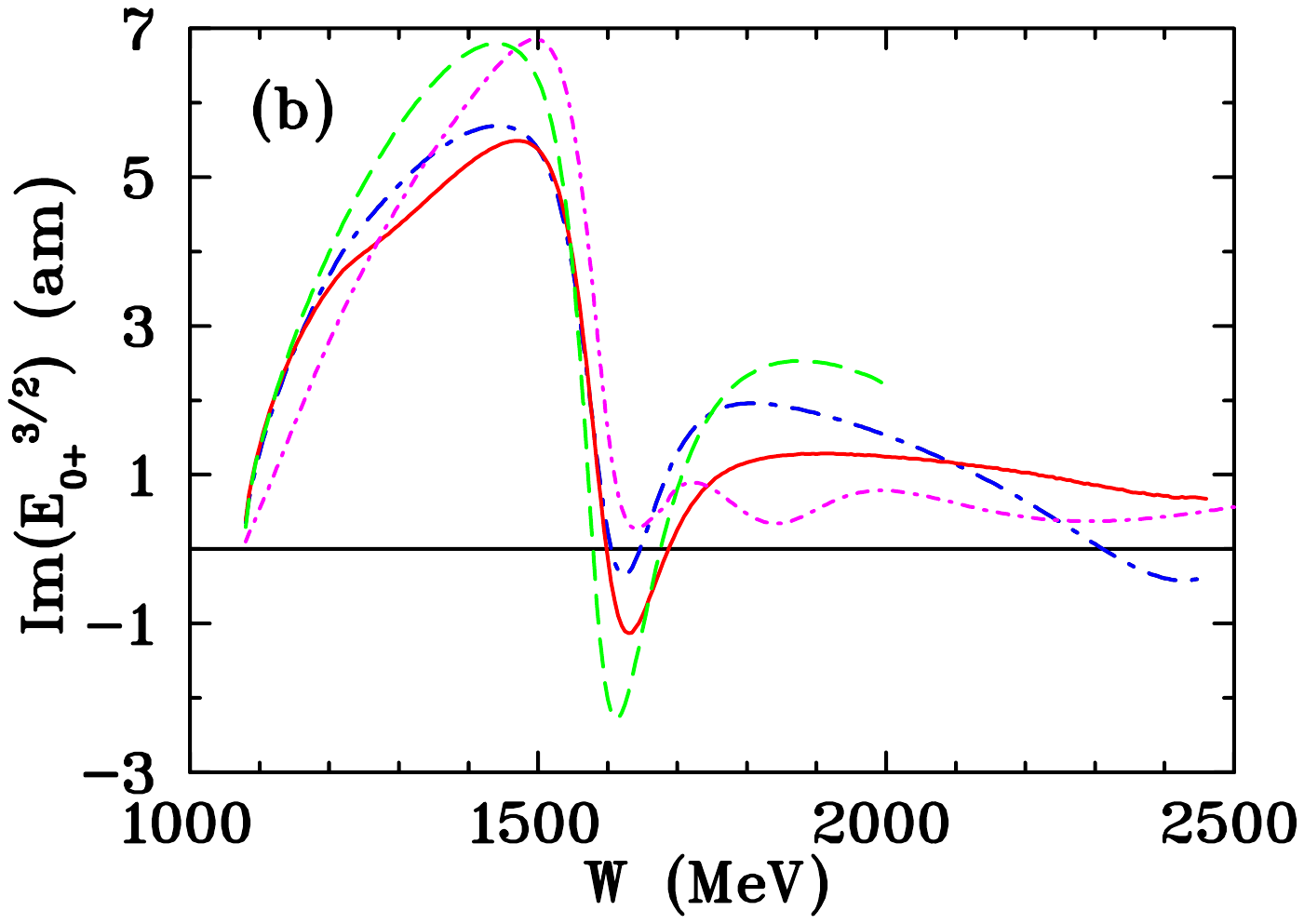}
}
\centering
{
    \includegraphics[width=0.32\textwidth,angle=90,keepaspectratio]{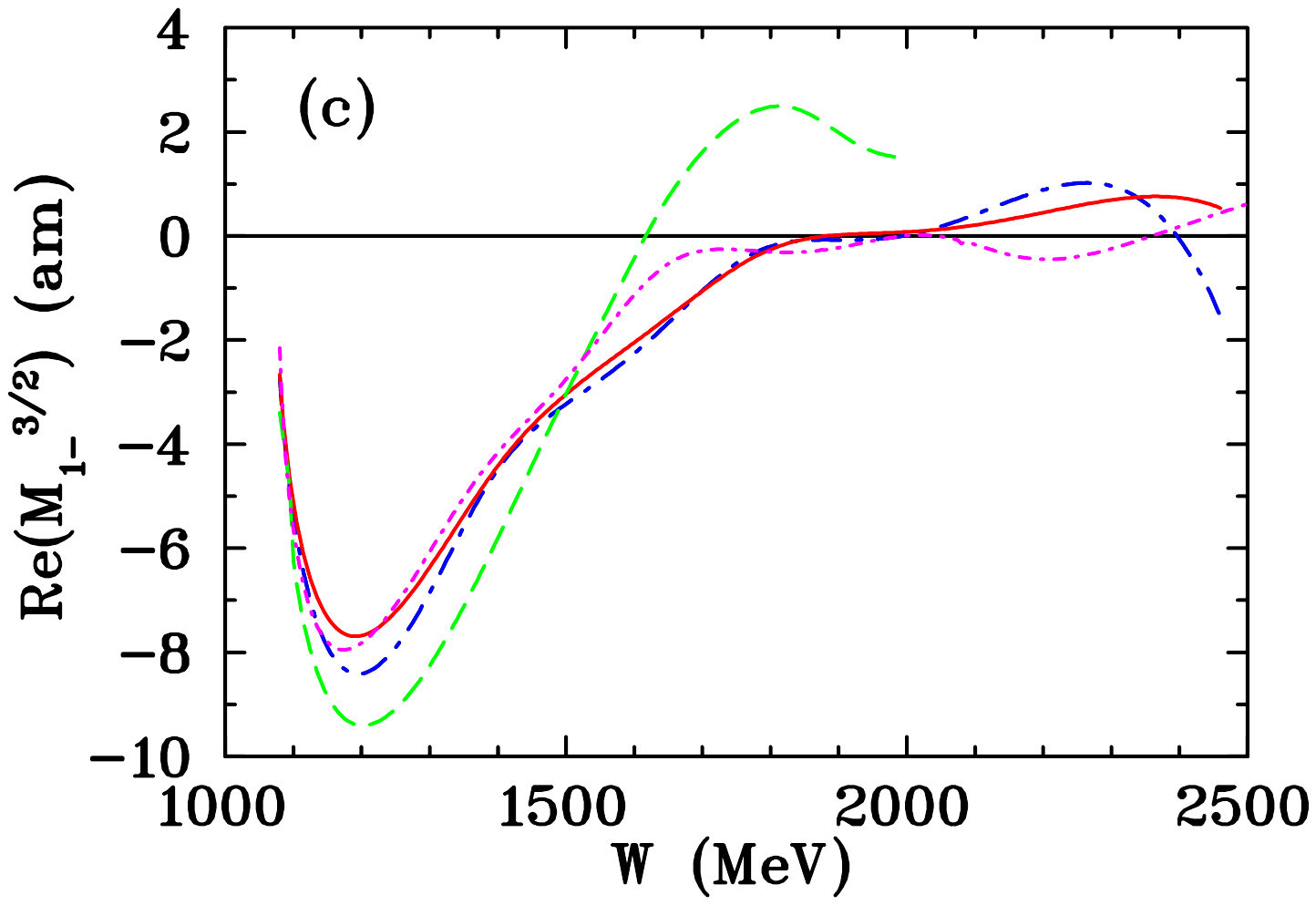}
    \includegraphics[width=0.32\textwidth,angle=90,keepaspectratio]{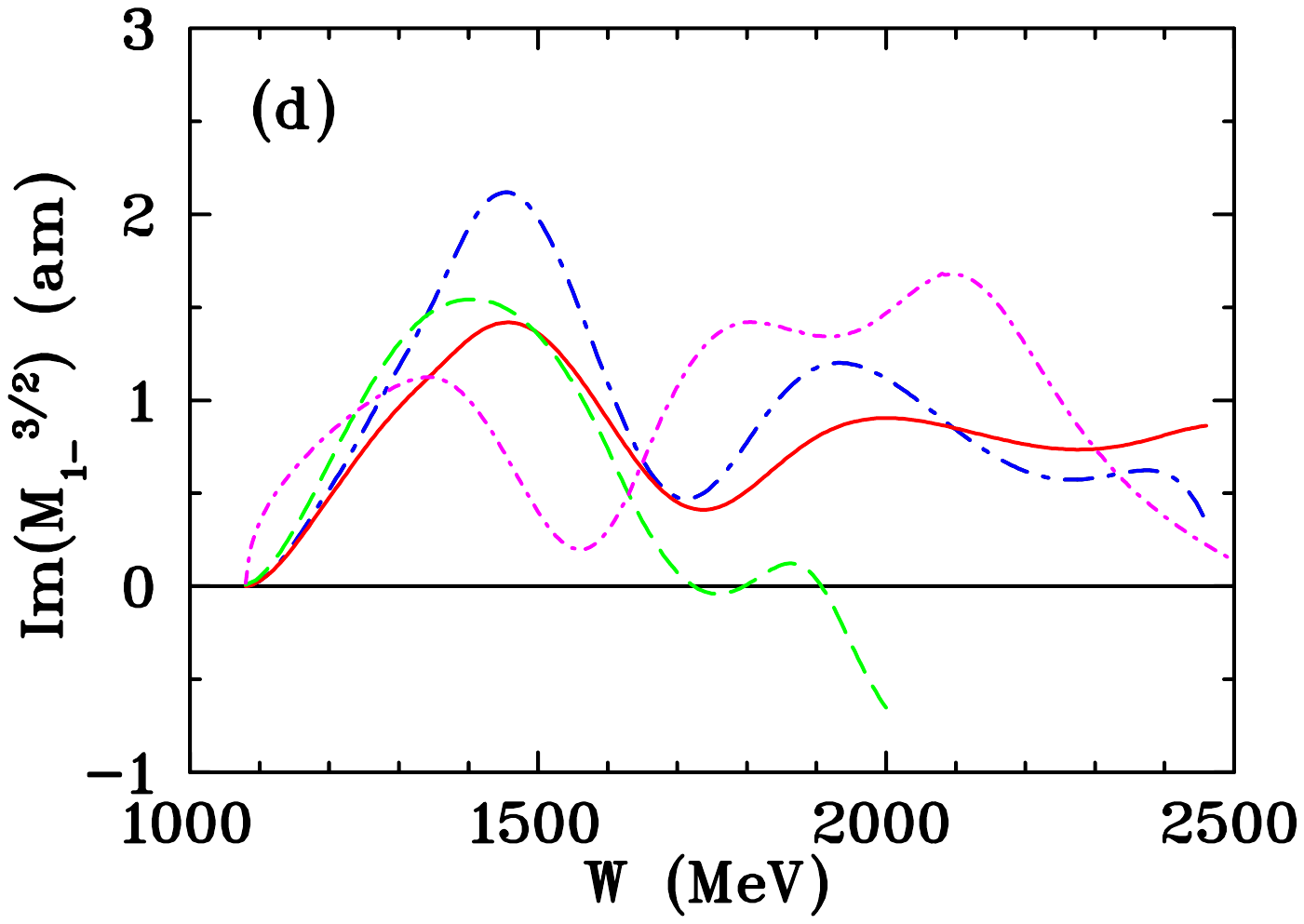}
}
\centering
{
    \includegraphics[width=0.32\textwidth,angle=90,keepaspectratio]{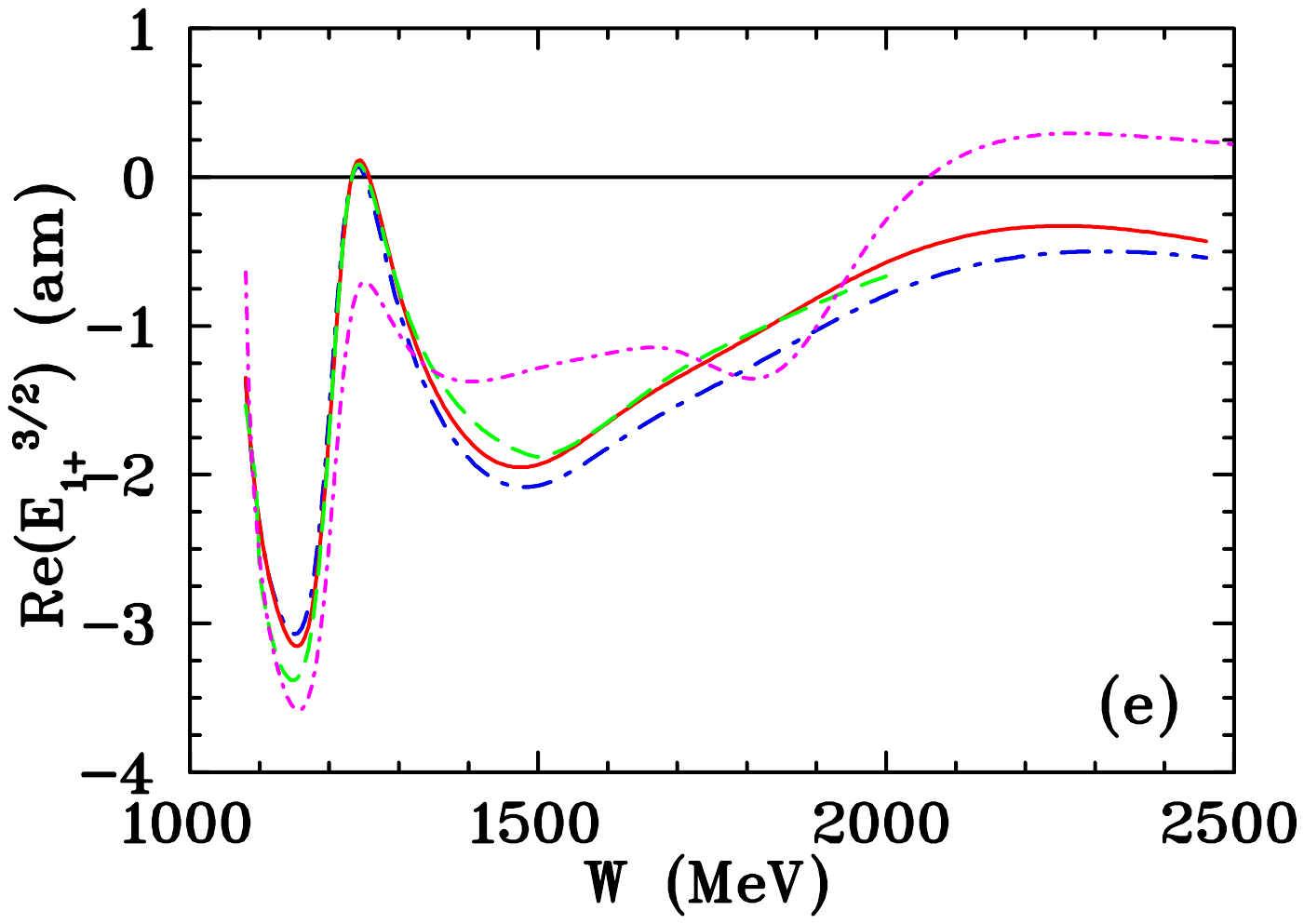}
    \includegraphics[width=0.32\textwidth,angle=90,keepaspectratio]{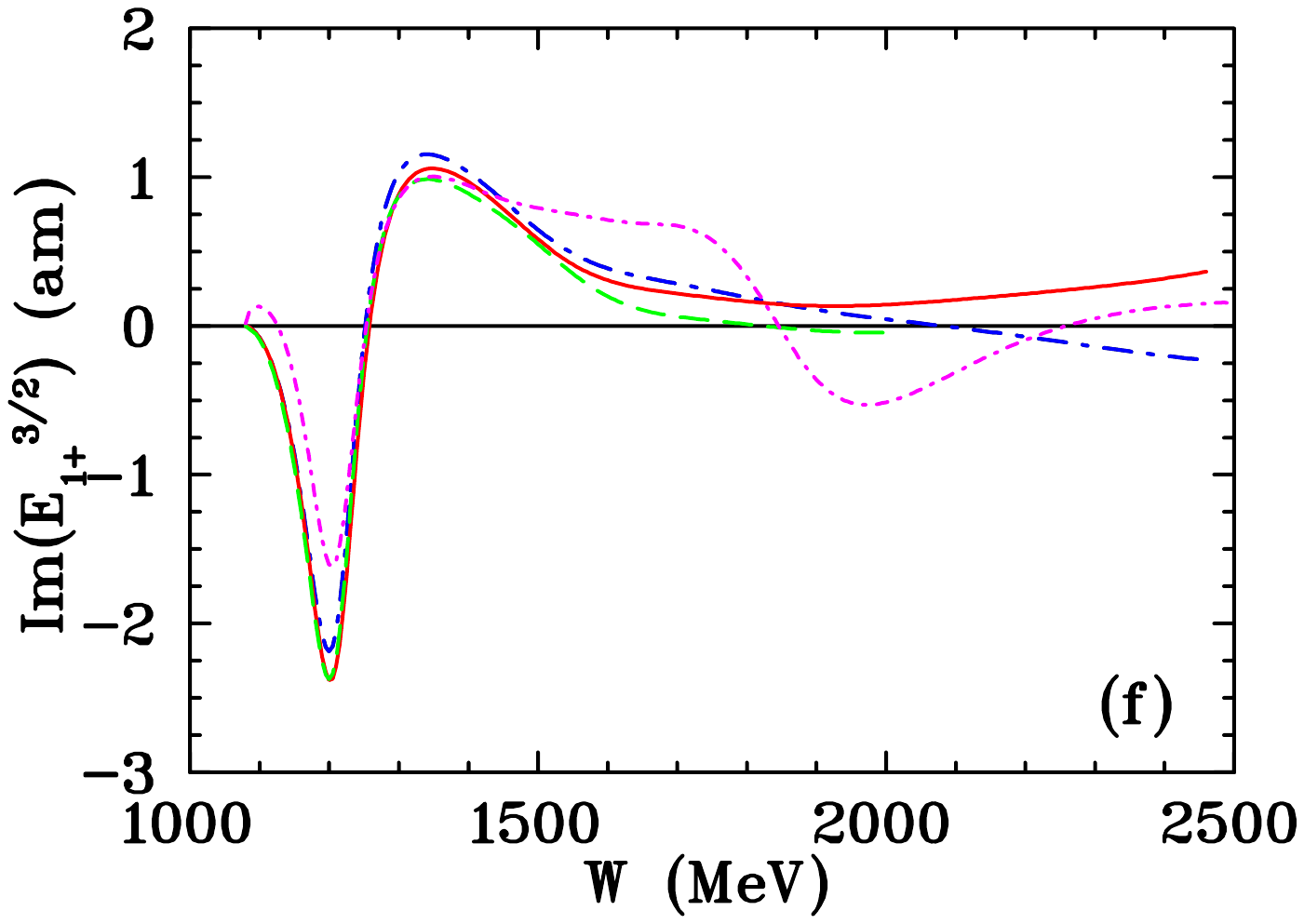}
}
\centering
{
    \includegraphics[width=0.32\textwidth,angle=90,keepaspectratio]{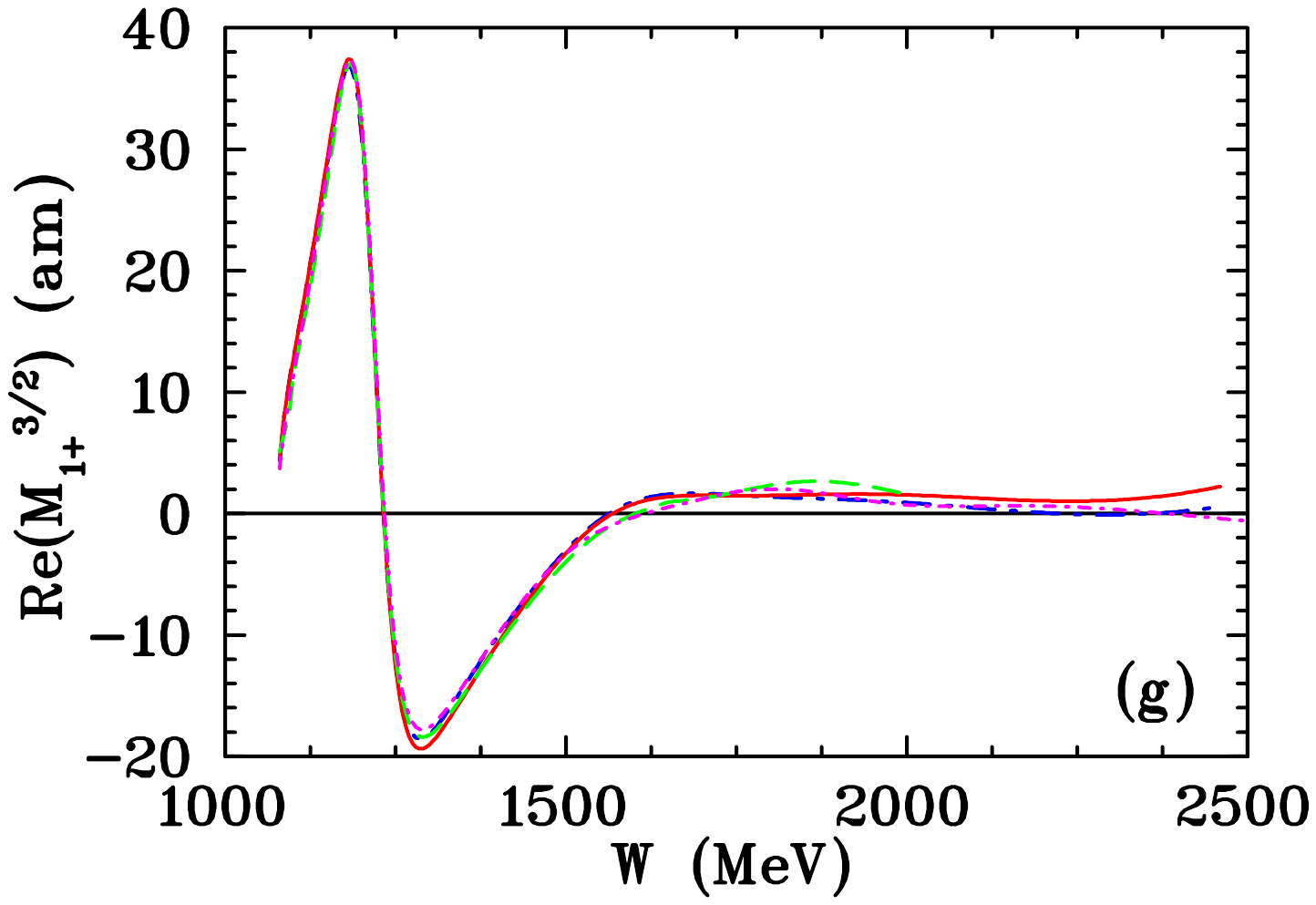}
    \includegraphics[width=0.32\textwidth,angle=90,keepaspectratio]{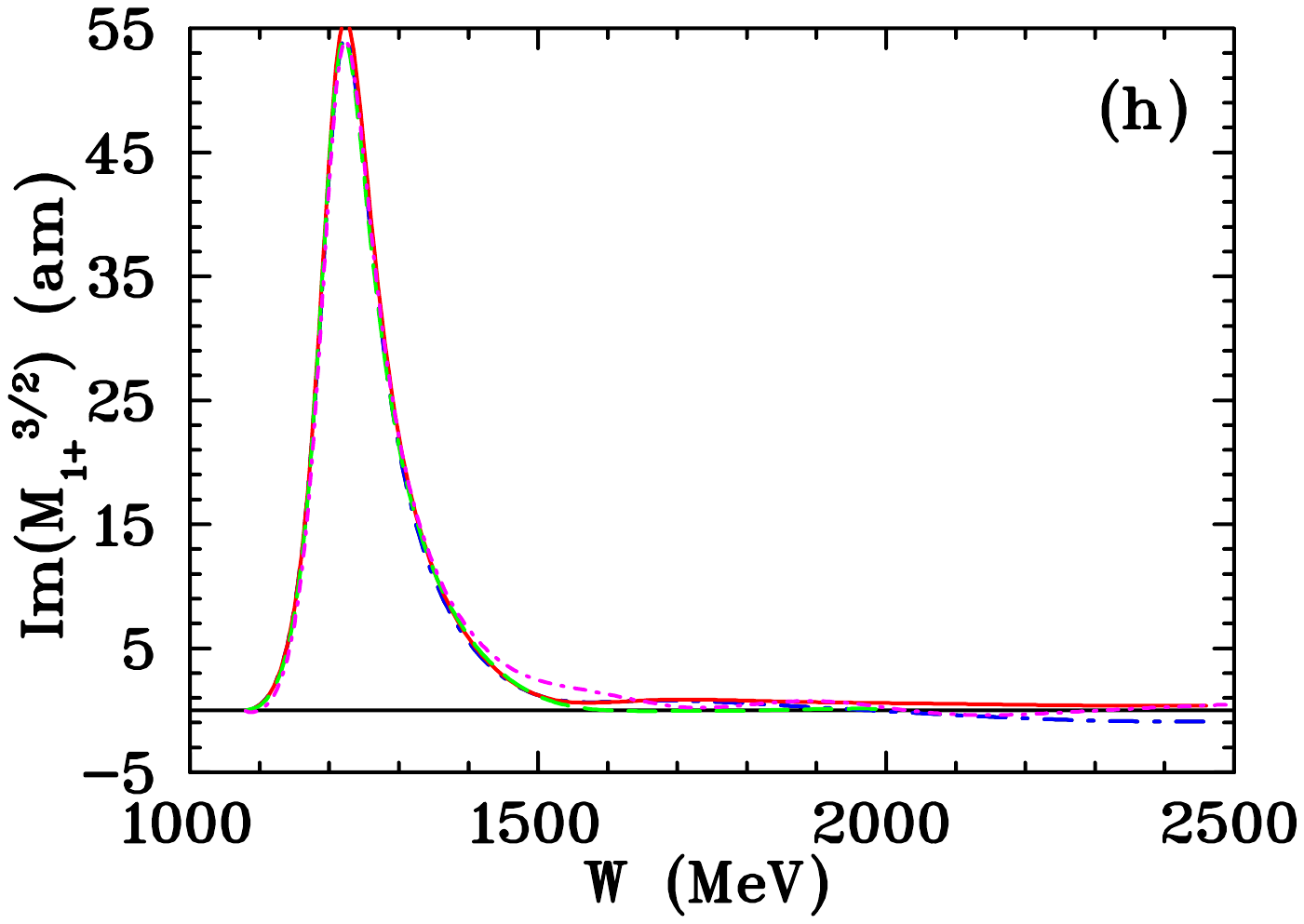}
}

\caption{Comparison $I = 3/2$ multipole amplitudes (orbital momentum $l = 0, 1$) from threshold to $W = 2.5~\mathrm{GeV}$ ($E_\gamma = 2.7~\mathrm{GeV}$). For the amplitudes, the subscript $l\pm$ gives the value of $j = l \pm 1/2$, and the superscript gives the isospin index. Notation for solutions is given in the caption of Table~\ref{tab:tbl2}. New SAID SM22 fit is shown by red solid curves. Previous SAID CM12~\cite{Workman:2012jf} (MAID2007~\cite{Drechsel:2007if}, terminates at $W = 2~\mathrm{GeV}$) predictions show by blue dash-dotted (green dashed) curves. BG2019~\cite{CBELSATAPS:2014wvh} predictions show by magenta short dash-dotted curves. 
}
\label{fig:amp1a}
\end{figure*}

\begin{figure*}[hbt!]
\vspace{0.4cm}
\centering
{
    \includegraphics[width=0.32\textwidth,angle=90,keepaspectratio]{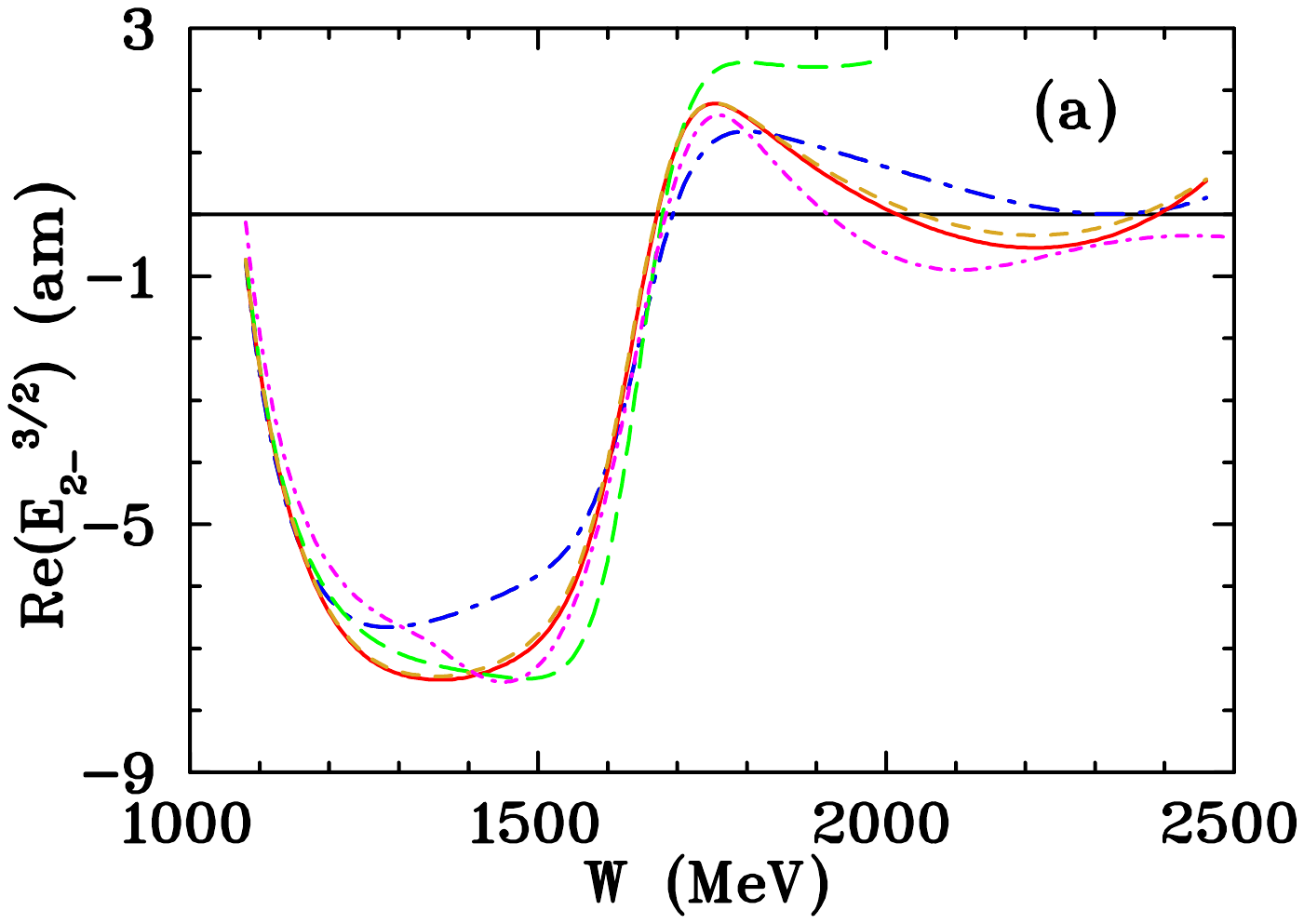}
    \includegraphics[width=0.32\textwidth,angle=90,keepaspectratio]{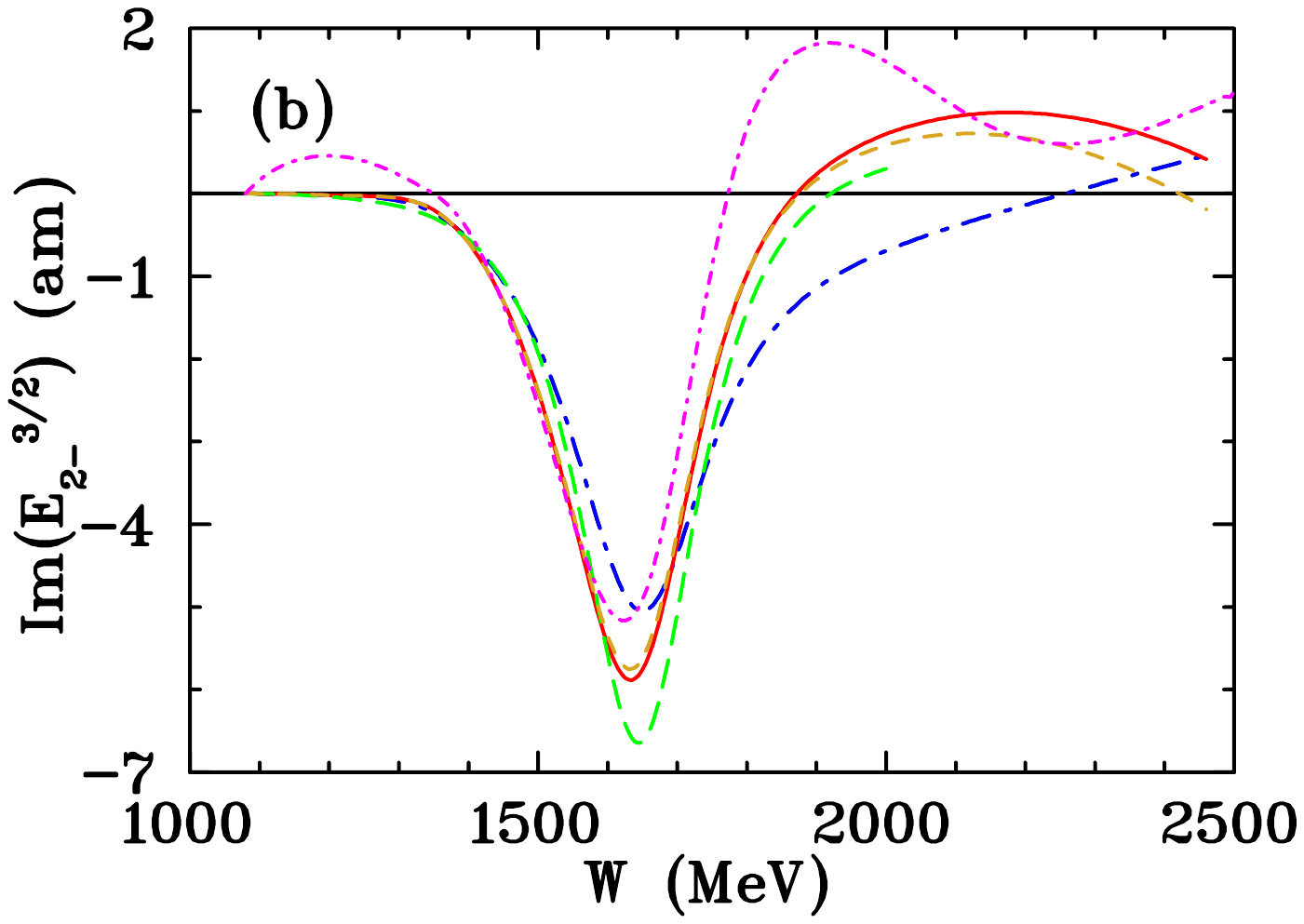}
}
\centering
{
    \includegraphics[width=0.32\textwidth,angle=90,keepaspectratio]{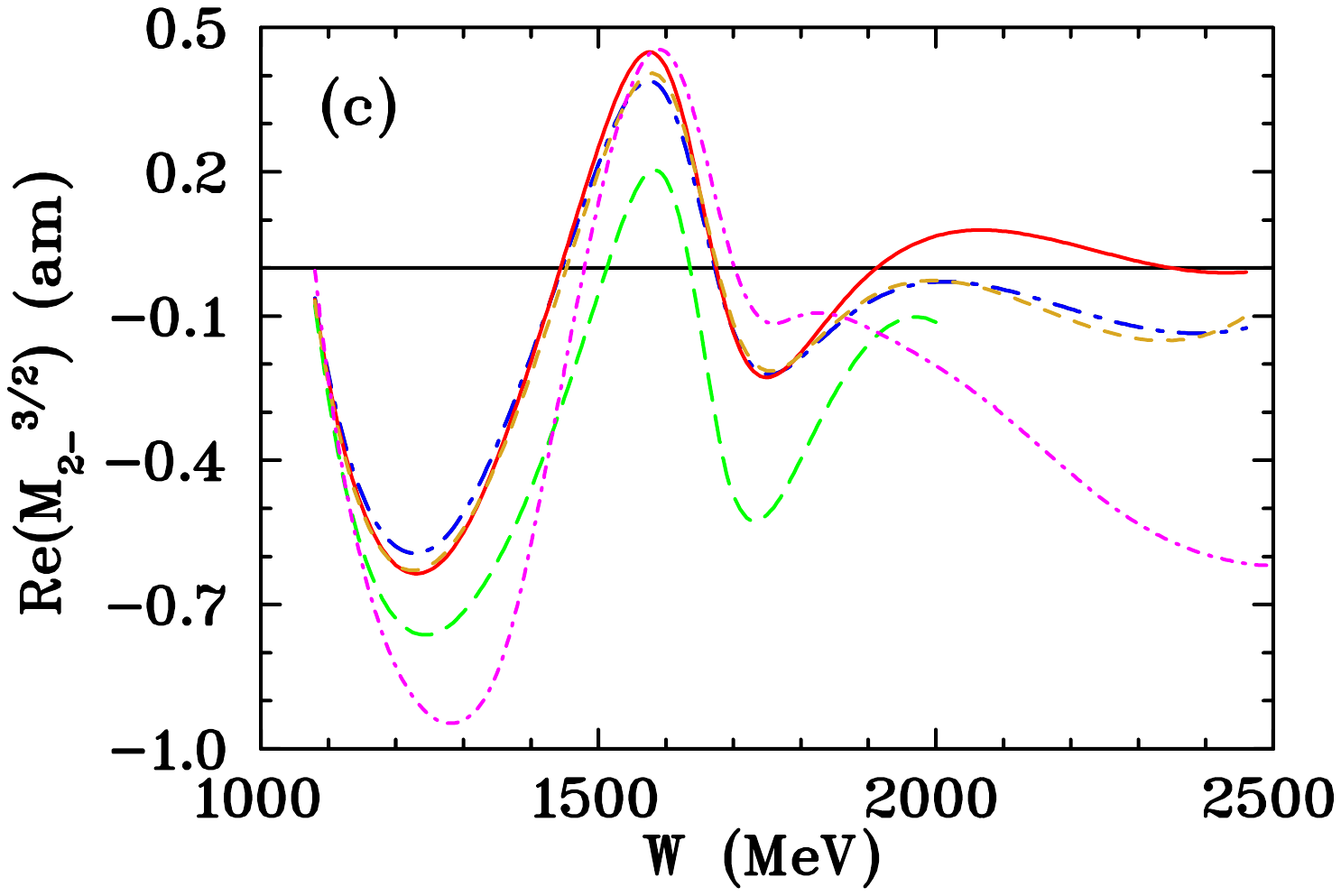}
    \includegraphics[width=0.32\textwidth,angle=90,keepaspectratio]{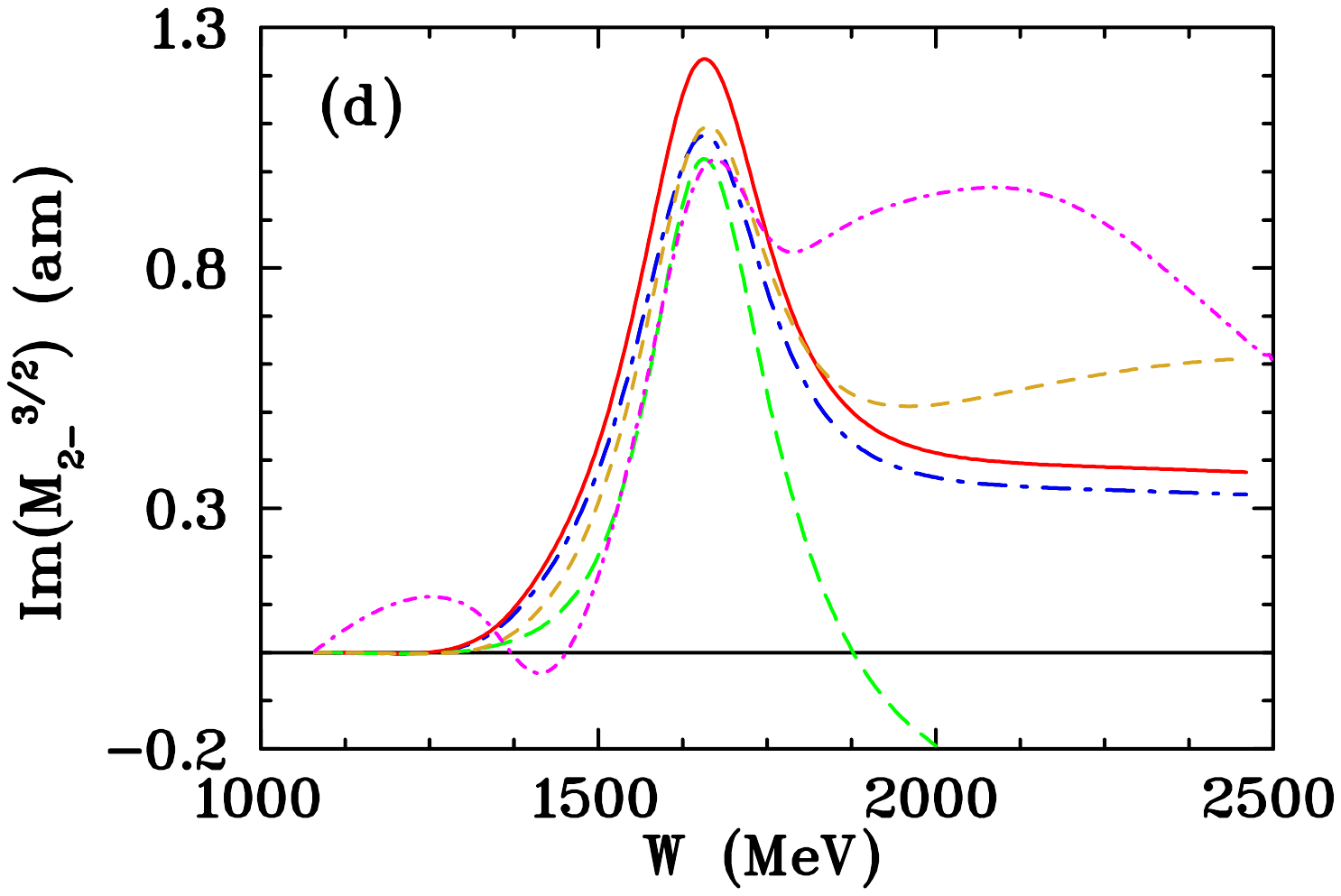}
}
\centering
{
    \includegraphics[width=0.32\textwidth,angle=90,keepaspectratio]{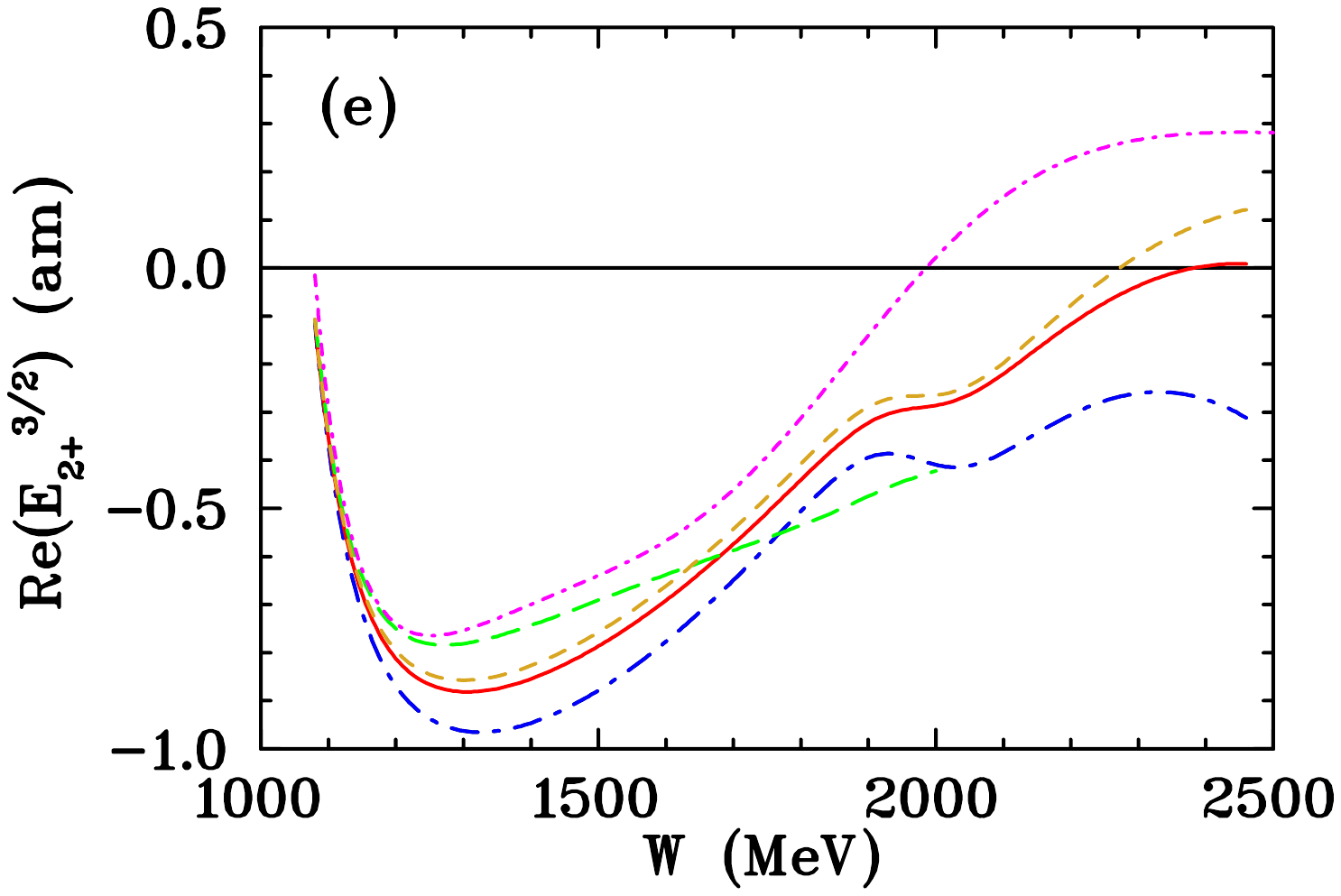}
    \includegraphics[width=0.32\textwidth,angle=90,keepaspectratio]{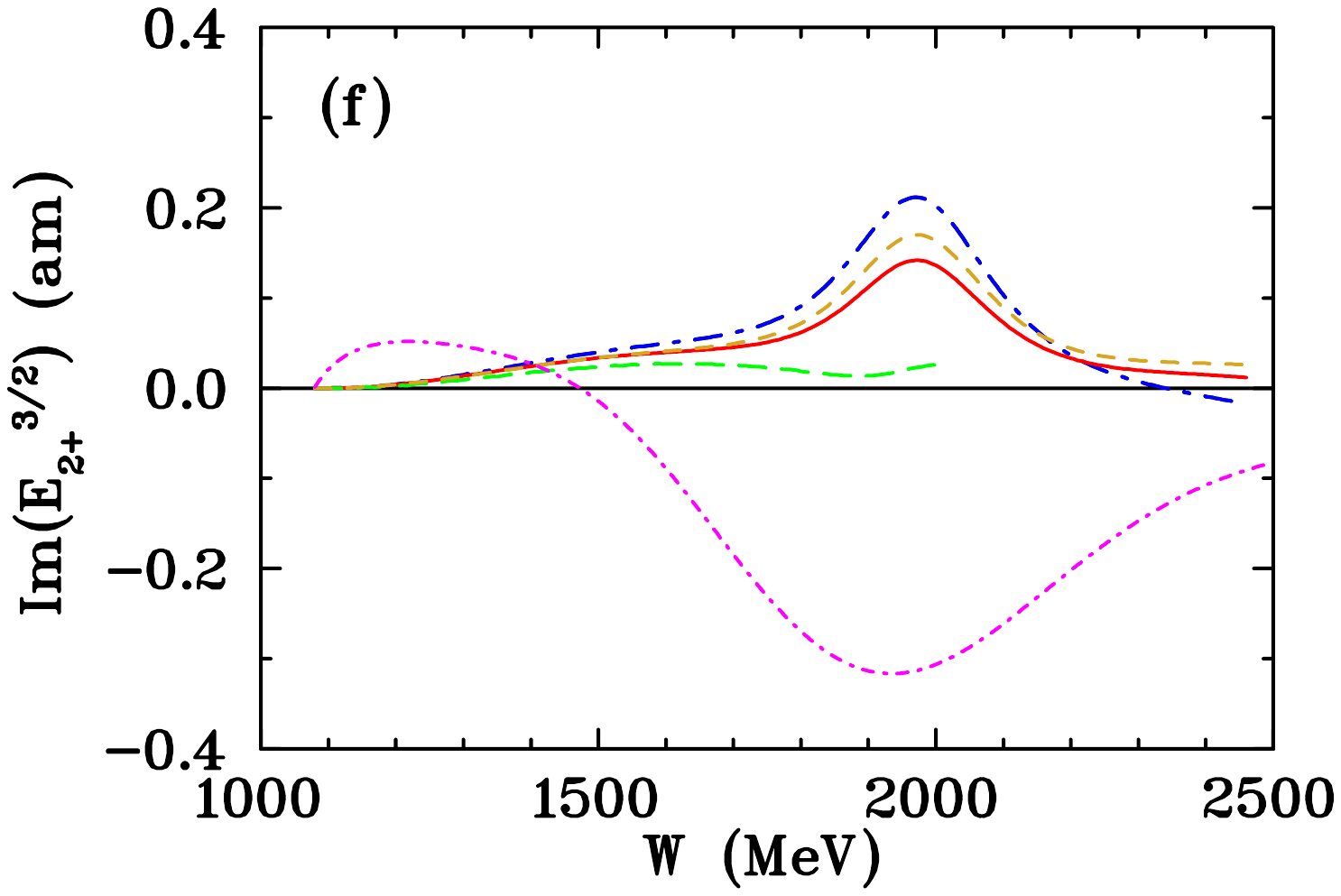}
}
\centering
{
    \includegraphics[width=0.32\textwidth,angle=90,keepaspectratio]{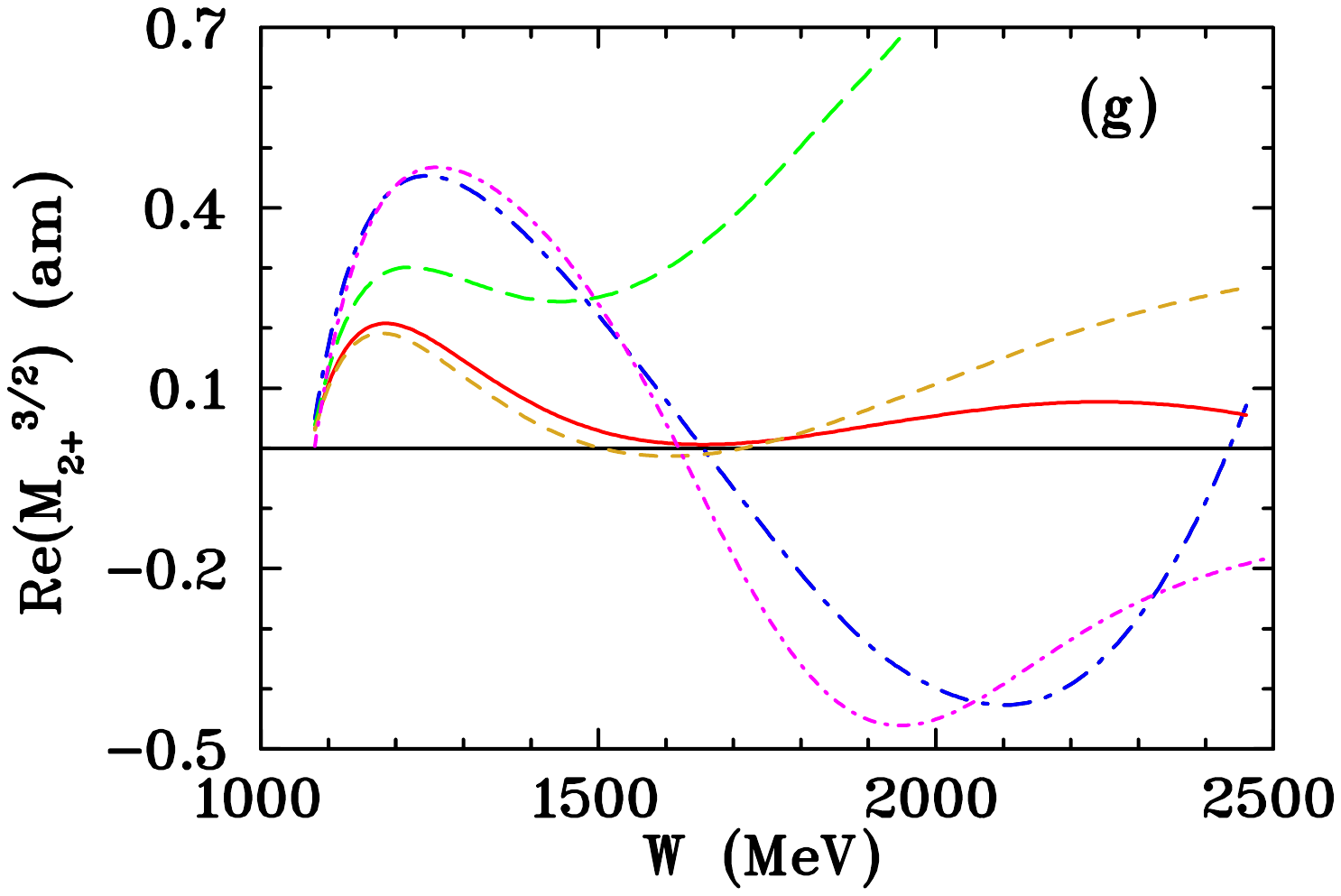}
    \includegraphics[width=0.32\textwidth,angle=90,keepaspectratio]{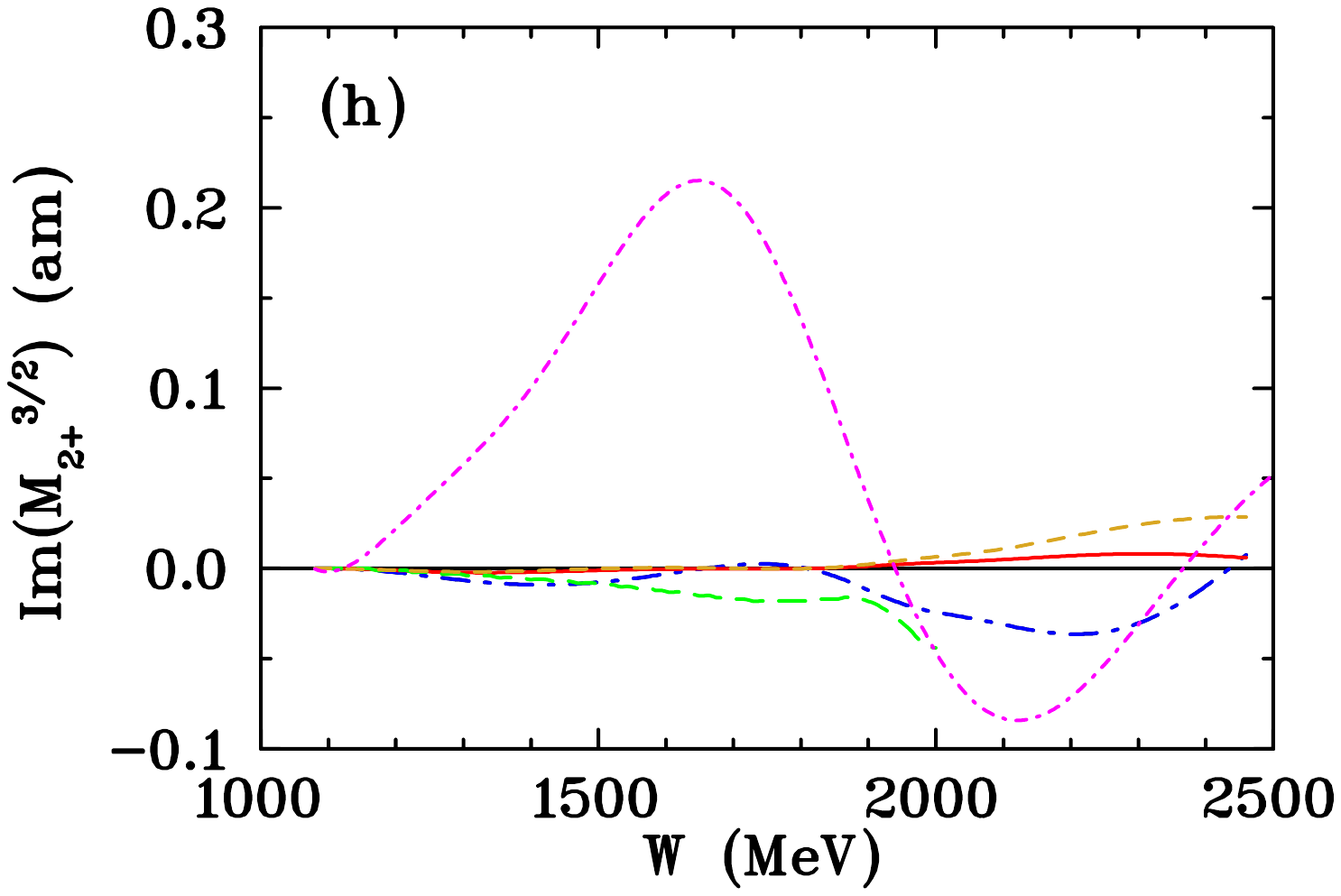}
}

\caption{Comparison $I = 3/2$ multipole amplitudes (orbital momentum $l = 2$) from threshold to $W = 2.5~\mathrm{GeV}$. Notation of the solutions is the  same as in Fig.~\ref{fig:amp1a}. Additionally, the WM22 fit is shown by yellow dashed curves.}  
\label{fig:amp2a}
\end{figure*}

\begin{figure*}[hbt!]
%\vspace{0.4cm}
\centering
{
    \includegraphics[width=0.32\textwidth,angle=90,keepaspectratio]{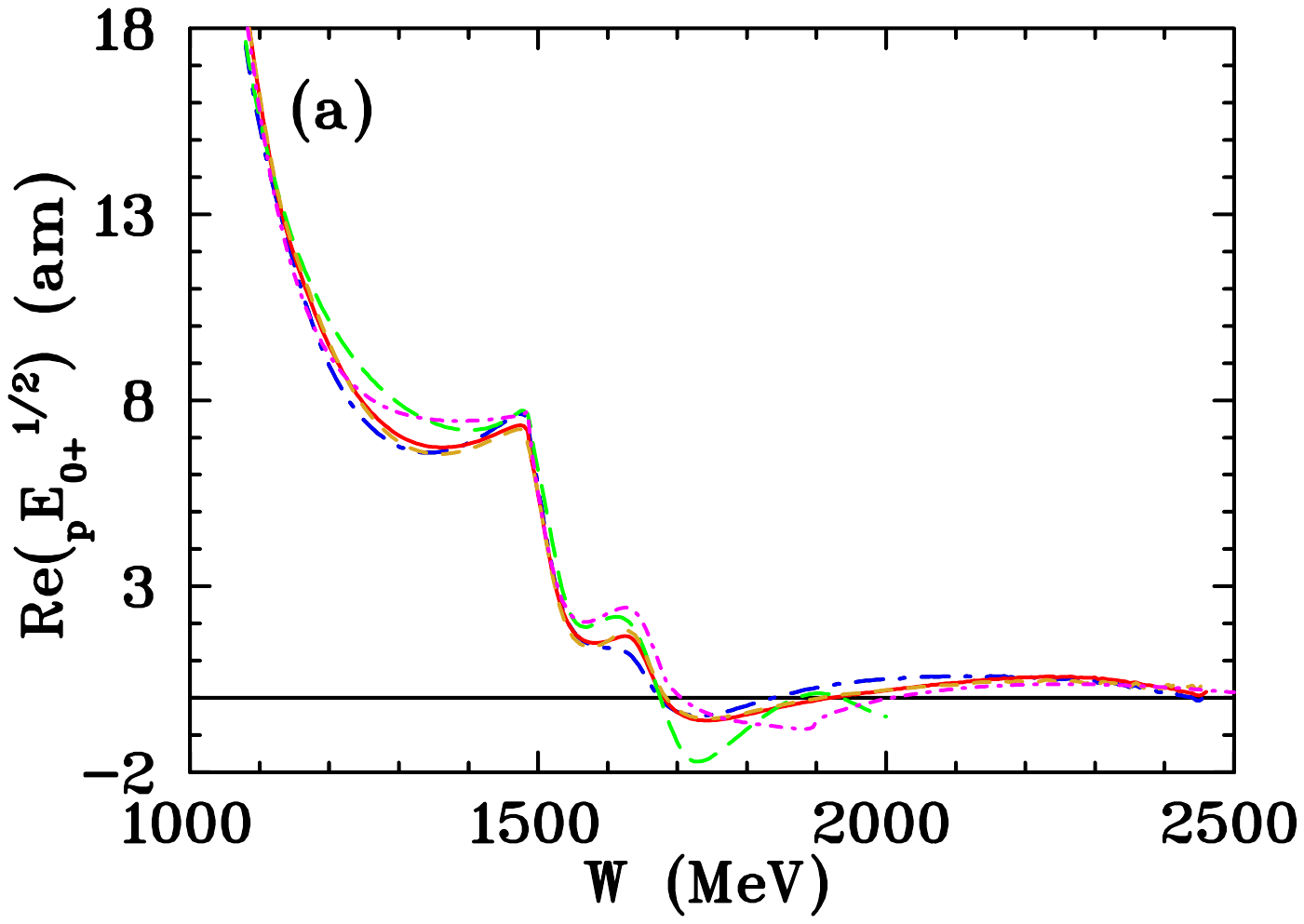}
    \includegraphics[width=0.32\textwidth,angle=90,keepaspectratio]{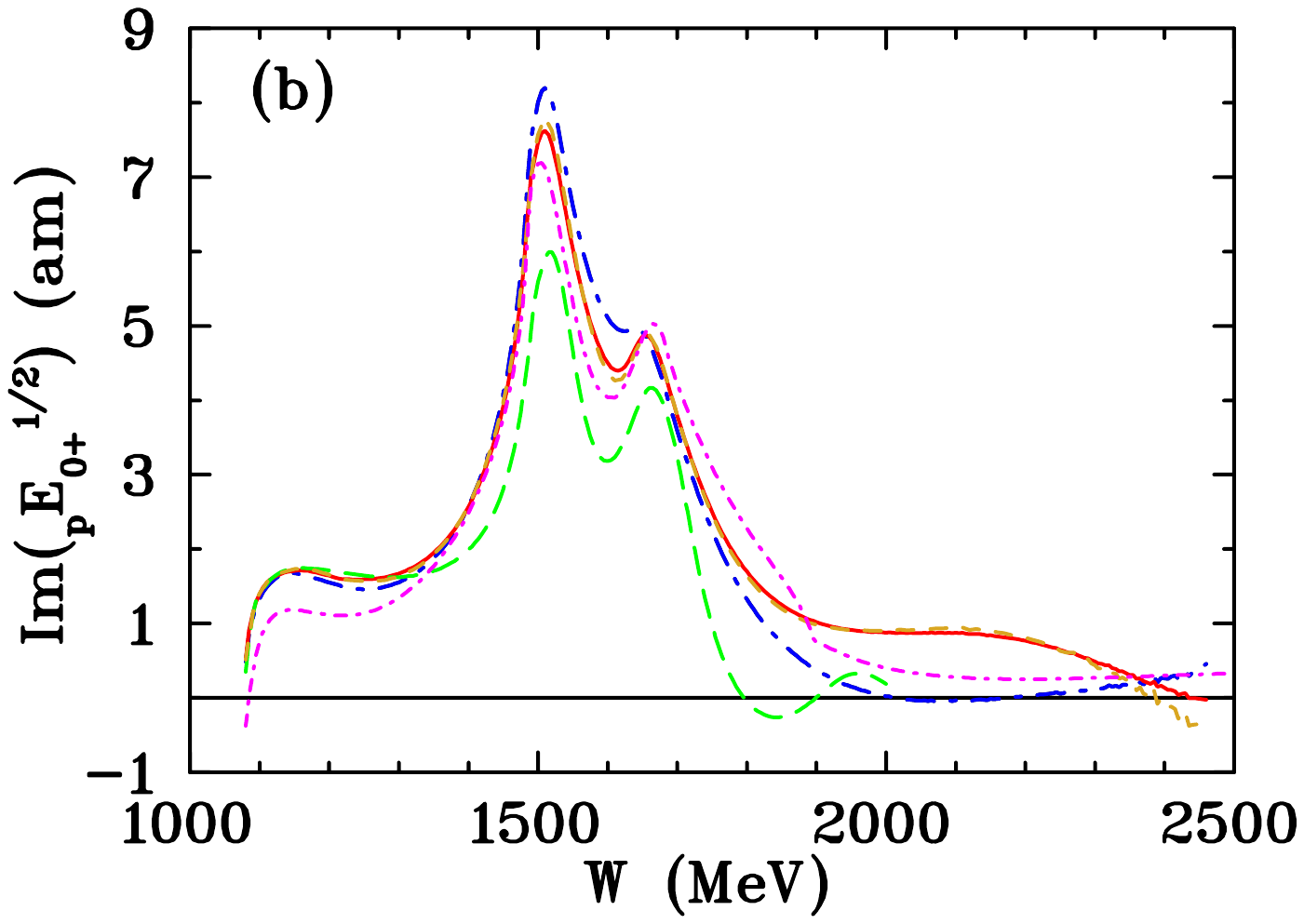}
}
\centering
{
    \includegraphics[width=0.32\textwidth,angle=90,keepaspectratio]{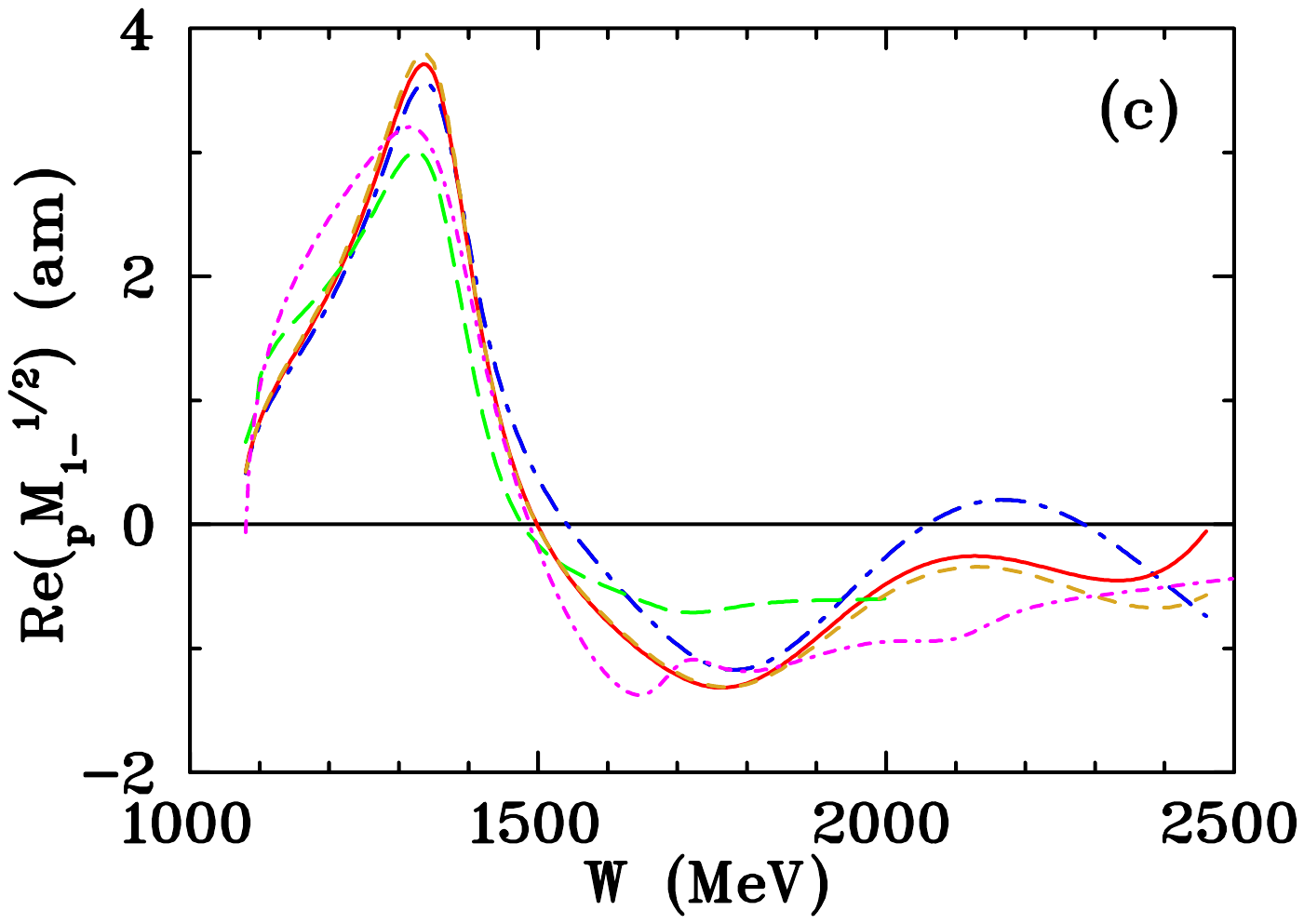}
    \includegraphics[width=0.32\textwidth,angle=90,keepaspectratio]{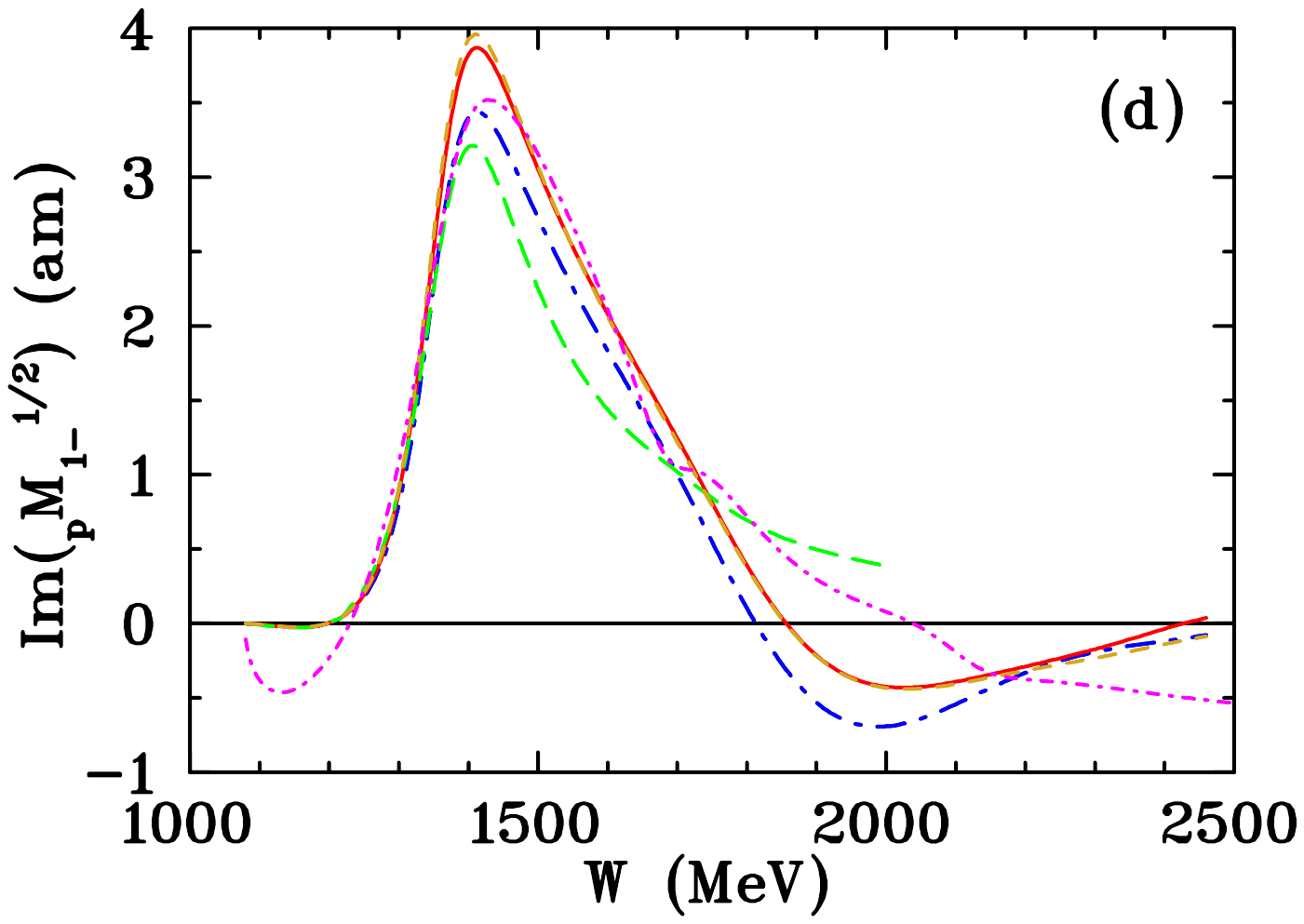}
}
\centering
{
    \includegraphics[width=0.32\textwidth,angle=90,keepaspectratio]{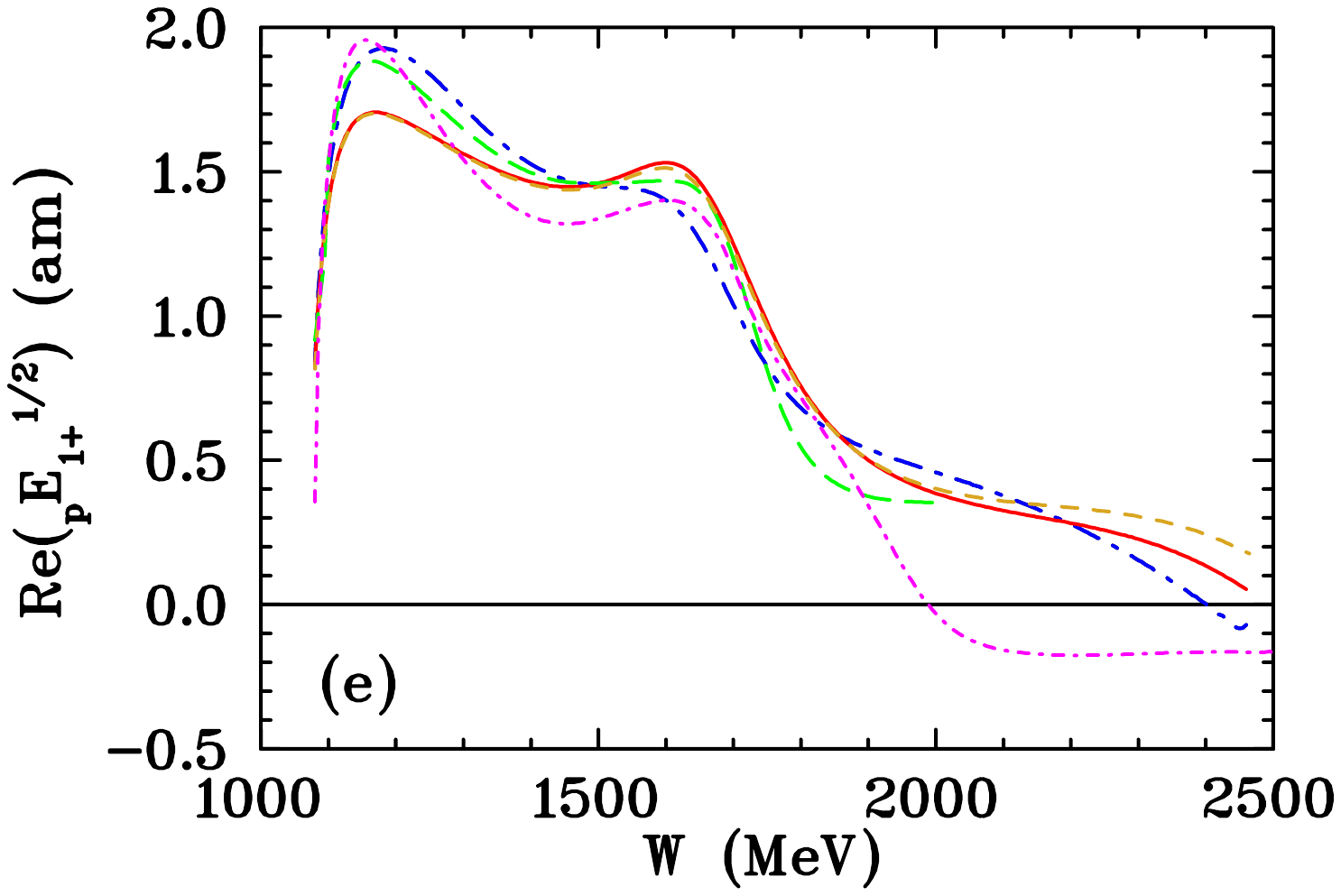}
    \includegraphics[width=0.32\textwidth,angle=90,keepaspectratio]{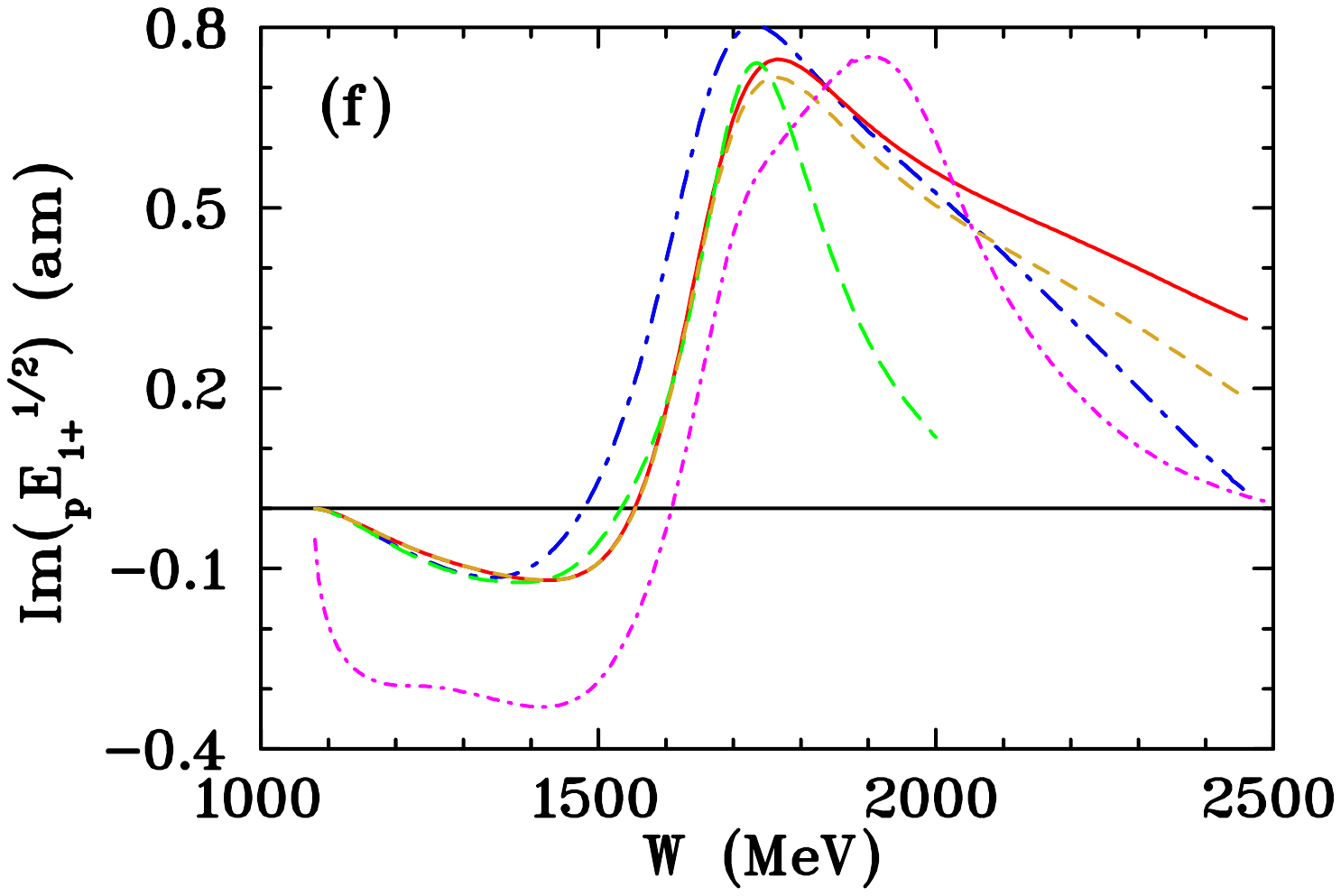}
}
\centering
{
    \includegraphics[width=0.32\textwidth,angle=90,keepaspectratio]{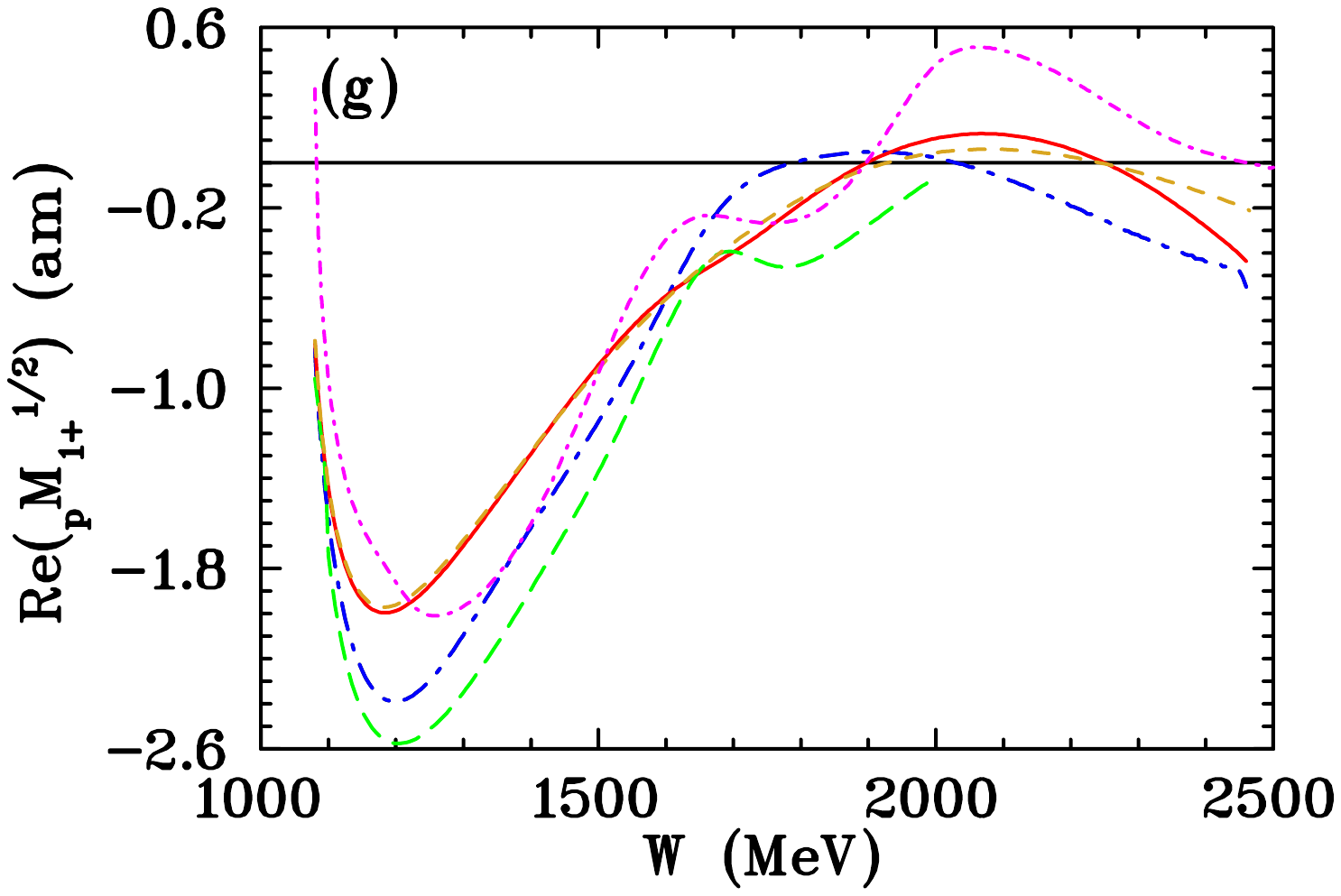}
    \includegraphics[width=0.32\textwidth,angle=90,keepaspectratio]{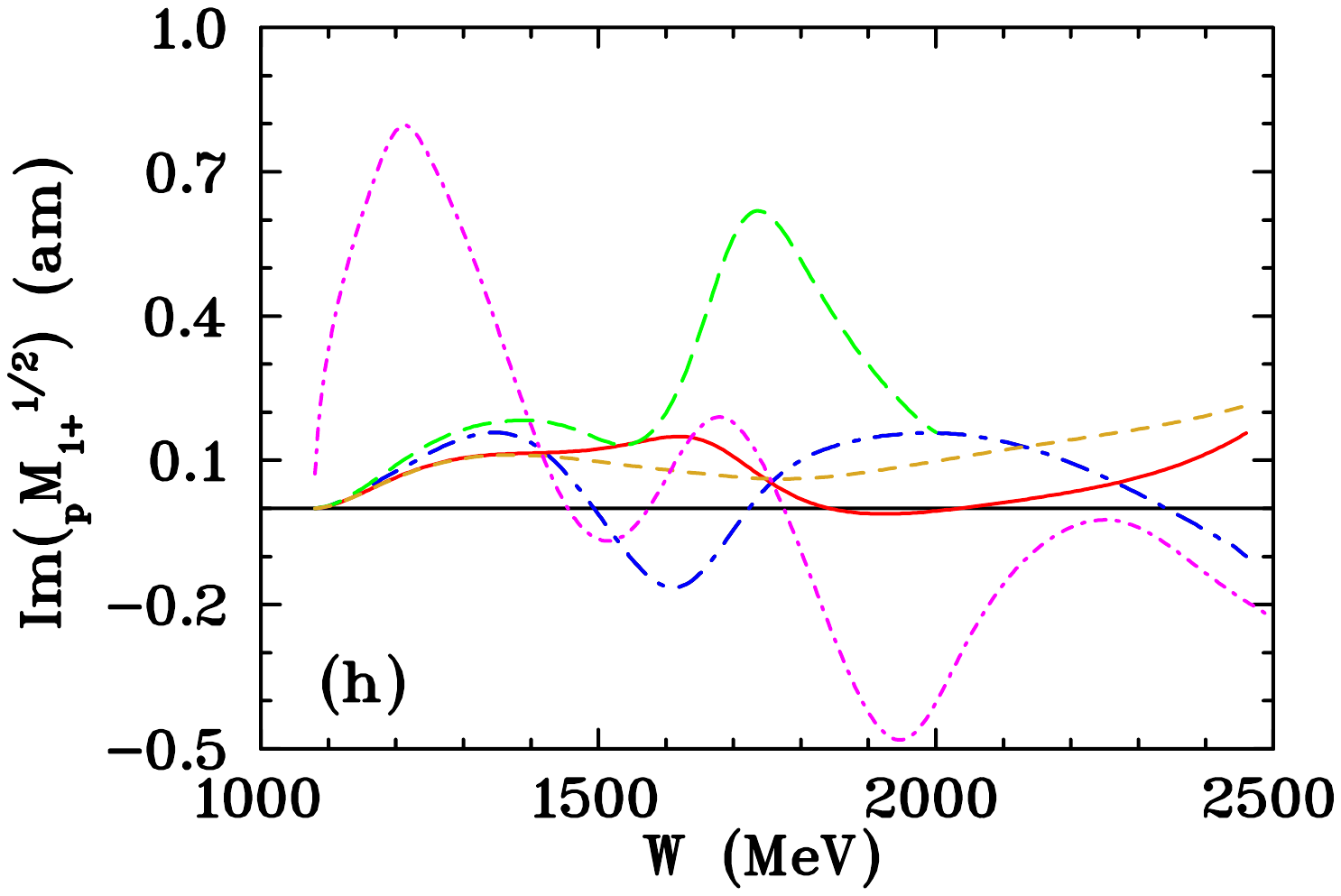}
}

\caption{Comparison proton $I = 1/2$ multipole amplitudes (orbital momentum $l = 0, 1$) from threshold to $W = 2.5~\mathrm{GeV}$ ($E_\gamma = 2.7~\mathrm{GeV}$). Notation of the solutions is the same as in Fig.~\ref{fig:amp1a}. Additionally, WM22 fit is shown by yellow dashed curves.} 
\label{fig:amp4a}
\end{figure*}

\begin{figure*}[hbt!]
%\vspace{0.4cm}
\centering
{
    \includegraphics[width=0.32\textwidth,angle=90,keepaspectratio]{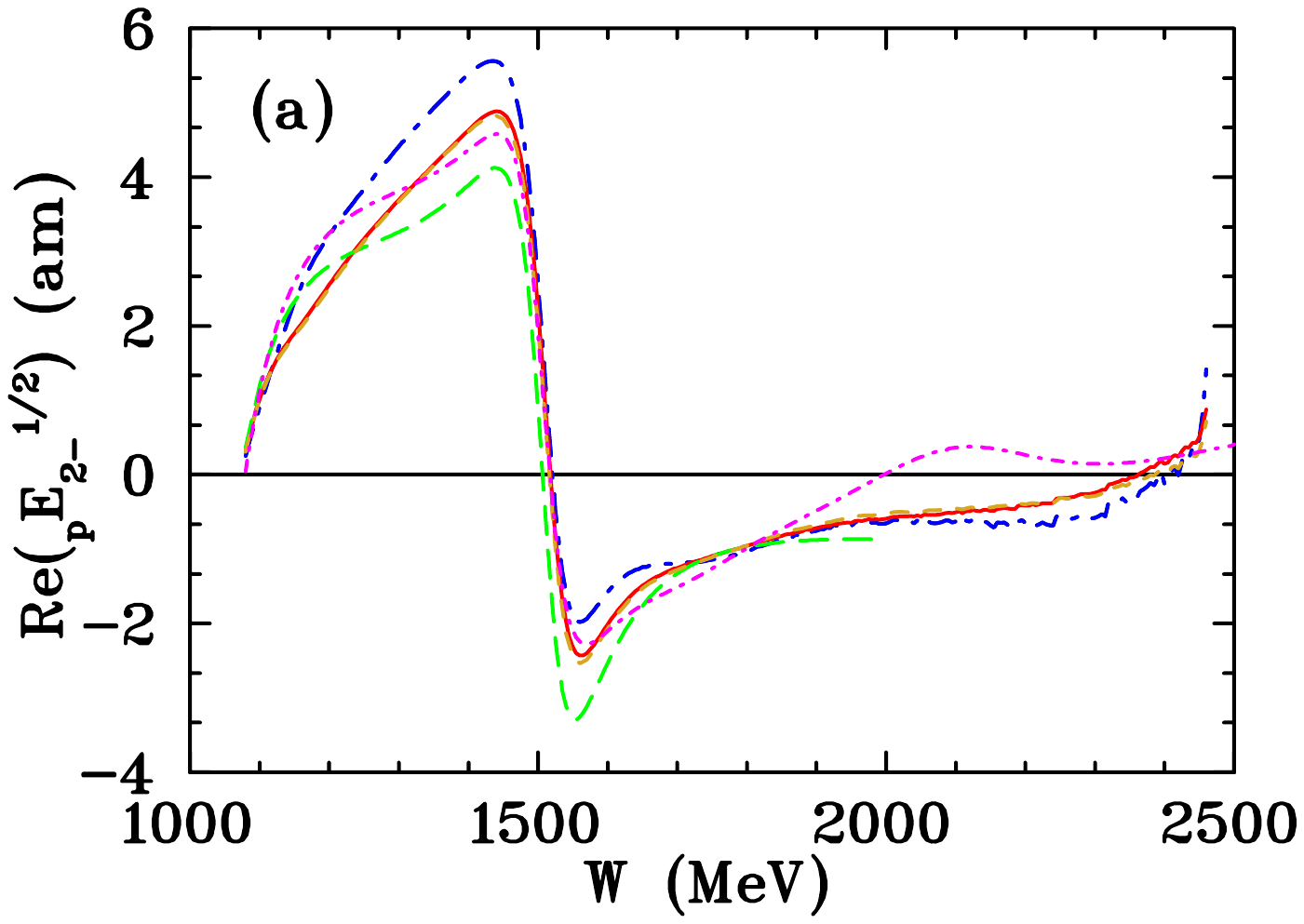}
    \includegraphics[width=0.32\textwidth,angle=90,keepaspectratio]{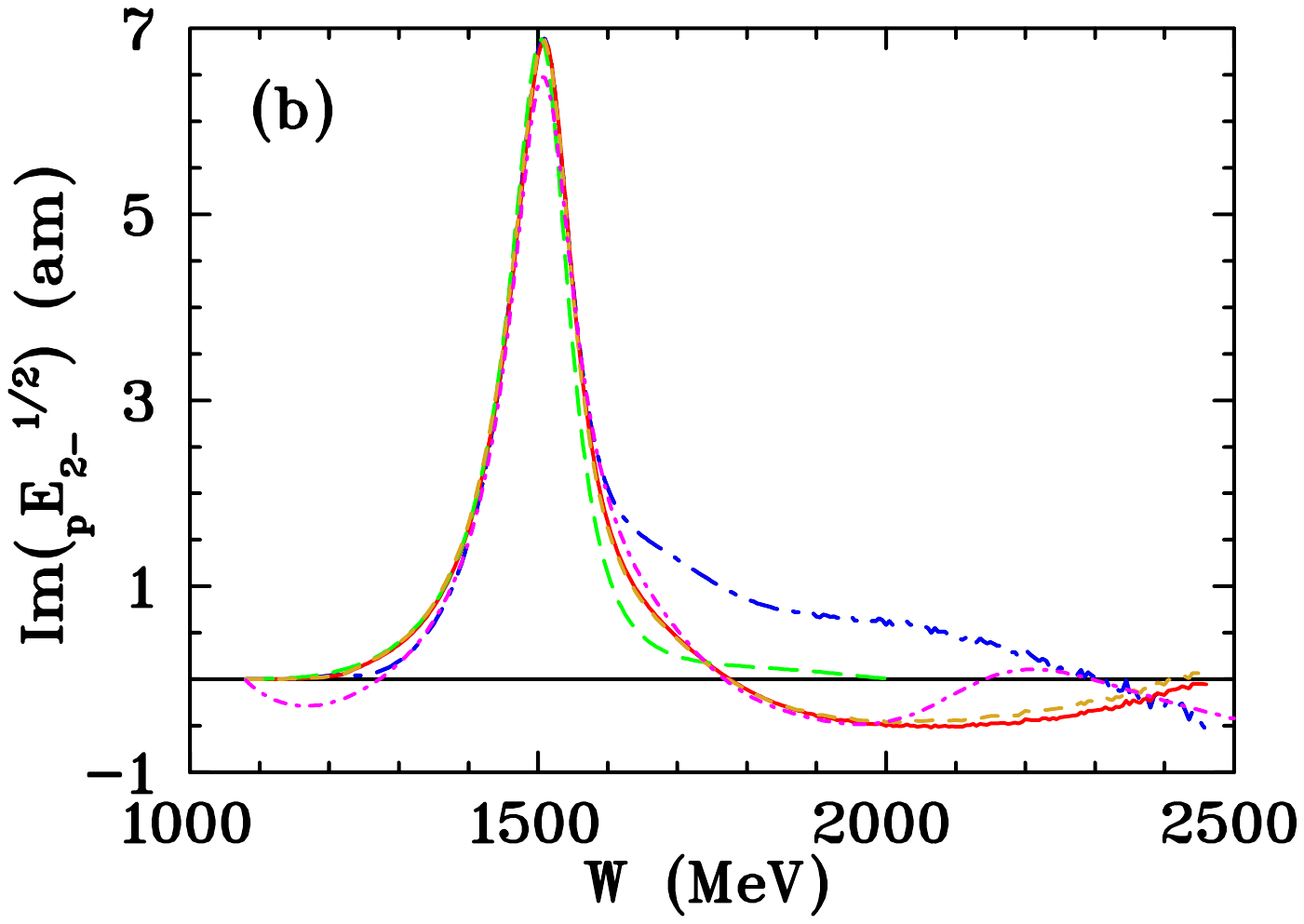}
}
\centering
{
    \includegraphics[width=0.32\textwidth,angle=90,keepaspectratio]{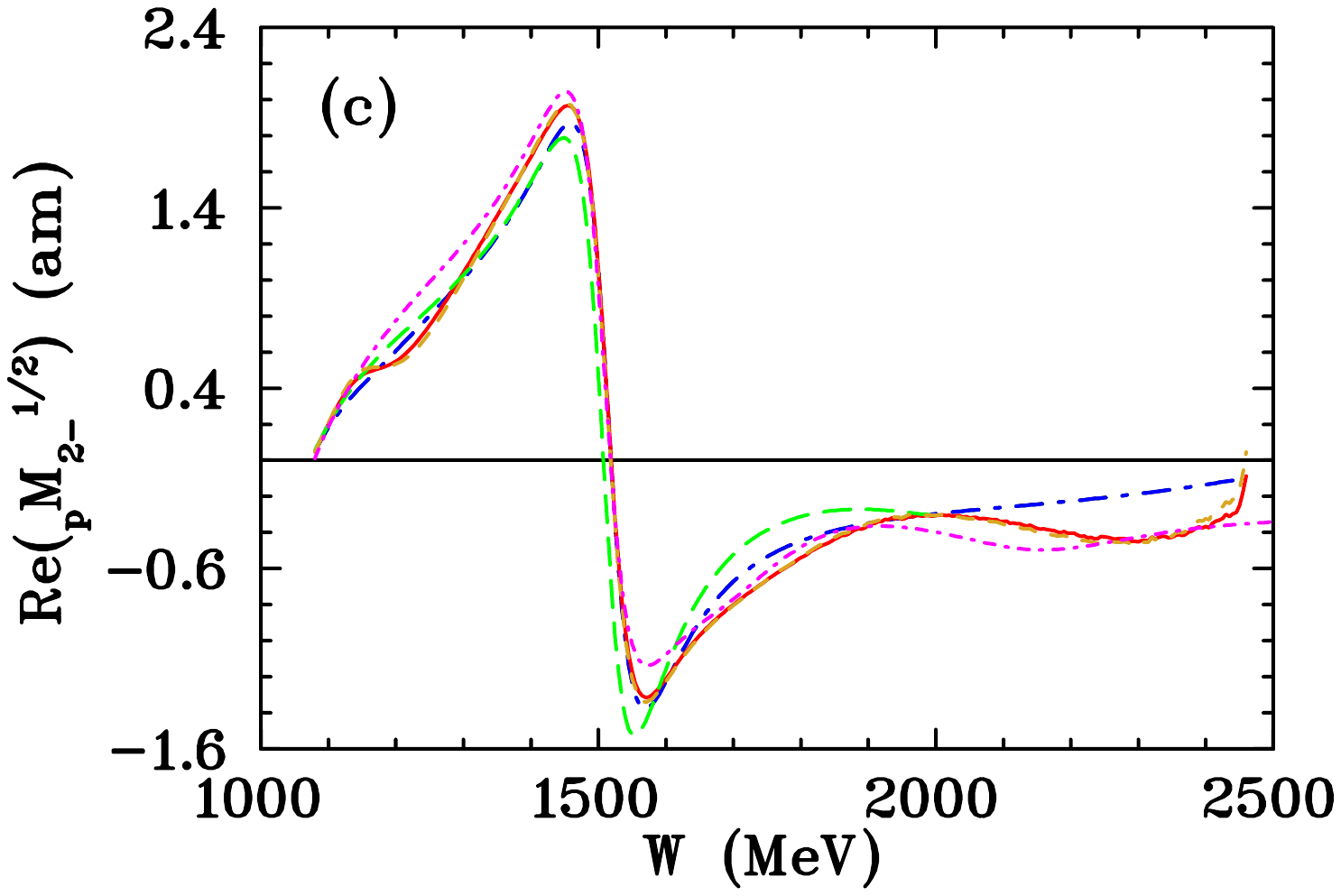}
    \includegraphics[width=0.32\textwidth,angle=90,keepaspectratio]{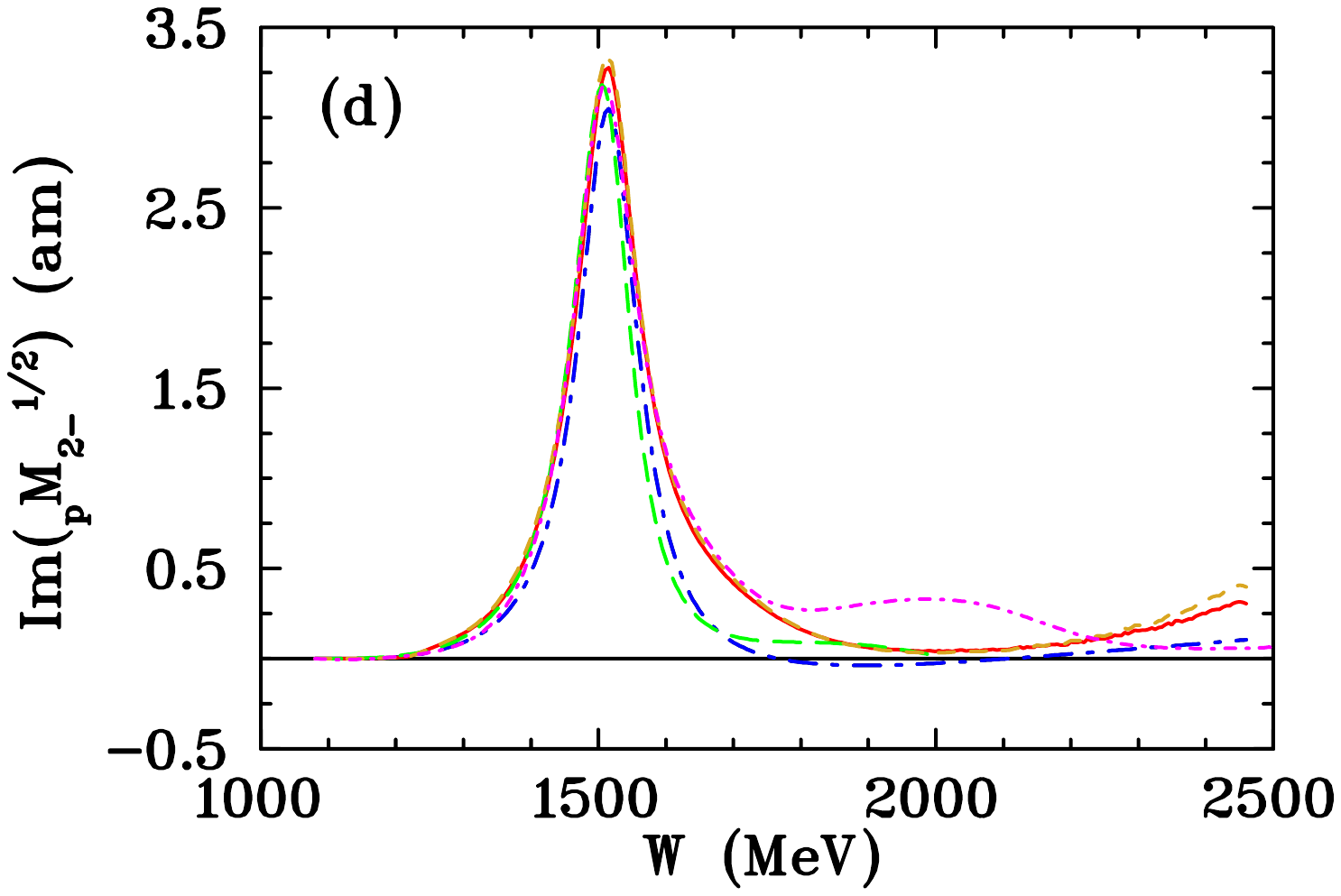}
}
\centering
{
    \includegraphics[width=0.32\textwidth,angle=90,keepaspectratio]{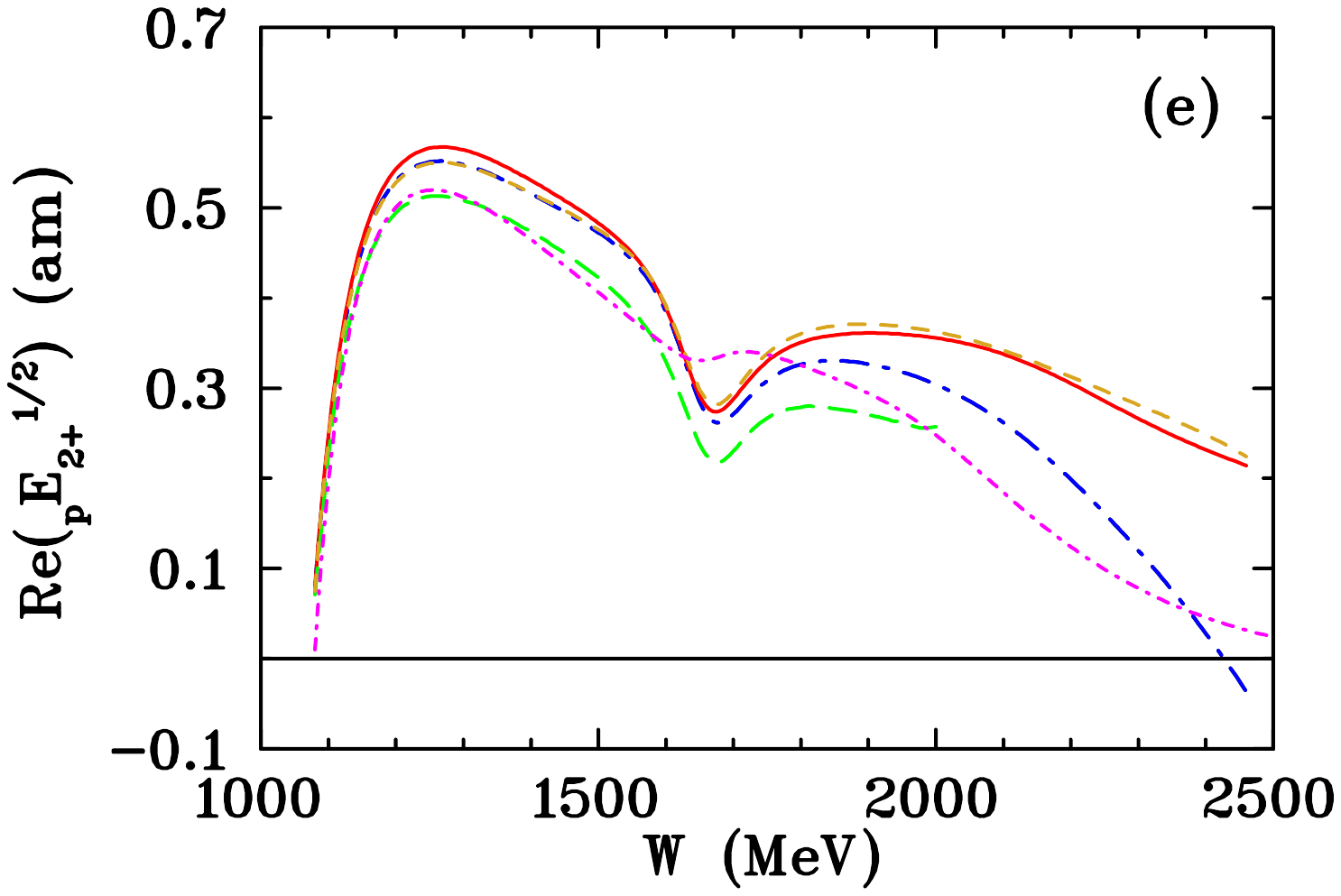}
    \includegraphics[width=0.32\textwidth,angle=90,keepaspectratio]{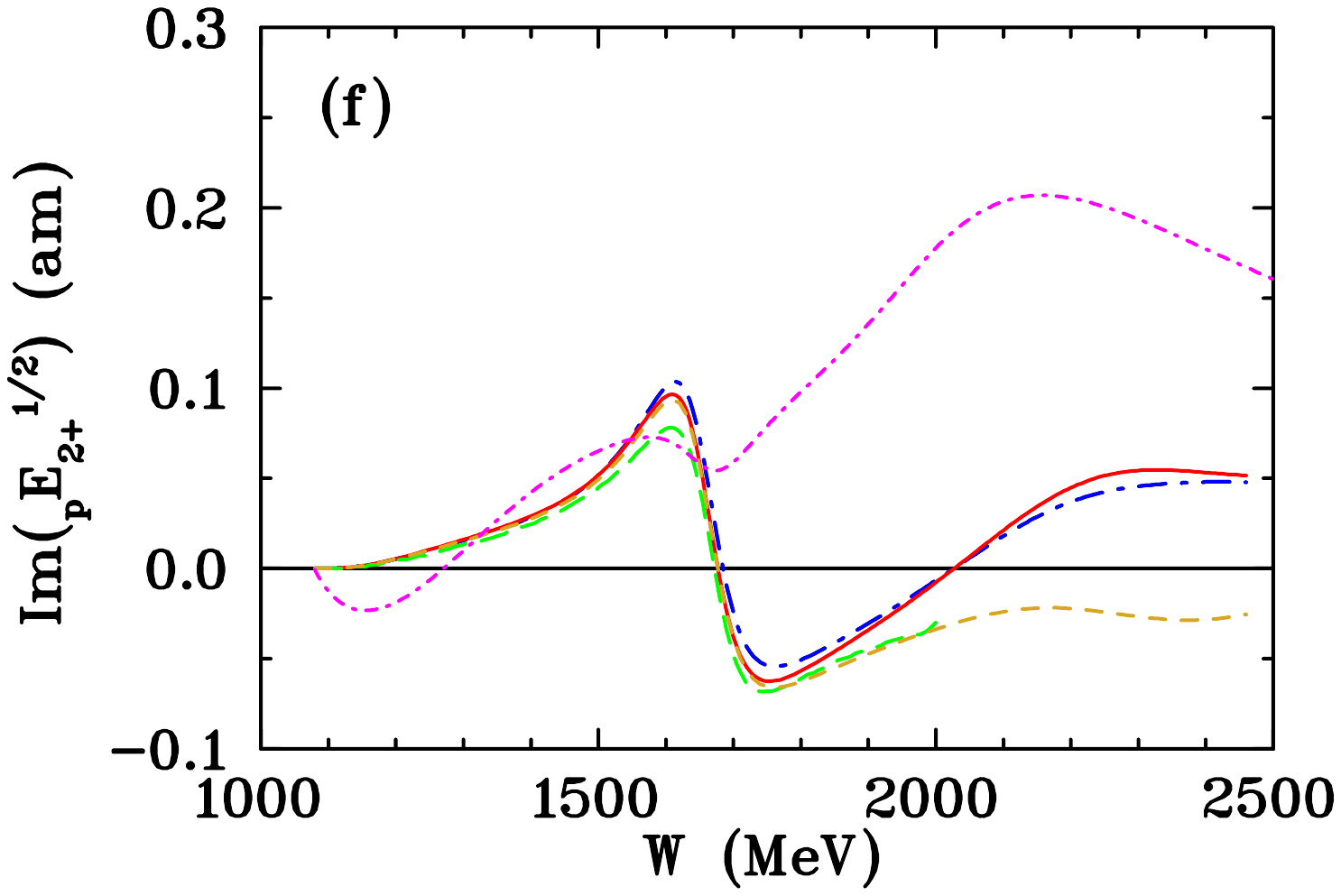}
}
\centering
{
    \includegraphics[width=0.32\textwidth,angle=90,keepaspectratio]{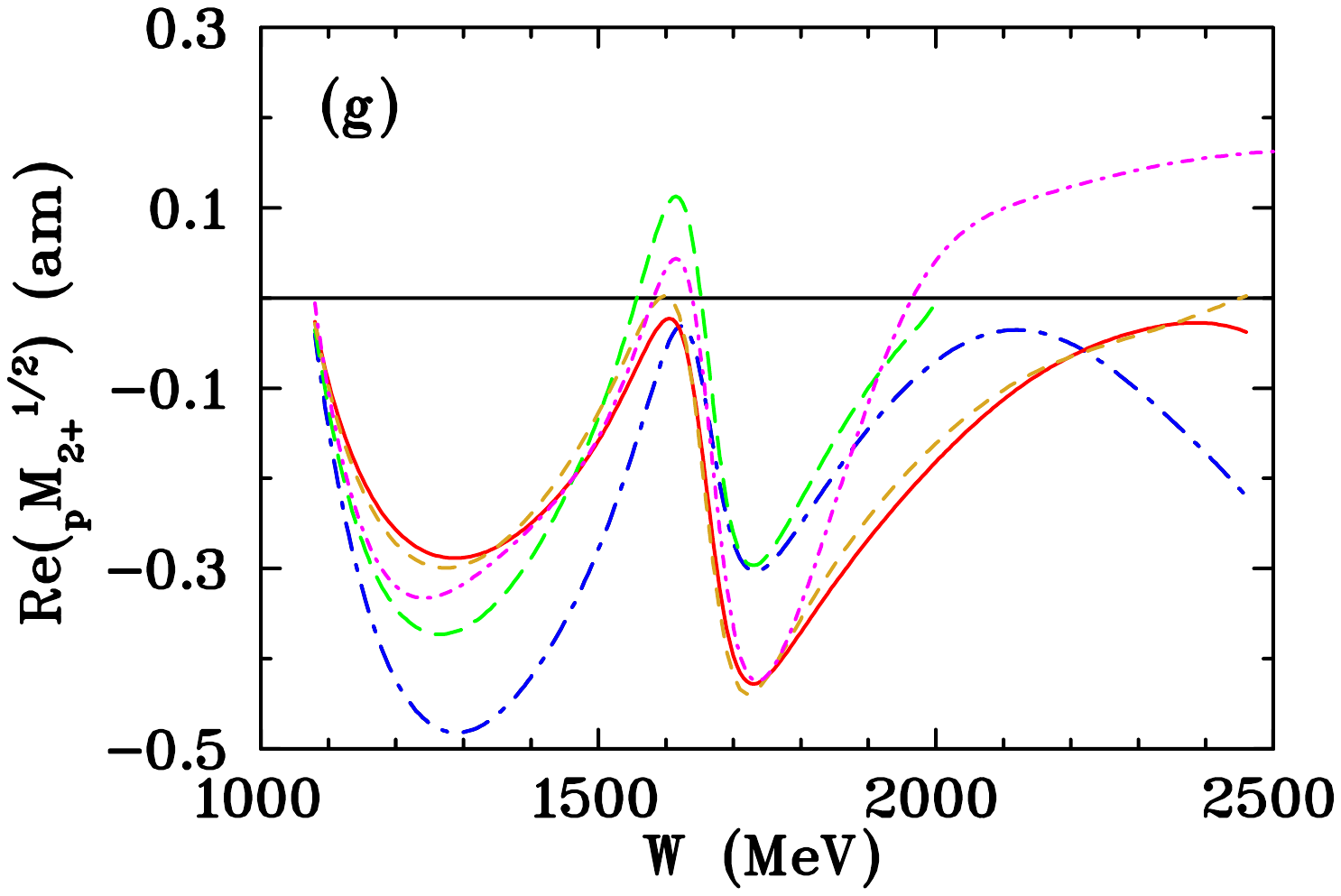}
    \includegraphics[width=0.32\textwidth,angle=90,keepaspectratio]{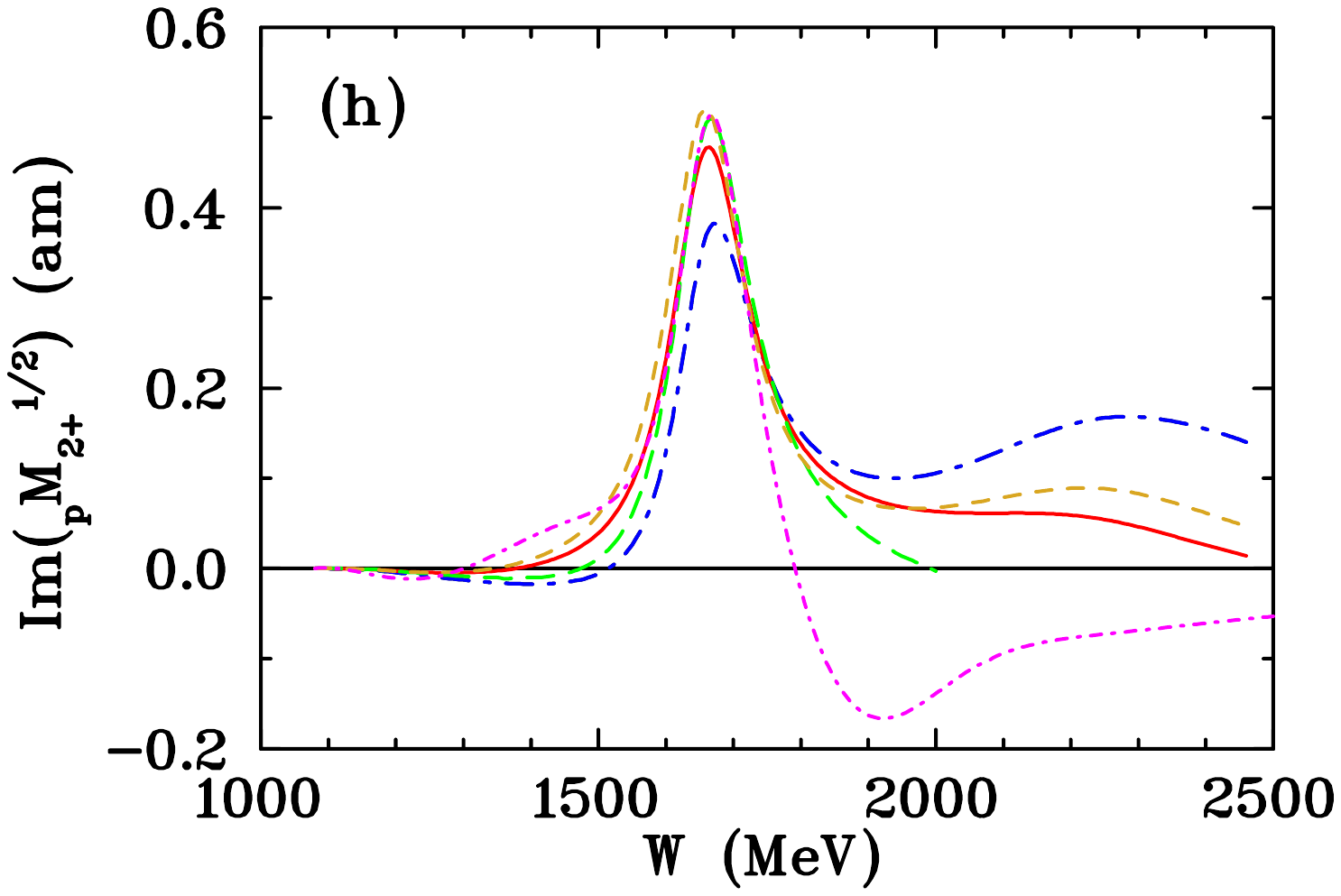}
}

\caption{Comparison of proton $I = 1/2$ multipole amplitudes (orbital momentum $l = 2$) from threshold to $W = 2.5~\mathrm{GeV}$. Notation of the solutions is the same as in Fig.~\ref{fig:amp1a}. For the amplitudes, the subscript $p$ denotes a proton target,
Additionally, WM22 fit shown by yellow dashed curves.}  
\label{fig:amp5a}
\end{figure*}

\begin{figure*}[hbt!]
%\vspace{0.4cm}
\centering
{
    \includegraphics[width=0.32\textwidth,angle=90,keepaspectratio]{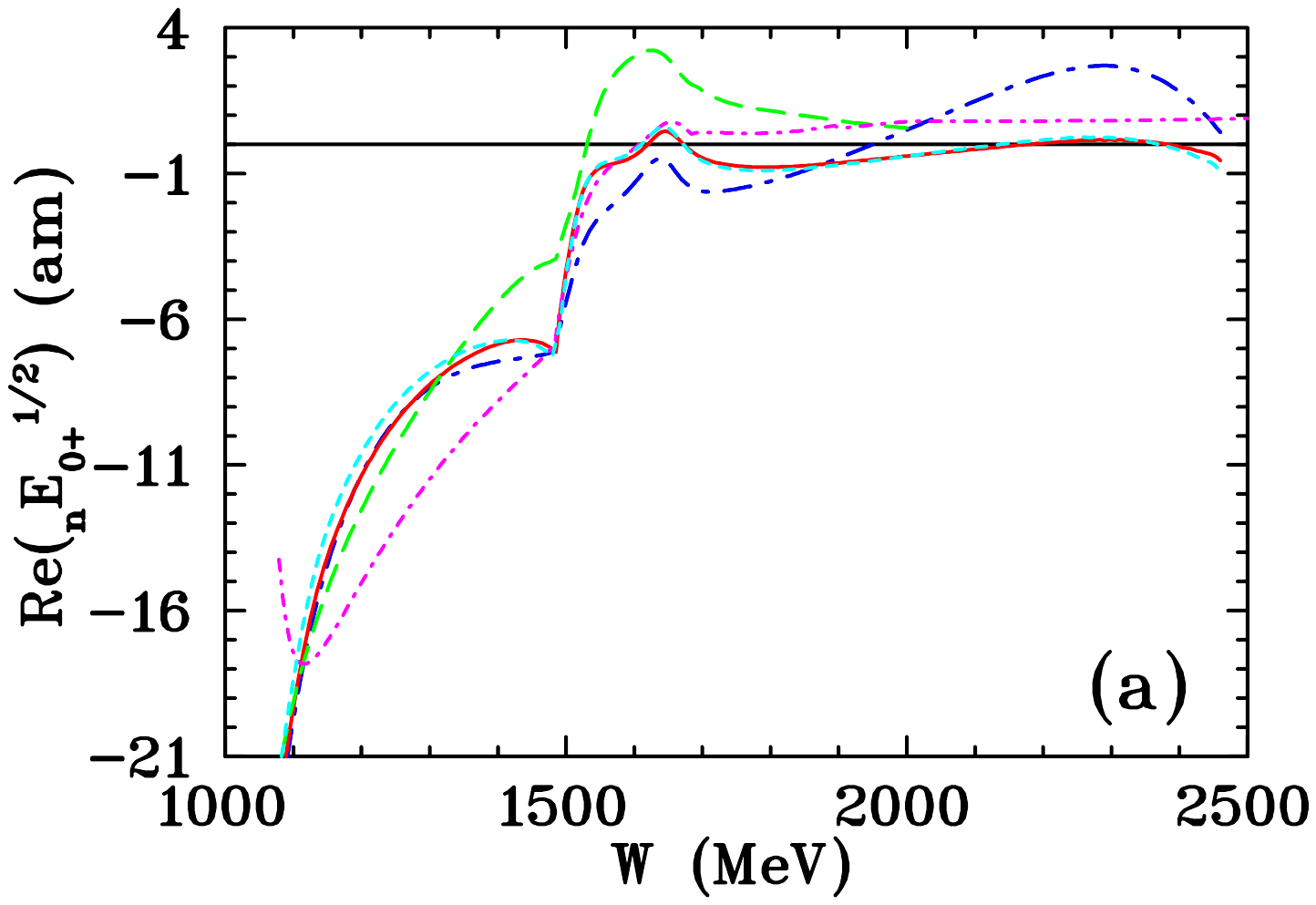}
    \includegraphics[width=0.32\textwidth,angle=90,keepaspectratio]{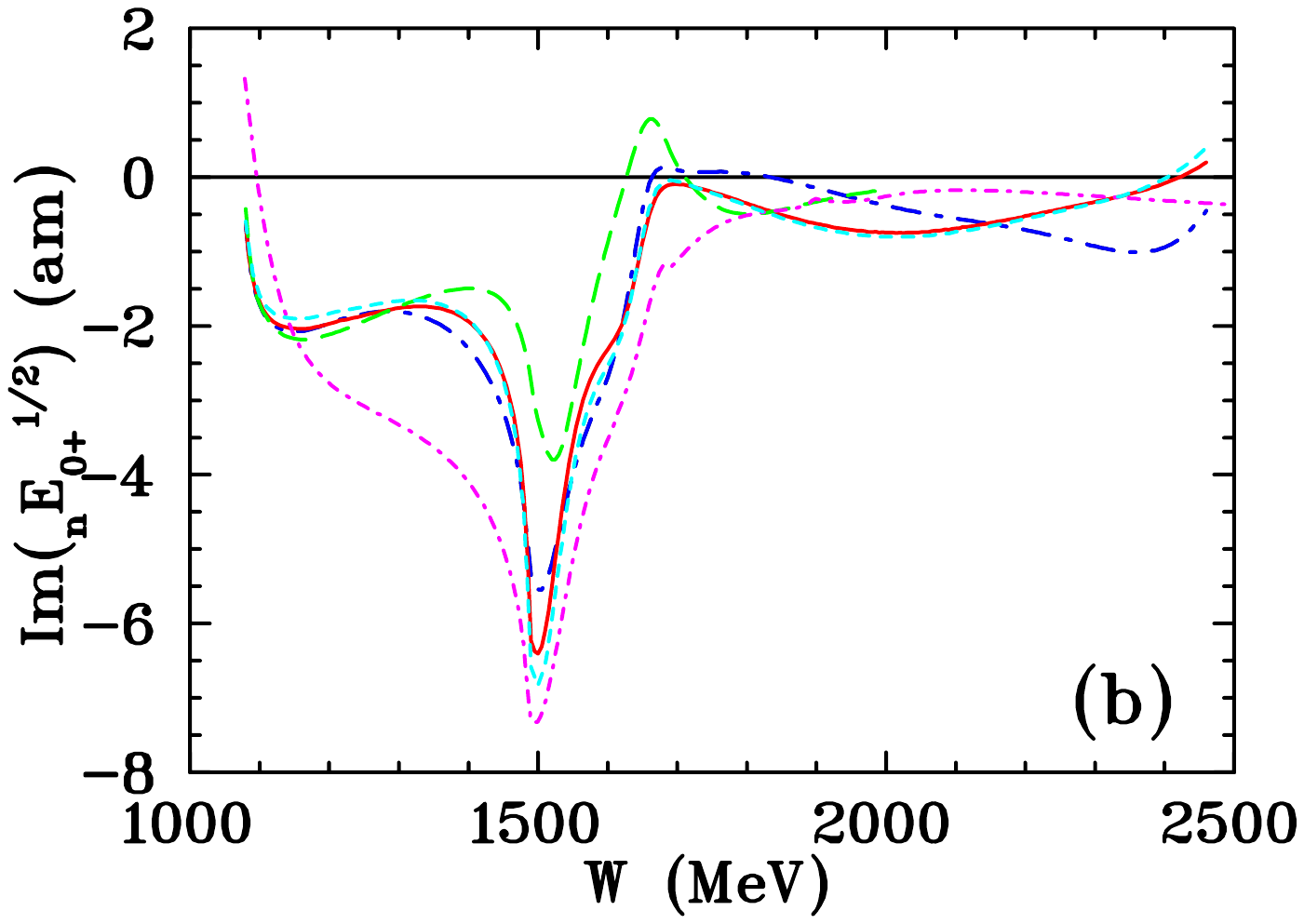}
}
\centering
{
    \includegraphics[width=0.32\textwidth,angle=90,keepaspectratio]{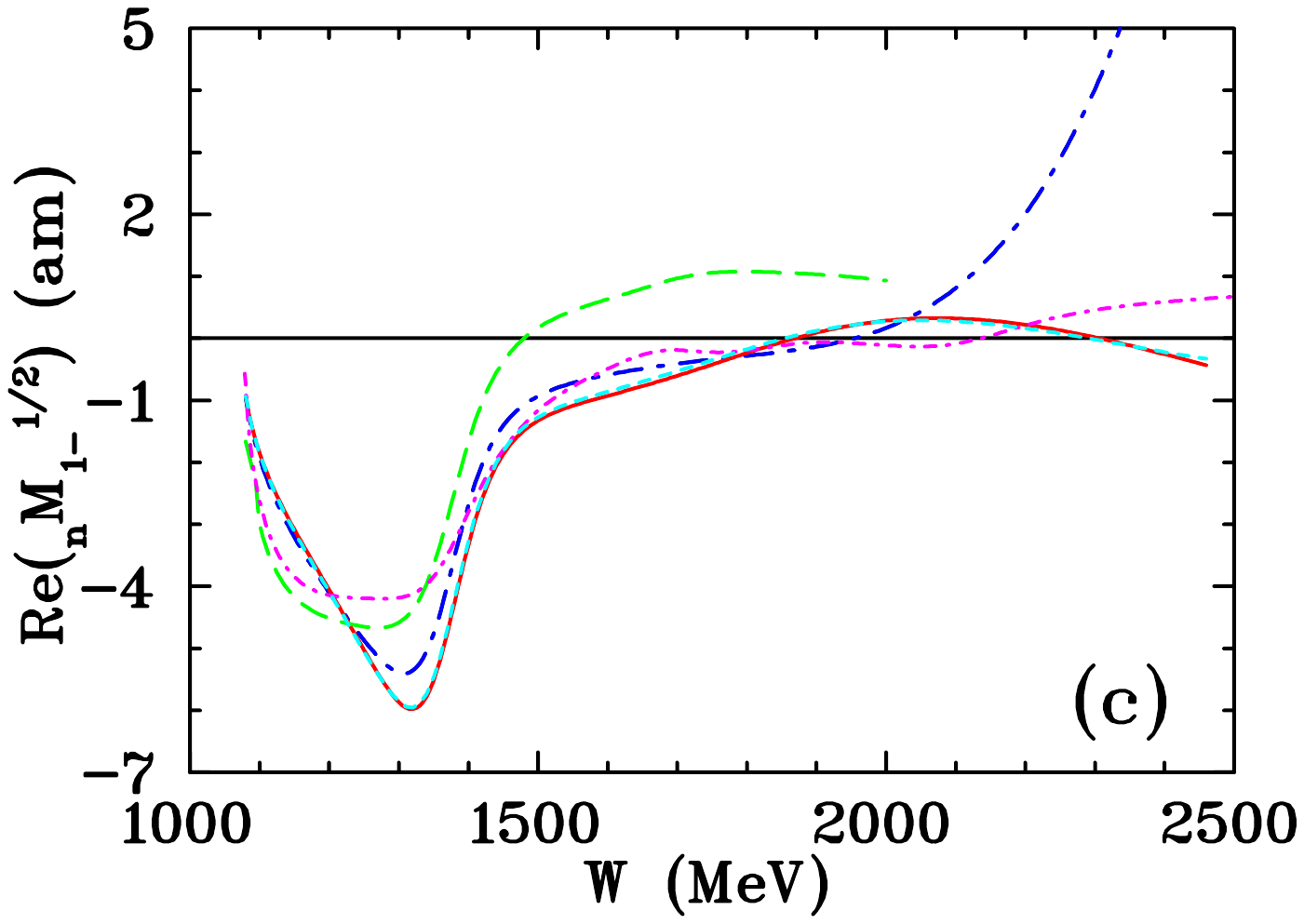}
    \includegraphics[width=0.32\textwidth,angle=90,keepaspectratio]{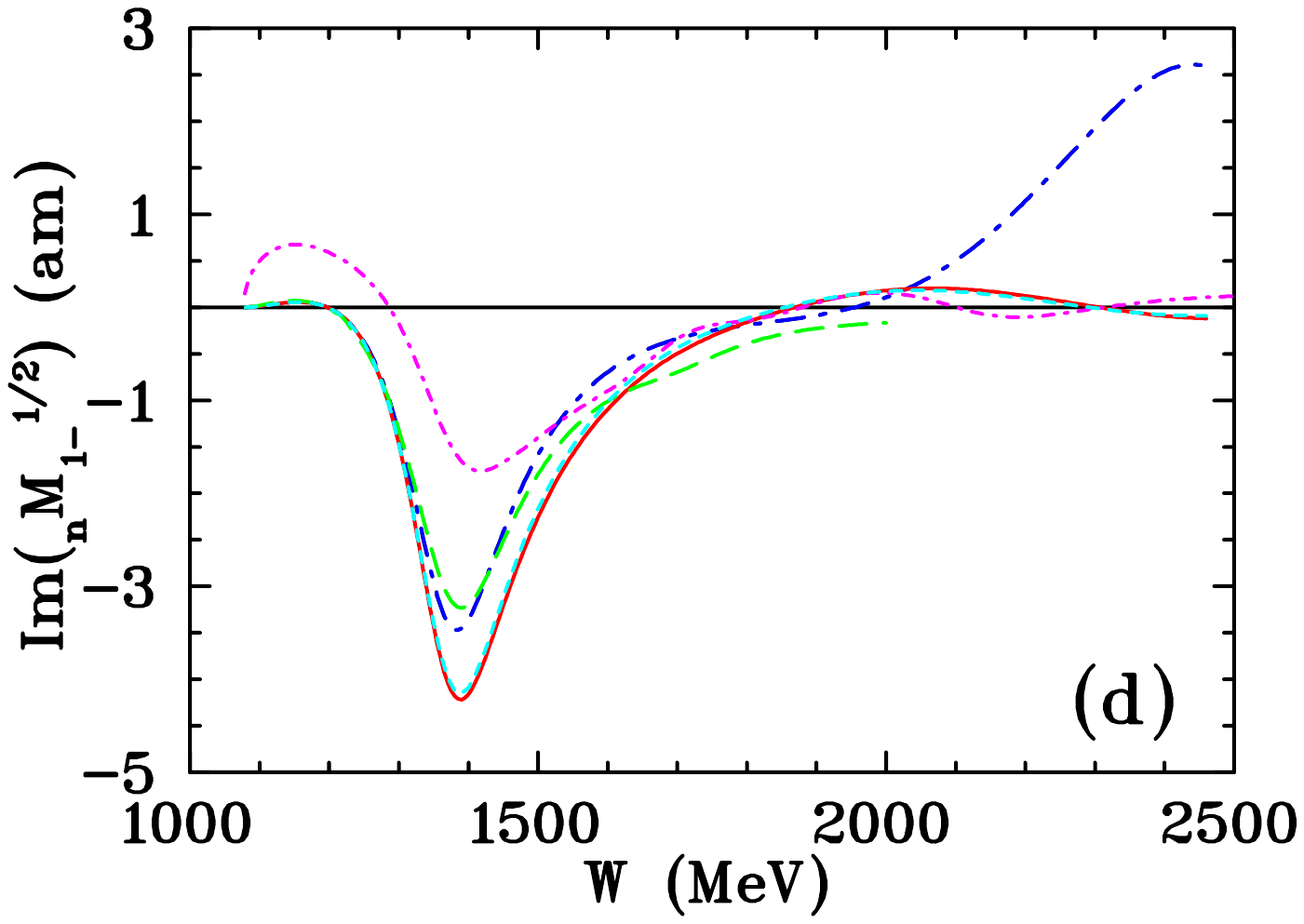}
}
\centering
{
    \includegraphics[width=0.32\textwidth,angle=90,keepaspectratio]{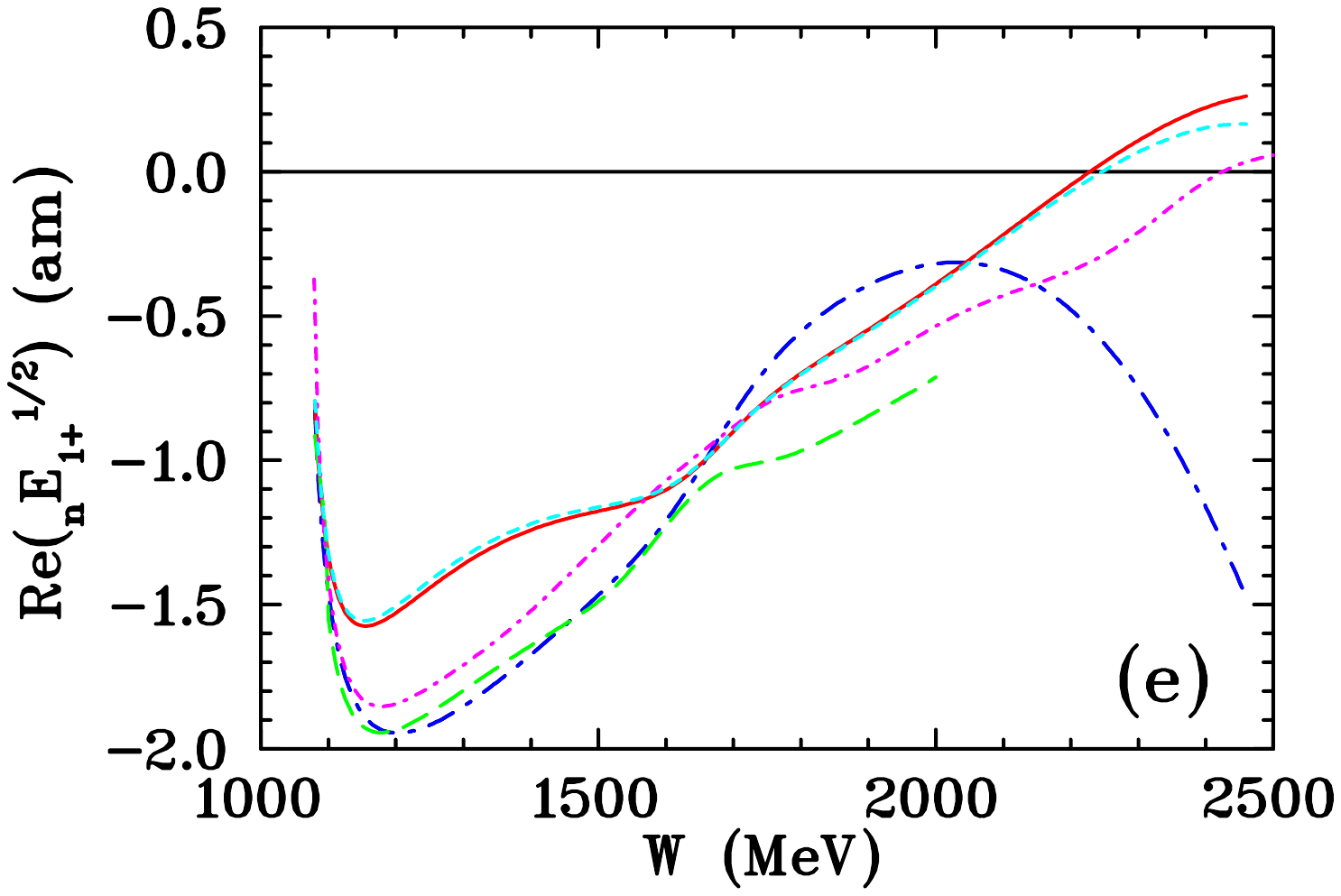}
    \includegraphics[width=0.32\textwidth,angle=90,keepaspectratio]{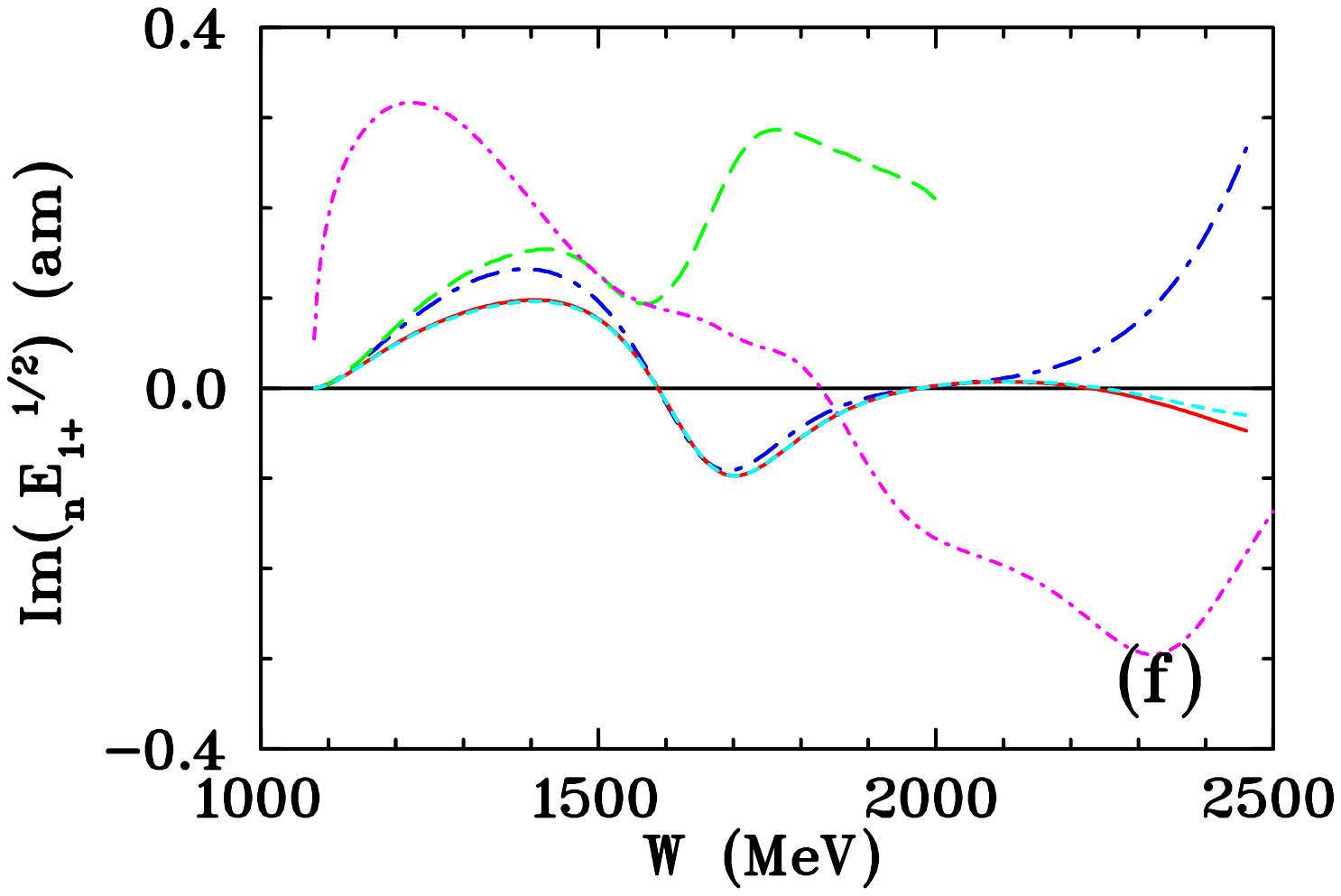}
}
\centering
{
    \includegraphics[width=0.32\textwidth,angle=90,keepaspectratio]{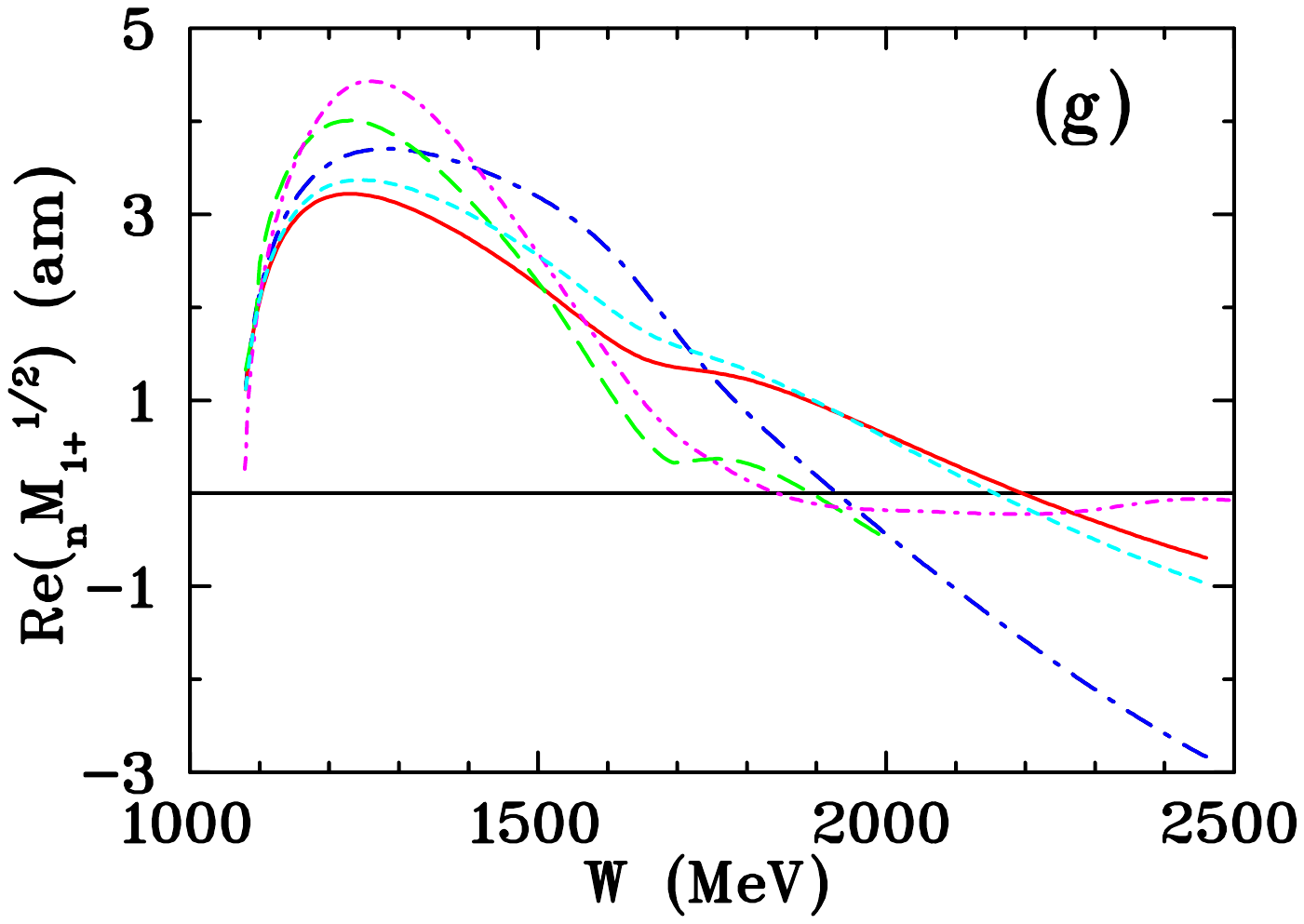}
    \includegraphics[width=0.32\textwidth,angle=90,keepaspectratio]{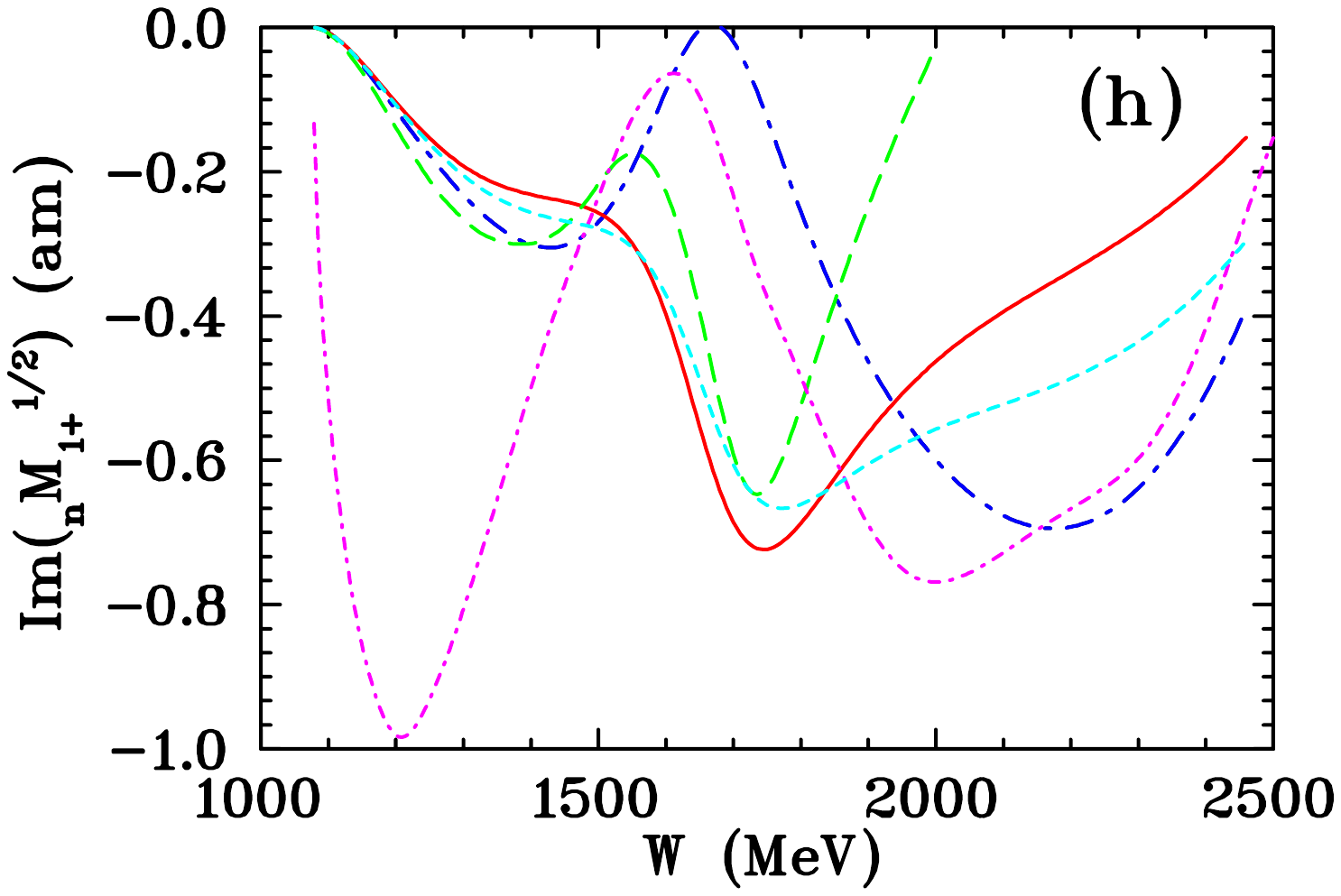}
}

\caption{Comparison of neutron $I = 1/2$ multipole amplitudes (orbital momentum $l = 0, 1$) from threshold to $W = 2.5~\mathrm{GeV}$ ($E_\gamma = 2.7~\mathrm{GeV}$). Notation of the solutions is the same as in Fig.~\ref{fig:amp1a}. Additionally, cyan short-dashed curves are SM44 fits.}   
\label{fig:amp7a}
\end{figure*}

\begin{figure*}[hbt!]
%\vspace{0.4cm}
\centering
{
    \includegraphics[width=0.32\textwidth,angle=90,keepaspectratio]{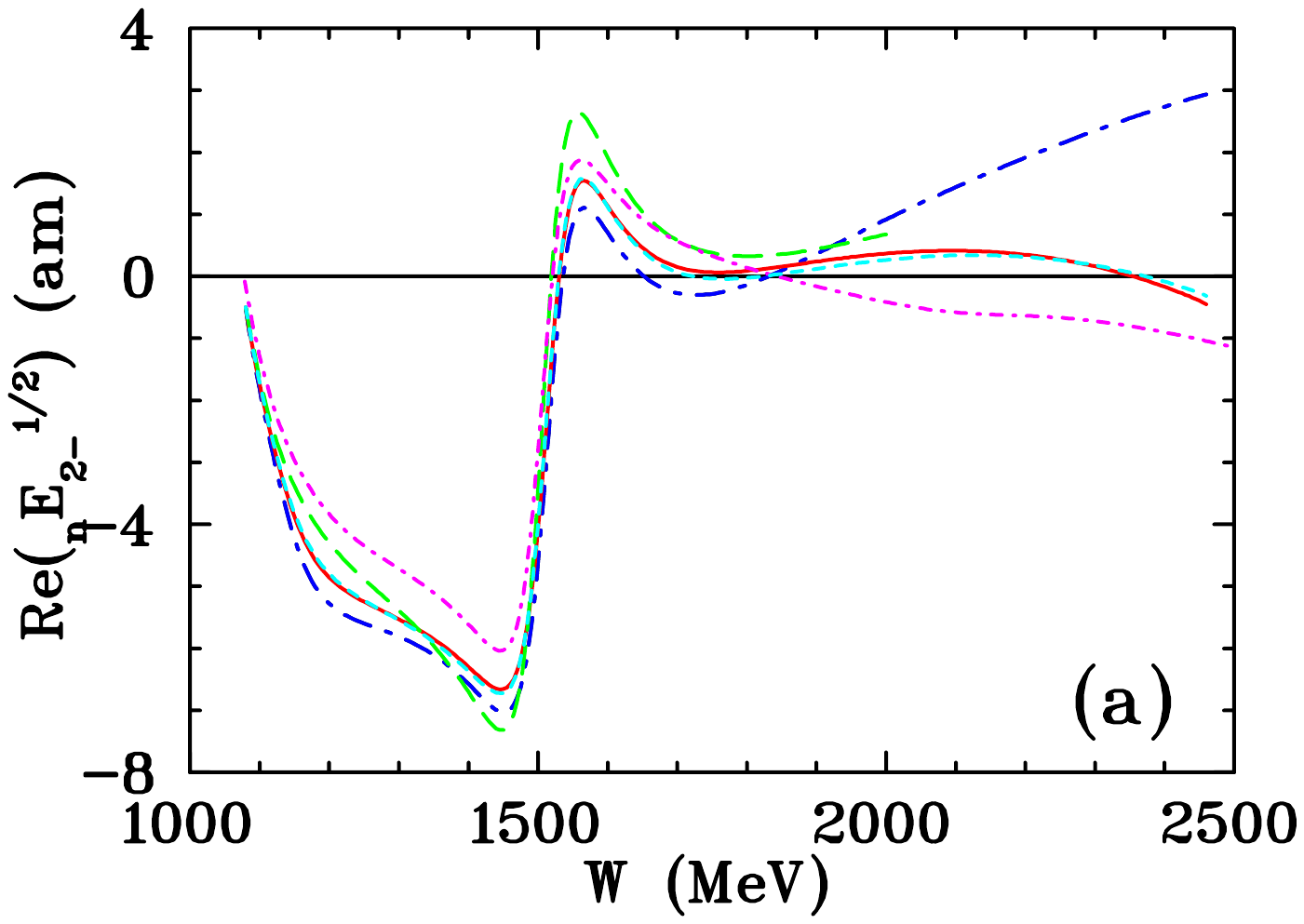}
    \includegraphics[width=0.32\textwidth,angle=90,keepaspectratio]{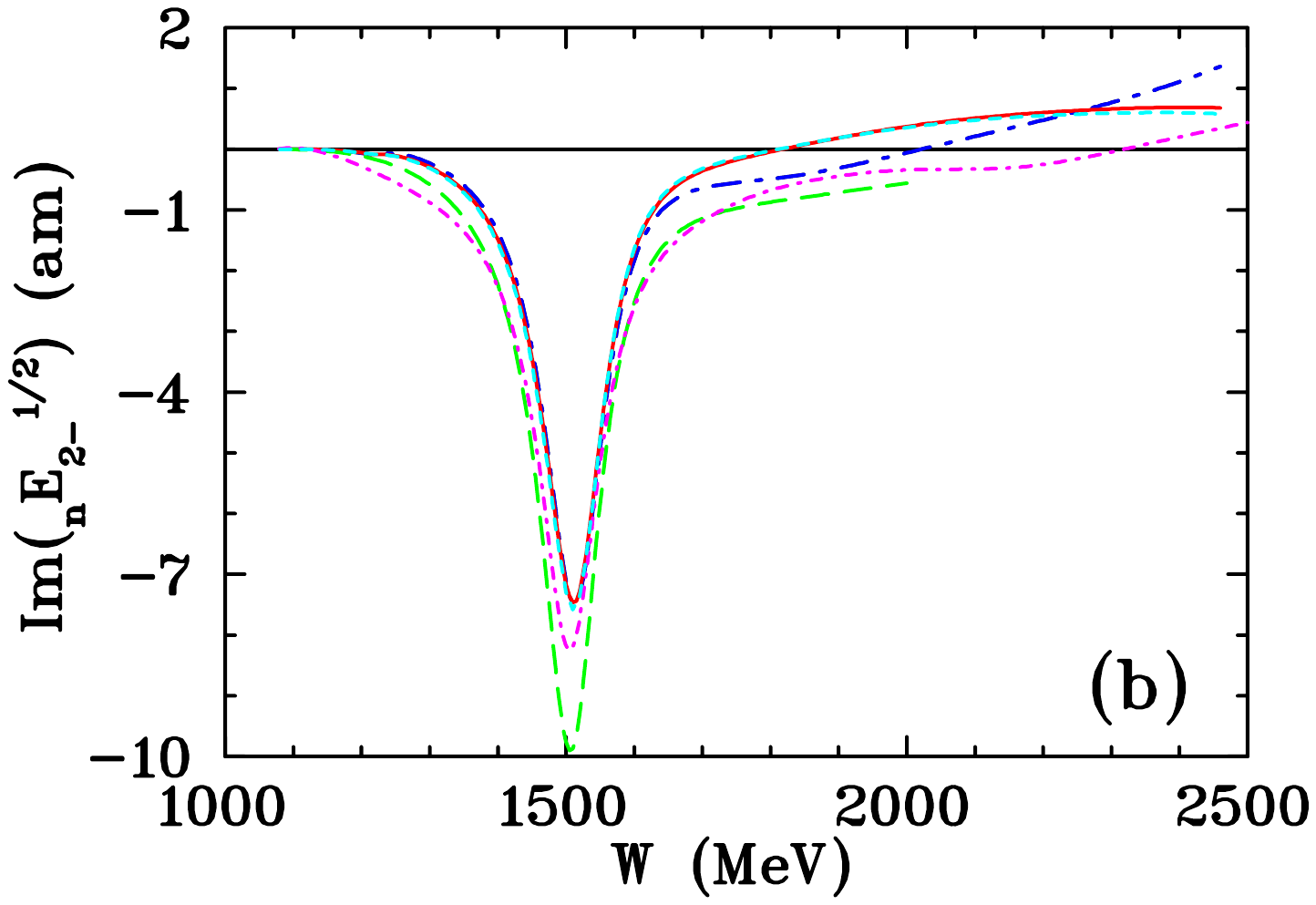}
}
\centering
{
    \includegraphics[width=0.32\textwidth,angle=90,keepaspectratio]{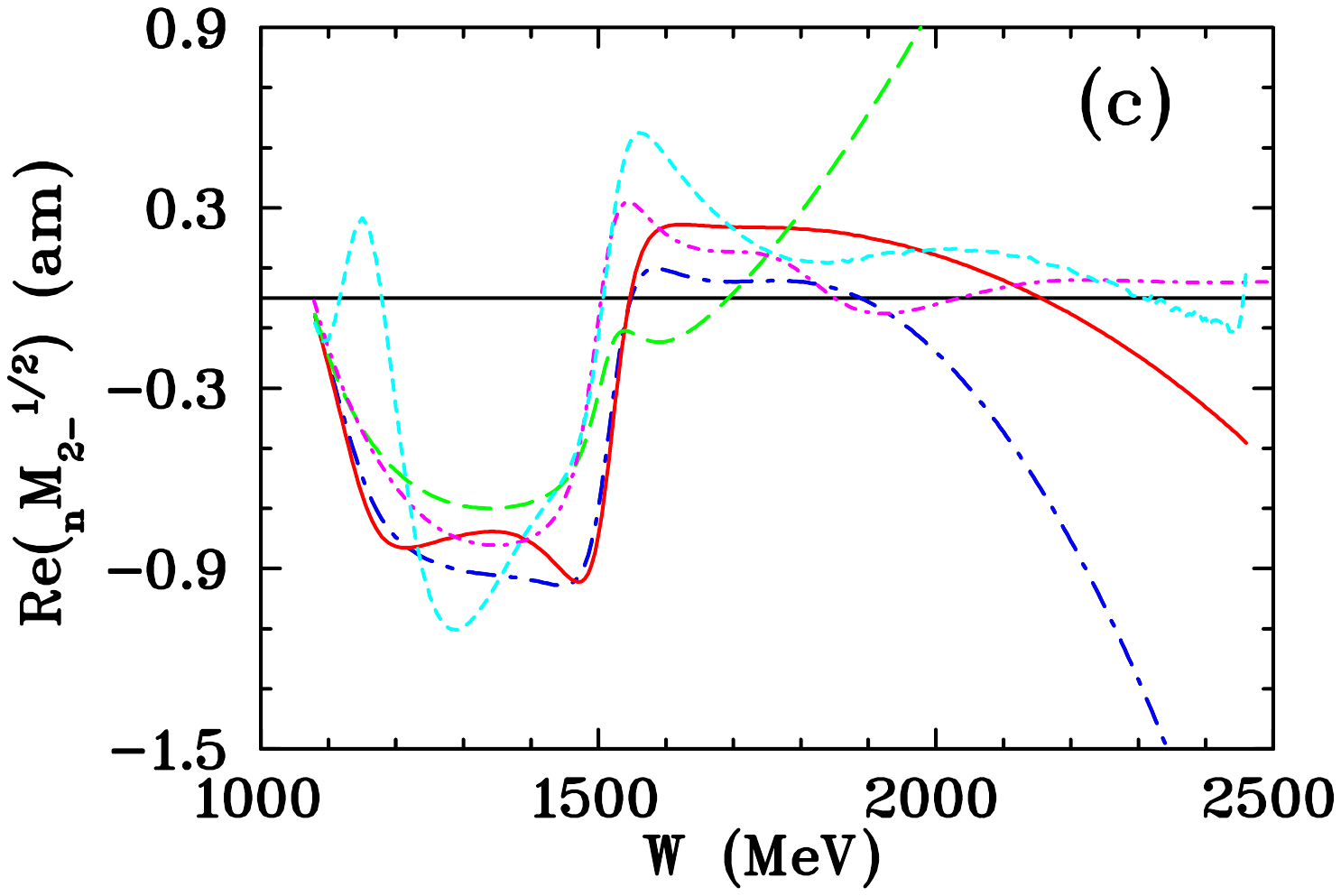}
    \includegraphics[width=0.32\textwidth,angle=90,keepaspectratio]{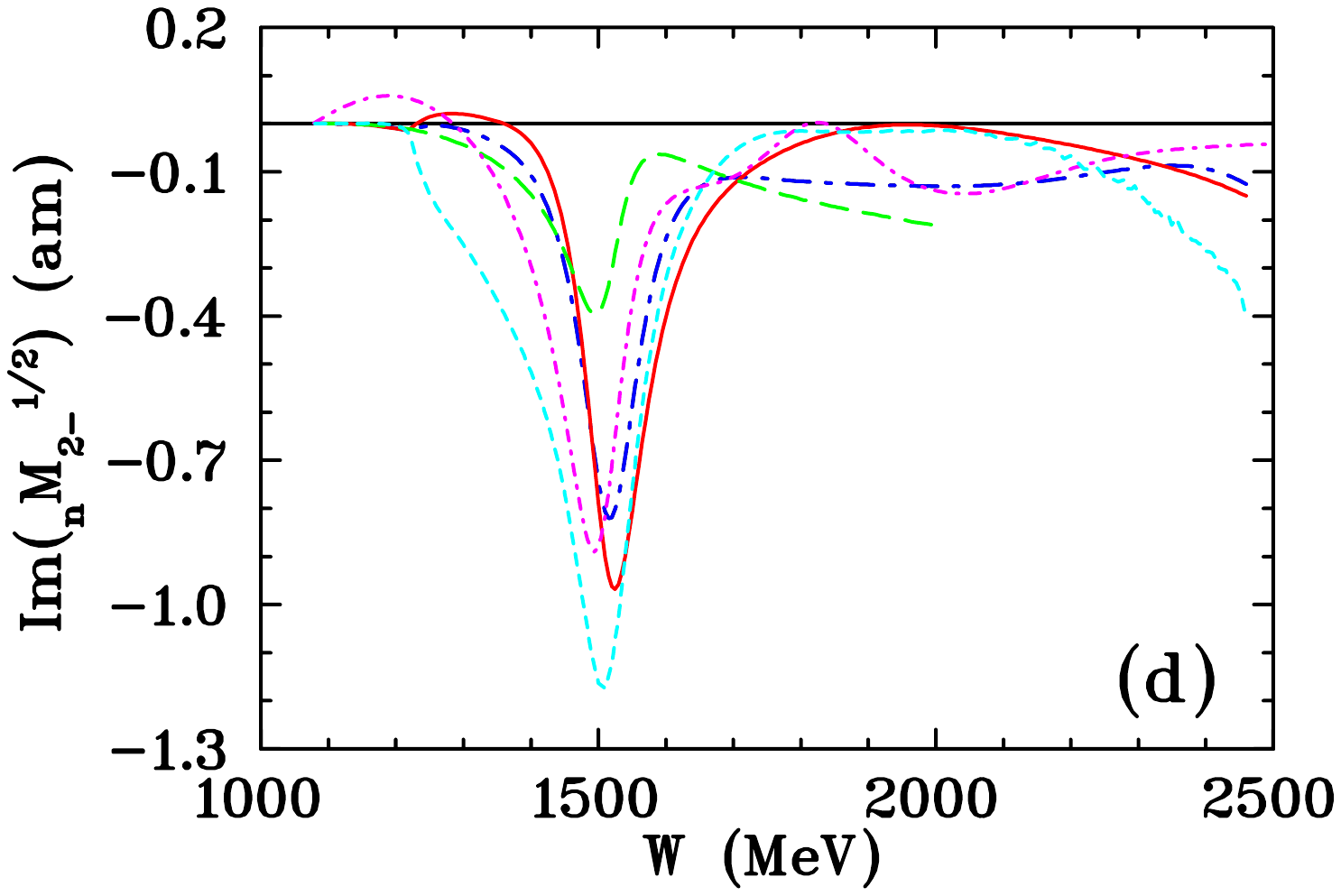}
}
\centering
{
    \includegraphics[width=0.32\textwidth,angle=90,keepaspectratio]{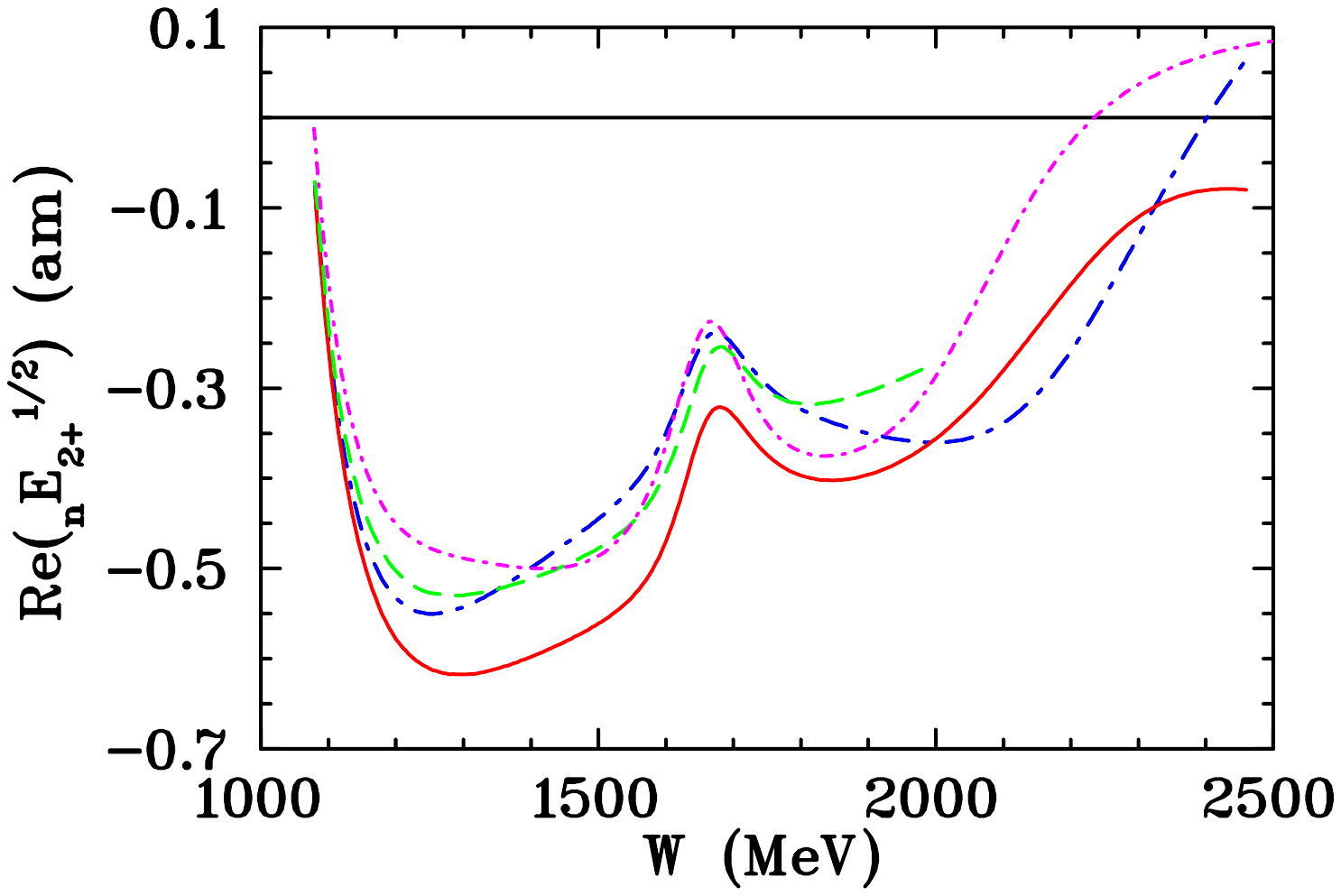}
    \includegraphics[width=0.32\textwidth,angle=90,keepaspectratio]{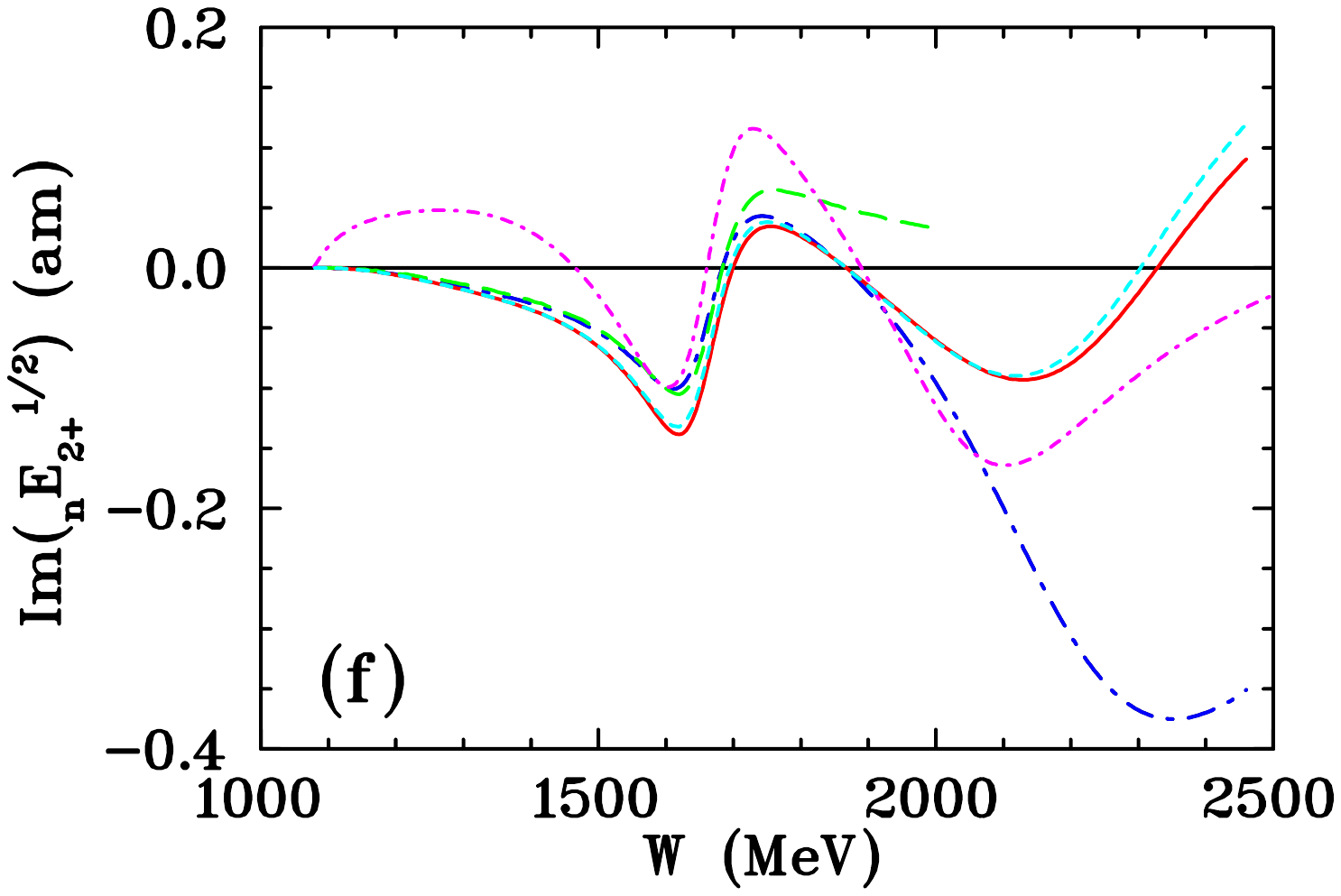}
}
\centering
{
    \includegraphics[width=0.32\textwidth,angle=90,keepaspectratio]{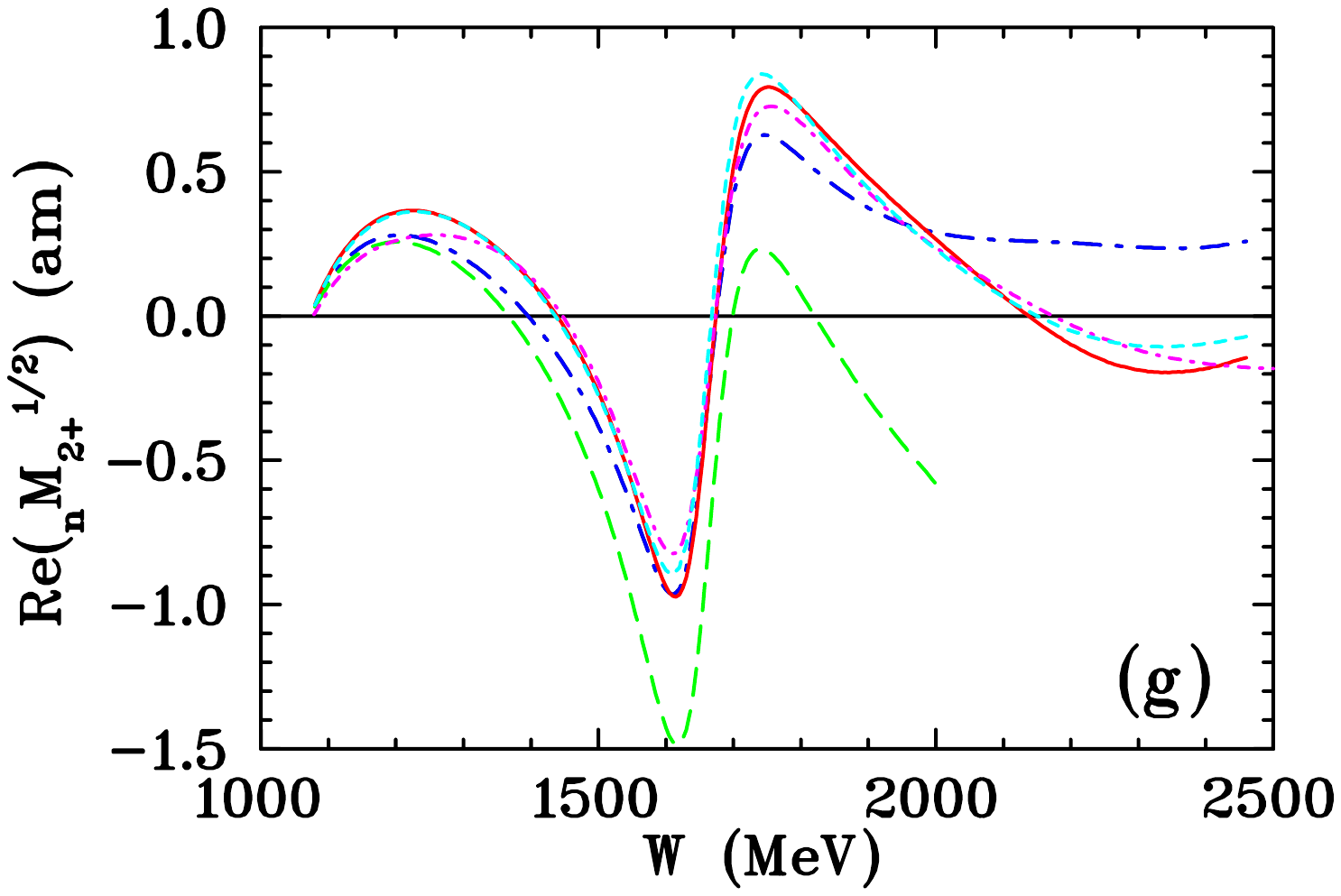}
    \includegraphics[width=0.32\textwidth,angle=90,keepaspectratio]{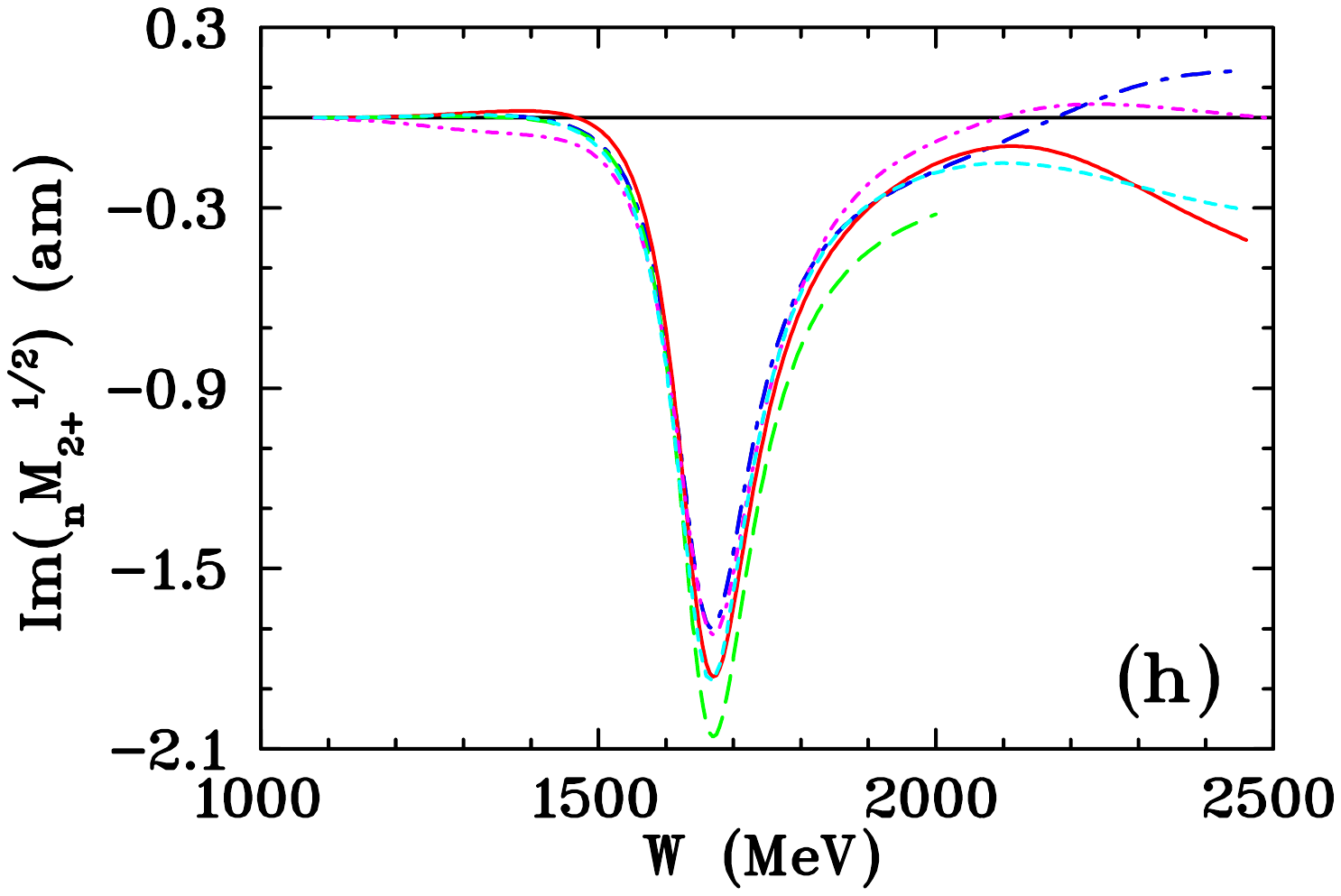}
}

\caption{Comparison neutron $I = 1/2$ multipole amplitudes (orbital momentum $l = 2$) from threshold to $W = 2.5~\mathrm{GeV}$. Notation of the solutions is the same as in Fig.~\ref{fig:amp7a}.
}  
\label{fig:amp8a}
\end{figure*}

Comparisons of the present SAID $I = 3/2$ and $I = 1/2$ multipoles amplitudes from threshold to $W = 2.5~\mathrm{GeV}$ ($E_\gamma = 2.7~\mathrm{GeV}$) shown in Figs.~\ref{fig:amp1a} - \ref{fig:amp8a}.
Also included, for comparison, are the BnGa and MAID multipoles.

\begin{figure*}[hbt!]
%\vspace{0.4cm}
\centering
{
    \includegraphics[width=0.32\textwidth,angle=90,keepaspectratio]{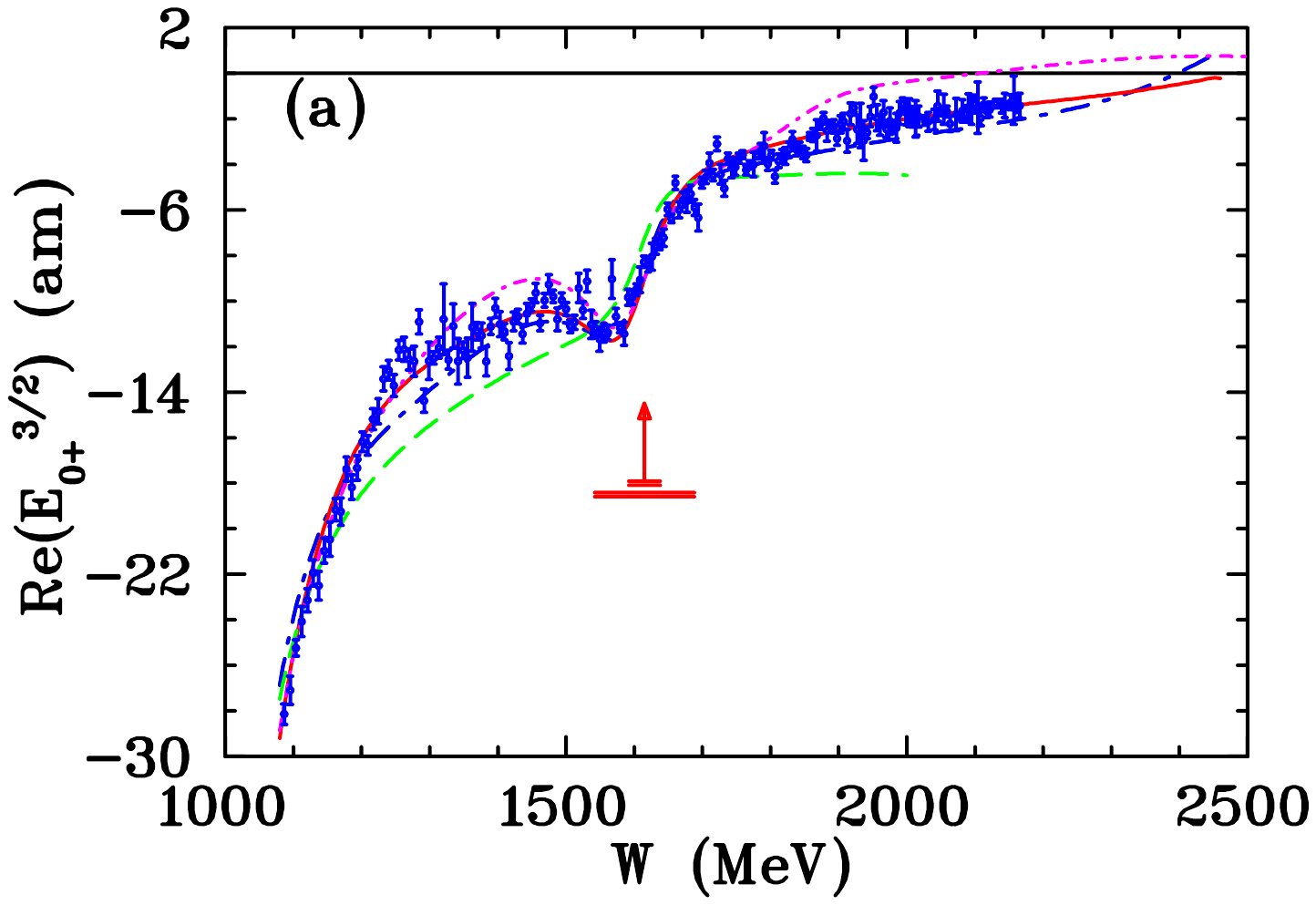}
    \includegraphics[width=0.32\textwidth,angle=90,keepaspectratio]{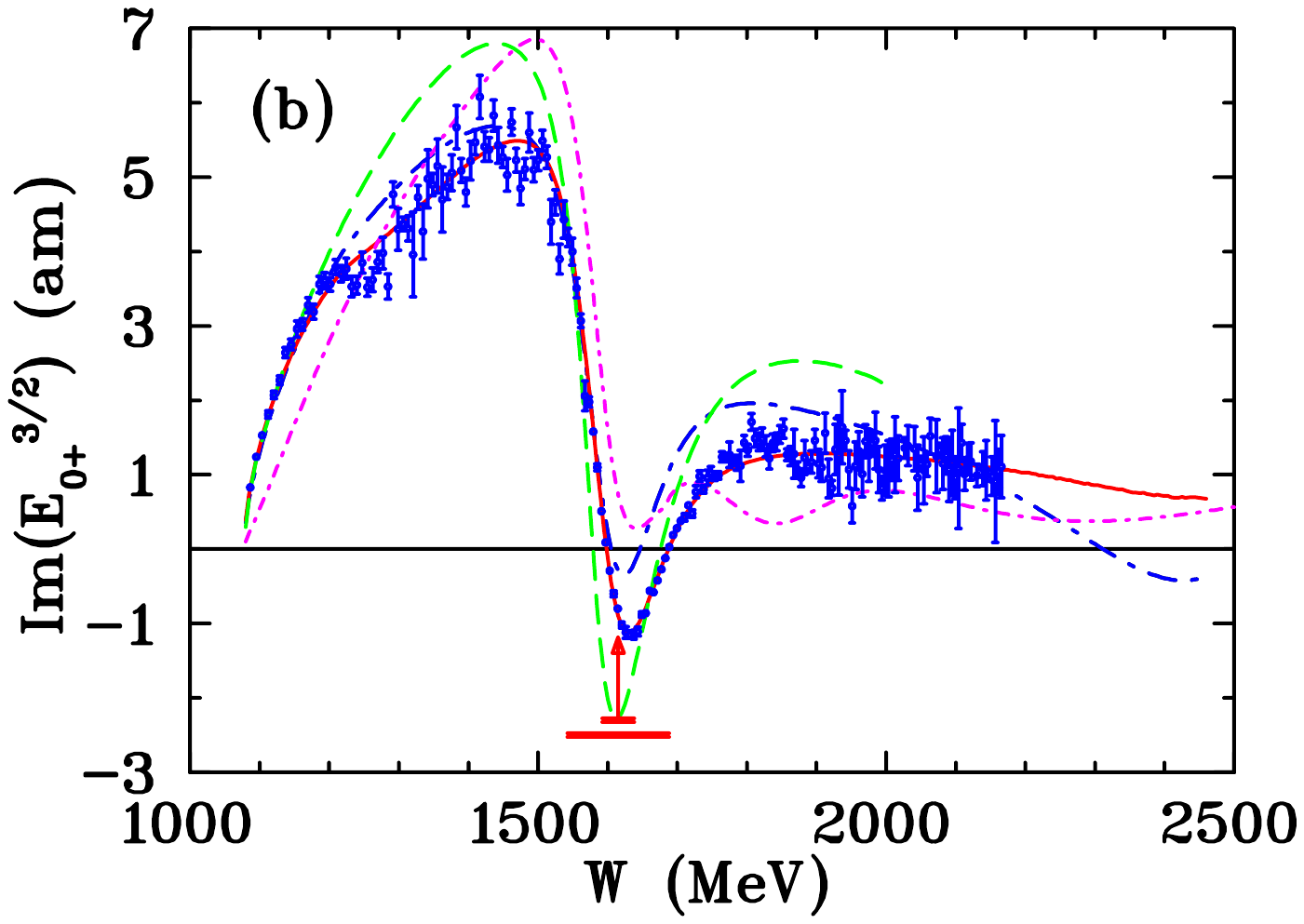}
}
\centering
{
    \includegraphics[width=0.32\textwidth,angle=90,keepaspectratio]{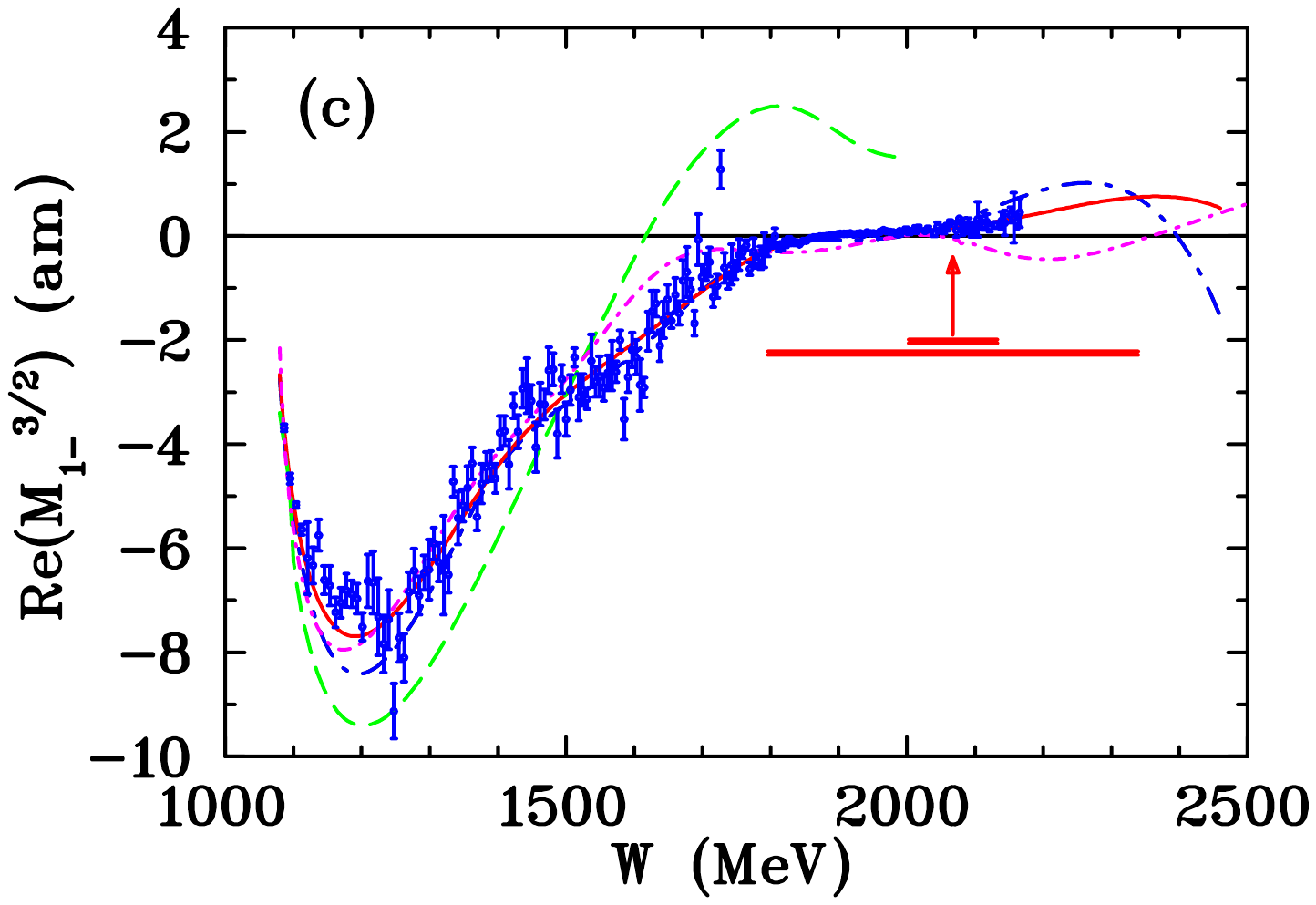}
    \includegraphics[width=0.32\textwidth,angle=90,keepaspectratio]{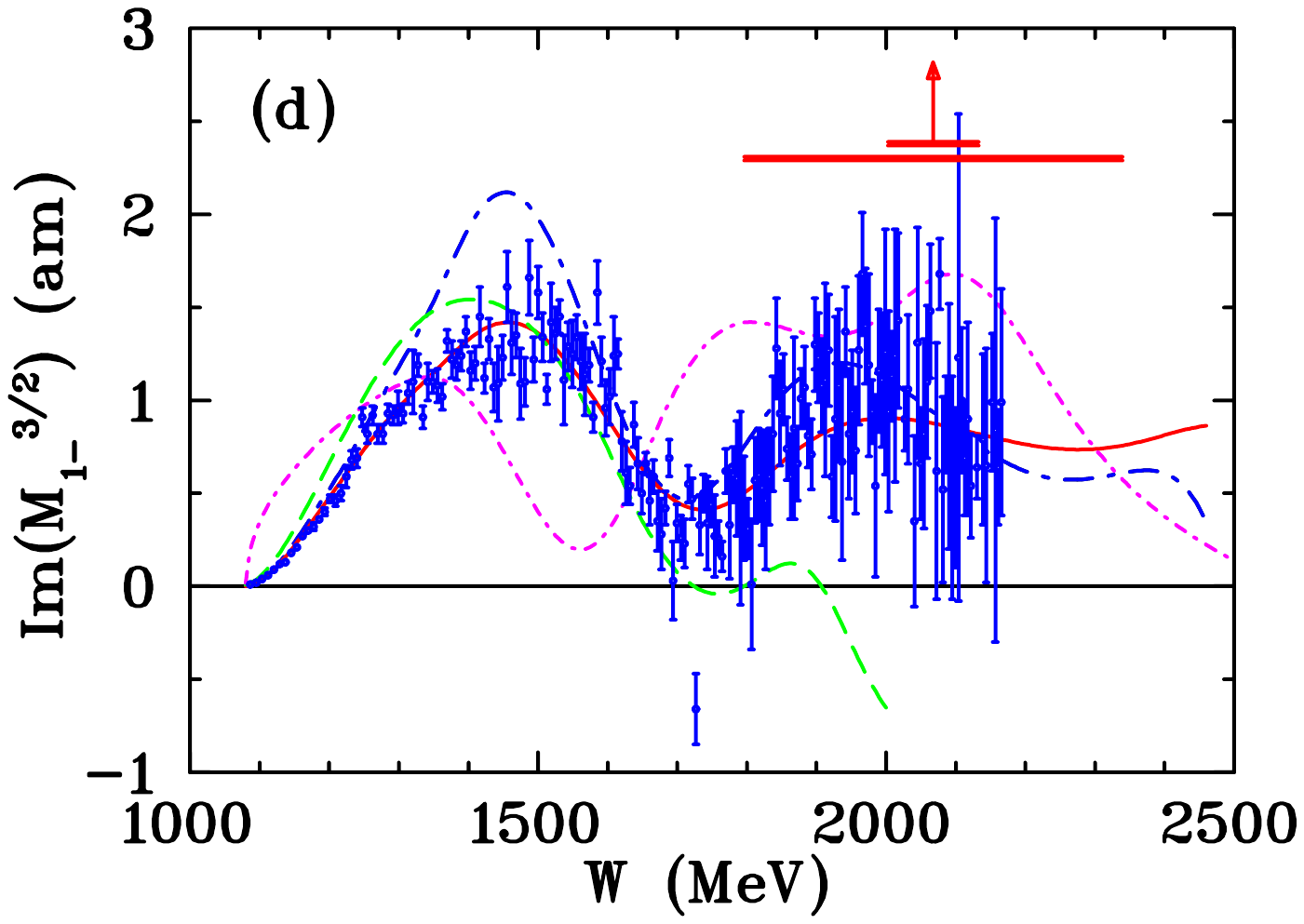}
}
\centering
{
    \includegraphics[width=0.32\textwidth,angle=90,keepaspectratio]{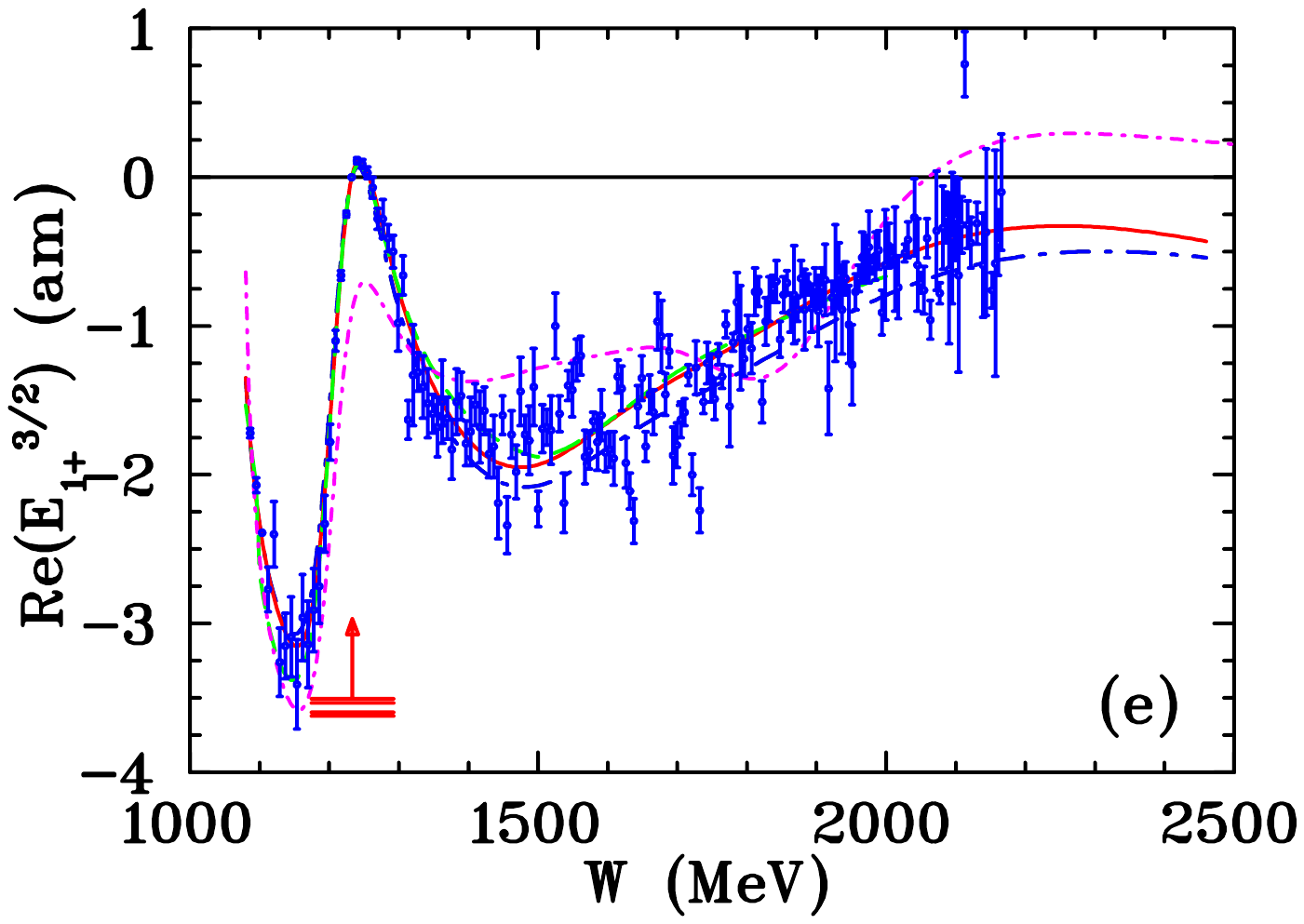}
    \includegraphics[width=0.32\textwidth,angle=90,keepaspectratio]{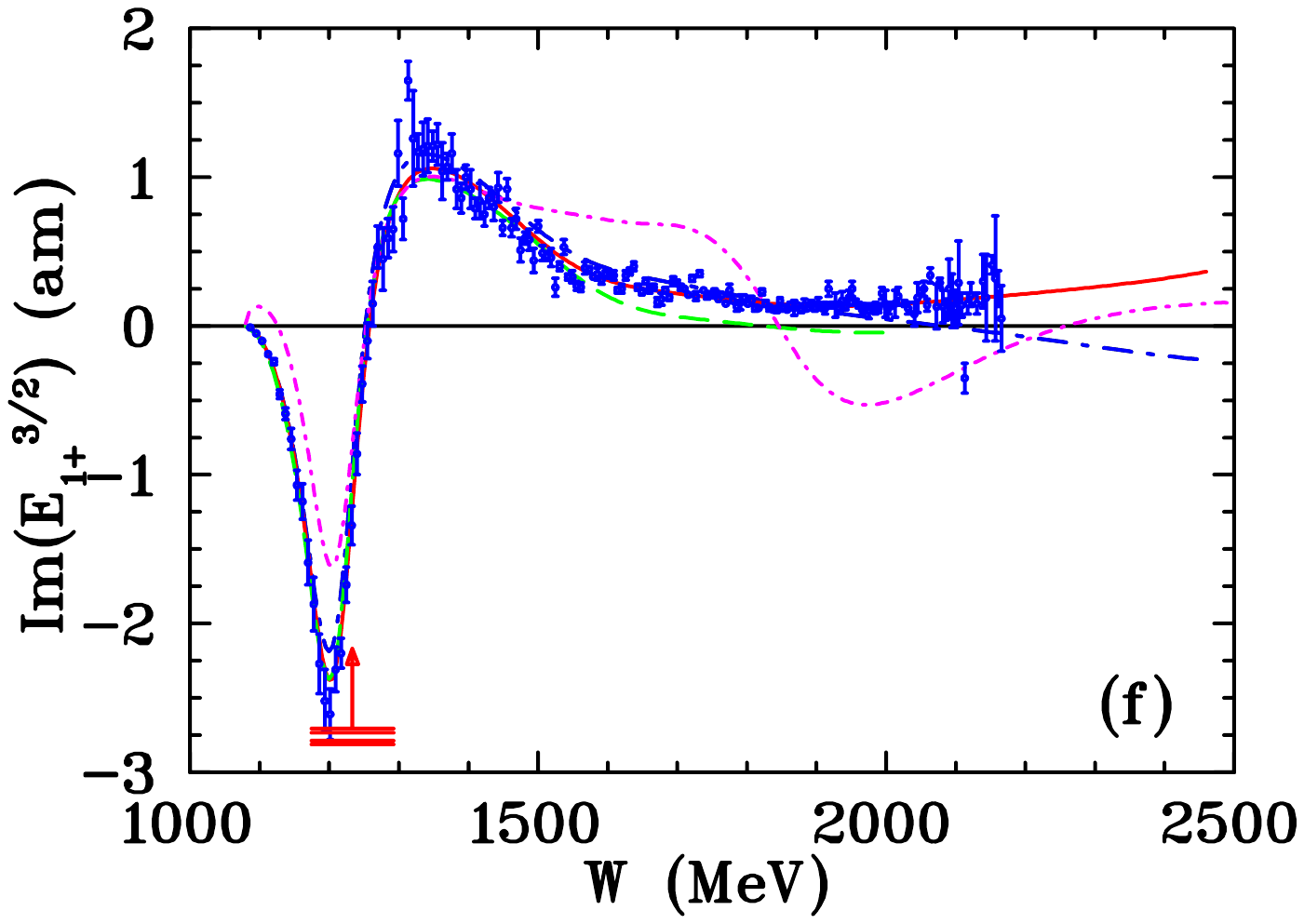}
}
\centering
{
    \includegraphics[width=0.32\textwidth,angle=90,keepaspectratio]{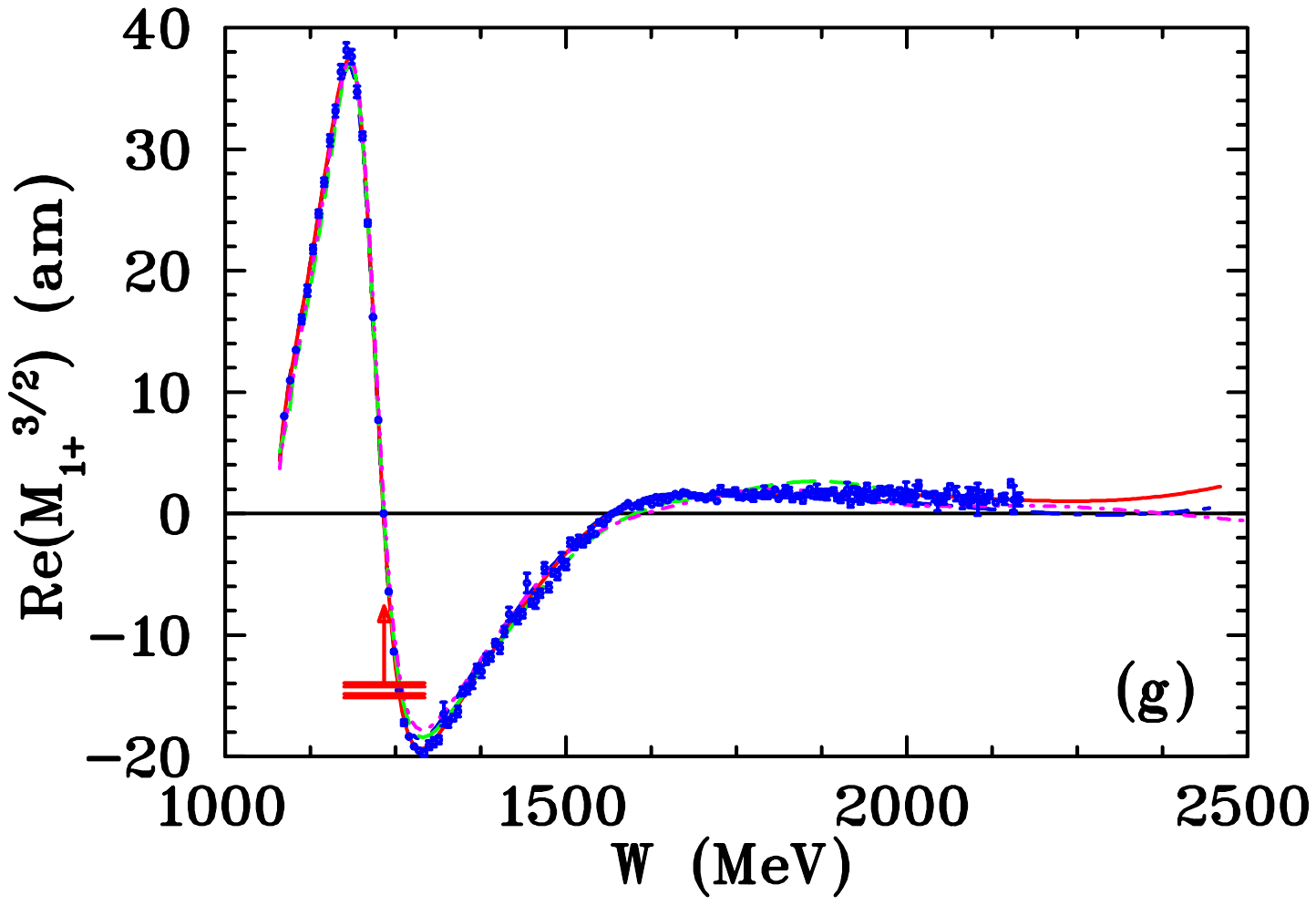}
    \includegraphics[width=0.32\textwidth,angle=90,keepaspectratio]{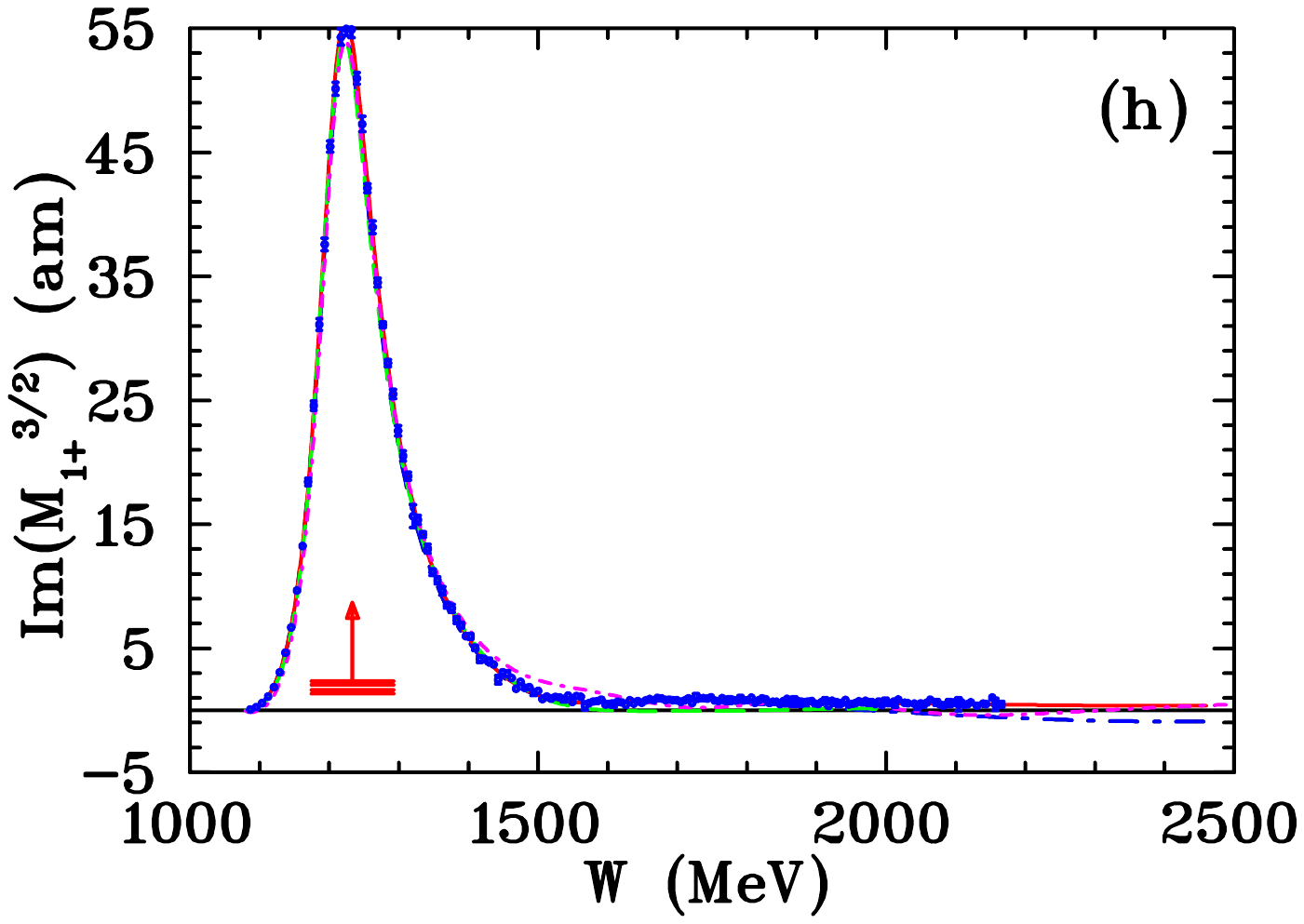}
}

\caption{Comparison of $I = 3/2$ multipole amplitudes (orbital momentum $l = 0, 1$) from threshold to $W = 2.5~\mathrm{GeV}$ ($E_\gamma = 2.7~\mathrm{GeV}$). Notation for solutions is given in the caption of Table~\ref{tab:tbl2}. For the amplitudes, the subscript $n$ denotes a neutron target, New SAID SM22 fit is shown by red solid curves. Previous SAID CM12~\cite{Workman:2012jf} (MAID2007~\cite{Drechsel:2007if}, terminates at $W = 2~\mathrm{GeV}$) predictions show by blue dash-dotted (green dashed) curves. BG2019~\cite{CBELSATAPS:2014wvh} predictions show by magenta short dash-dotted curves.  SE associated with SM22 shown as blue open circles. Vertical arrows indicate resonance energies, $W_R$, and horizontal bars show full ($\Gamma$) and partial ($\Gamma_{\pi N}$) widths associated with the SAID $\pi N$ solution SP06 (Breit-Wigner parameters)~\cite{Arndt:2006bf}.
}  
\label{fig:amp1}
\end{figure*}

\begin{figure*}[hbt!]
\vspace{0.4cm}
\centering
{
    \includegraphics[width=0.32\textwidth,angle=90,keepaspectratio]{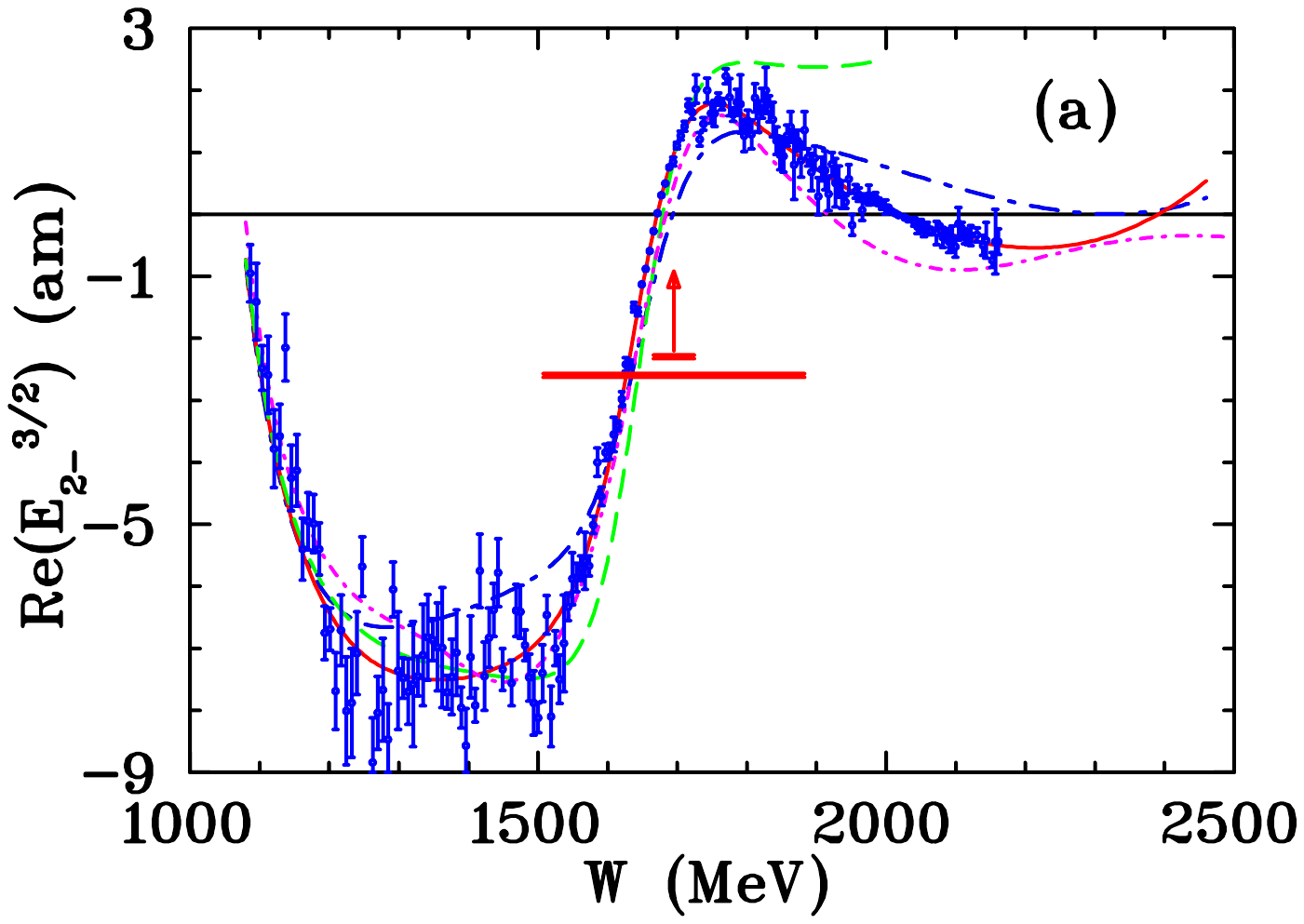}
    \includegraphics[width=0.32\textwidth,angle=90,keepaspectratio]{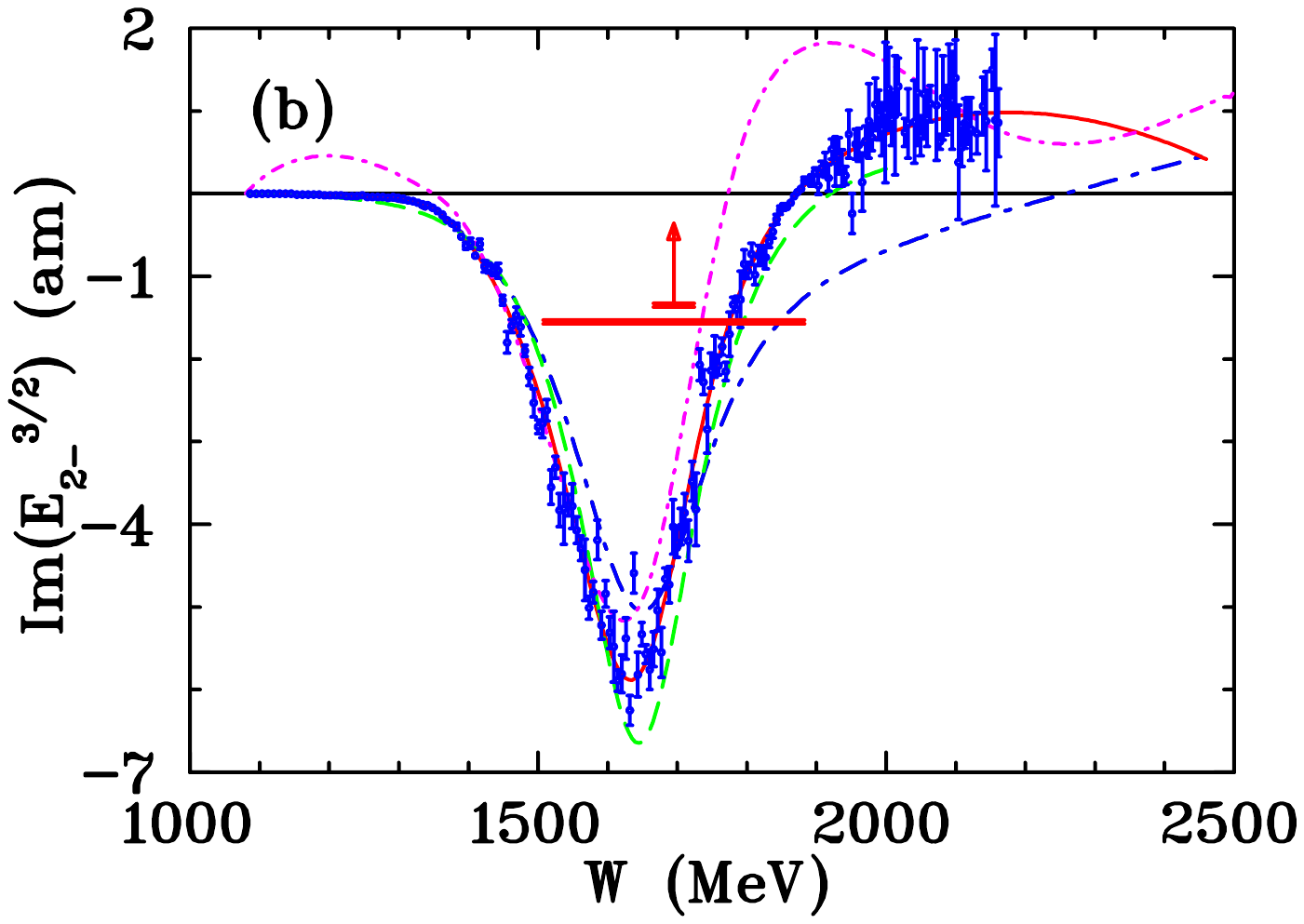}
}
\centering
{
    \includegraphics[width=0.32\textwidth,angle=90,keepaspectratio]{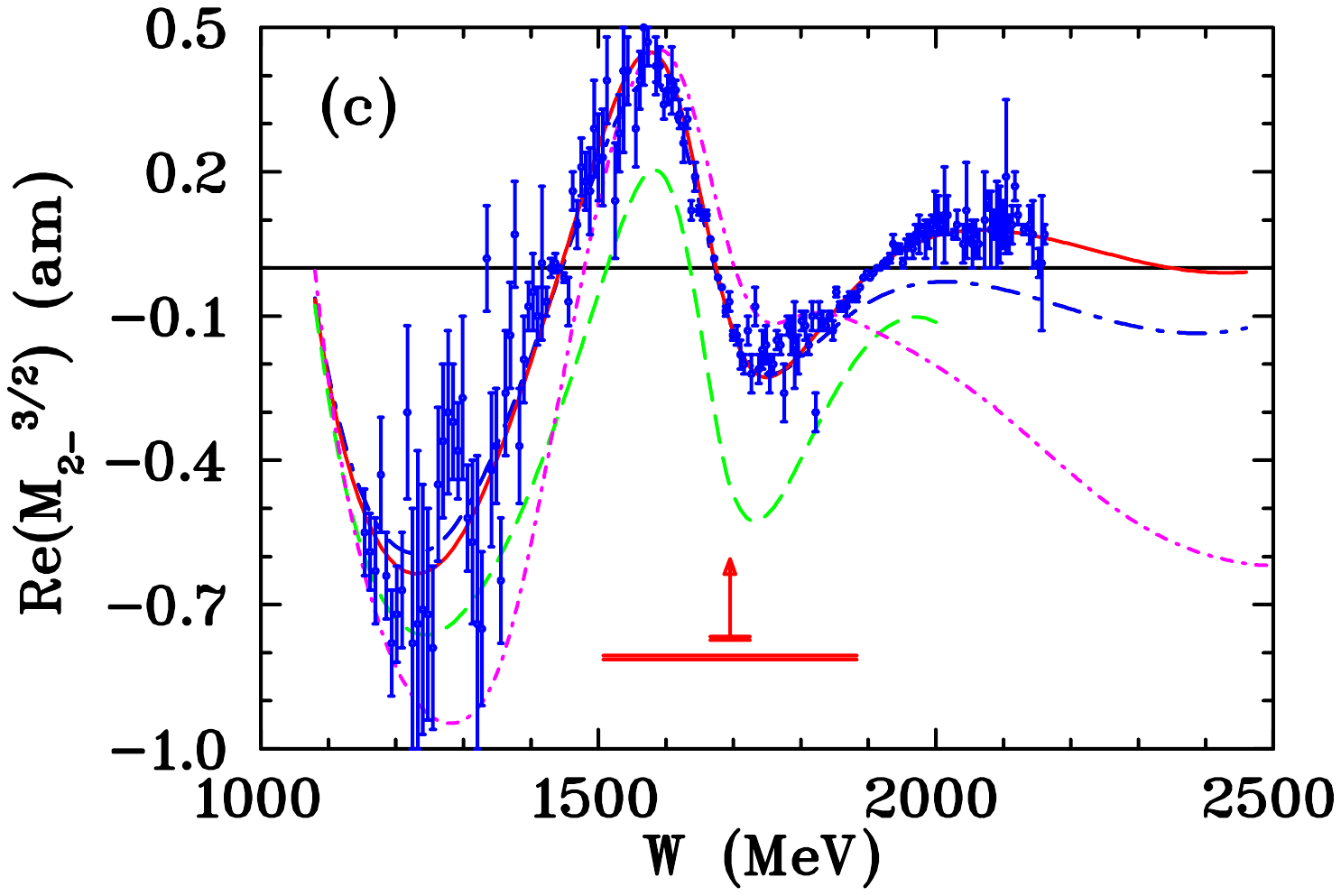}
    \includegraphics[width=0.32\textwidth,angle=90,keepaspectratio]{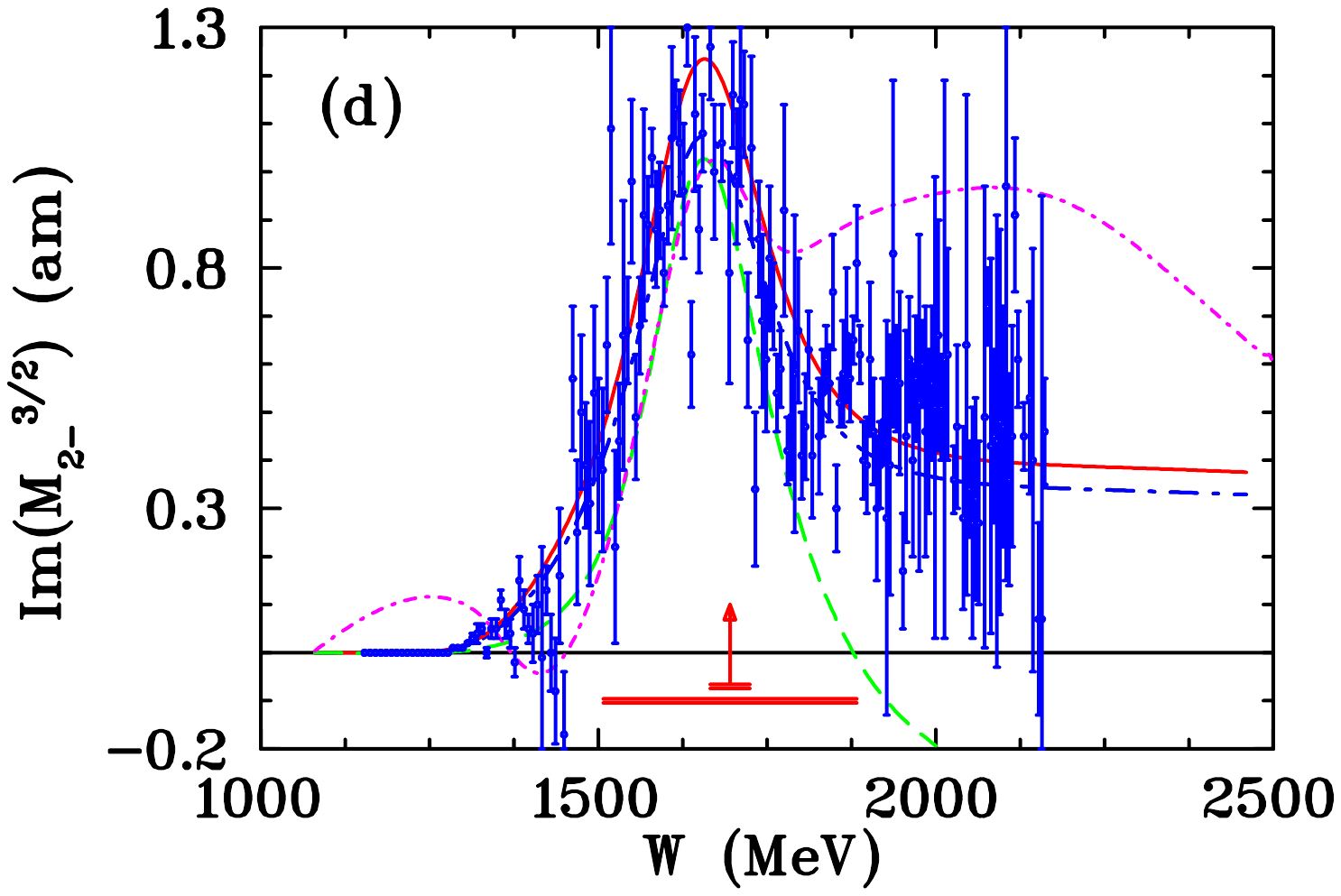}
}
\centering
{
    \includegraphics[width=0.32\textwidth,angle=90,keepaspectratio]{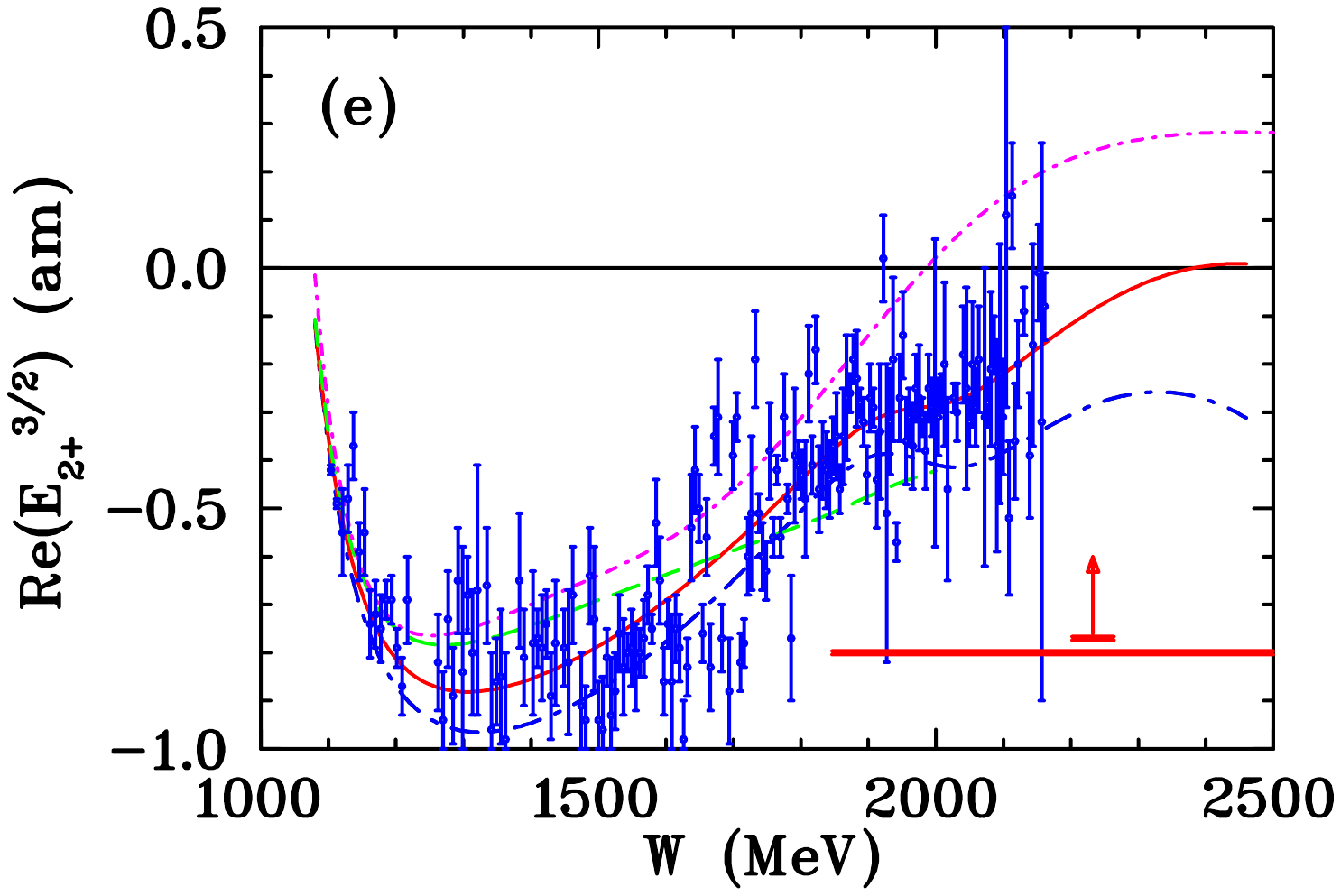}
    \includegraphics[width=0.32\textwidth,angle=90,keepaspectratio]{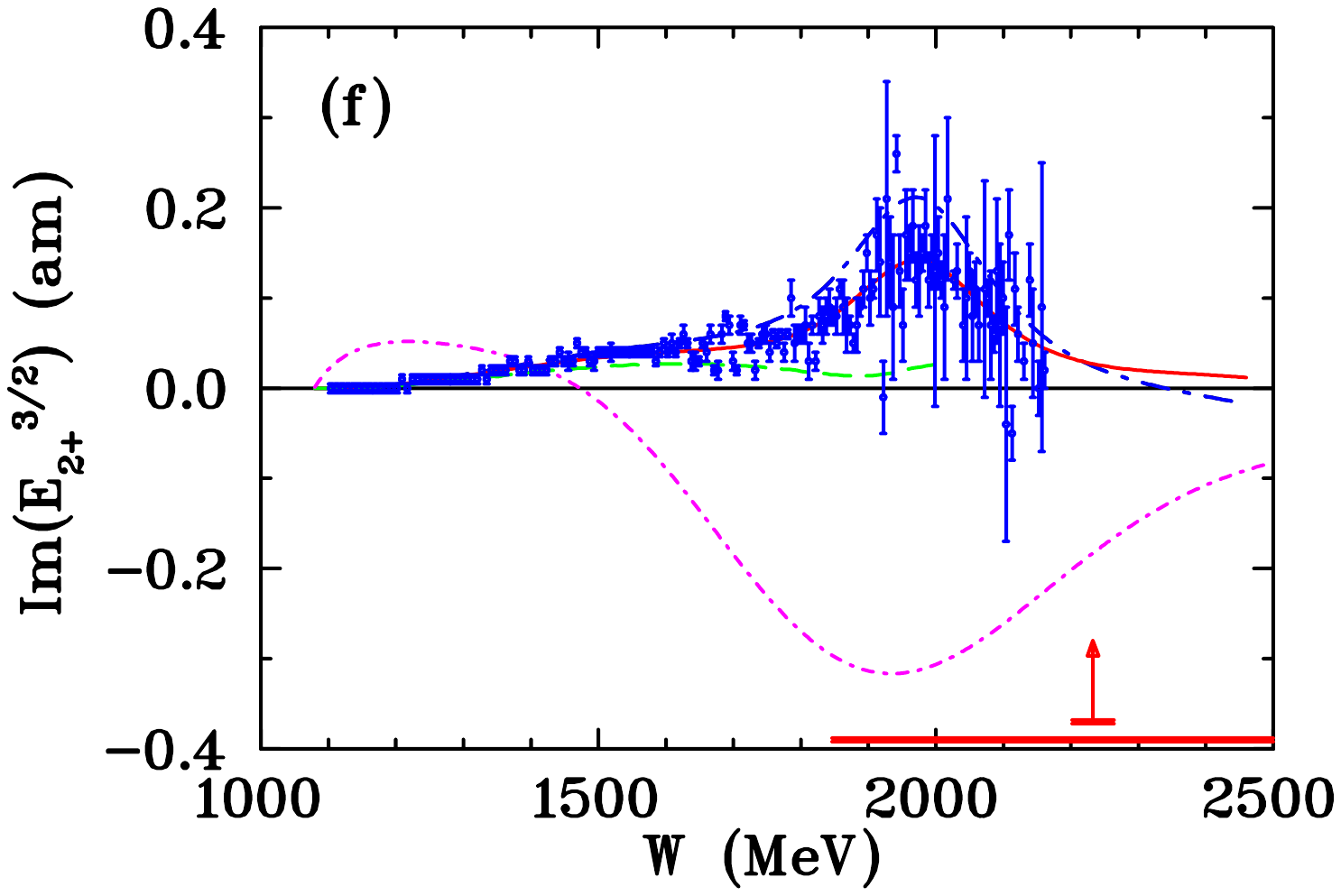}
}
%\centering
%{
%    \includegraphics[width=0.38\textwidth,angle=0,keepaspectratio]{./v4ta.png}
%    \includegraphics[width=0.38\textwidth,angle=0,keepaspectratio]{./v4tb.png}
%}

\caption{Comparison $I = 3/2$ multipole amplitudes (orbital momentum $l = 2$) from threshold to $W = 2.5~\mathrm{GeV}$. Notation of the solutions and data is the same as in Fig.~\ref{fig:amp1}.
}  
\label{fig:amp2}
\end{figure*}

\begin{figure*}[hbt!]
%\vspace{0.4cm}
\centering
{
    \includegraphics[width=0.32\textwidth,angle=90,keepaspectratio]{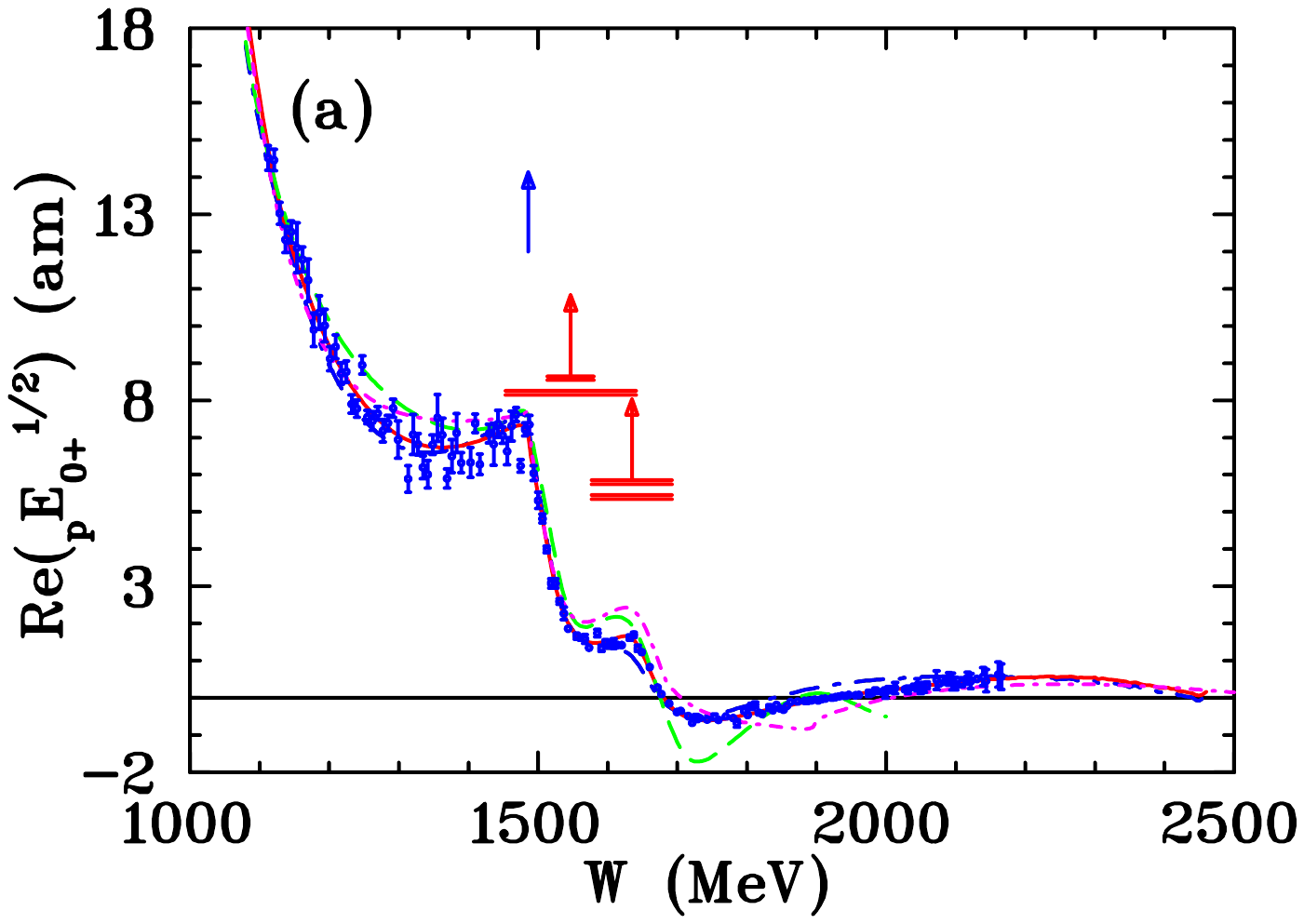}
    \includegraphics[width=0.32\textwidth,angle=90,keepaspectratio]{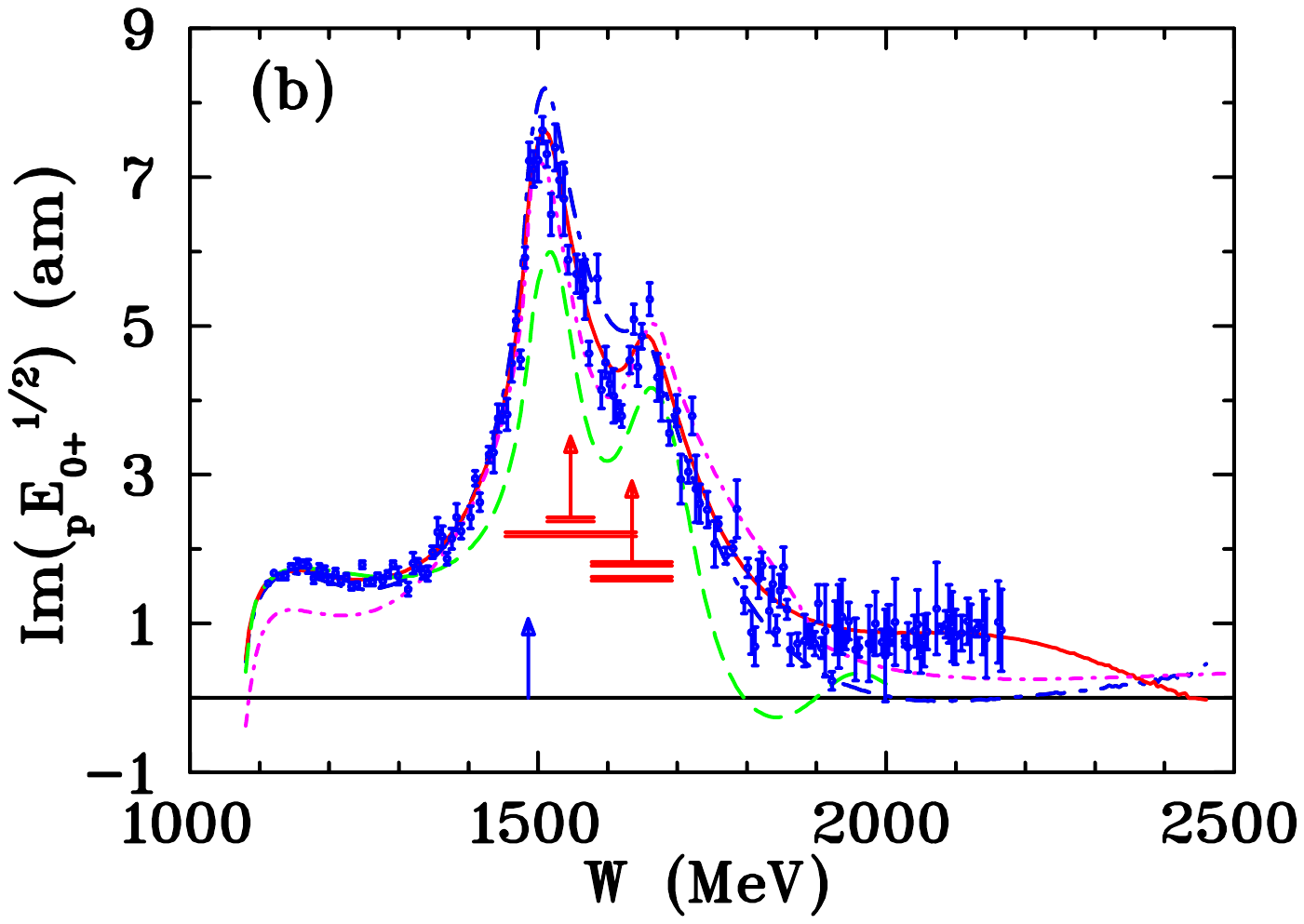}
}
\centering
{
    \includegraphics[width=0.32\textwidth,angle=90,keepaspectratio]{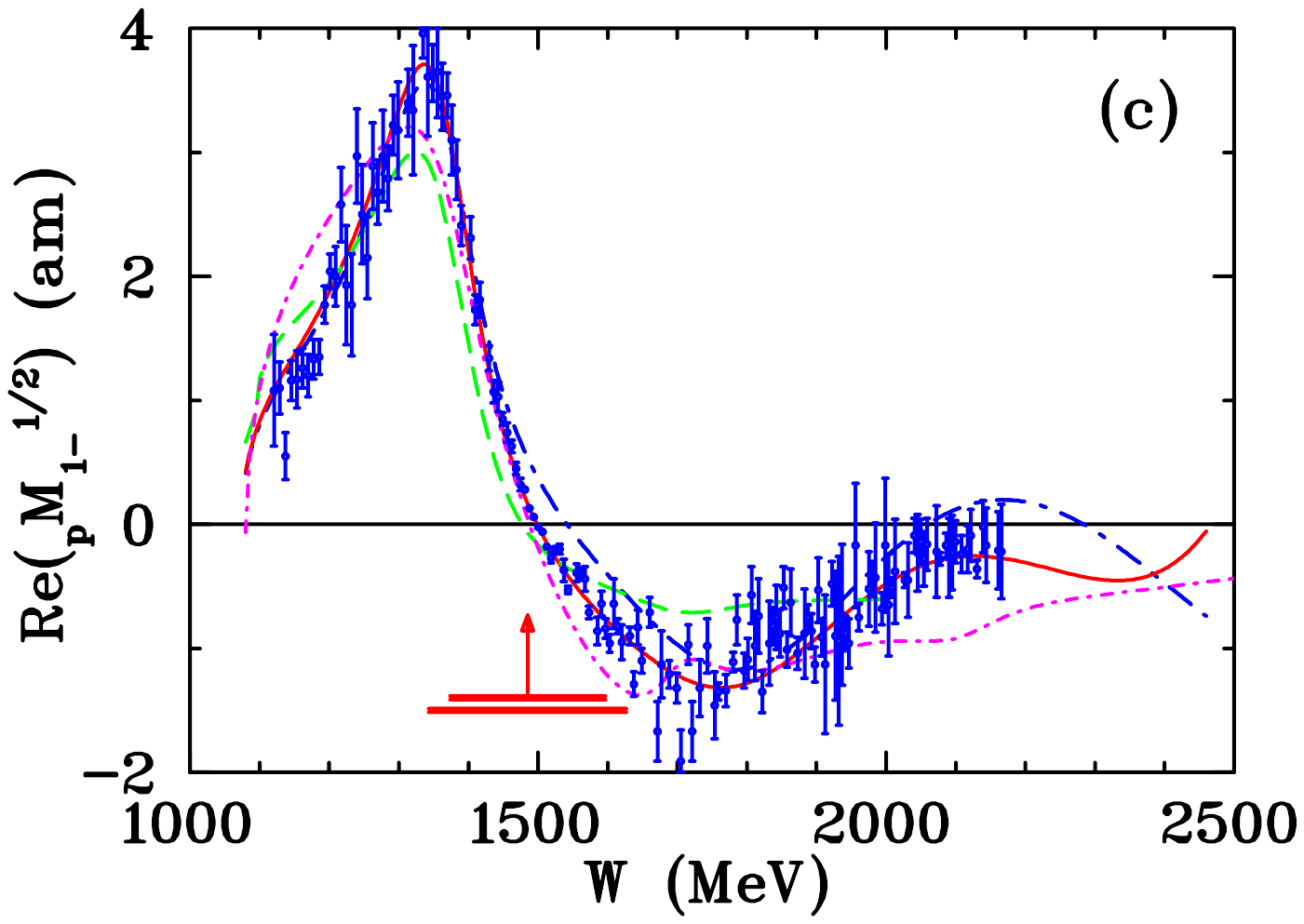}
    \includegraphics[width=0.32\textwidth,angle=90,keepaspectratio]{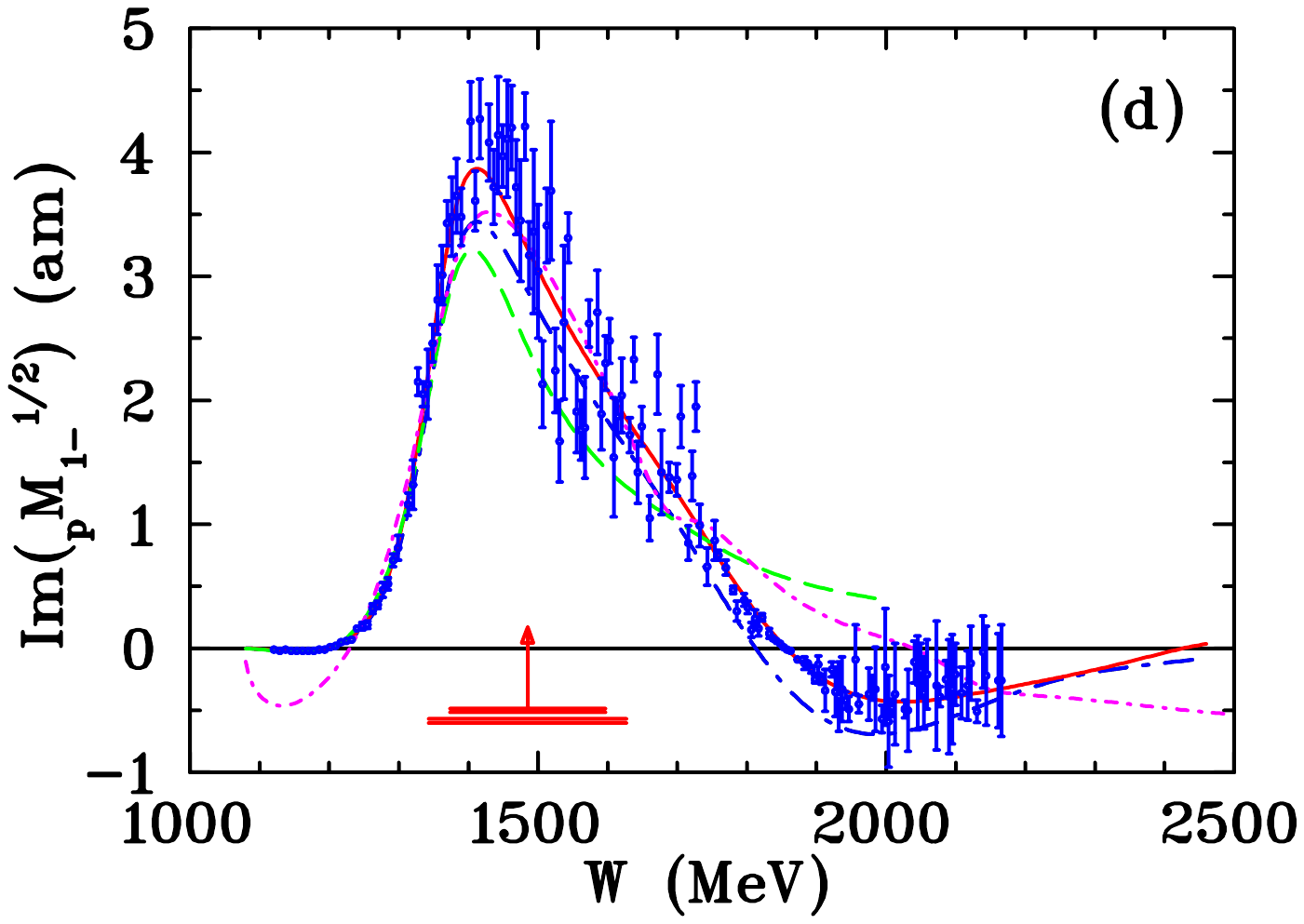}
}
\centering
{
    \includegraphics[width=0.32\textwidth,angle=90,keepaspectratio]{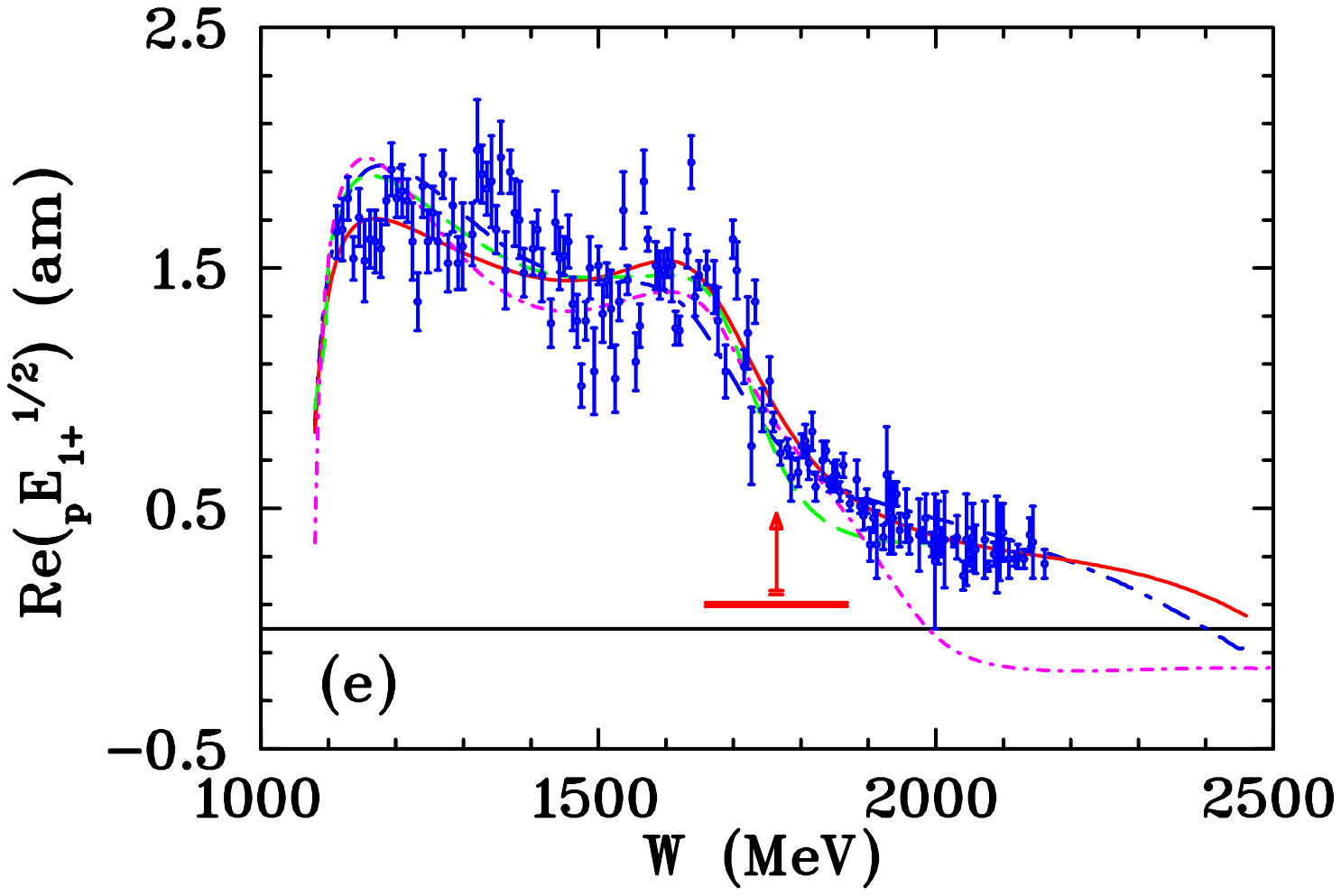}
    \includegraphics[width=0.32\textwidth,angle=90,keepaspectratio]{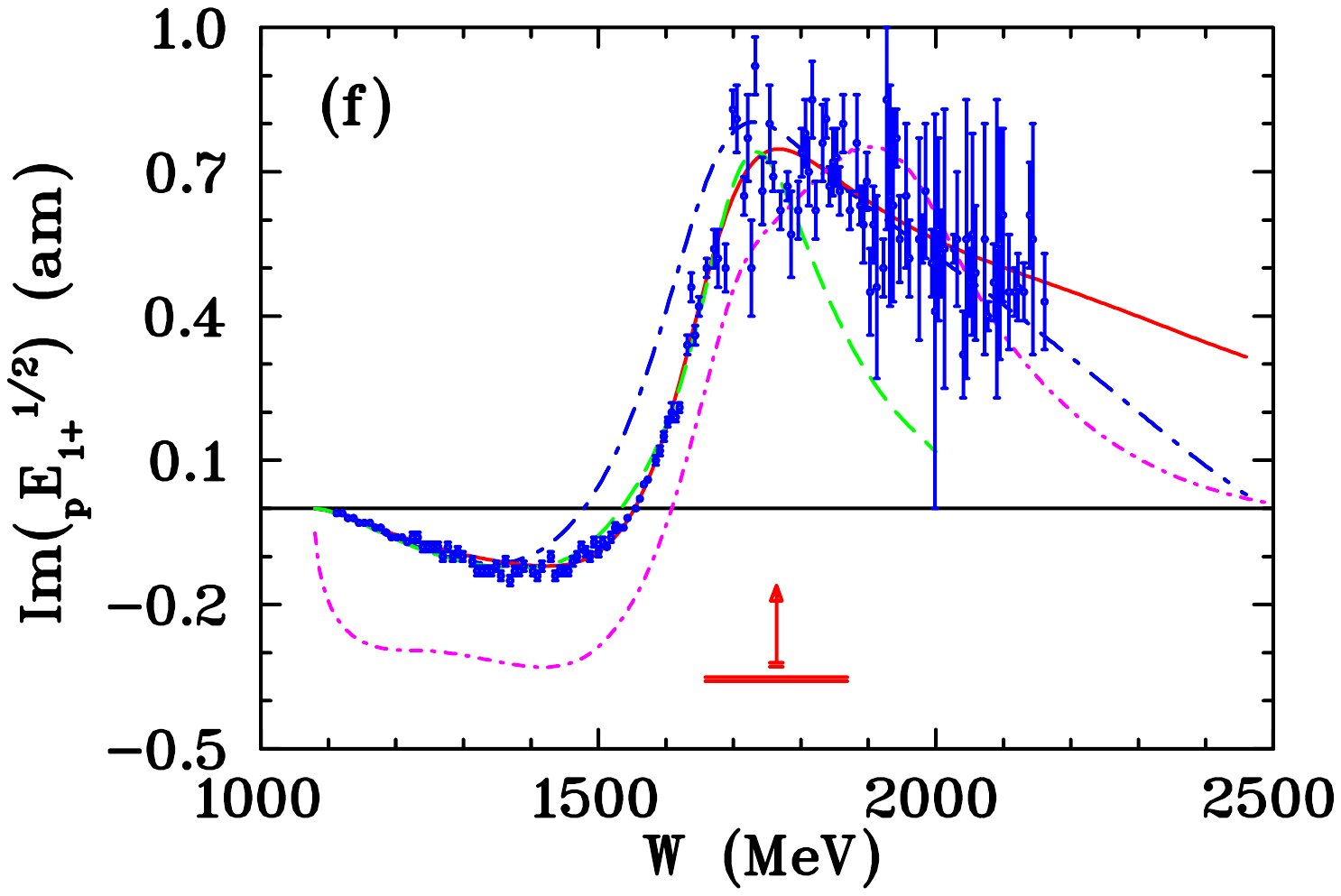}
}
\centering
{
    \includegraphics[width=0.32\textwidth,angle=90,keepaspectratio]{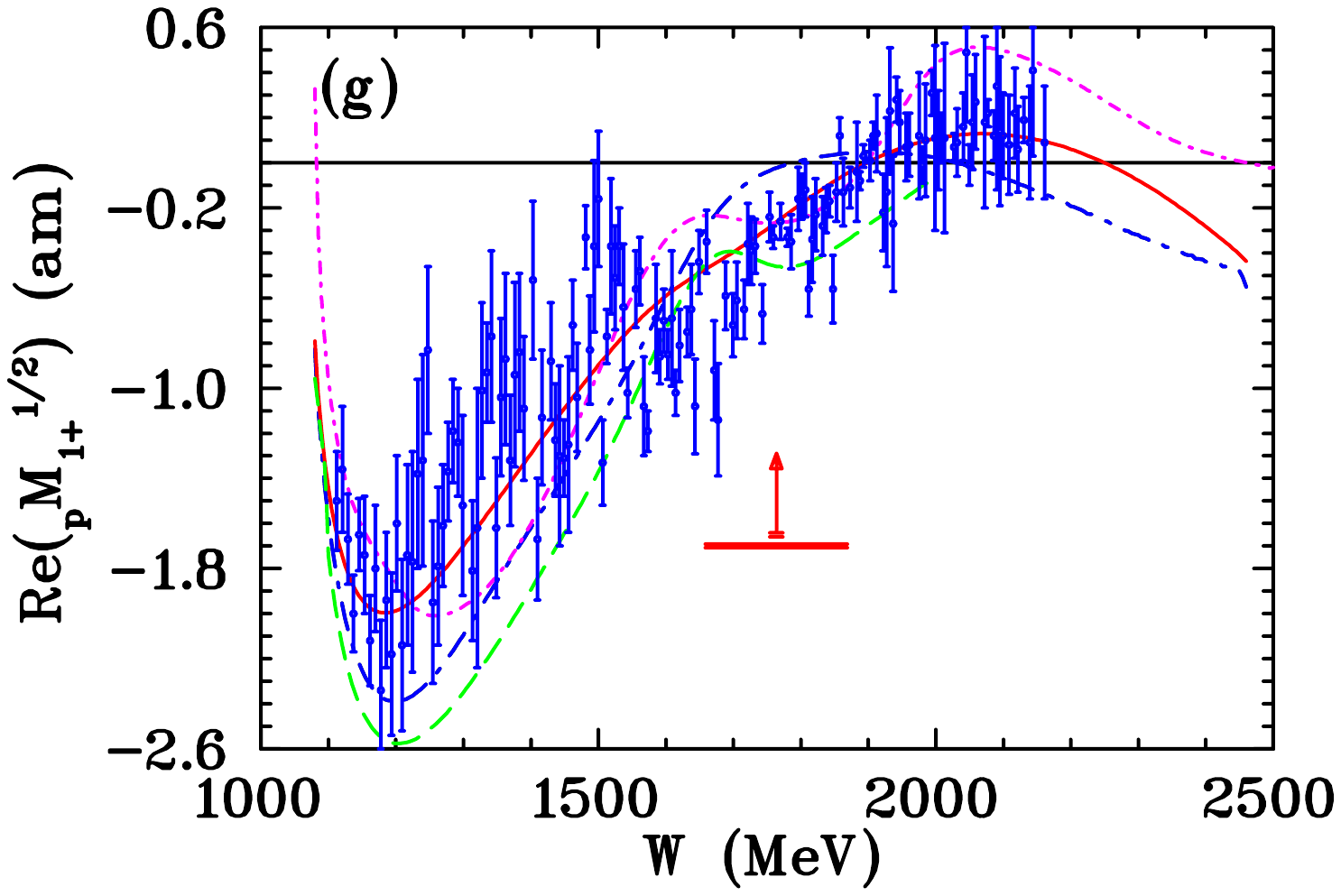}
    \includegraphics[width=0.32\textwidth,angle=90,keepaspectratio]{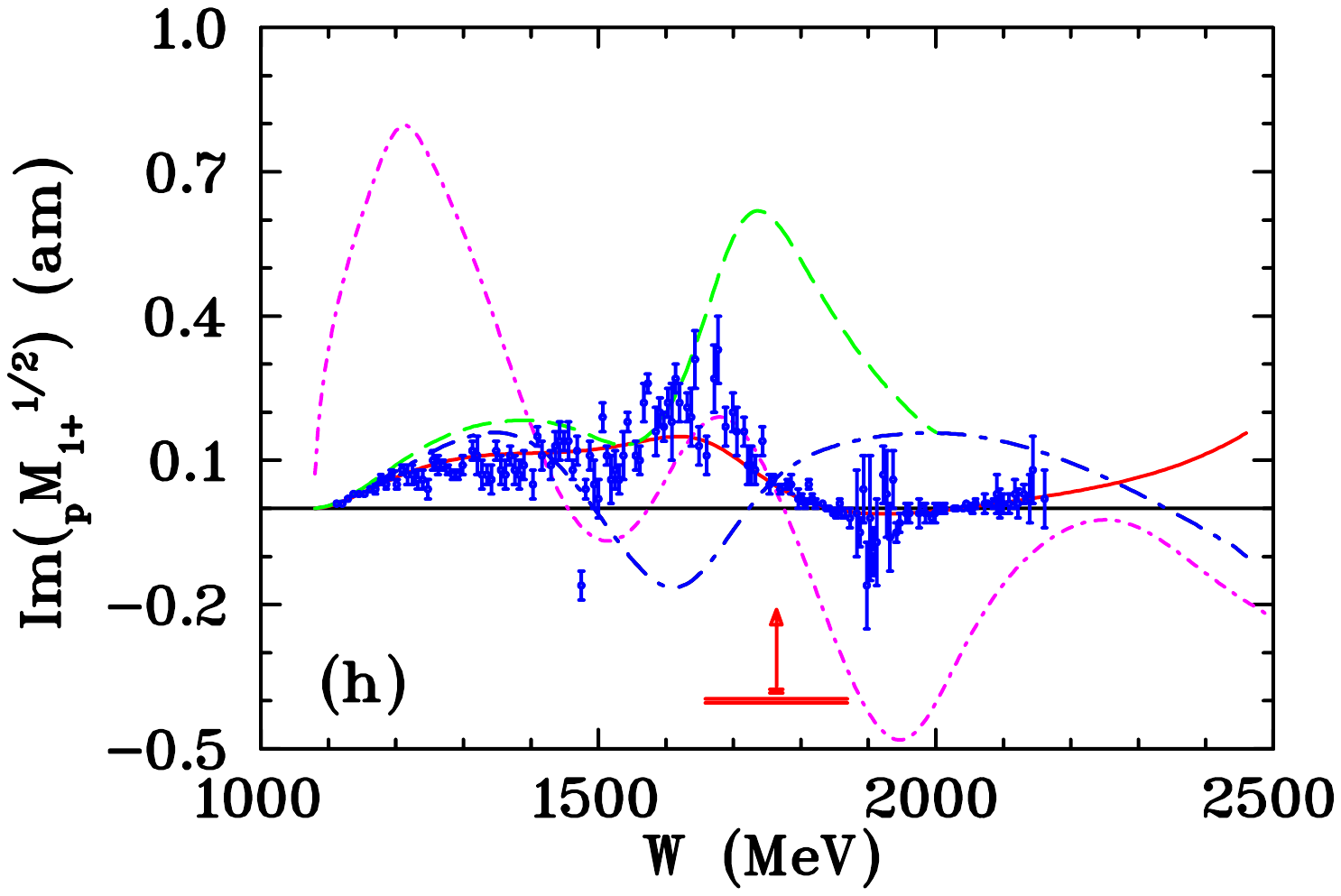}
}

\caption{Comparison of proton $I = 1/2$ multipole amplitudes (orbital momentum $l = 0, 1$) from threshold to $W = 2.5~\mathrm{GeV}$ ($E_\gamma = 2.7~\mathrm{GeV}$).  Notation of the solutions is the same as in Fig.~\ref{fig:amp1}. The blue vertical arrows for (a) and (b) indicate the $\eta$ production threshold.}  
\label{fig:amp4}
\end{figure*}

\begin{figure*}[hbt!]
%\vspace{0.4cm}
\centering
{
    \includegraphics[width=0.32\textwidth,angle=90,keepaspectratio]{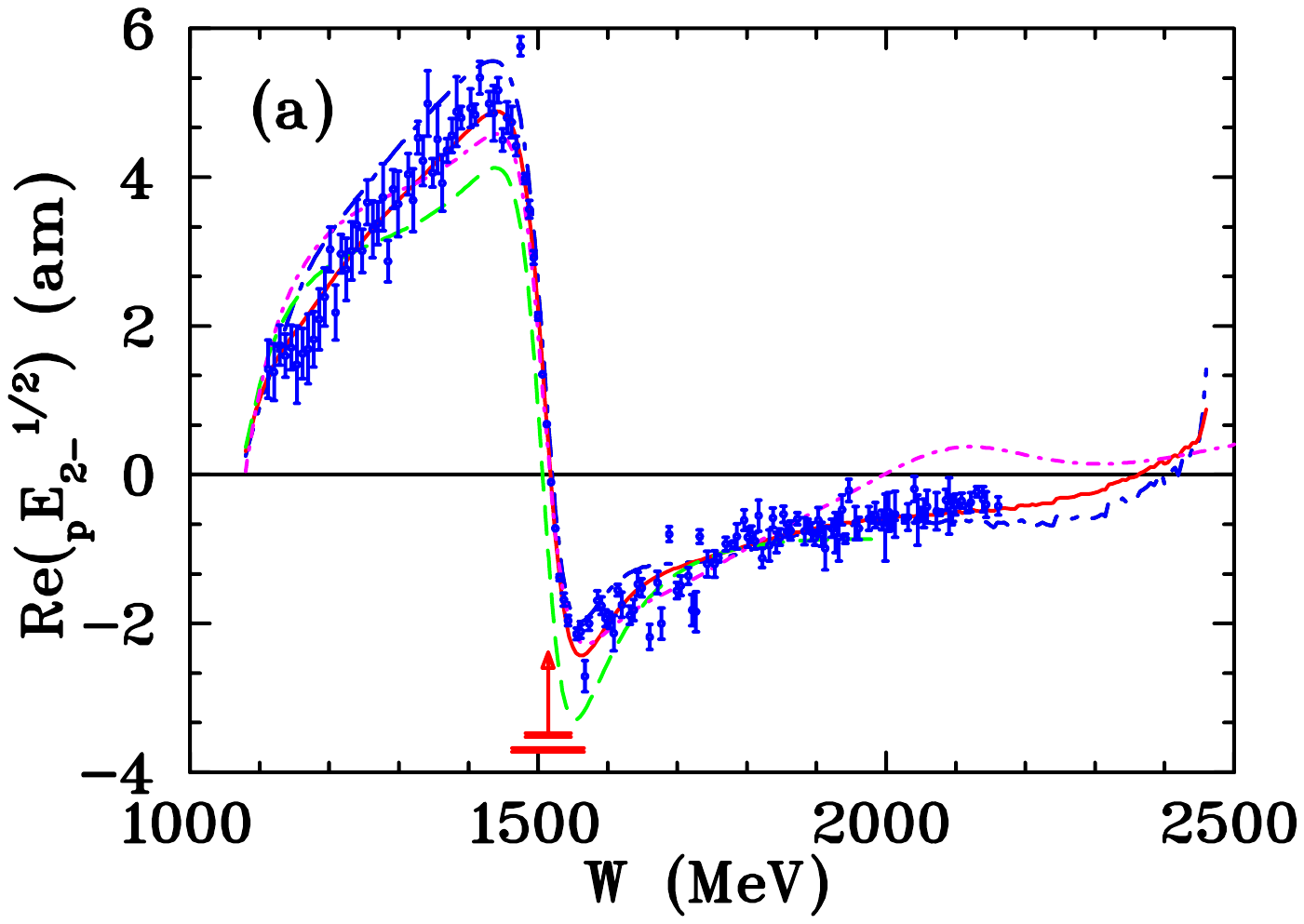}
    \includegraphics[width=0.32\textwidth,angle=90,keepaspectratio]{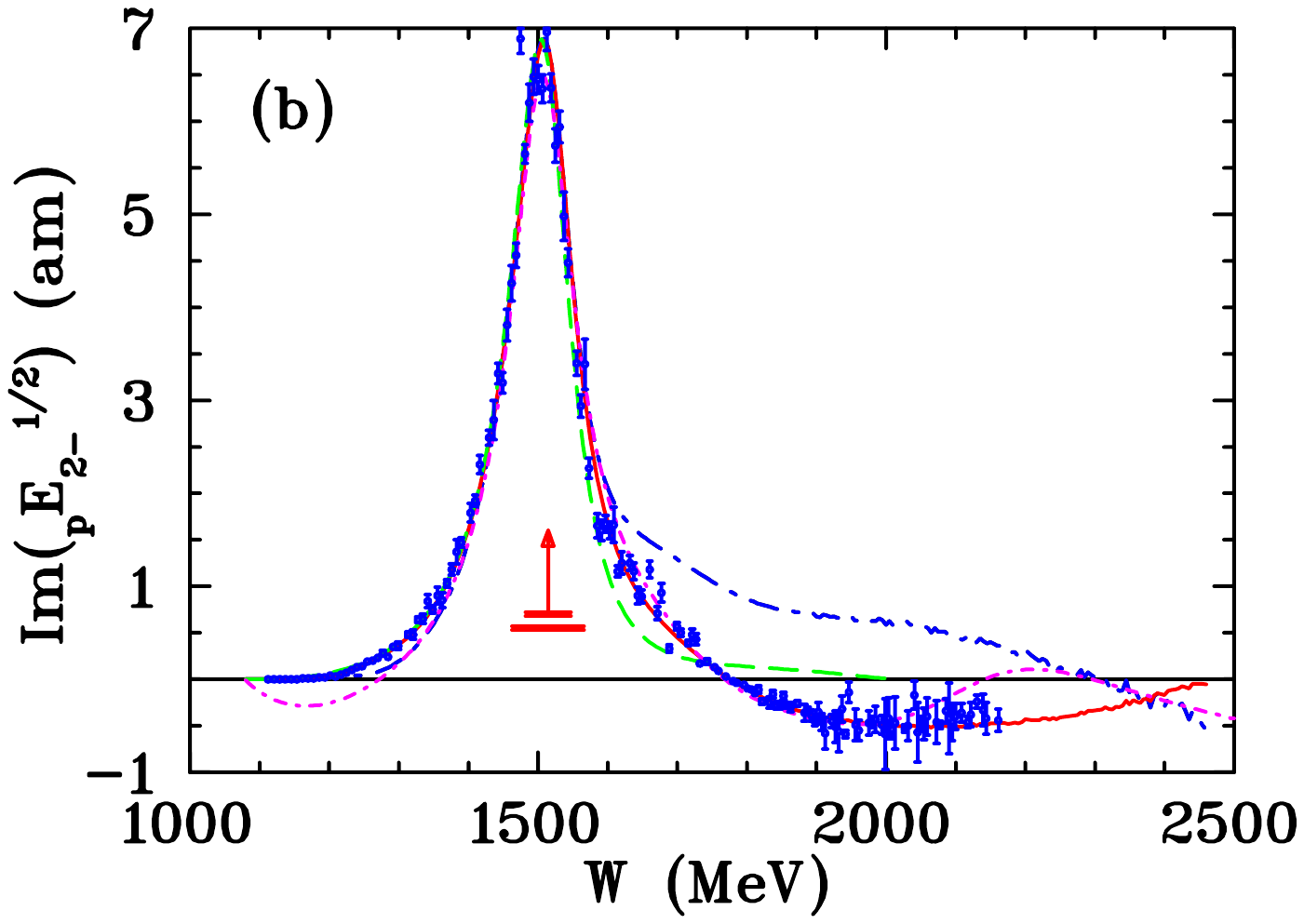}
}
\centering
{
    \includegraphics[width=0.32\textwidth,angle=90,keepaspectratio]{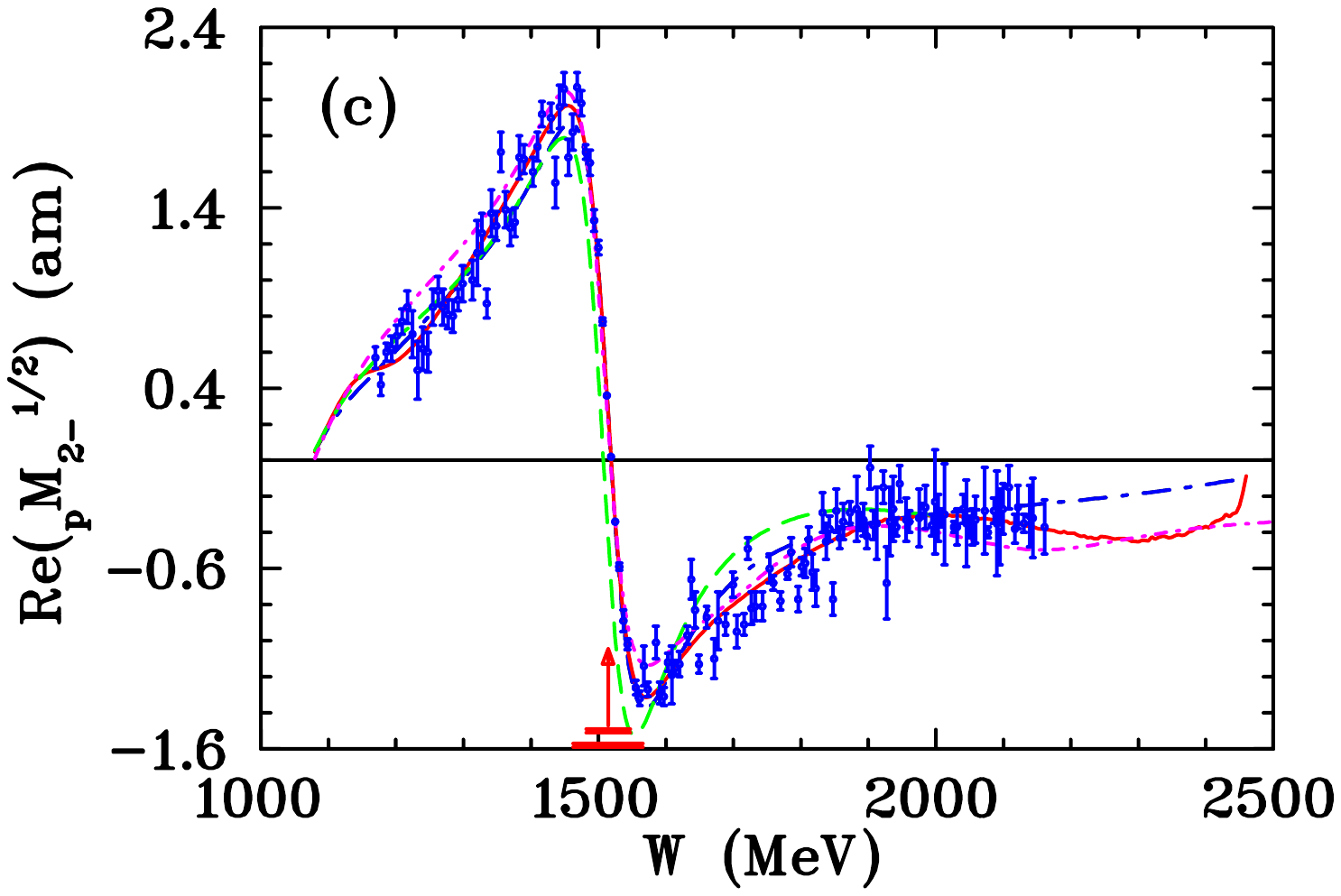}
    \includegraphics[width=0.32\textwidth,angle=90,keepaspectratio]{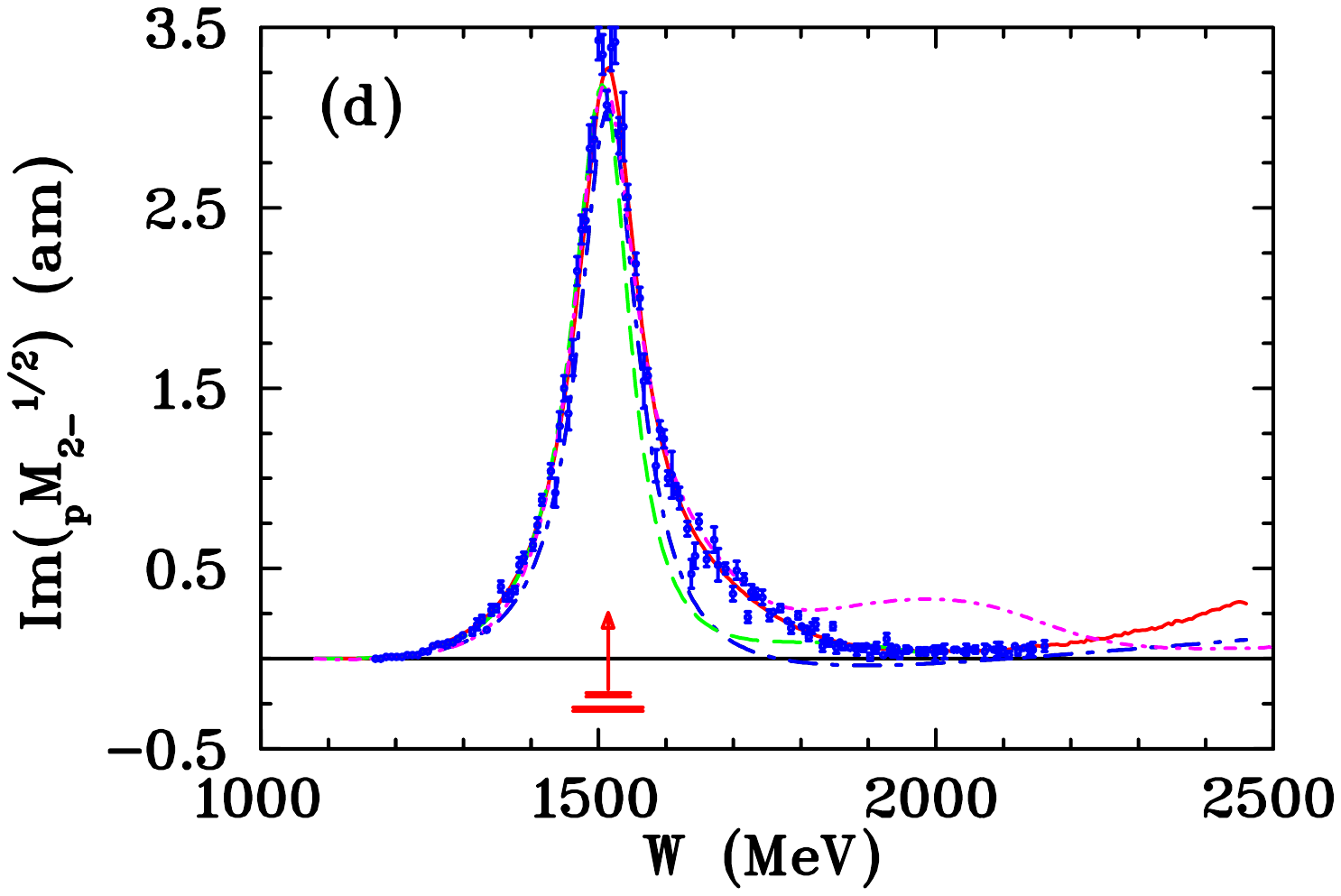}
}
\centering
{
    \includegraphics[width=0.32\textwidth,angle=90,keepaspectratio]{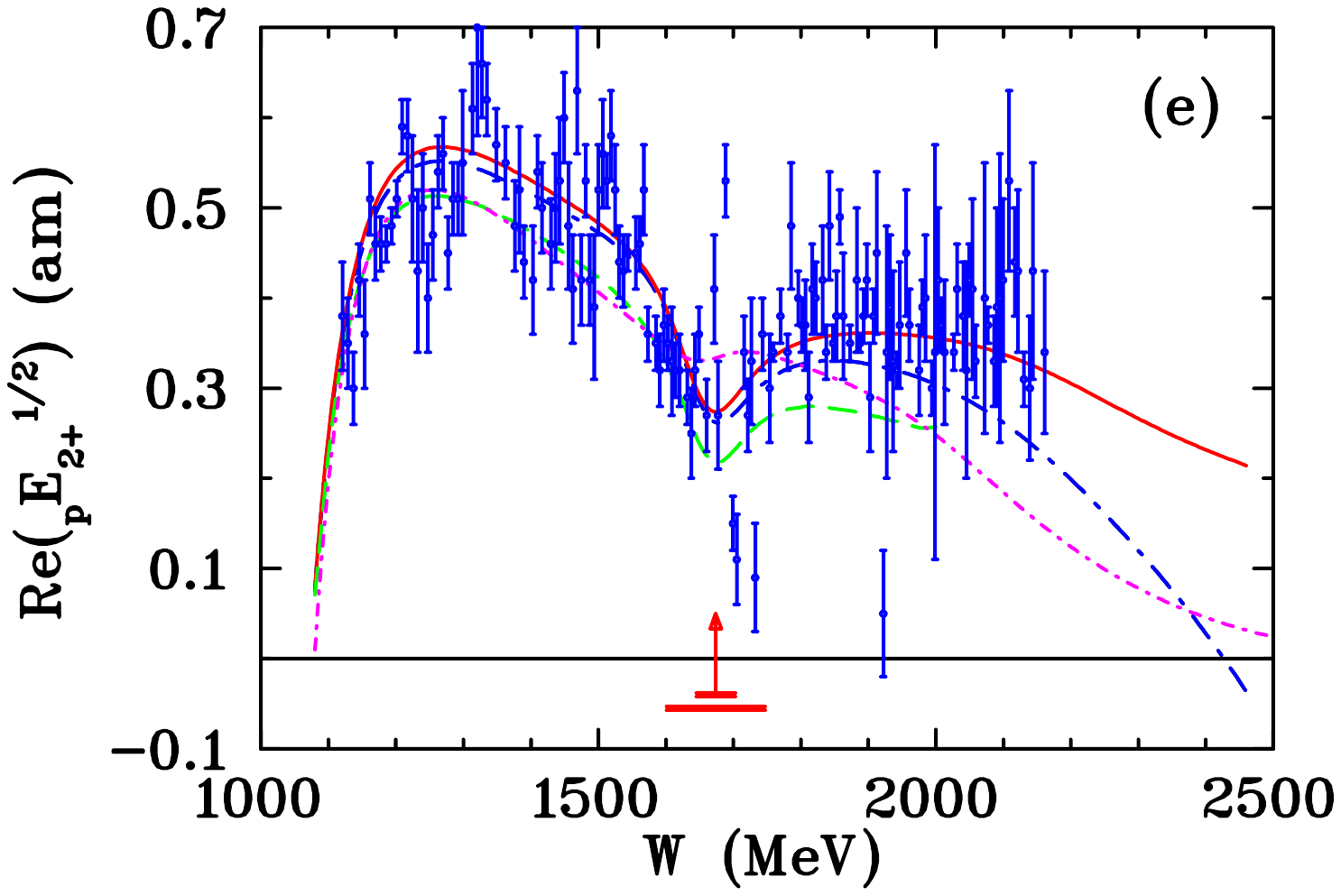}
    \includegraphics[width=0.32\textwidth,angle=90,keepaspectratio]{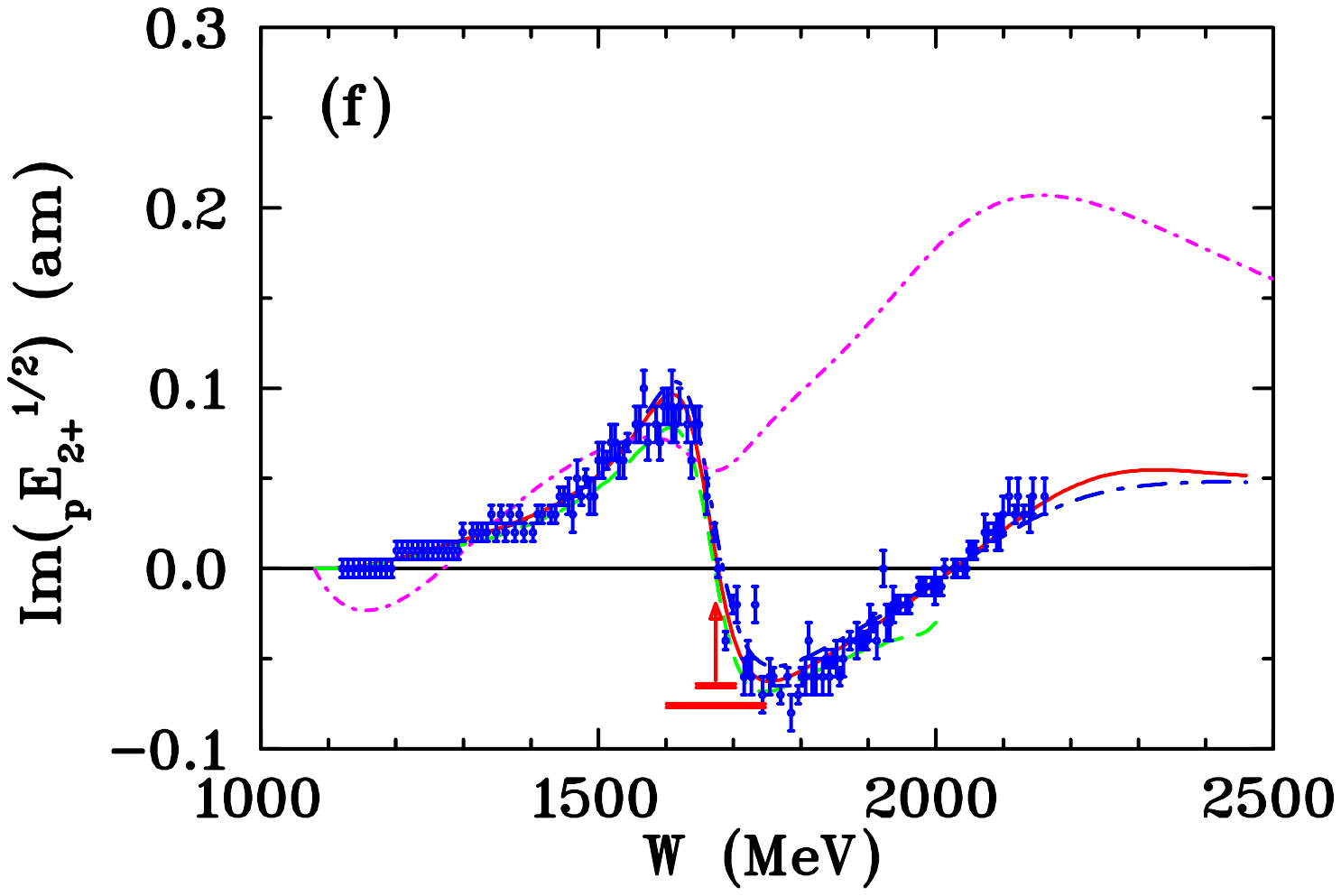}
}
\centering
{
    \includegraphics[width=0.32\textwidth,angle=90,keepaspectratio]{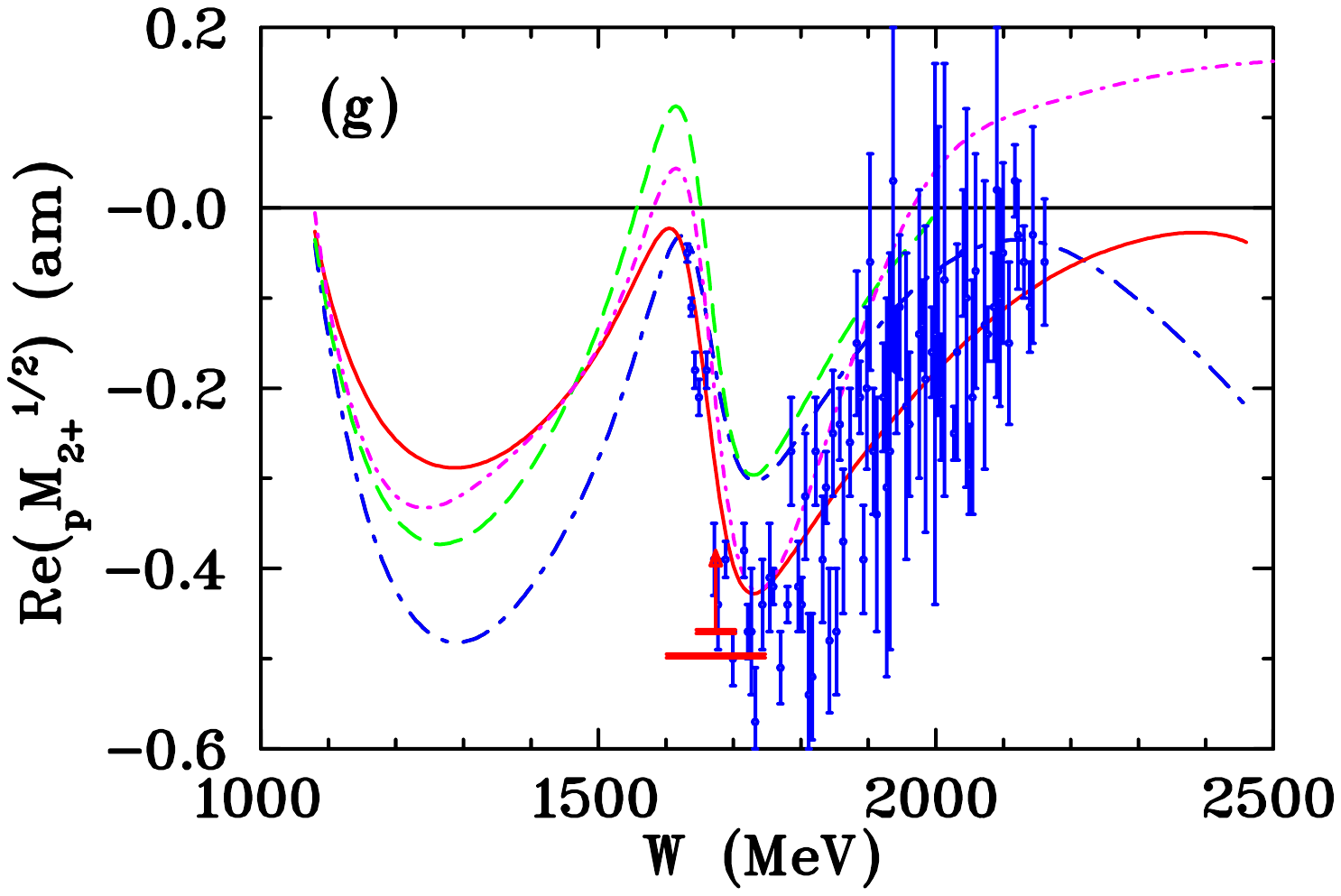}
    \includegraphics[width=0.32\textwidth,angle=90,keepaspectratio]{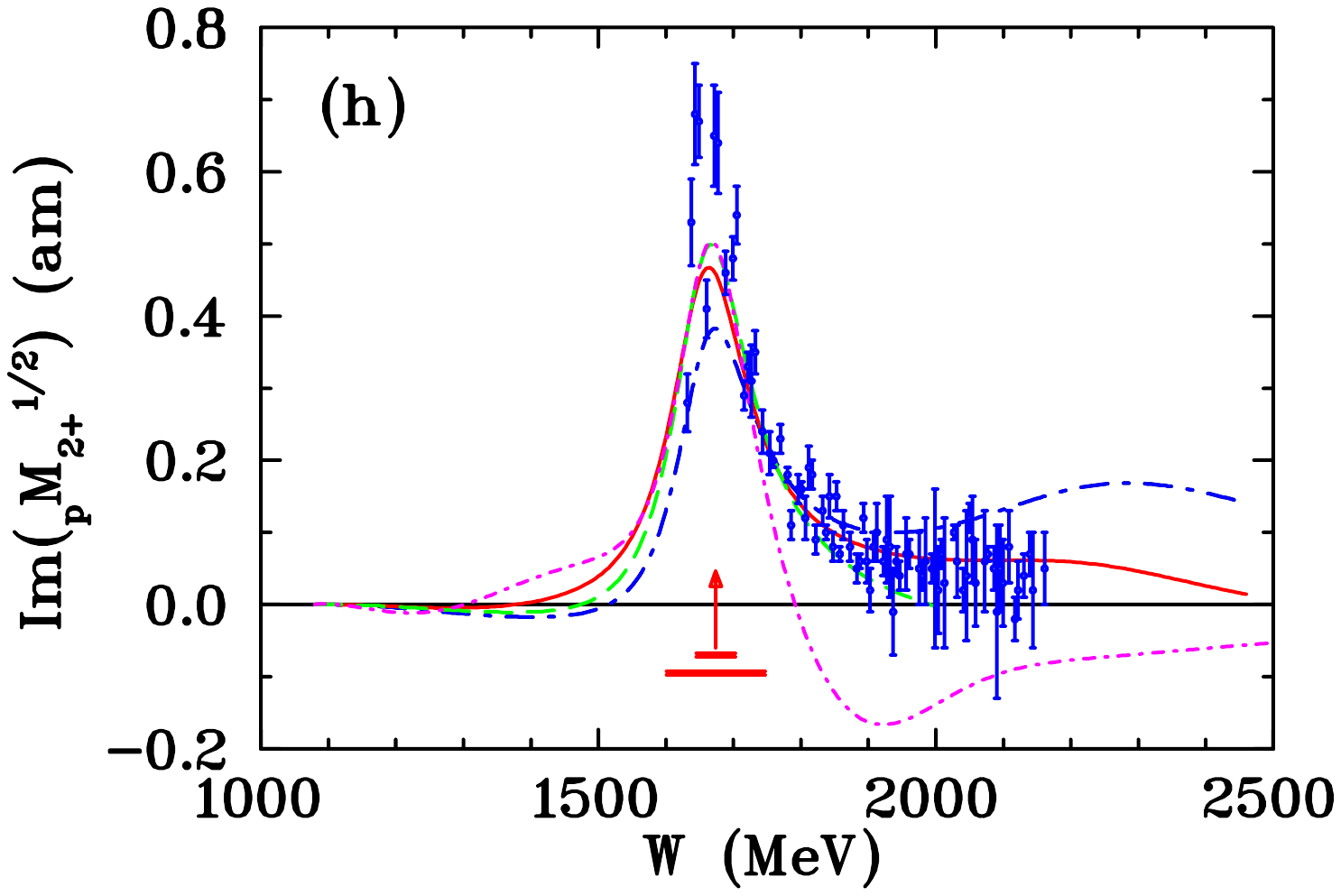}
}

\caption{Comparison proton $I = 1/2$ multipole amplitudes (orbital momentum $l = 2$) from threshold to $W = 2.5~\mathrm{GeV}$. Notation of the solutions is the same as in Fig.~\ref{fig:amp1}.
}  
\label{fig:amp5}
\end{figure*}

\begin{figure*}[hbt!]
%\vspace{0.4cm}
\centering
{
    \includegraphics[width=0.32\textwidth,angle=90,keepaspectratio]{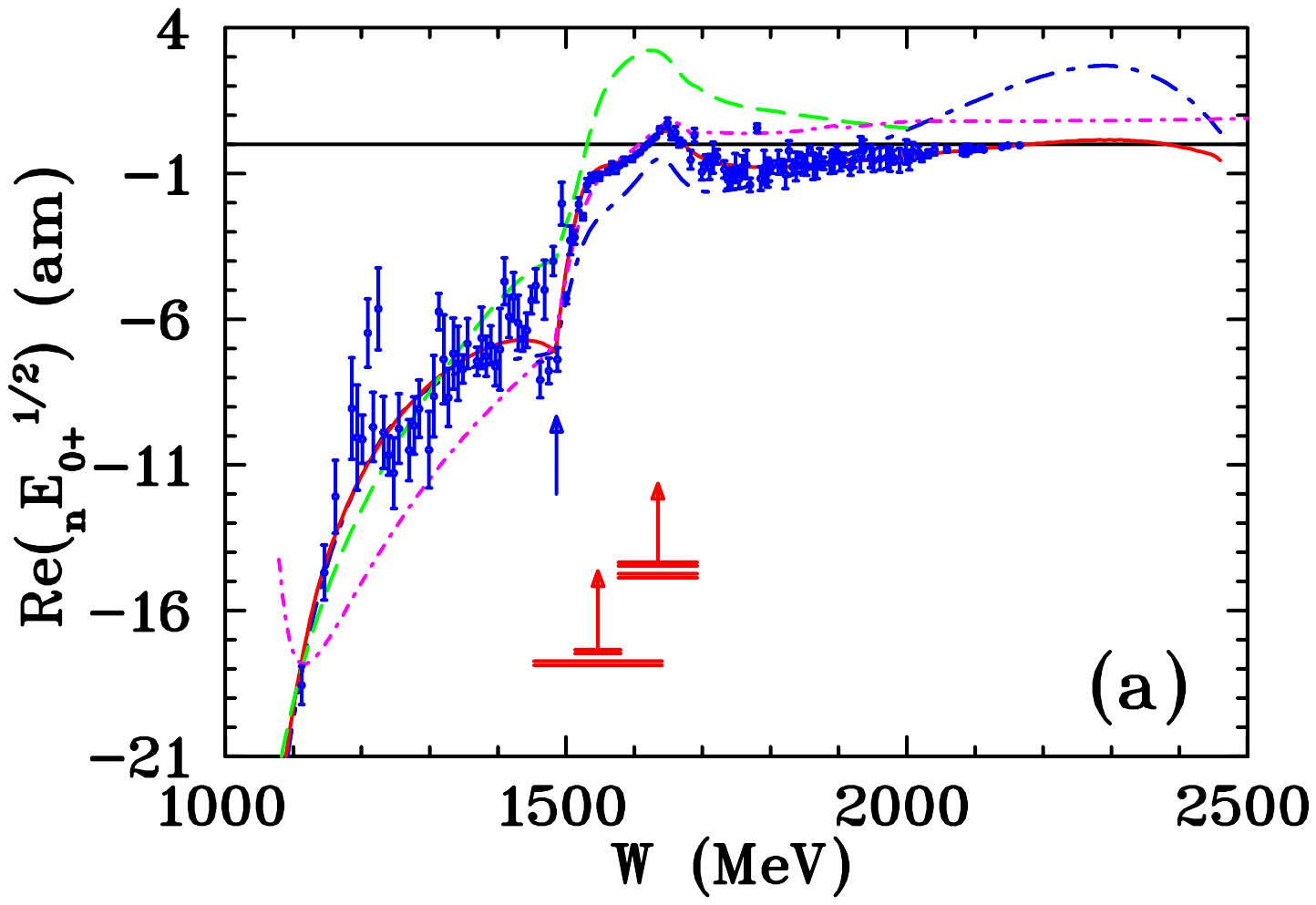}
    \includegraphics[width=0.32\textwidth,angle=90,keepaspectratio]{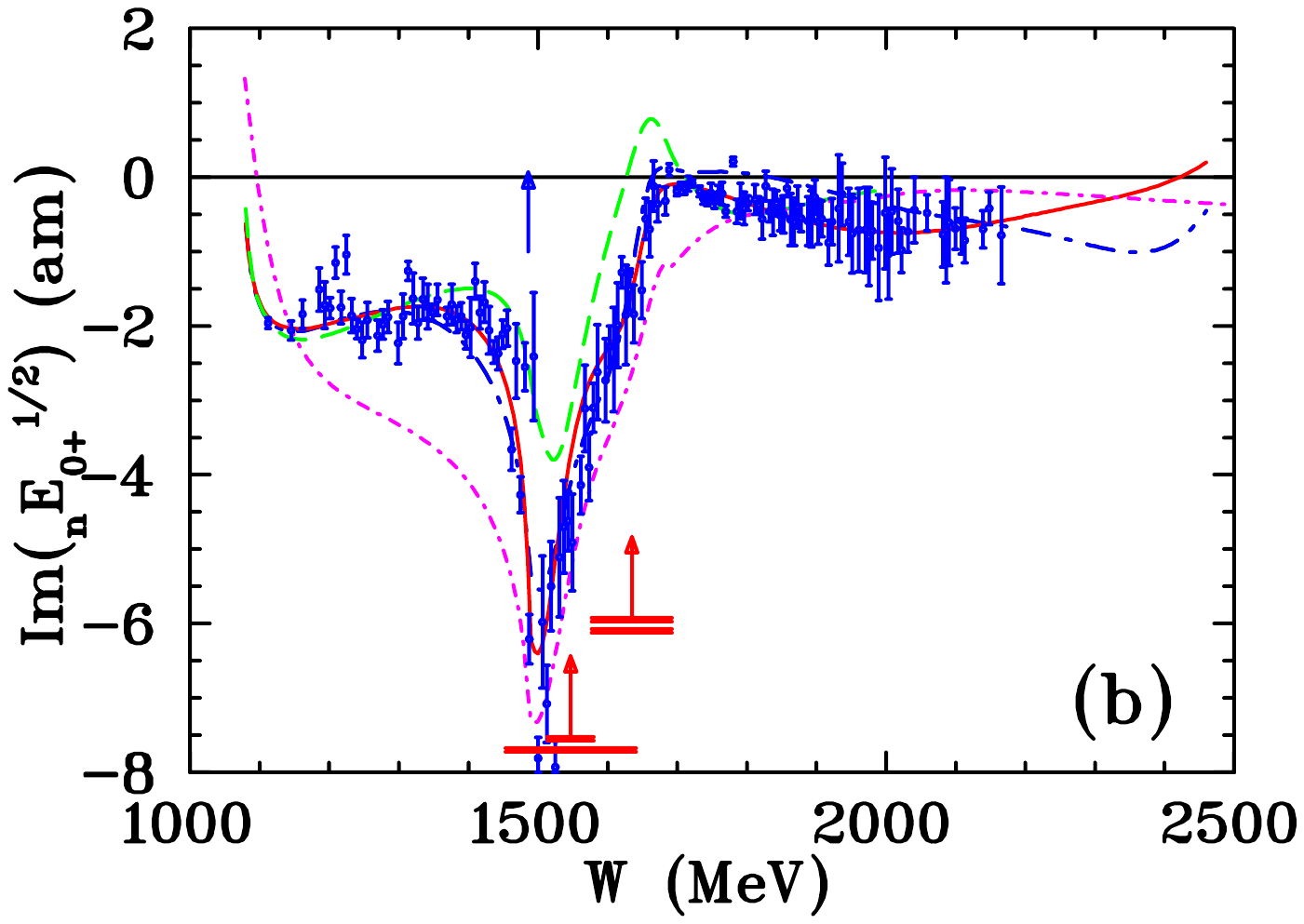}
}
\centering
{
    \includegraphics[width=0.32\textwidth,angle=90,keepaspectratio]{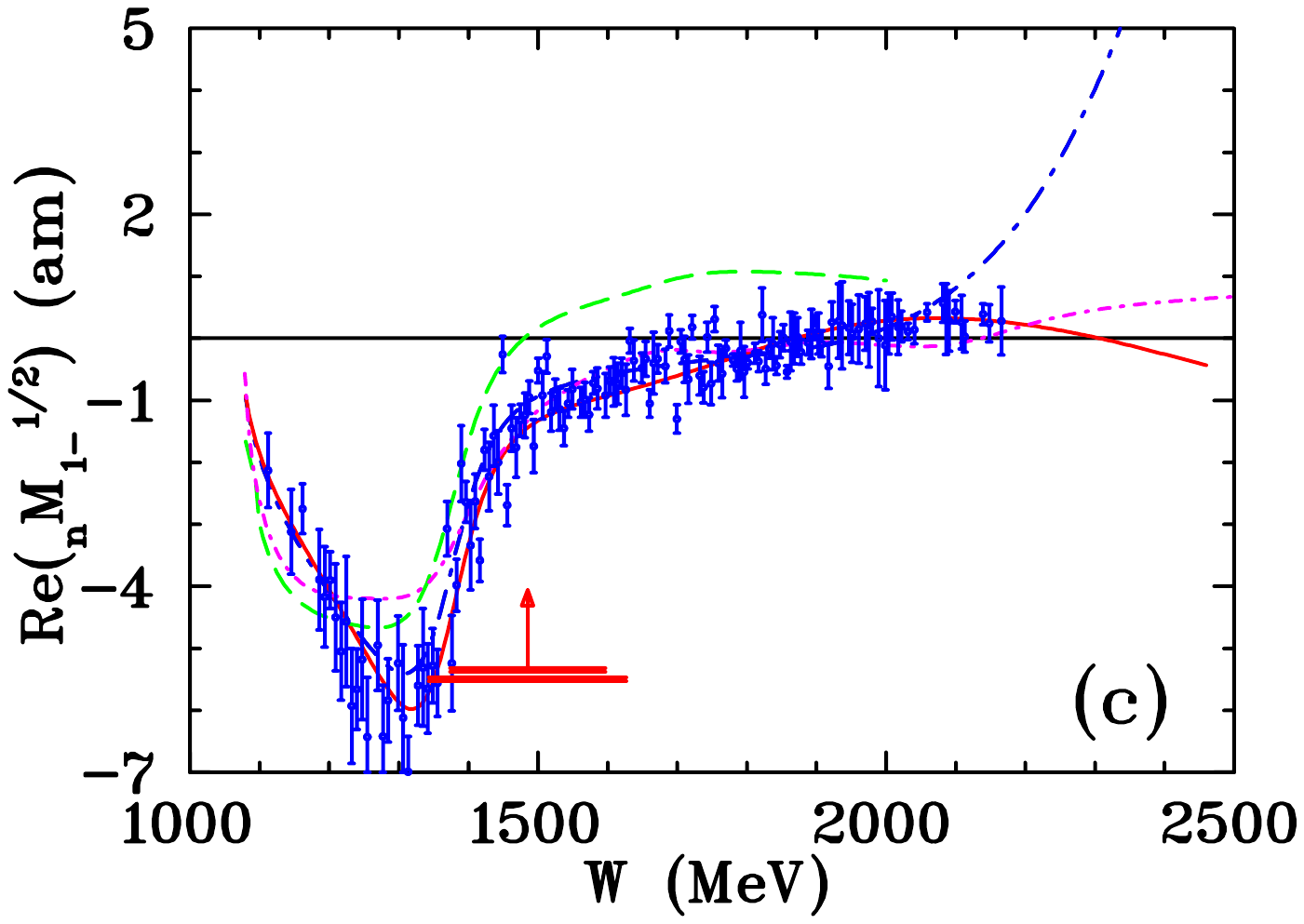}
    \includegraphics[width=0.32\textwidth,angle=90,keepaspectratio]{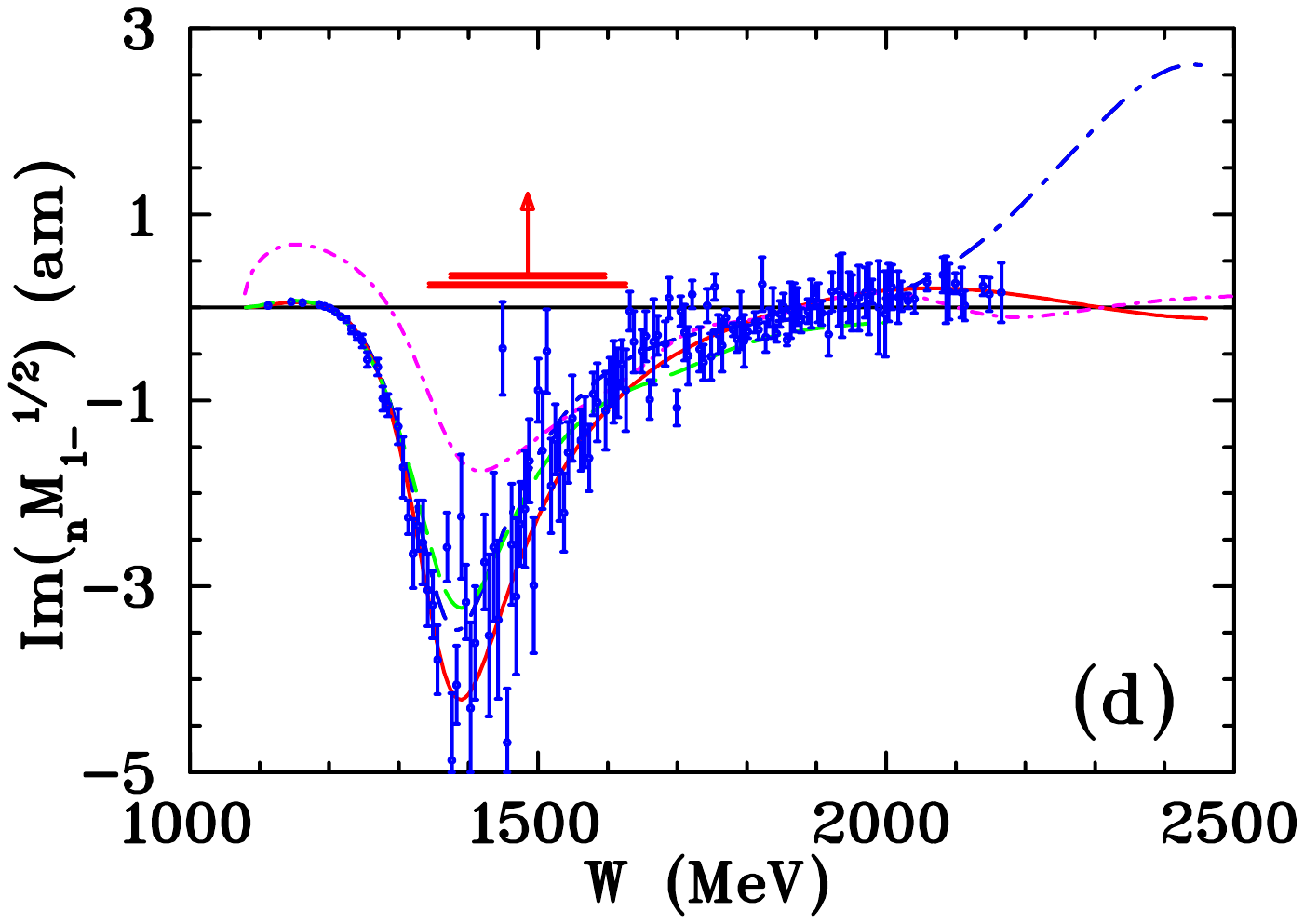}
}
\centering
{
    \includegraphics[width=0.32\textwidth,angle=90,keepaspectratio]{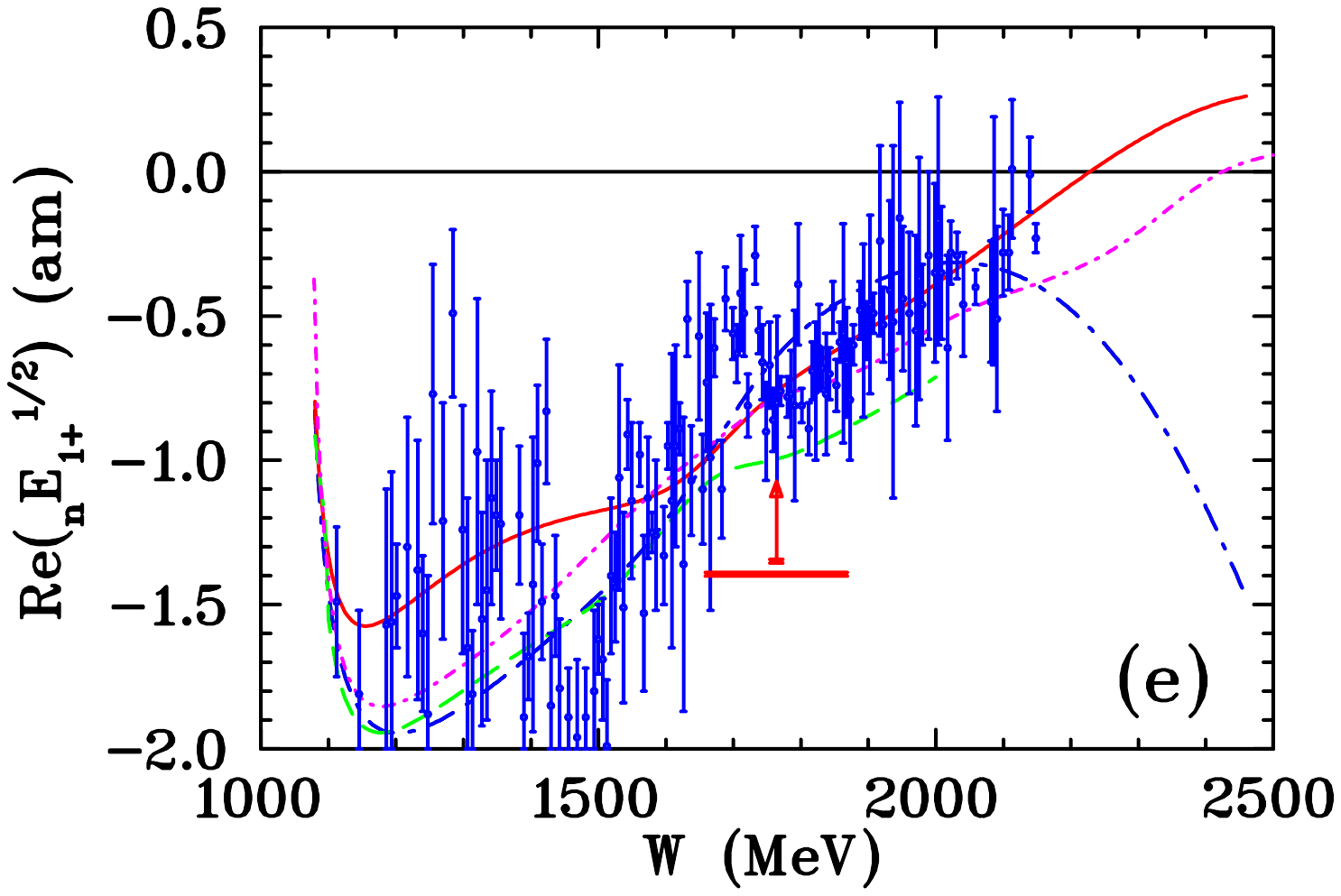}
    \includegraphics[width=0.32\textwidth,angle=90,keepaspectratio]{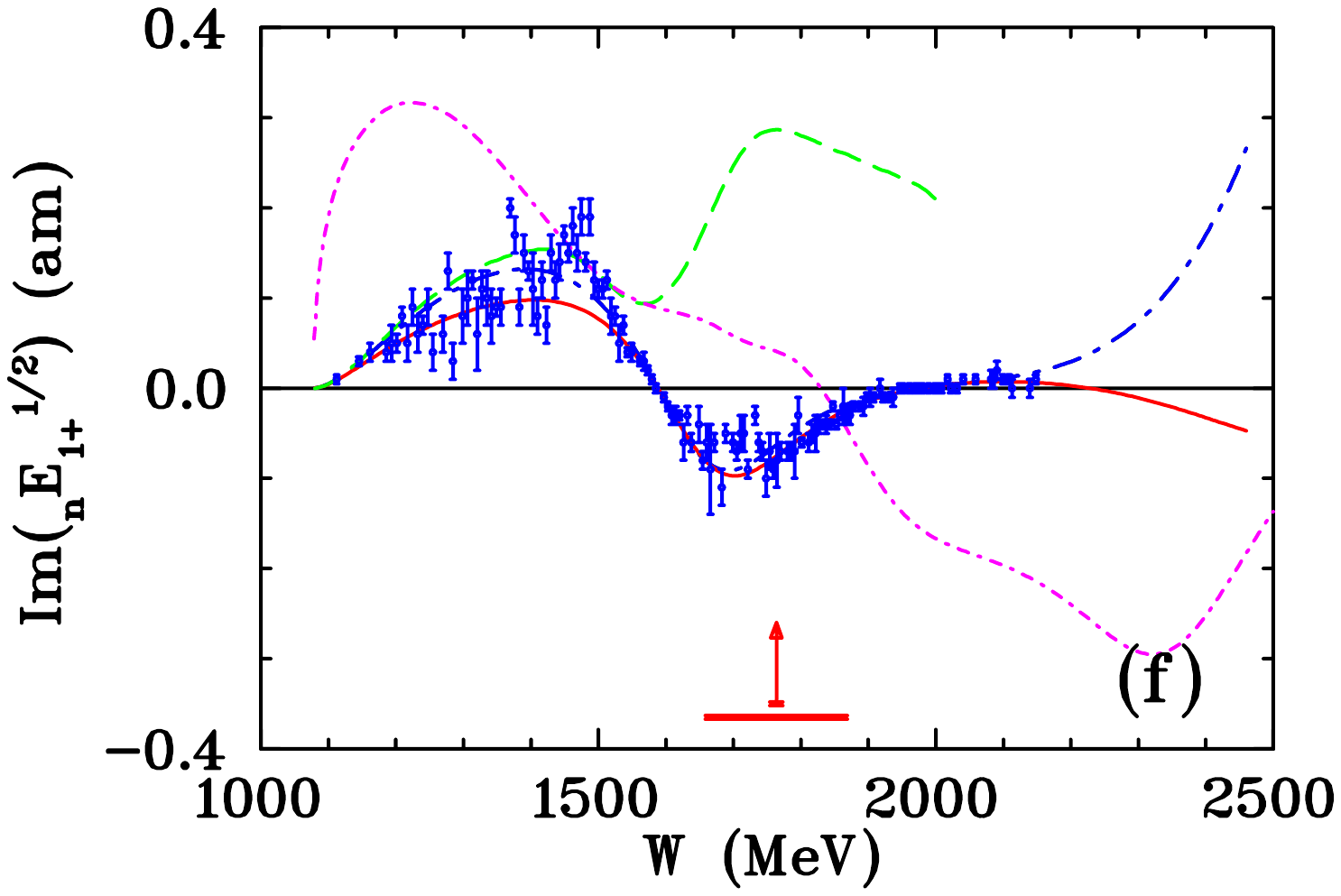}
}
\centering
{
    \includegraphics[width=0.32\textwidth,angle=90,keepaspectratio]{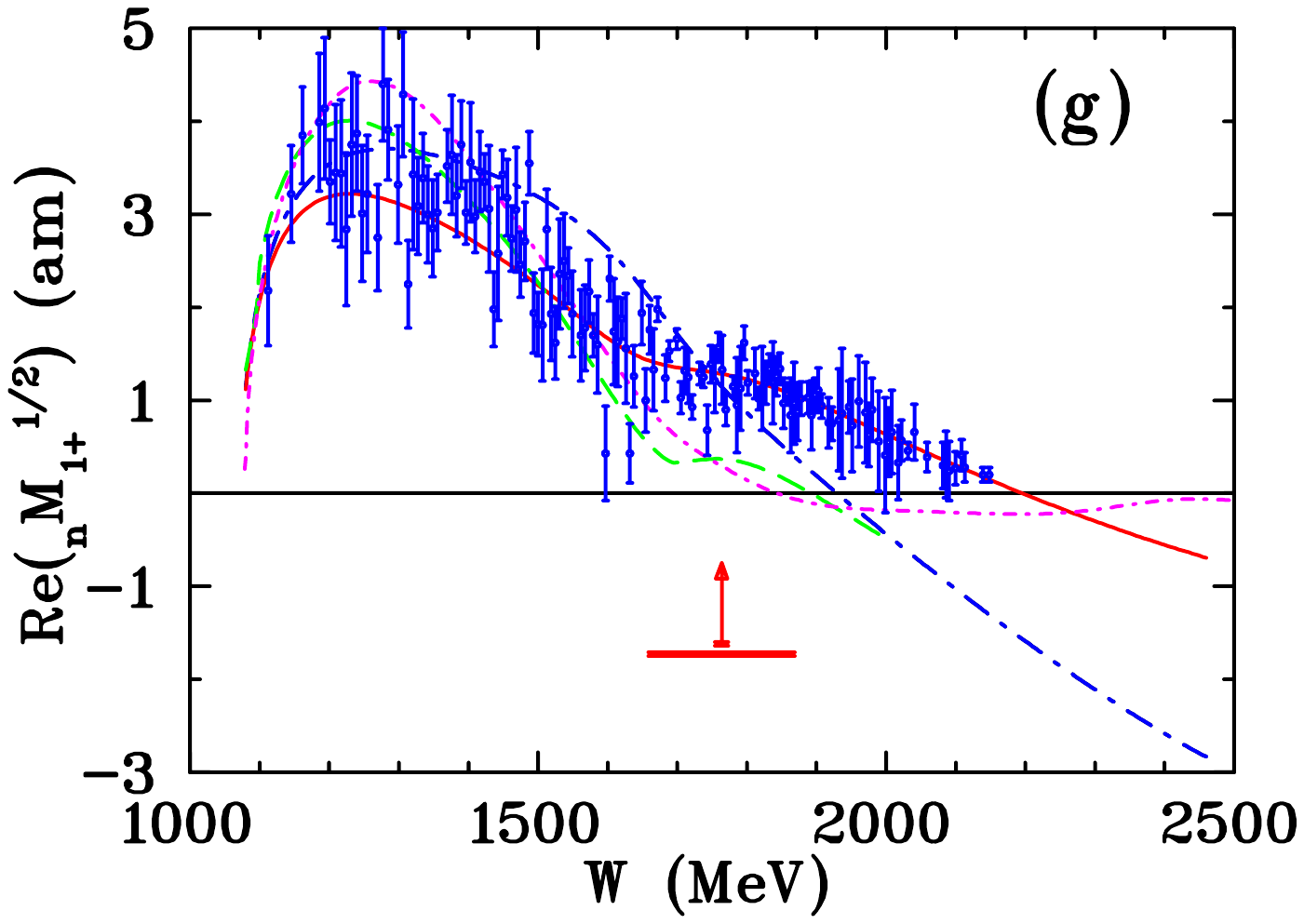}
    \includegraphics[width=0.32\textwidth,angle=90,keepaspectratio]{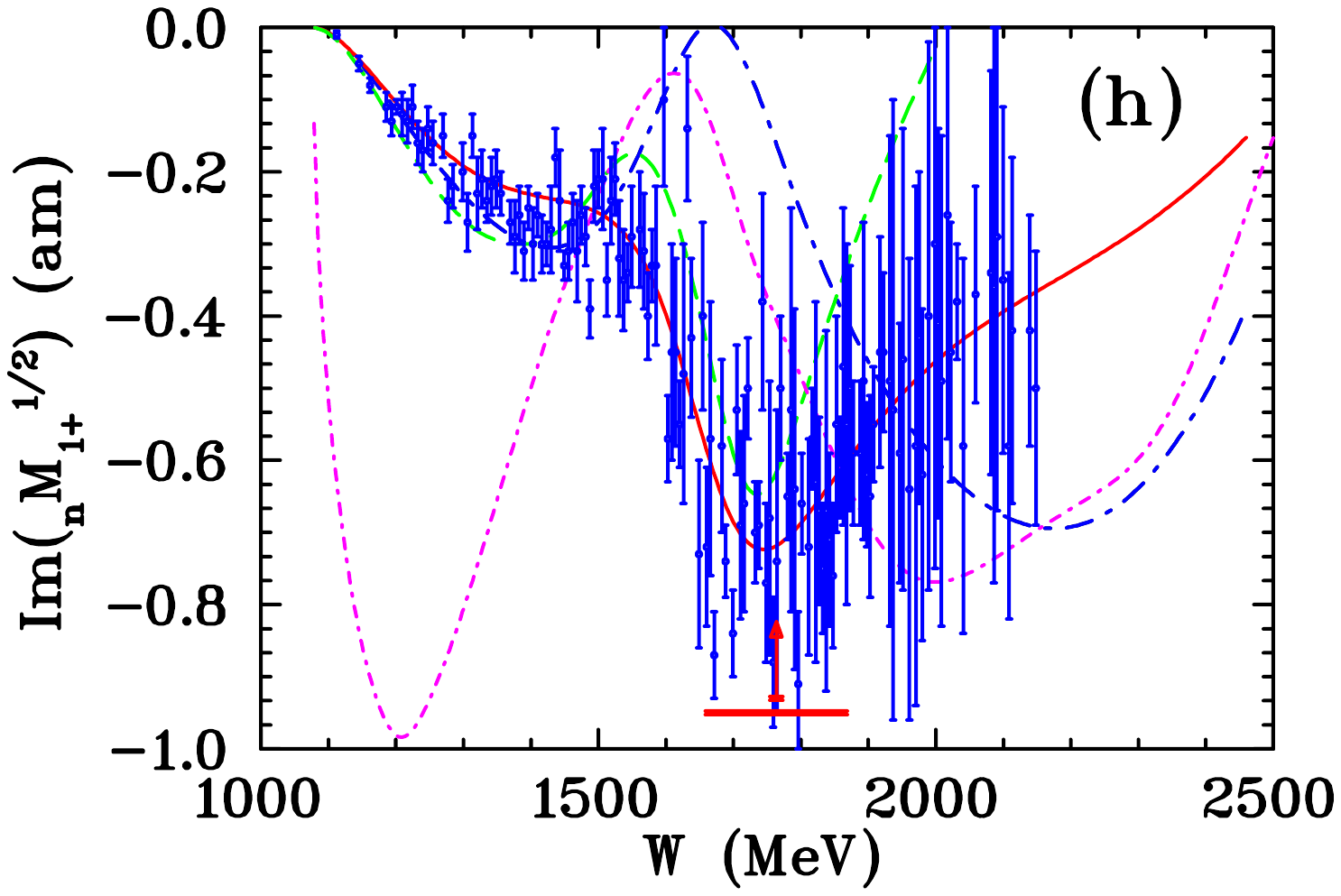}
}

\caption{Comparison neutron $I = 1/2$ multipole amplitudes (orbital momentum $l = 0, 1$) from threshold to $W = 2.5~\mathrm{GeV}$ ($E_\gamma = 2.7~\mathrm{GeV}$). For the amplitudes, the subscript $n$ denotes a neutron target, the subscript $l\pm$ gives the value of $j = l\pm 1/2$, and the superscript gives the isospin index. Notation of the solutions is the same as in Fig.~\ref{fig:amp1}. The blue vertical arrows for (a) and (b) indicate the $\eta$ production threshold.
}  
\label{fig:amp7}
\end{figure*}

\begin{figure*}[hbt!]
%\vspace{0.4cm}
\centering
{
    \includegraphics[width=0.32\textwidth,angle=90,keepaspectratio]{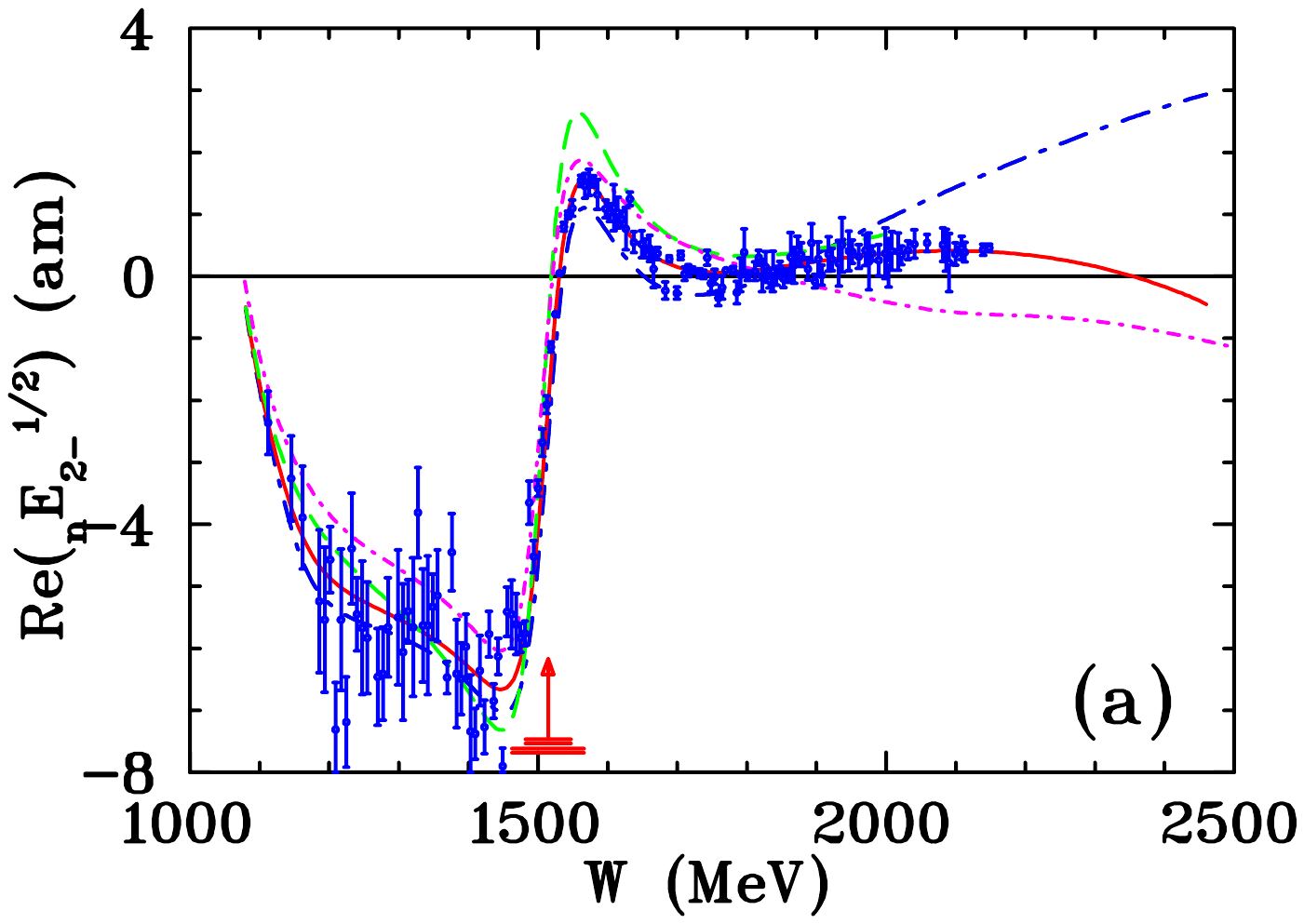}
    \includegraphics[width=0.32\textwidth,angle=90,keepaspectratio]{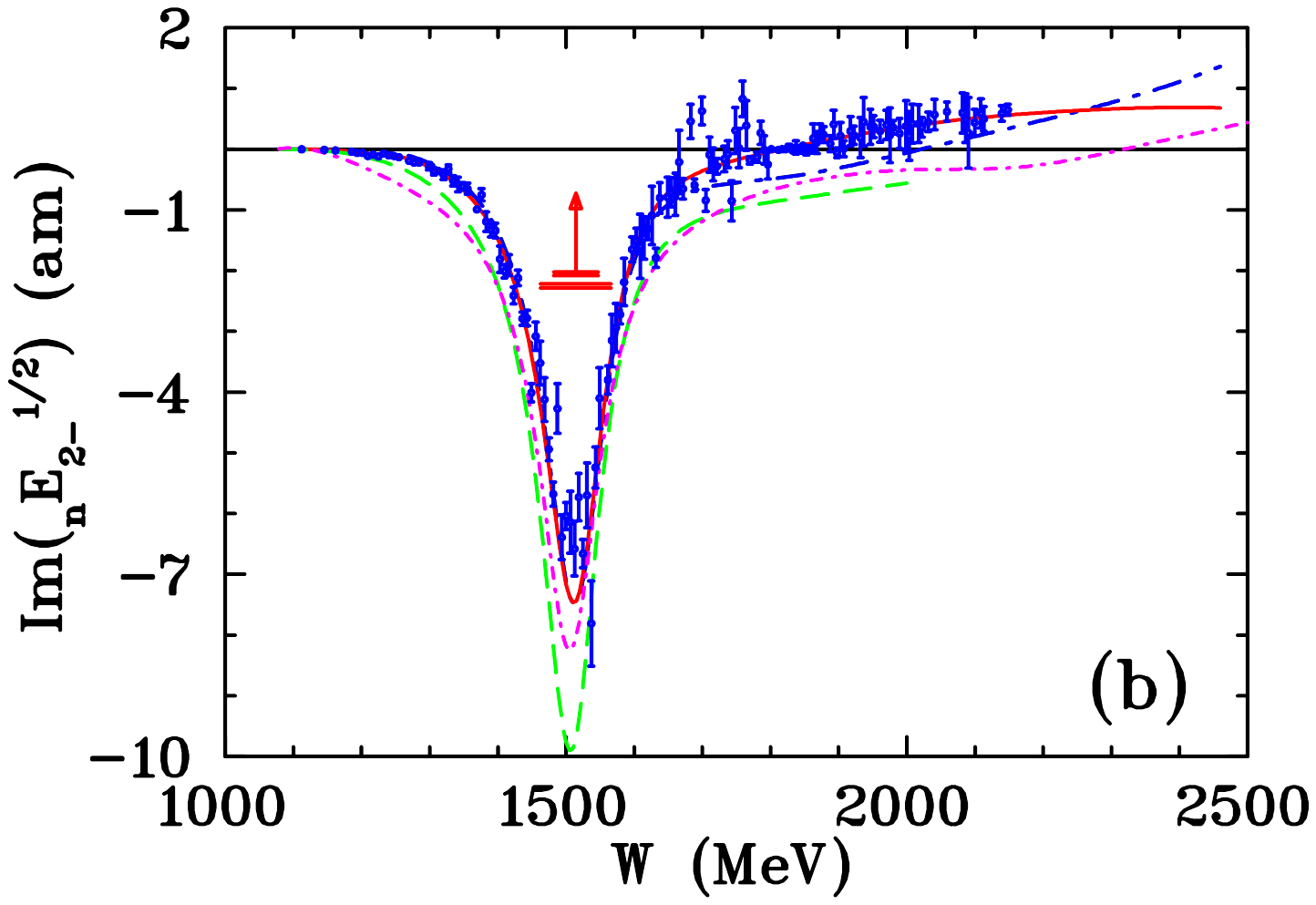}
}
\centering
{
    \includegraphics[width=0.32\textwidth,angle=90,keepaspectratio]{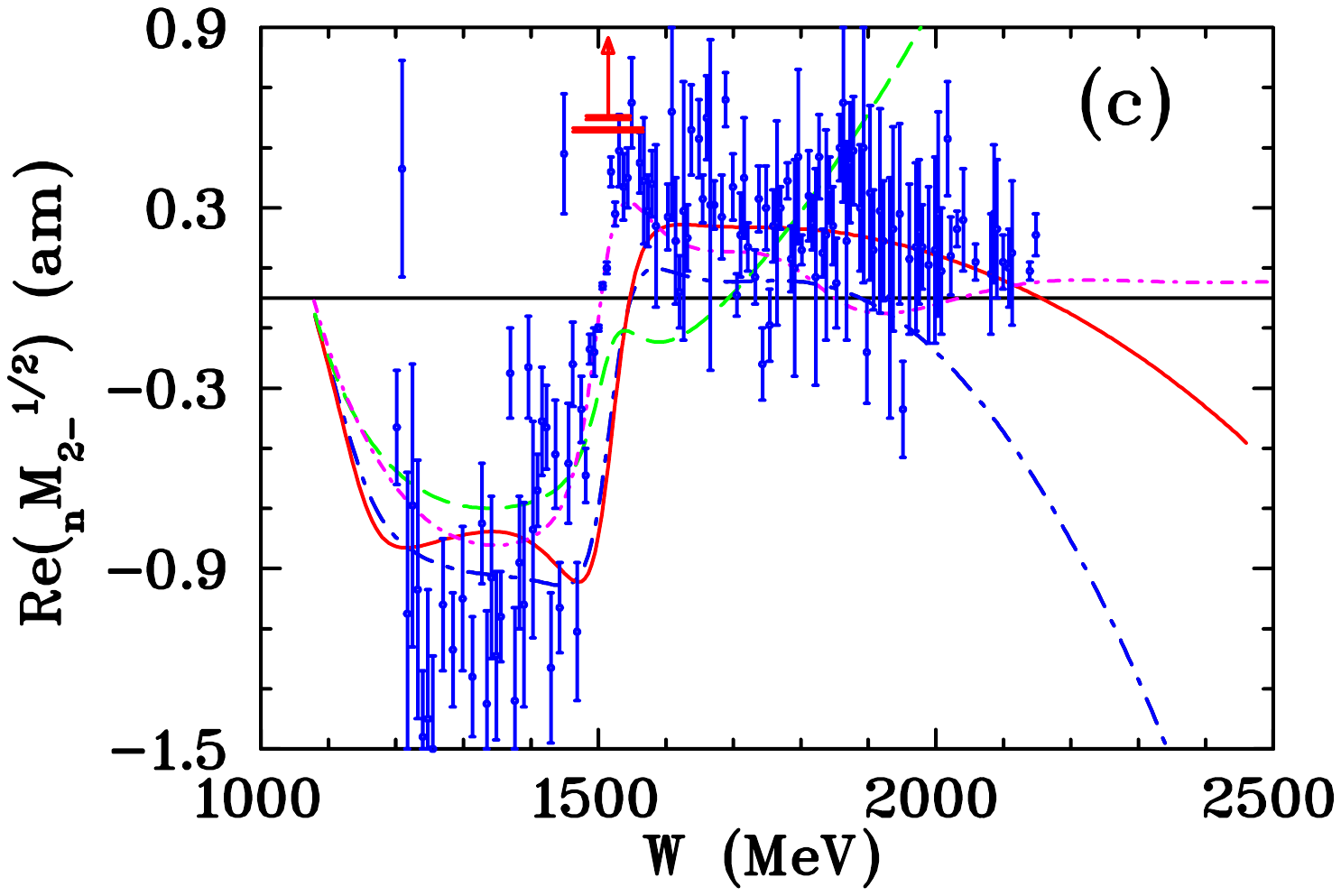}
    \includegraphics[width=0.32\textwidth,angle=90,keepaspectratio]{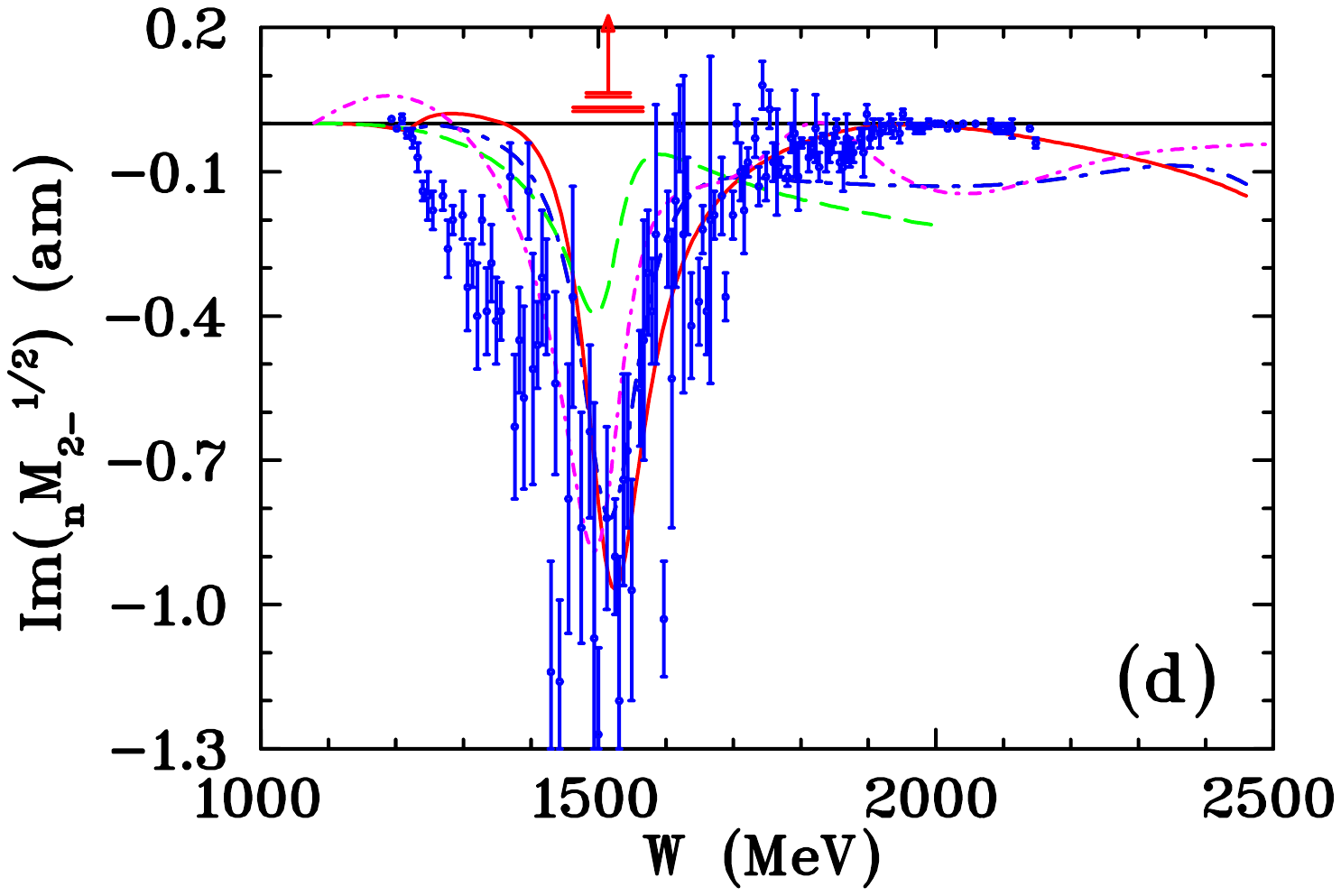}
}
\centering
{
    \includegraphics[width=0.32\textwidth,angle=90,keepaspectratio]{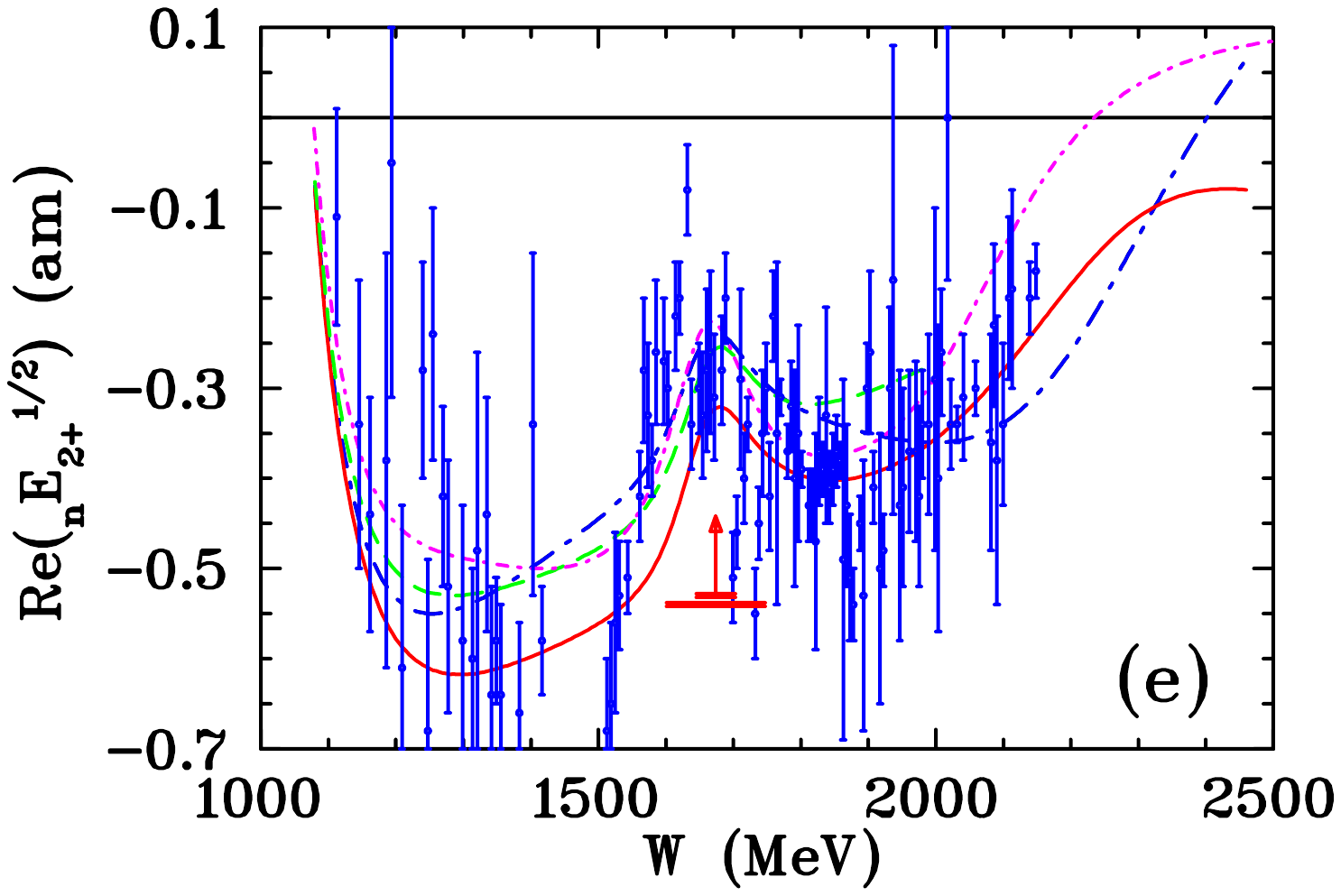}
    \includegraphics[width=0.32\textwidth,angle=90,keepaspectratio]{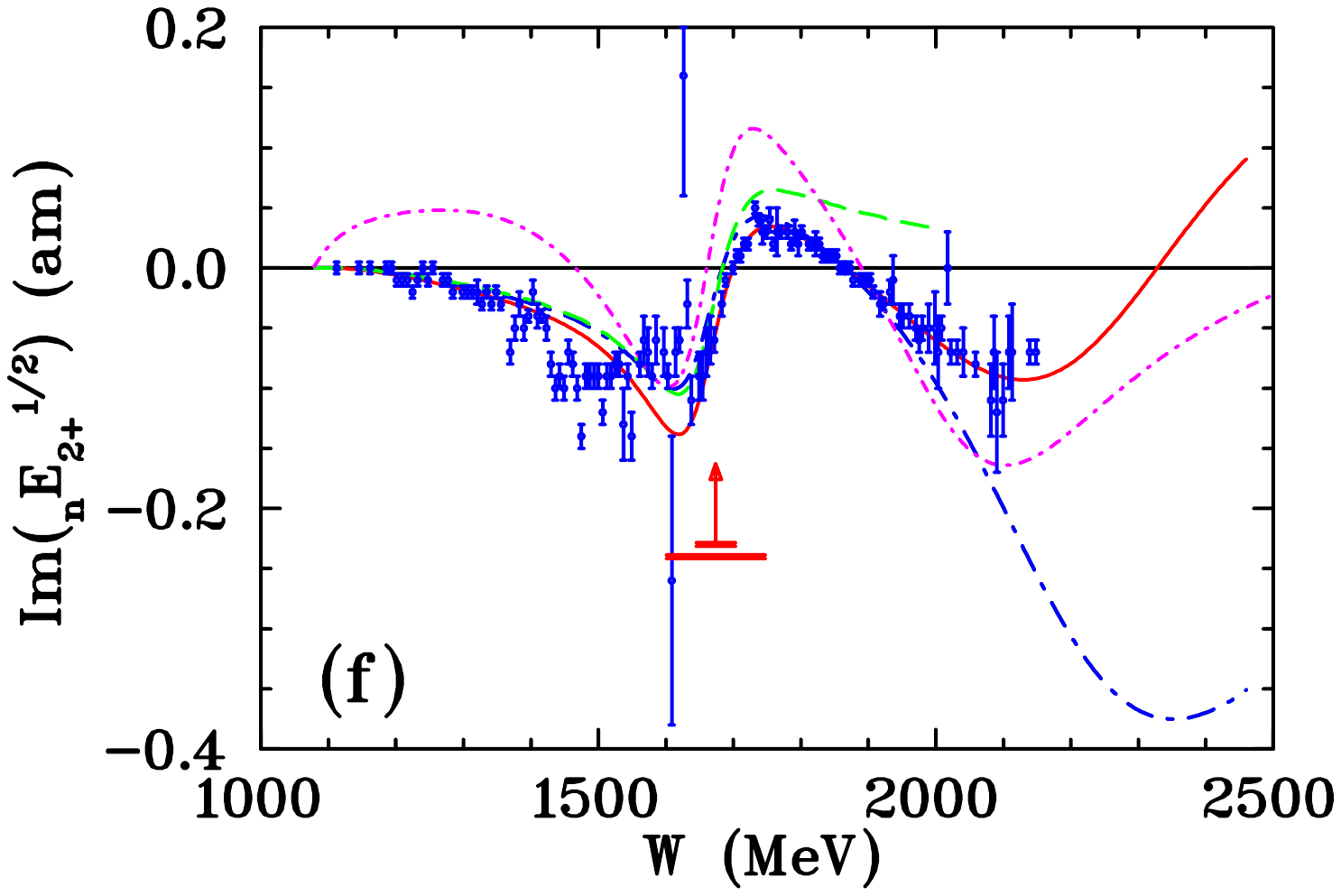}
}
\centering
{
    \includegraphics[width=0.32\textwidth,angle=90,keepaspectratio]{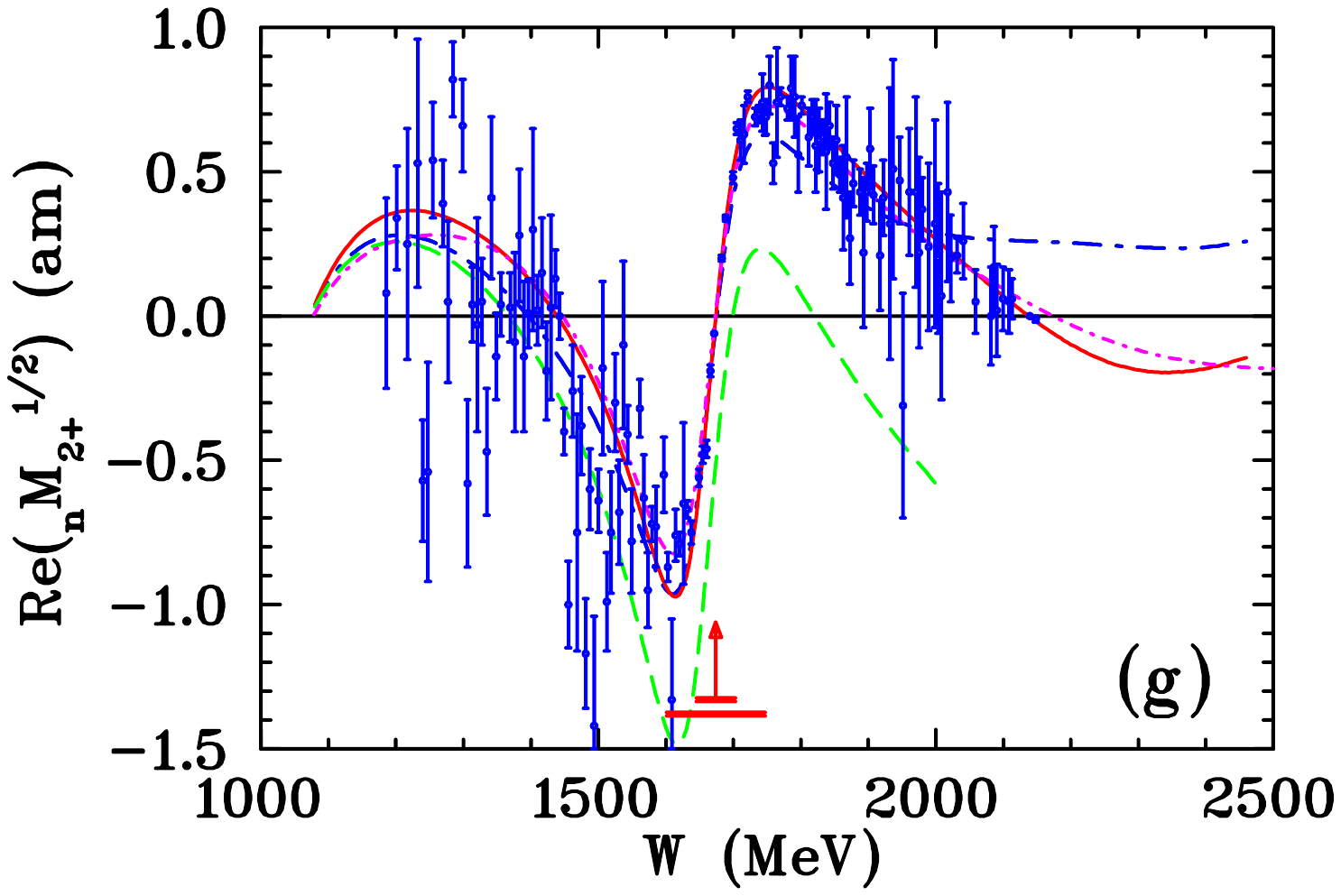}
    \includegraphics[width=0.32\textwidth,angle=90,keepaspectratio]{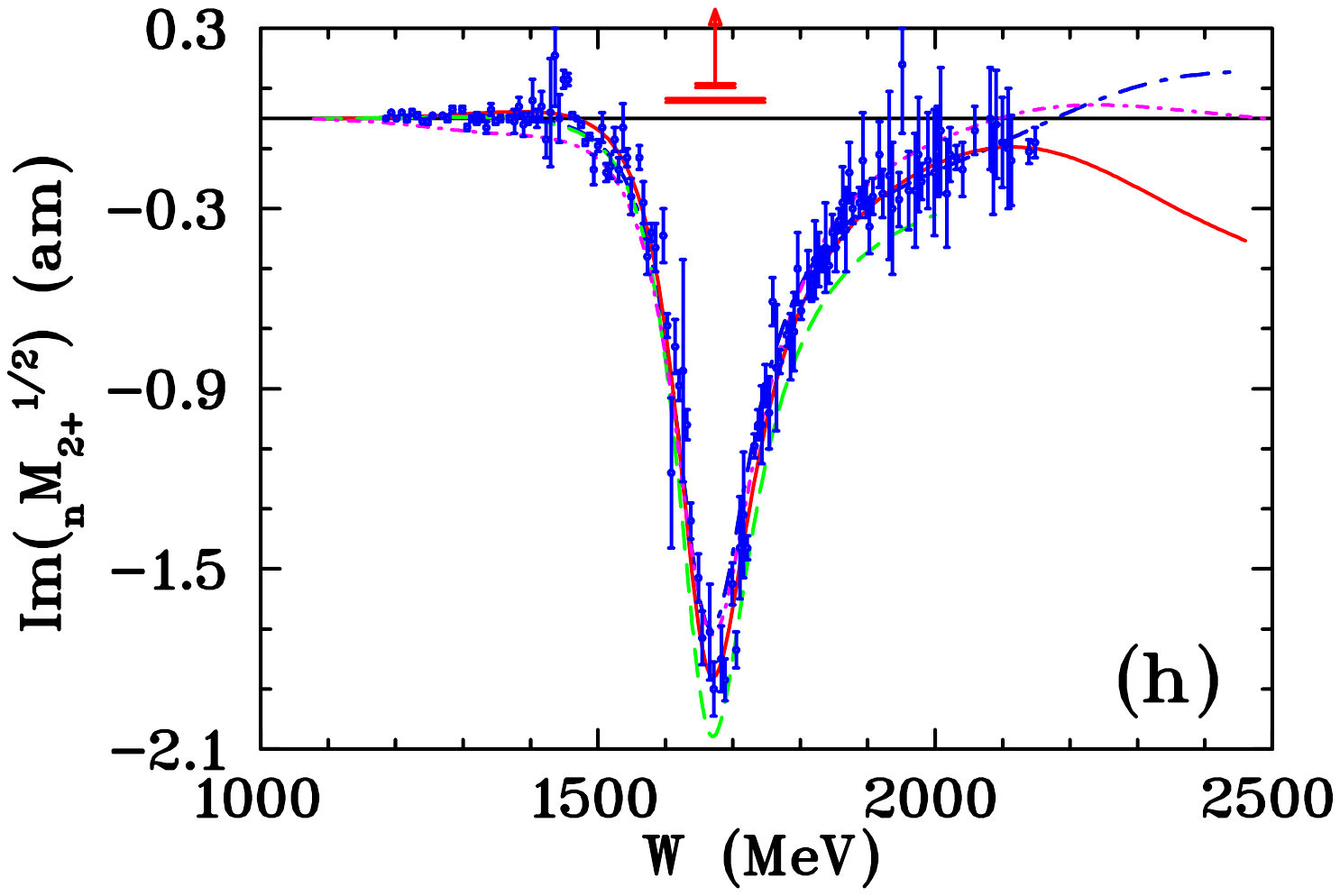}
}

\caption{Comparison of neutron $I = 1/2$ multipole amplitudes (orbital momentum $l = 2$) from threshold to $W = 2.5~\mathrm{GeV}$. Notation of the solutions is the same as in Fig.~\ref{fig:amp7}.
}  
\label{fig:amp8}
\end{figure*}

Comparisons of the present $I = 3/2$ and $I = 1/2$ ED and SE multipole amplitudes from threshold to $W = 2.5~\mathrm{GeV}$ ($E_\gamma = 2.7~\mathrm{GeV}$) shown on  Figs.~\ref{fig:amp1} - \ref{fig:amp8}.

%----------------------------------------------------------------------xxxx - no WM22
%------------------------------------------------------------
%----------------------------------------------------------------------xxx
%---------------------------------------------------------------------
%--------------------------------------------------------------------
%---------------------------------------------------------------------
%----------------------------------------------------------------------
%---------------------------------------------------------------------
%---------------------------------------------------------------------
%---------------------------------------------------------------------
%----------------------------------------------------------------------
%---------------------------------------------------------------------

%----------------------------------------------------------------------
%------------------------------------------------------------
%----------------------------------------------------------------------xxx- Figs 10g & 10h
%---------------------------------------------------------------------
%----------------------------------------------------------------------XXX-Fig. 11
%\begin{figure*}[hbtp]
%\vspace{0.4cm}
%\centering
%{
%    \includegraphics[width=0.38\textwidth,angle=0,keepaspectratio]{./v4ya.png}
%    \includegraphics[width=0.38\textwidth,angle=0,keepaspectratio]{./v4yb.png}
%}
%\centering
%{
%    \includegraphics[width=0.38\textwidth,angle=0,keepaspectratio]{./v4za.png}
%    \includegraphics[width=0.38\textwidth,angle=0,keepaspectratio]{./v4zb.png}
%}
%\centering
%{
%    \includegraphics[width=0.38\textwidth,angle=0,keepaspectratio]{./v45a.png}
%    \includegraphics[width=0.38\textwidth,angle=0,keepaspectratio]{./v45b.png}
%}
%\centering
%{
%    \includegraphics[width=0.38\textwidth,angle=0,keepaspectratio]{./v46a.png}
%    \includegraphics[width=0.38\textwidth,angle=0,keepaspectratio]{./v46b.png}
%}
%
%\caption{Comparison $I = 3/2$ multipole amplitudes (orbital momentum $l = 3$) from threshold to $W = 2.5~\mathrm{GeV}$. Notation of the solutions is the same as in Fig.~\ref{fig:amp1}.
%}  
%\label{fig:amp3}
%\end{figure*}
%---------------------------------------------------------------------
%--------------------------------------------------------------------
%---------------------------------------------------------------------
%----------------------------------------------------------------------
%---------------------------------------------------------------------
%----------------------------------------------------------------------XXX-Fig. 14
%\begin{figure*}[hbtp]
%\vspace{0.4cm}
%\centering
%{
%    \includegraphics[width=0.38\textwidth,angle=0,keepaspectratio]{./v41a.png}
%    \includegraphics[width=0.38\textwidth,angle=0,keepaspectratio]{./v41b.png}
%}
%\centering
%{
%    \includegraphics[width=0.38\textwidth,angle=0,keepaspectratio]{./v42a.png}
%    \includegraphics[width=0.38\textwidth,angle=0,keepaspectratio]{./v42b.png}
%}
%\centering
%{
%    \includegraphics[width=0.38\textwidth,angle=0,keepaspectratio]{./v47a.png}
%    \includegraphics[width=0.38\textwidth,angle=0,keepaspectratio]{./v47b.png}
%}
%\centering
%{
%    \includegraphics[width=0.38\textwidth,angle=0,keepaspectratio]{./v48a.png}
%    \includegraphics[width=0.38\textwidth,angle=0,keepaspectratio]{./v48b.png}
%}
%
%\caption{Comparison proton $I = 1/2$ multipole amplitudes (orbital momentum $l = 3$) from threshold to $W = 2.5~\mathrm{GeV}$. Notation of the solutions is the same as in Fig.~\ref{fig:amp1}.
%}  
%\label{fig:amp6}
%\end{figure*}
%---------------------------------------------------------------------

%---------------------------------------------------------------------
%---------------------------------------------------------------------
%----------------------------------------------------------------------
\clearpage
\section{Resonance Couplings}
\label{Sec:res}
Following the notation of Refs.~\cite{Workman:2013rca, Svarc:2014sqa}, the $(\gamma,\pi)$ T-matrix element for helicity $h$ is given by
\begin{equation}
    T_{\gamma,\pi}^{h}=\sqrt{2\,k\,q}\; \mathcal{A}^{h}_\alpha\; C\,, \label{uniamp}
\end{equation}
where $\alpha$ denotes the partial wave and $k, q$ are the center-of-mass (c.m.) momenta of the photon and the pion.
%{\red ($\mathcal{A}_\alpha^h$ needsalso to be defined!) }
The factor $C$ is $\sqrt{2/3}$ for isospin $3/2$ and $-\sqrt{3}$ for isospin $1/2$. The helicity
multipoles $\mathcal{A}^{h}_\alpha$ are given in terms of electric and magnetic multipoles
\begin{eqnarray}
    \mathcal{A}_{\ell +}^{1/2}     & = & -{1\over 2} \left[ (\ell + 2) {E}_{\ell +} + \ell {M}_{\ell +} \right] \>, \label{helimult1}
\\
    \mathcal{A}_{\ell +}^{3/2}     & = & {1\over 2} \sqrt{\ell (\ell + 2)} \left[ {E}_{\ell +} -{M}_{\ell +} \right] \>, \label{helimult2}
\\
    \mathcal{A}_{(\ell +1)-}^{1/2} & = & -{1\over 2} \left[\ell {E}_{(\ell +1)-} - (\ell +2 )
    {M}_{(\ell +1)-} \right] \>, \label{helimult3}
\\
    \mathcal{A}_{(\ell +1)-}^{3/2} & = & -{1\over 2} \sqrt{\ell (\ell +2)} \left[ {E}_{(\ell +1)-} + {M}_{(\ell +1) -} \right] \>, \label{helimult4}
\end{eqnarray}
with $J=\ell+1/2$ for ``$+$'' multipoles and $J = (\ell+1)-1/2$ for ``$-$'' multipoles, all having the
same total spin $J\,$.

In Tables~\ref{tab:tblR2} to \ref{tab:tblR11}, we list the pole positions together with the photo-decay amplitudes
\begin{eqnarray}
    A_h &=& C\,\sqrt{\frac{q_p}{k_p}\frac{2\pi(2J+1)W_p}{m_N Res_{\pi N}}}\,\mbox{Res}\,\mathcal{A}_\alpha^h \>,
\end{eqnarray}
where the subscript $p$ denotes quantities evaluated at the pole position and $m_N$ is the nucleon mass. In Ref.~\cite{Svarc:2014sqa}, the elastic residues, $Res_{\pi N}$, and the pole positions, $W_p=M_p-i\Gamma_p/2$, were taken from the GWU SAID PWA, SP06~\cite{Arndt:2006bf} and each multipole was fitted separately, using the Laurent plus Pietarinen (L+P) method~\cite{Svarc:2014sqa}, to determine the corresponding residues.

Here, we have made a coupled multipole fit of all partial-wave amplitudes associated with particular resonances, including the pion-nucleon elastic scattering amplitudes. Thus, for example, the L+P fit of Ref.~\cite{Svarc:2014sqa} for the $E_{2-}^{1/2}$ multipole has been expanded to a simultaneous fit of the $D_{13}$ elastic amplitude, $E_{2-}^{1/2}$ and $M_{2-}^{1/2}$ (proton target), plus $E_{2-}^{1/2}$ and $M_{2-}^{1/2}$ (neutron target), yielding more self-consistent results. 

As in Ref.~\cite{Svarc:2014sqa}, the fitted partial waves are $S_{11}$, $P_{11}$, $D_{13}$, $F_{15}$, $P_{33}$, $D_{33}$, and $F_{37}$ with pion-nucleon partial waves taken from Ref.~\cite{Workman:2012hx}.
%The standard L+P method is detailed in %\textcolor{red}{Refs.~\cite{Svarc:2013laa, %Svarc:2015usk}}.

%-----------------------------------------------------------------
%\begin{table}[htb!]
%
%\centering \protect\caption{Photoproduction multipoles, partial waves, and associated resonances. Two multipoles lead to one spin-parity wave.}
%
%\vspace{2mm}
%{%
%\begin{tabular}{|c|c|c|c|c|c|}
%\hline
%Multipoles &         & Partial waves &          & Resonances     \tabularnewline
%\hline
%$E_0^+$    &         & $S_{11}$      &          & $N(1535)1/2^-$      \tabularnewline
%           &         &               &          & $N(1650)1/2^-$      \tabularnewline
%           & $M_1^-$ & $P_{11}$      &          & $N(1440)1/2^+$      \tabularnewline
%$E_1^+$    & $M_1^+$ & $P_{13}$      &          & $N(1720)3/2^+$      \tabularnewline
%           &         &               & $P_{33}$ & $\Delta(1232)3/2^+$ \tabularnewline
%$E_2^-$    & $M_2^=$ & $D_{13}$      &          & $N(1520)3/2^-$      \tabularnewline
%           &         &               & $D_{33}$ & $\Delta(1700)3/2^-$  \tabularnewline
%$E_2^+$    & $M_2^+$ & $D_{15}$      &          & $N(1675)5/2^-$       \tabularnewline
%$E_3^-$    & $M_3^-$ & $F_{15}$      &          & $N(1680)5/2^+$       \tabularnewline
%           &         &               & $F_{35}$ & $\Delta(1905)5/2^+$  \tabularnewline
%$E_3^+$    & $M_3^+$ &               & $F_{37}$ & $\Delta(1950)7/2^+$  \tabularnewline
%\hline
%\end{tabular}} \label{tab:tblR1}
%\end{table}

%---------------------------------------------------------
\begin{table*}[htb!]

\centering \protect\caption{Photon-decay helicity amplitudes at the pole for $p\gamma$ and $n\gamma$ decays. Fit to pion-nucleon elastic amplitude $S_{11}$ and multipole $E_{0+}^{1/2}$. Complex quantities given as modulus and phase. Results from present study (first row), PR2014~\cite{Svarc:2014sqa} (second row), and BnGa~\cite{CBELSATAPS:2015kka} (for proton couplings) and \cite{Anisovich:2017afs} (for neutron couplings) (third row).}

\vspace{2mm}
{%
\begin{tabular}{|c|c|c|c|c|c|c|}
\hline
Resonance
&  $Re W_p$             
&  $-2 Im W_p$      
& $|pA^{1/2}|$    
& $p\phi A^{1/2}$ 
& $|nA^{1/2}|$    
& $n\phi A^{1/2}$ \tabularnewline
&  (GeV)             
&  (GeV)         
&  ($GeV^{-1/2}$) 
&  (deg)         
&  ($GeV^{-1/2}$)  
&  (deg) \tabularnewline
\hline
$N(1535)1/2^-$
& 1.500$\pm$0.001    
& 0.096$\pm$0.006 
& 0.079$\pm$0.012 
& -11.4$\pm$1.7 
& 0.067$\pm$0.009 
& -174$\pm$22 \tabularnewline  
& 1.501$\pm$0.006    
& 0.095$\pm$0.011 
& 0.074$\pm$0.010 
& -17$\pm$11      
&          
&  \tabularnewline
& 1.500$\pm$0.004           
& 0.128$\pm$0.009      
& 0.114$\pm$0.008       
& 10$\pm$5        
& 0.088$\pm$0.004          
& -175$\pm$4 \tabularnewline
\hline
\hline
$N(1650)1/2^-$
& 1.650$\pm$0.001    
& 0.110$\pm$0.008 
& 0.042$\pm$0.001 
& -12.5$\pm$0.4 
& 0.026$\pm$0.005 
& -72$\pm$13 \tabularnewline
& 1.655$\pm$0.011    
& 0.127$\pm$0.017 
& 0.041$\pm$0.006 
& 16$\pm$27       
&                
& \tabularnewline
& 1.652$\pm$0.007           
& 0.102$\pm$0.008      
& 0.032$\pm$0.006       
& -2$\pm$11        
&  0.016$\pm$0.004         
&  -28$\pm$10 \tabularnewline
\hline
\end{tabular}} \label{tab:tblR2}
\end{table*}

%---------------------------------------------------------
\begin{table*}[htb!]

\centering \protect\caption{Photon-decay helicity amplitudes at the pole for $p\gamma$ and $n\gamma$ decays. Fit to pion-nucleon elastic amplitude $P_{11}$ and multipole $M_{1-}^{1/2}$. Complex quantities given as modulus and phase. Results from present study (first row), PR2014~\cite{Svarc:2014sqa} (second row), and BnGa~\cite{CBELSATAPS:2015kka} (for proton couplings) and \cite{Anisovich:2017afs} (for neutron couplings) (third row).}

\vspace{2mm}
{%
\begin{tabular}{|c|c|c|c|c|c|c|}
\hline
Resonance
&   $Re W_p$             
&   $-2 Im W_p$      
& $|pA^{1/2}|$    
& $p\phi A^{1/2}$ 
& $|nA^{1/2}|$    
& $n\phi A^{1/2}$ \tabularnewline
& (GeV)             
&   (GeV)         
&  ($GeV^{-1/2}$) 
&   (deg)         
& ($GeV^{-1/2}$)  
&  (deg) \tabularnewline
\hline
$N(1440)1/2^+$
& 1.358$\pm$0.003    
& 0.192$\pm$0.005 
& 0.062$\pm$0.004 
& 160$\pm$11  
& 0.080$\pm$0.005 
& 1.25$\pm$0.08 \tabularnewline 
& 1.360$\pm$0.005    
& 0.183$\pm$0.019 
& 0.055$\pm$0.003 
& 167$\pm$11      
&                 
&  \tabularnewline 
& 1.369$\pm$0.003           
& 0.189$\pm$0.005      
& 0.044$\pm$0.005       
& 140$\pm$8        
& 0.041$\pm$0.005         
& 23$\pm$10 \tabularnewline
\hline
\end{tabular}} \label{tab:tblR3}
\end{table*}

%---------------------------------------------------------
\begin{table*}[htb!]

\centering \protect\caption{Photon-decay helicity amplitudes at the pole for $p\gamma$ and $n\gamma$ decays. Fit to pion-nucleon elastic amplitude $P_{13}$ and multipoles $E_{1+}^{1/2}$ and $M_{1+}^{1/2}$. Complex quantities given as modulus and phase. Results from present study (first row), PR2014~\cite{Svarc:2014sqa} (second row), and BnGa~\cite{CBELSATAPS:2015kka} (for proton couplings) and \cite{Anisovich:2017afs} (for neutron couplings) (third row).}

\vspace{2mm}
{%
\begin{tabular}{|c|c|c|c|c|c|c|c|c|c|c|}
\hline
Resonance
& $Re W_p$             
& $-2 Im W_p$ 
& $|pA^{1/2}|$ 
& $p\phi A^{1/2}$ 
& $|pA^{3/2}|$ 
& $p\phi A^{3/2}$ 
& $|nA_{1/2}|$ 
& $n\phi A^{1/2}$ 
& $|nA^{3/2}|$ 
& $nA\phi ^{3/2}$ \tabularnewline
& (GeV)
& (GeV)
& ($GeV^{-1/2}$) 
& (deg) 
& ($GeV^{-1/2}$) 
& (deg) 
& ($GeV^{-1/2}$)
&  (deg) 
& ($GeV^{-1/2}$) 
& (deg)\tabularnewline
\hline
$N(1720)3/2^+$
& 1.670$\pm$0.001
& 0.280$\pm$0.002 
& 0.057$\pm$0.027 
& -42$\pm$19 
& 0.071$\pm$0.033
& -8$\pm$4
& 0.056$\pm$0.021 
& -21$\pm$8 
& 0.065$\pm$0.024
&  169$\pm$64 \tabularnewline
& 1.651$\pm$0.009 
& 0.311$\pm$0.045 
& 0.059$\pm$0.002 
& -14$\pm$8 
& 0.045$\pm$0.005
& -151$\pm$11
& 
& 
& 
& \tabularnewline
& 1.670$\pm$0.025           
& 0.430$\pm$0.100
& 0.115$\pm$0.045       
& 0$\pm$35
& 0.140$\pm$0.040
& 65$\pm$35
& 0.025$_{-0.015}^{+0.040}$        
& 105$\pm$35 
& 0.100$\pm$0.035
& -80$\pm$35 \tabularnewline
\hline
\end{tabular}} \label{tab:tblR4}
\end{table*}

%---------------------------------------------------------
\begin{table*}[htb!]

\centering \protect\caption{Photon-decay helicity amplitudes at the pole for $p\gamma$ and $n\gamma$ decays. Fit to pion-nucleon elastic amplitude $D_{13}$ and multipoles $E_{2-}^{1/2}$ and $M_{2-}^{1/2}$. Complex quantities given as modulus and phase. Results from present study (first row), PR2014~\cite{Svarc:2014sqa} (second row), and BnGa~\cite{CBELSATAPS:2015kka} (for proton couplings) and \cite{Anisovich:2017afs} (for neutron couplings) (third row).}

\vspace{2mm}
{%
\begin{tabular}{|c|c|c|c|c|c|c|c|c|c|c|}
\hline
Resonance
& $Re W_p$             
& $-2 Im W_p$ 
& $|pA^{1/2}|$ 
& $p\phi A^{1/2}$ 
& $|pA^{3/2}|$ 
& $p\phi A^{3/2}$ 
& $|nA_{1/2}|$ 
& $n\phi A^{1/2}$ 
& $|nA^{3/2}|$ 
& $nA\phi ^{3/2}$ \tabularnewline
& (GeV)
& (GeV)
& ($GeV^{-1/2}$) 
& (deg) 
& ($GeV^{-1/2}$) 
& (deg) 
& ($GeV^{-1/2}$)
&  (deg) 
& ($GeV^{-1/2}$) 
& (deg)\tabularnewline
\hline
$N(1520)3/2^-$
& 1.511$\pm$0.001
& 0.116$\pm$0.002 
& 0.029$\pm$0.001 
& 156$\pm$8 
& 0.144$\pm$0.007
& 4.0$\pm$0.2
& 0.044$\pm$0.004 
& -175$\pm$15 
& 0.121$\pm$0.010
& -170$\pm$14 \tabularnewline
& 1.514$\pm$0.001 
& 0.109$\pm$0.005 
& 0.028$\pm$0.001 
& 154$\pm$7 
& 0.133$\pm$0.006
& 13$\pm$2
& 
& 
& 
& \tabularnewline
& 1.507$\pm$0.002           
& 0.111$\pm$0.003
& 0.023$\pm$0.004       
& 174$\pm$5
& 0.131$\pm$0.006
& 4$\pm$4
& 0.045$\pm$0.005         
& 175$\pm$4 
& 0.119$\pm$0.005
& -175$\pm$4 \tabularnewline
\hline
\end{tabular}} \label{tab:tblR5}
\end{table*}

%---------------------------------------------------------
\begin{table*}[htb!]

\centering \protect\caption{Photon-decay helicity amplitudes at the pole for $p\gamma$ and $n\gamma$ decays. Fit to pion-nucleon elastic amplitude $D_{15}$ and multipoles $E_{2-}^{1/2}$ and $M_{2-}^{1/2}$. Complex quantities given as modulus and phase. Results from present study (first row), PR2014~\cite{Svarc:2014sqa} (second row), and BnGa~\cite{CBELSATAPS:2015kka} (for proton couplings) and \cite{Anisovich:2017afs} (for neutron couplings) (third row).}

\vspace{2mm}
{%
\begin{tabular}{|c|c|c|c|c|c|c|c|c|c|c|}
\hline
Resonance
& $Re W_p$             
& $-2 Im W_p$ 
& $|pA^{1/2}|$ 
& $p\phi A^{1/2}$ 
& $|pA^{3/2}|$ 
& $p\phi A^{3/2}$ 
& $|nA_{1/2}|$ 
& $n\phi A^{1/2}$ 
& $|nA^{3/2}|$ 
& $nA\phi ^{3/2}$ \tabularnewline
& (GeV)
& (GeV)
& ($GeV^{-1/2}$) 
& (deg) 
& ($GeV^{-1/2}$) 
& (deg) 
& ($GeV^{-1/2}$)
&  (deg) 
& ($GeV^{-1/2}$) 
& (deg)\tabularnewline
\hline
$N(1675)5/2^-$
& 1.658$\pm$0.003
& 0.141$\pm$0.005 
& 0.020$\pm$0.006 
& 165$\pm$43 
& 0.020$\pm$0.005
& 23$\pm$6
& 0.123$\pm$0.027 
& -19$\pm$4 
& 0.084$\pm$0.018
& -170$\pm$38 \tabularnewline
& 1.657$\pm$0.005 
& 0.141$\pm$0.011 
& 0.015$\pm$0.002 
& 25$\pm$12 
& 0.019$\pm$0.002
& -40$\pm$8
& 
& 
& 
& \tabularnewline
& 1.655$\pm$0.004           
& 0.147$\pm$0.005
& 0.022$\pm$0.003       
& -12$\pm$7
& 0.028$\pm$0.006
& -17$\pm$6
& 0.053$\pm$0.004         
& 177$\pm$5 
& 0.073$\pm$0.005
& 168$\pm$5 \tabularnewline
\hline
\end{tabular}} \label{tab:tblR6}
\end{table*}

%---------------------------------------------------------
\begin{table*}[htb!]

\centering \protect\caption{Photon-decay helicity amplitudes at the pole for $p\gamma$ and $n\gamma$ decays. Fit to pion-nucleon elastic amplitude $F_{15}$ and multipoles $E_{3-}^{1/2}$ and $M_{3-}^{1/2}$. Complex quantities given as modulus and phase. Results from present study (first row), PR2014~\cite{Svarc:2014sqa} (second row), and BnGa~\cite{CBELSATAPS:2015kka} (for proton couplings) and \cite{Anisovich:2017afs} (for neutron couplings) (third row).}

\vspace{2mm}
{%
\begin{tabular}{|c|c|c|c|c|c|c|c|c|c|c|}
\hline
Resonance
& $Re W_p$             
& $-2 Im W_p$ 
& $|pA^{1/2}|$ 
& $p\phi A^{1/2}$ 
& $|pA^{3/2}|$ 
& $p\phi A^{3/2}$ 
& $|nA_{1/2}|$ 
& $n\phi A^{1/2}$ 
& $|nA^{3/2}|$ 
& $nA\phi ^{3/2}$ \tabularnewline
& (GeV)
& (GeV)
& ($GeV^{-1/2}$) 
& (deg) 
& ($GeV^{-1/2}$) 
& (deg) 
& ($GeV^{-1/2}$)
&  (deg) 
& ($GeV^{-1/2}$) 
& (deg)\tabularnewline
\hline
$N(1680)5/2^+$
& 1.672$\pm$0.017
& 0.113$\pm$0.004 
& 0.020$\pm$0.002 
& 141$\pm$25 
& 0.126$\pm$0.011
& -1.1$\pm$0.1
& 0.037$\pm$0.006 
& -15$\pm$3
& 0.040$\pm$0.007
& -176$\pm$29 \tabularnewline
& 1.674$\pm$0.003 
& 0.113$\pm$0.005 
& 0.014$\pm$0.005 
& 130$\pm$20 
& 0.123$\pm$0.004
& -6$\pm$3
& 
& 
& 
& \tabularnewline
& 1.678$\pm$0.005           
& 0.113$\pm$0.004      
& 0.013$\pm$0.003      
& 160$\pm$17       
& 0.135$\pm$0.005
& 1$\pm$3
& 0.032$\pm$0.003         
& -7$\pm$5 
& 0.063$\pm$0.004
& 170$\pm$5 \tabularnewline
\hline
\end{tabular}} \label{tab:tblR7}
\end{table*}
%---------------------------------------------------------
\begin{table*}[htb!]

\centering \protect\caption{Photon-decay helicity amplitudes at the pole for $p\gamma$ decay. Fit to pion-nucleon elastic amplitude $P_{33}$ and multipoles $E_{1+}^{3/2}$ and $M_{1+}^{3/2}$. Complex quantities given as modulus and phase. Results from present study (first row), PR2014~\cite{Svarc:2014sqa} (second row), and BnGa~\cite{Anisovich:2011fc} (for $\Delta(1232)3/2^+$) and \cite{CBELSATAPS:2015kka} (for $\Delta(1620)3/2^+$) (third row).}

\vspace{2mm}
{%
\begin{tabular}{|c|c|c|c|c|c|c|}
\hline
Resonance
& $Re W_p$             
& $-2 Im W_p$ 
& $|pA^{1/2}|$ 
& $p\phi A^{1/2}$ 
& $|pA^{3/2}|$ 
& $p\phi A^{3/2}$ \tabularnewline
& (GeV)
& (GeV)
& ($GeV^{-1/2}$) 
& (deg) 
& ($GeV^{-1/2}$) 
& (deg) \tabularnewline
\hline
$\Delta(1232)3/2^+$
& 1.210$\pm$0.001
& 0.995$\pm$0.001 
& 0.130$\pm$0.005
& 161$\pm$7 
& 0.263$\pm$0.012
& 171$\pm$8 \tabularnewline
& 1.211$\pm$0.001 
& 0.101$\pm$0.002 
& 0.129$\pm$0.002 
& 167$\pm$2 
& 0.259$\pm$0.002
& 179$\pm$2 \tabularnewline
& 1.210$\pm$0.001           
& 0.099$\pm$0.002      
& 0.131$\pm$0.004       
& 161$\pm$2       
& 0.254$\pm$0.005 
& 171$\pm$1 \tabularnewline
\hline
\end{tabular}} \label{tab:tblR8}
\end{table*}
%---------------------------------------------------------
\begin{table*}[htb!]

\centering \protect\caption{Photon-decay helicity amplitudes at the pole for $p\gamma$ decay. Fit to pion-nucleon elastic amplitude $D_{33}$ and multipoles $E_{2-}^{3/2}$ and $M_{2-}^{3/2}$. Complex quantities given as modulus and phase. Results from present study (first row), PR2014~\cite{Svarc:2014sqa} (second row), and BnGa~\cite{CBELSATAPS:2015kka} (third row).}

\vspace{2mm}
{%
\begin{tabular}{|c|c|c|c|c|c|c|}
\hline
Resonance
& $Re W_p$             
& $-2 Im W_p$ 
& $|pA^{1/2}|$ 
& $p\phi A^{1/2}$ 
& $|pA^{3/2}|$ 
& $p\phi A^{3/2}$ \tabularnewline
& (GeV)
& (GeV)
& ($GeV^{-1/2}$) 
& (deg) 
& ($GeV^{-1/2}$) 
& (deg) \tabularnewline
\hline
$\Delta(1700)3/2^-$
& 1.638$\pm$0.002
& 0.267$\pm$0.004 
& 0.147$\pm$0.004 
& 12.0$\pm$0.3 
& 0.173$\pm$0.004
& 25.8$\pm$0.6 \tabularnewline 
& 1.650$\pm$0.004 
& 0.255$\pm$0.011 
& 0.125$\pm$0.002
& 20$\pm$2
& 0.132$\pm$0.004
& 27$\pm$3 \tabularnewline
& 1.685$\pm$0.010           
& 0.300$\pm$0.015      
& 0.175$\pm$0.020       
& 50$\pm$10        
& 0.180$\pm$0.020
& 45$\pm$10 \tabularnewline
\hline
\end{tabular}} \label{tab:tblR9}
\end{table*}
%---------------------------------------------------------
\begin{table*}[htb!]

\centering \protect\caption{Photon-decay helicity amplitudes at the pole for $p\gamma$ decay. Fit to pion-nucleon elastic amplitude $F_{35}$ and multipoles $E_{3-}^{3/2}$ and $M_{3-}^{3/2}$. Complex quantities given as modulus and phase. Results from present study (first row), PR2014~\cite{Svarc:2014sqa} (second row), and BnGa~\cite{CBELSATAPS:2015kka} (third row).}

\vspace{2mm}
{%
\begin{tabular}{|c|c|c|c|c|c|c|}
\hline
Resonance
& $Re W_p$             
& $-2 Im W_p$ 
& $|pA^{1/2}|$ 
& $p\phi A^{1/2}$ 
& $|pA^{3/2}|$ 
& $p\phi A^{3/2}$ \tabularnewline
& (GeV)
& (GeV)
& ($GeV^{-1/2}$) 
& (deg) 
& ($GeV^{-1/2}$) 
& (deg) \tabularnewline
\hline
$\Delta(1905)5/2^+$
& 1.799$\pm$0.006
& 0.227$\pm$0.012 
& 0.051$\pm$0.006 
& 166$\pm$21 
& 0.009$\pm$0.001
& -171$\pm$22 \tabularnewline
& 1.817$\pm$0.007 
& 0.257$\pm$0.015 
& 0.015$\pm$0.002 
& -29$\pm$9 
& 0.038$\pm$0.001
& -174$\pm$2 \tabularnewline
& 1.800$\pm$0.006          
& 0.290$\pm$0.015    
& 0.025$\pm$0.005      
& -28$\pm$12       
& 0.050$\pm$0.004
& -175$\pm$10 \tabularnewline
\hline
\end{tabular}} \label{tab:tblR10}
\end{table*}

%---------------------------------------------------------
\begin{table*}[htb!]

\centering \protect\caption{Photon-decay helicity amplitudes at the pole for $p\gamma$ decay. Fit to pion-nucleon elastic amplitude $F_{37}$ and multipoles $E_{3+}^{3/2}$ and $M_{3+}^{3/2}$. Complex quantities given as modulus and phase. Results from present study (first row), PR2014~\cite{Svarc:2014sqa} (second row), and BnGa~\cite{CBELSATAPS:2015kka} (third row).}

\vspace{2mm}
{%
\begin{tabular}{|c|c|c|c|c|c|c|}
\hline
Resonance
& $Re W_p$             
& $-2 Im W_p$ 
& $|pA^{1/2}|$ 
& $p\phi A^{1/2}$ 
& $|pA^{3/2}|$ 
& $p\phi A^{3/2}$ \tabularnewline
& (GeV)
& (GeV)
& ($GeV^{-1/2}$) 
& (deg) 
& ($GeV^{-1/2}$) 
& (deg) \tabularnewline
\hline
$\Delta(1950)7/2^+$
& 1.883$\pm$0.002
& 0.240$\pm$0.005 
& 0.072$\pm$0.008 
& 179$\pm$20 
& 0.090$\pm$0.010
& 173$\pm$19 \tabularnewline
& 1.879$\pm$0.005 
& 0.231$\pm$0.009 
& 0.076$\pm$0.004 
& 175$\pm$4 
& 0.095$\pm$0.005
& -178$\pm$4 \tabularnewline
& 1.888$\pm$0.004          
& 0.245$\pm$0.008    
& 0.067$\pm$0.004      
& 170$\pm$5       
& 0.095$\pm$0.004
& 170$\pm$5 \tabularnewline
\hline
\end{tabular}} \label{tab:tblR11}
\end{table*}

%----------------------------------------------------------------------
\begin{figure*}[hbt!]
\vspace{0.4cm}
\centering
{
\includegraphics[width=0.37\textwidth]{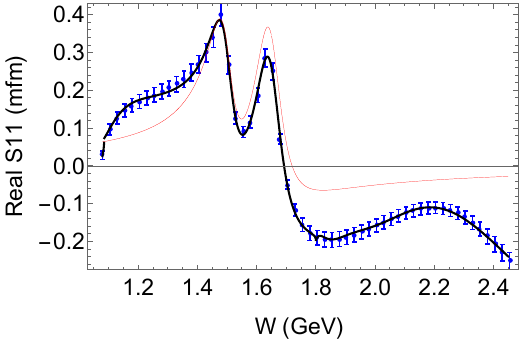} \hspace{0.5cm}
\includegraphics[width=0.37\textwidth]{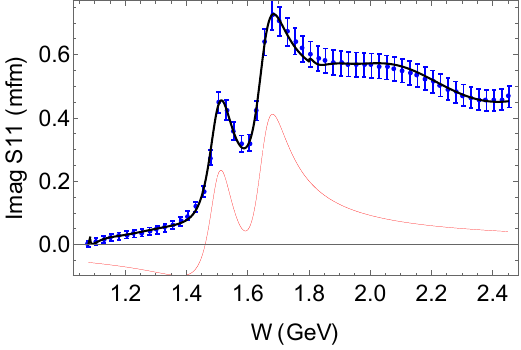}  \\
}
\centering
{
\includegraphics[width=0.37\textwidth]{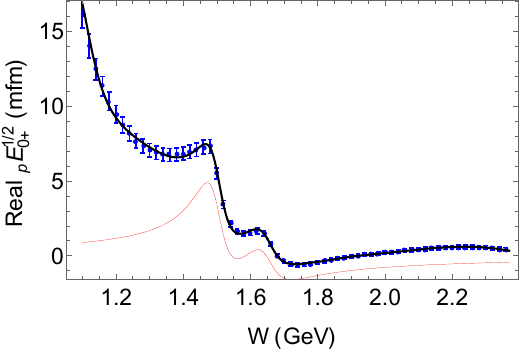} \hspace{0.5cm}
\includegraphics[width=0.37\textwidth]{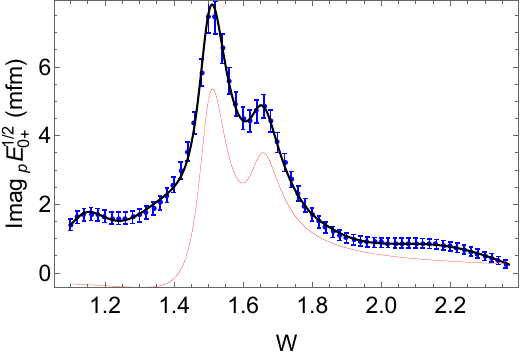}  \\
}
\centering
{
\includegraphics[width=0.37\textwidth]{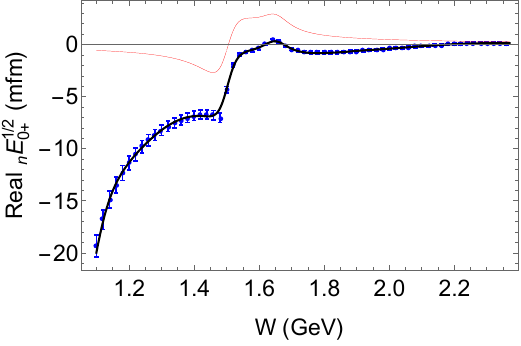} \hspace{0.5cm}
\includegraphics[width=0.37\textwidth]{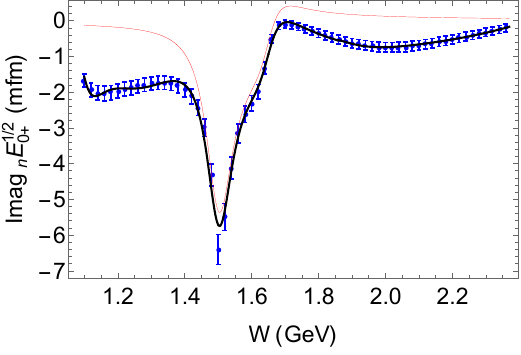}  \\
}

\caption{Samples of Laurent+Pietarinen (L+P) coupled fit of the $S_{11}$ $\pi N$ partial wave of the GWU-SAID fit WI08~\cite{Workman:2012hx} and the SM22 ED GWU-SAID multipole solutions. Blue symbols are the GWU-SAID solutions, solid black curves are the L+P coupled-multipole fit, and thin red curves are the resonant contribution in the L+P coupled-multipole fit.
}
\label{fig:a1}
\end{figure*}
%----------------------------------------------------------------------
\begin{figure*}[hbt!]
\vspace{0.4cm}
\centering
{
\includegraphics[width=0.37\textwidth]{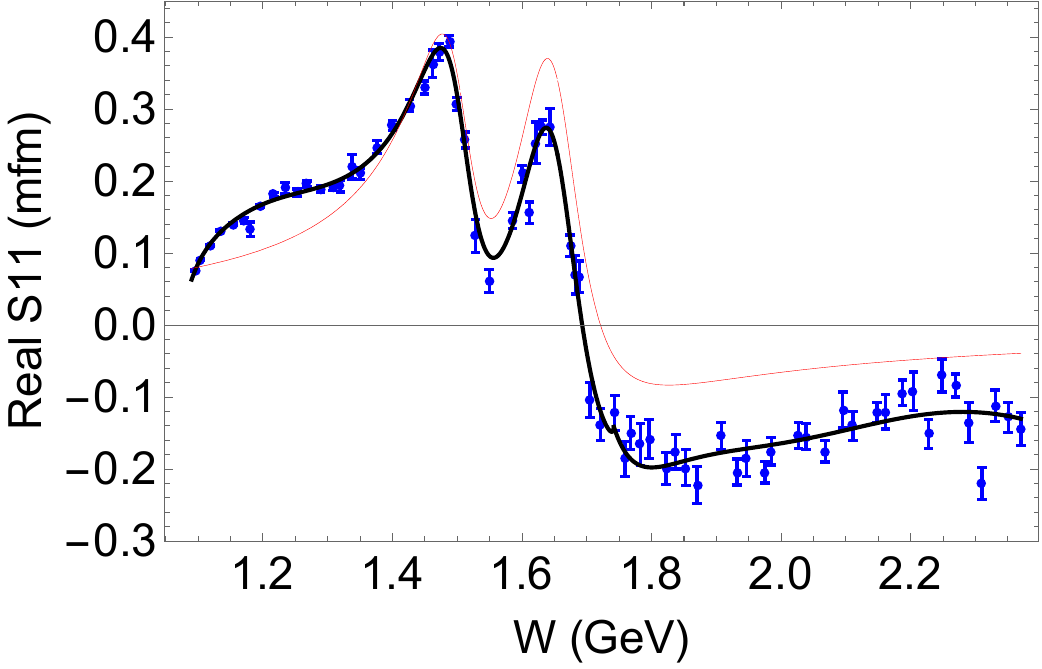} \hspace{0.5cm}
\includegraphics[width=0.37\textwidth]{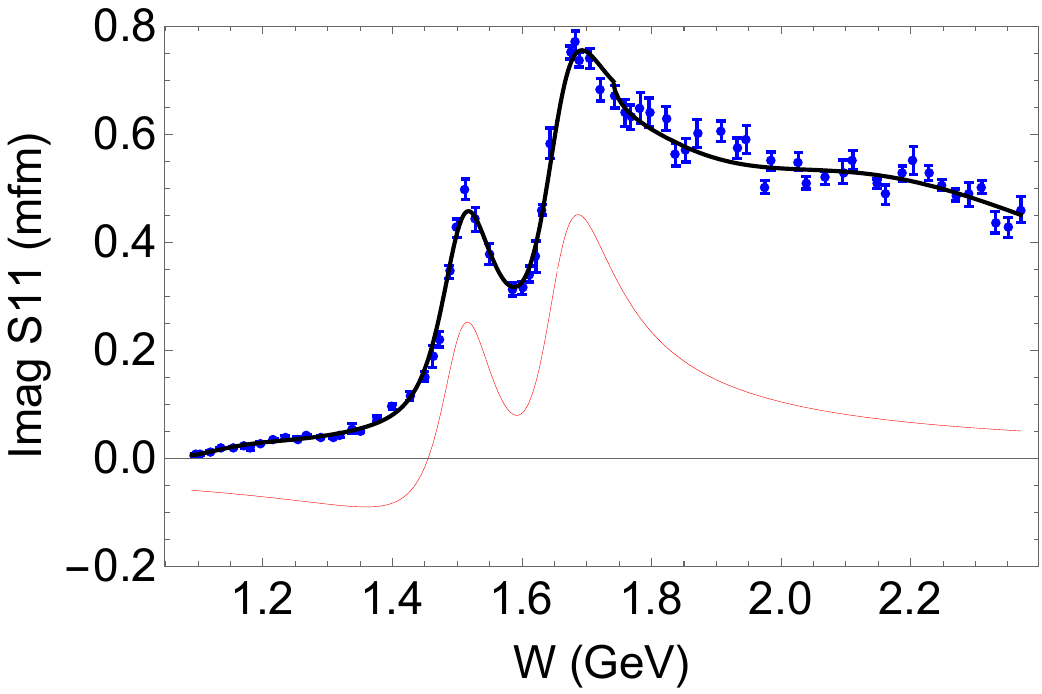}  \\
}
\centering
{
\includegraphics[width=0.37\textwidth]{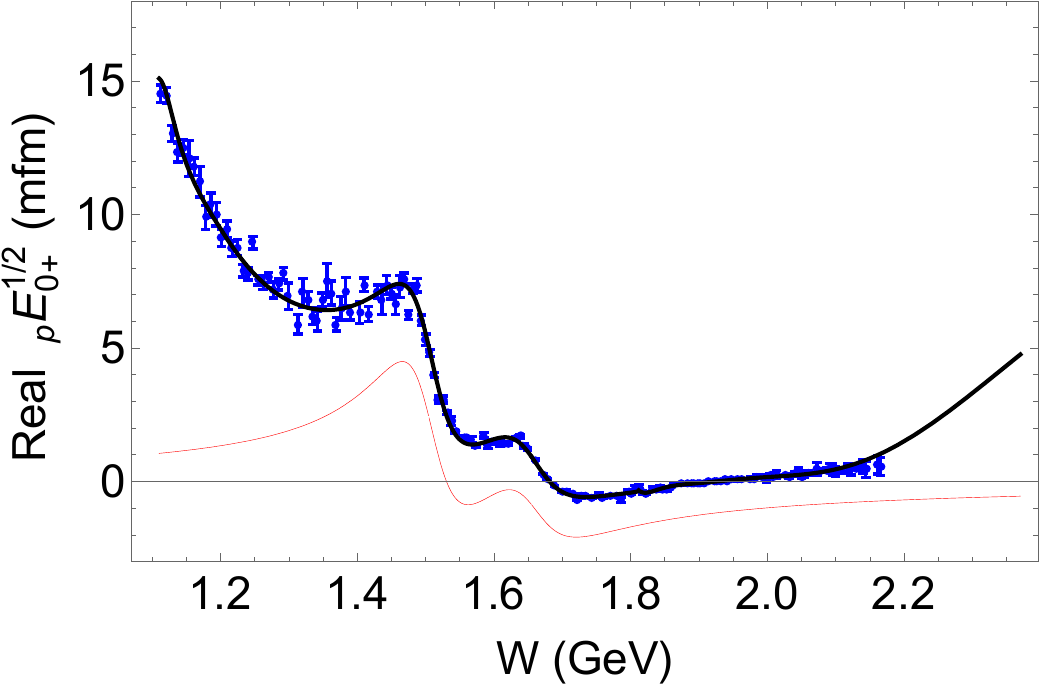} \hspace{0.5cm}
\includegraphics[width=0.37\textwidth]{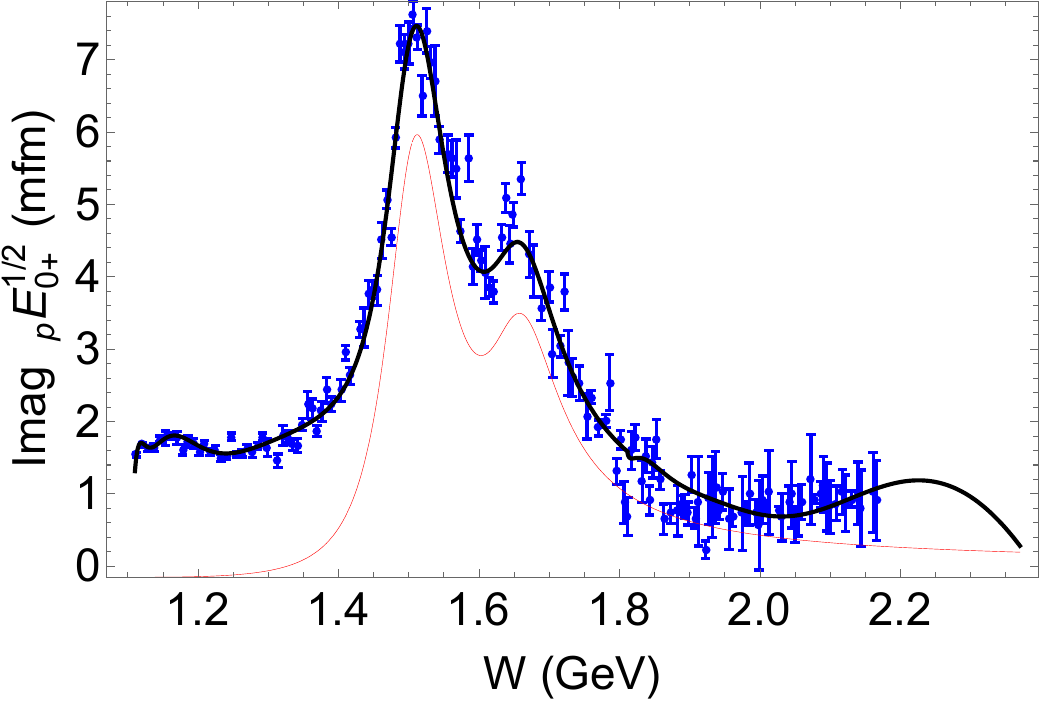}  \\
}
\centering
{
\includegraphics[width=0.37\textwidth]{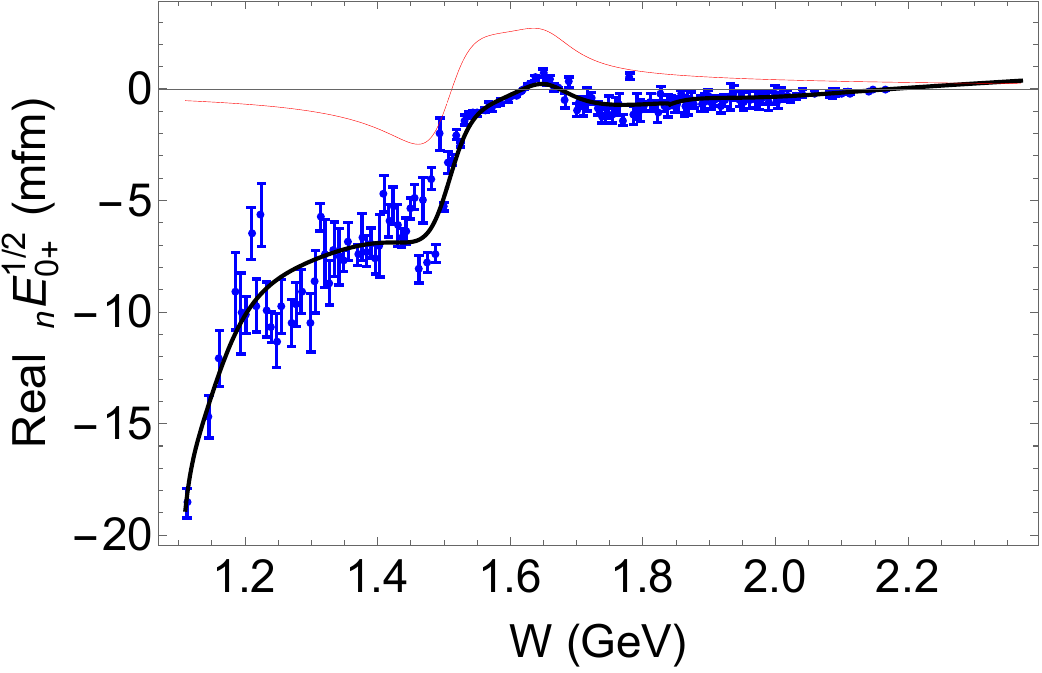} \hspace{0.5cm}
\includegraphics[width=0.37\textwidth]{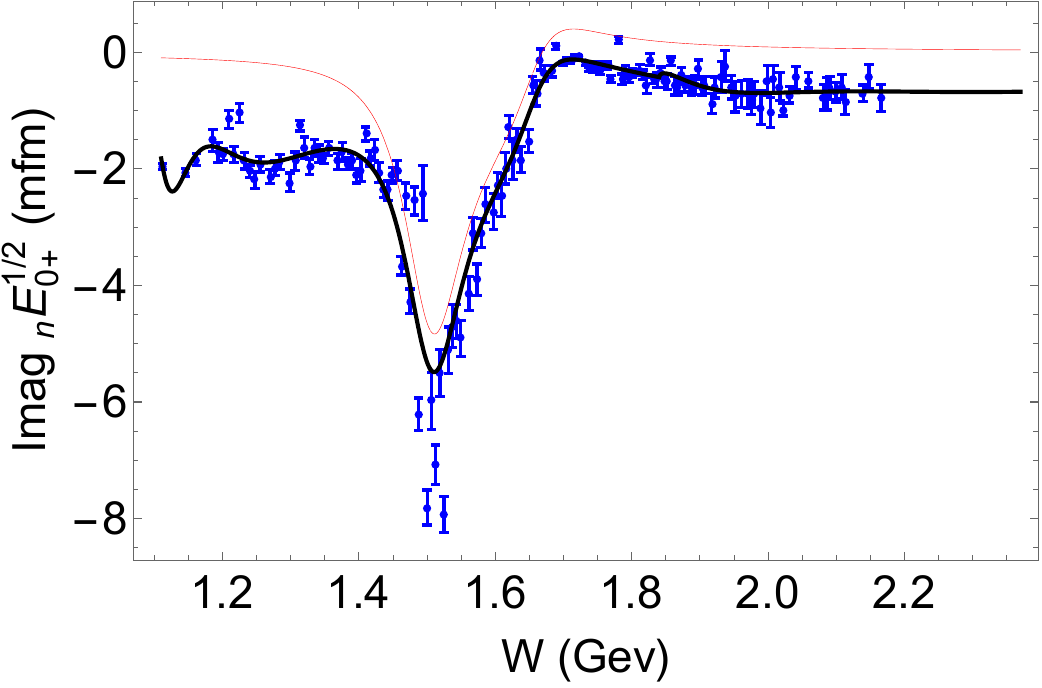}  \\
}

\caption{Samples of Laurent+Pietarinen (L+P) coupled fit of the $S_{11}$ $\pi N$ partial wave of the GWU-SAID fit WI08~\cite{Workman:2012hx} and SM22 SE4 GWU-SAID multipole solutions.  Notation of the solutions is the same as in Fig.~\ref{fig:a1}.
%Blue symbols are the GWU-SAID solutions, solid black curves are the L+P coupled-multipole fit, and thin red curves are the resonant contribution in the L+P coupled-multipole fit.
}
\label{fig:a2}
\end{figure*}
%----------------------------------------------------------------------
\begin{figure*}[hbt!]
\vspace{0.4cm}
\centering
{
\includegraphics[width=0.37\textwidth]{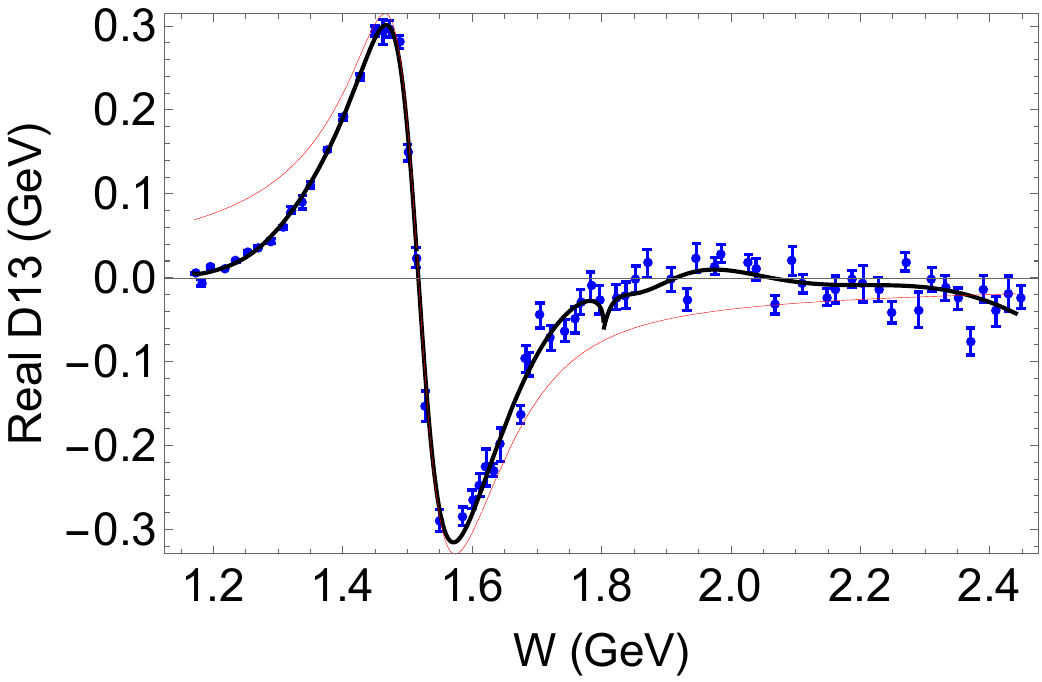} \hspace{0.5cm}
\includegraphics[width=0.37\textwidth]{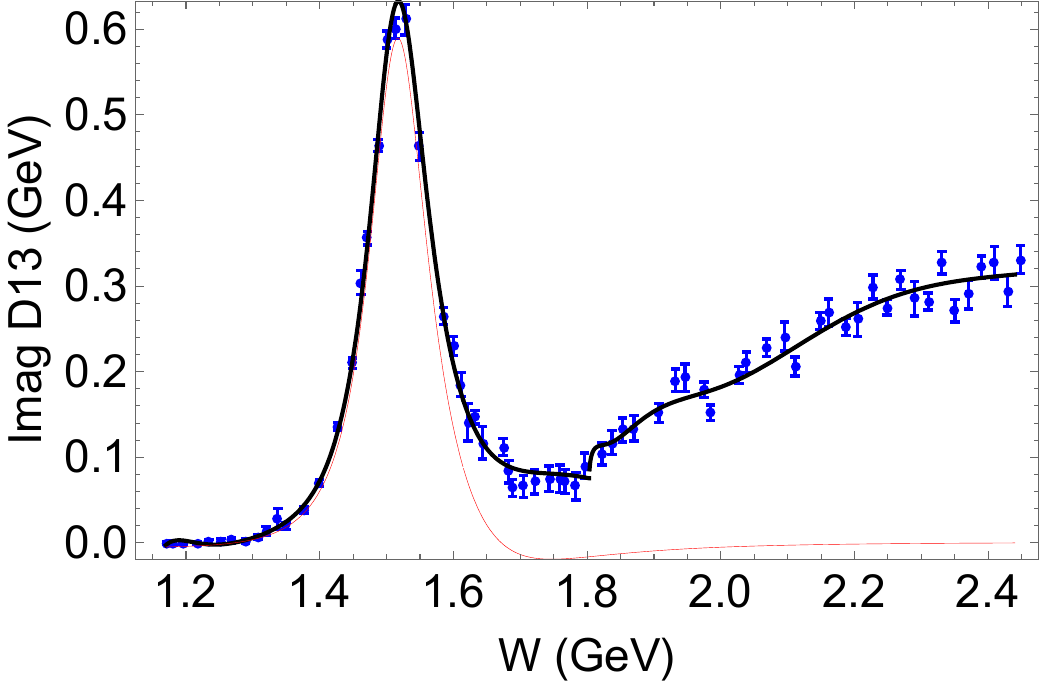}  \\
}
\centering
{
\includegraphics[width=0.37\textwidth]{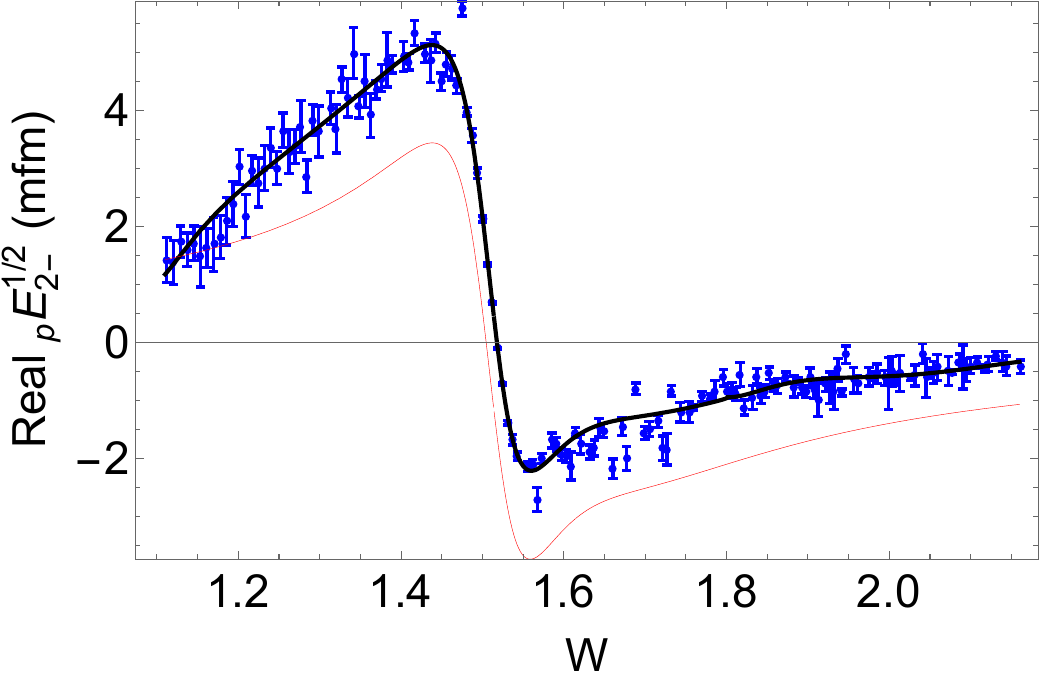} \hspace{0.5cm}
\includegraphics[width=0.37\textwidth]{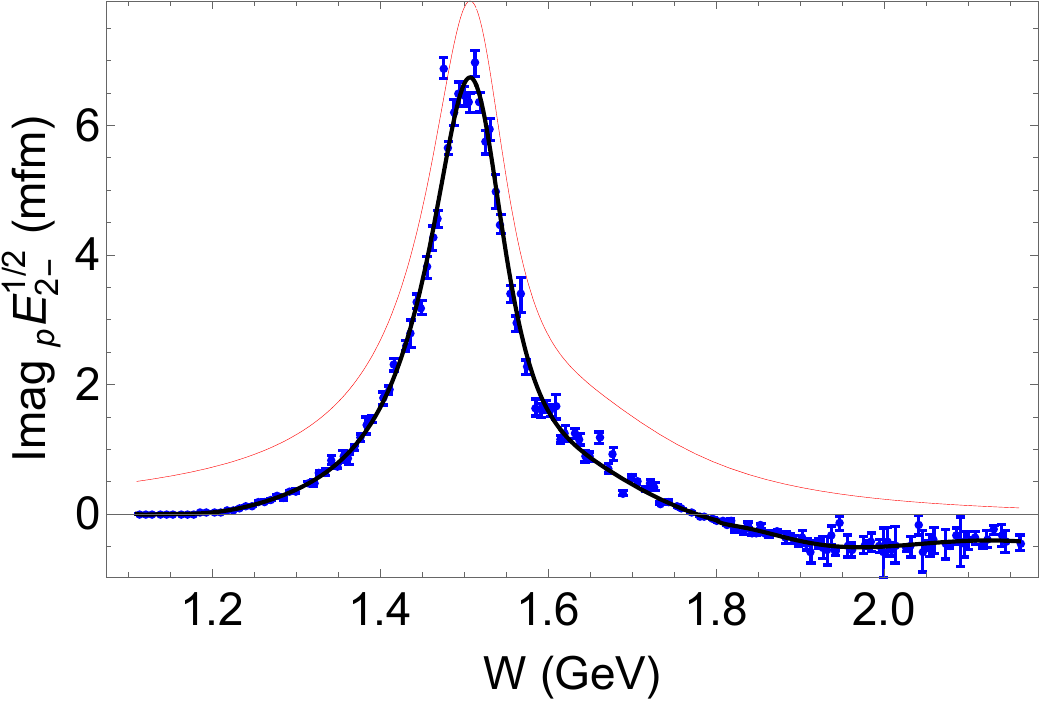}  \\
}
\centering
{
\includegraphics[width=0.37\textwidth]{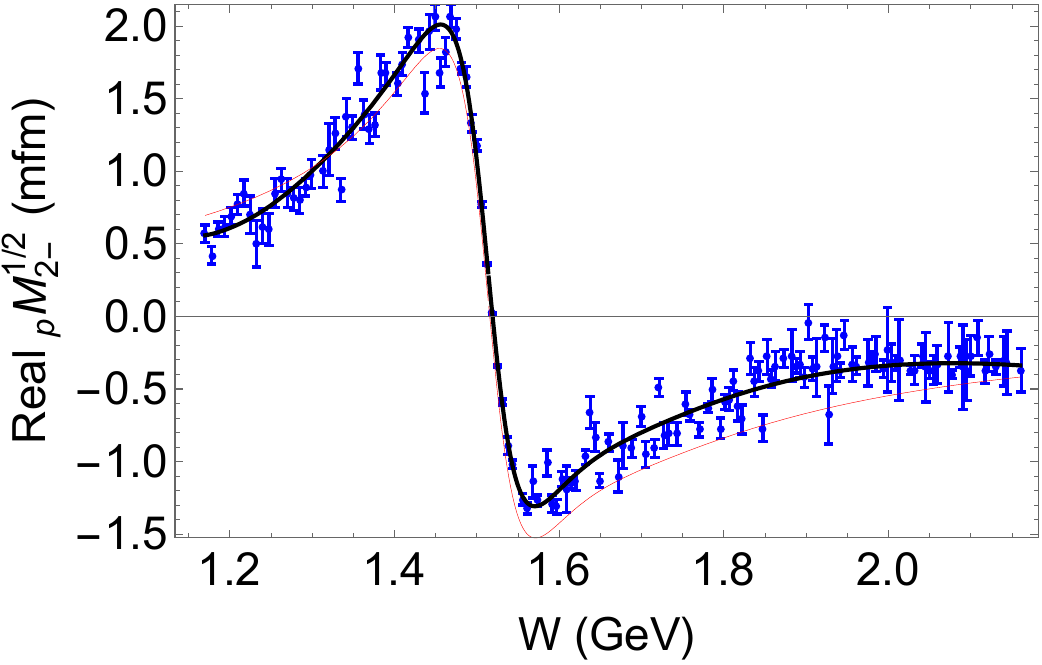} \hspace{0.5cm}
\includegraphics[width=0.37\textwidth]{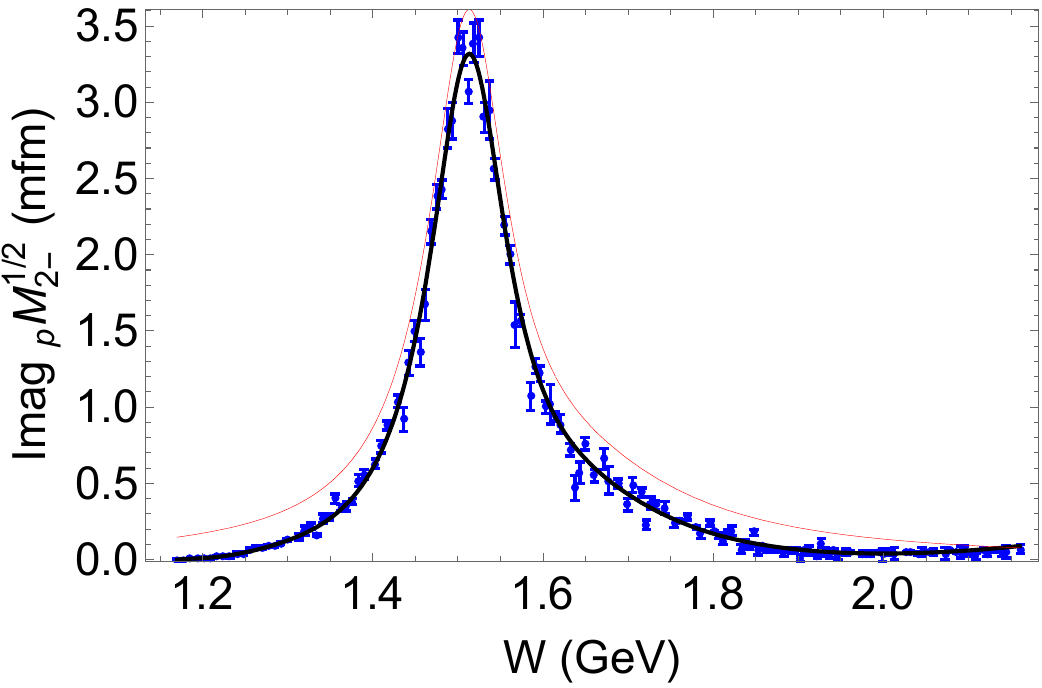}  \\
}
\centering
{
\includegraphics[width=0.37\textwidth]{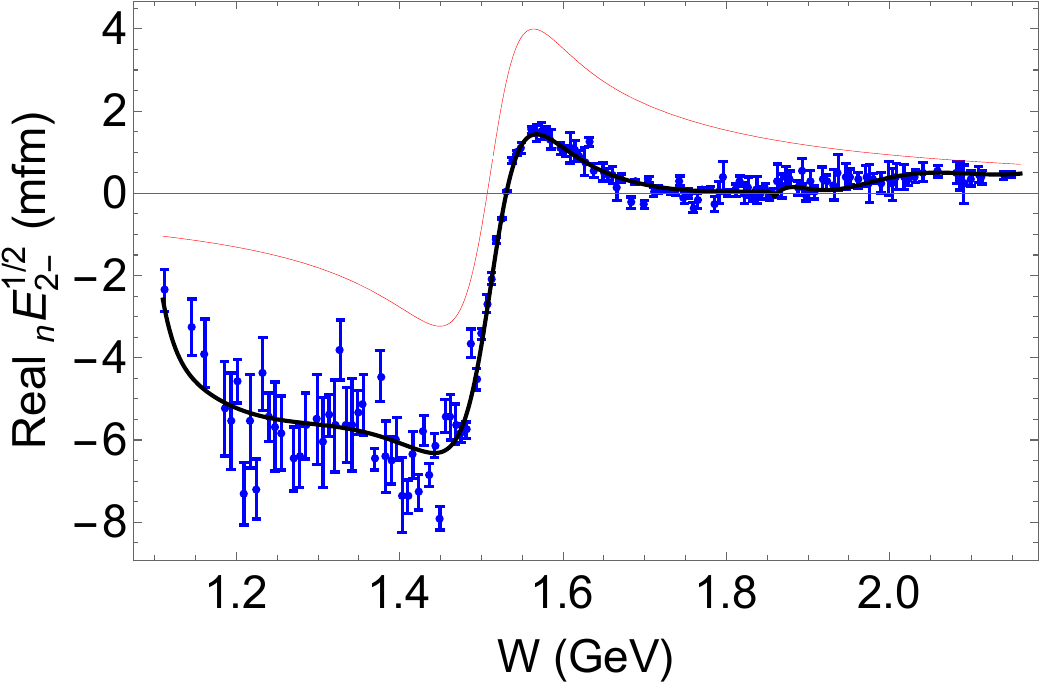} \hspace{0.5cm}
\includegraphics[width=0.37\textwidth]{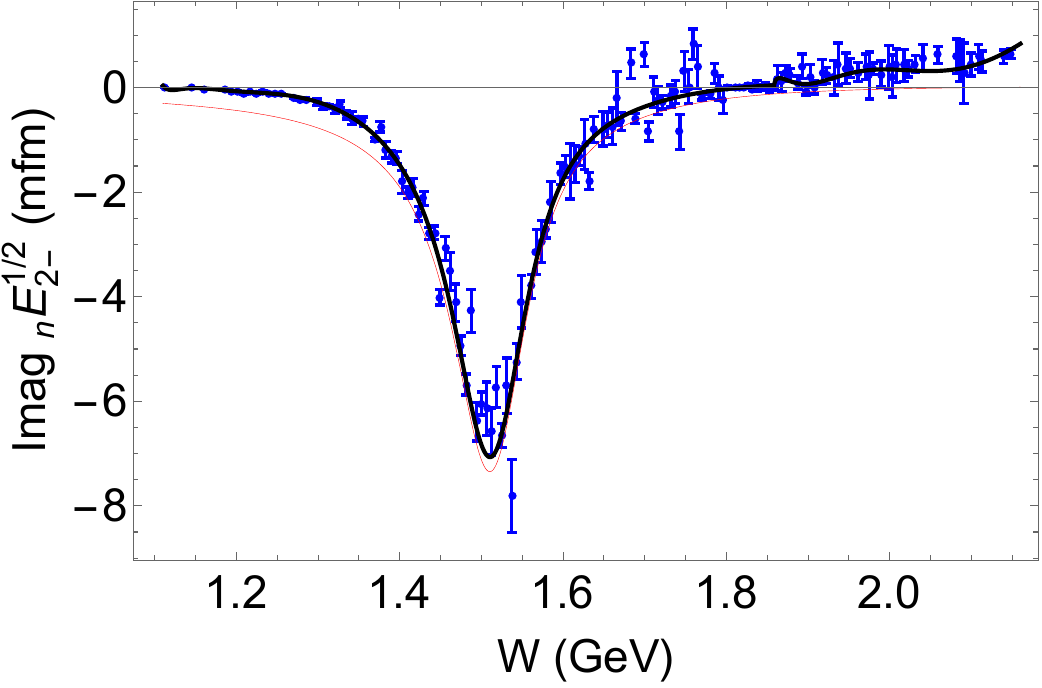}  \\
}
\centering
{
\includegraphics[width=0.37\textwidth]{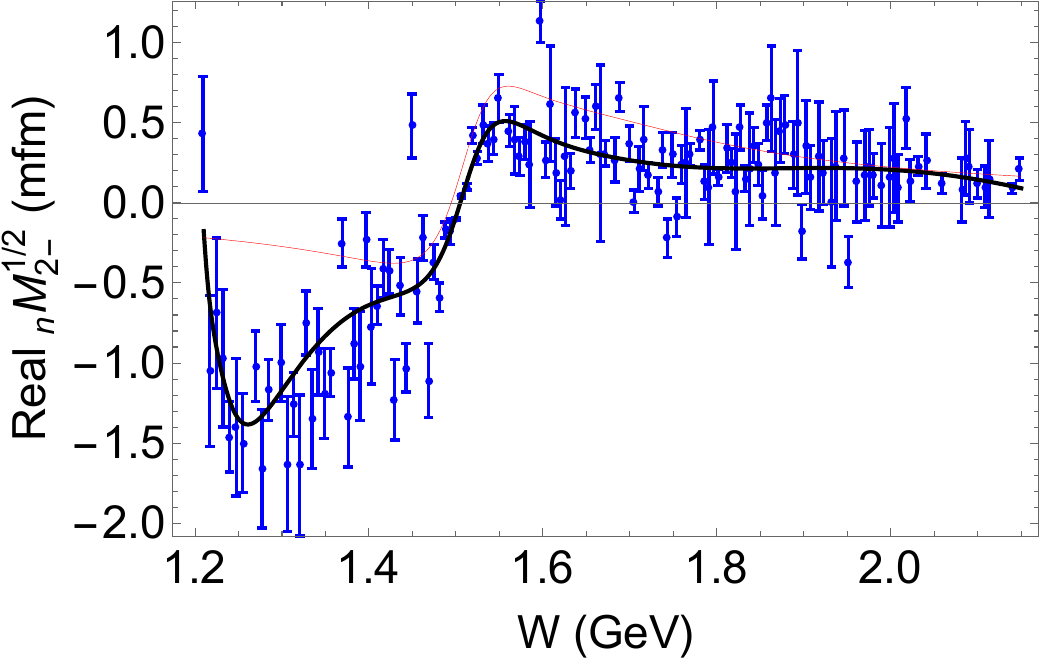} \hspace{0.5cm}
\includegraphics[width=0.37\textwidth]{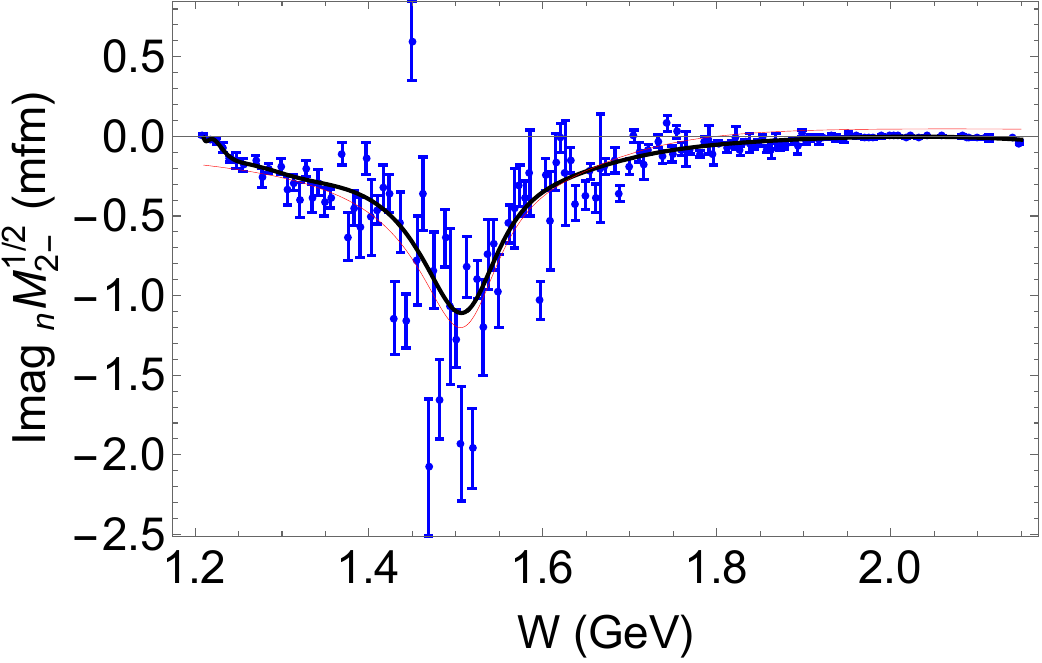}  \\
}

\caption{Samples of Laurent+Pietarinen (L+P) coupled fit of the $D_{13}$ $\pi N$ partial wave of the GWU-SAID fit WI08~\cite{Workman:2012hx} and SM22 SE4 GWU-SAID multipole solutions.  Notation of the solutions is the same as in Fig.~\ref{fig:a1}.
%Blue symbols are the GWU-SAID solutions, solid black curves are the L+P coupled-multipole fit, and thin red curves are the resonant contribution in the L+P coupled-multipole fit.
}
\label{fig:a3}
\end{figure*}
%----------------------------------------------------------------------
\begin{figure*}[hbt!]
\vspace{0.4cm}
\centering
{
\includegraphics[width=0.37\textwidth]{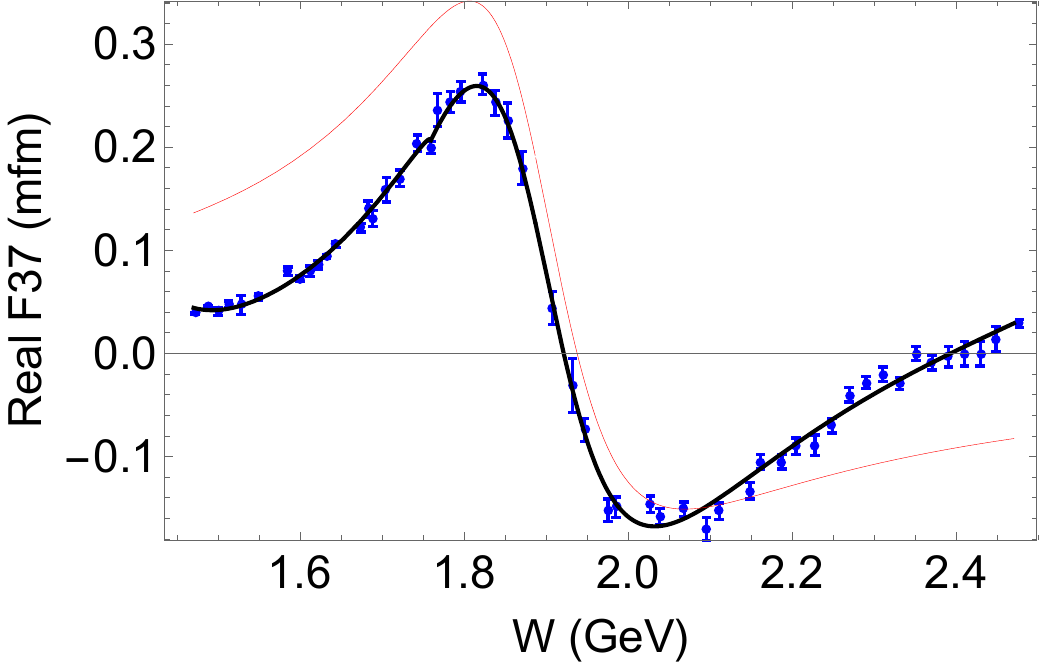} \hspace{0.5cm}
\includegraphics[width=0.37\textwidth]{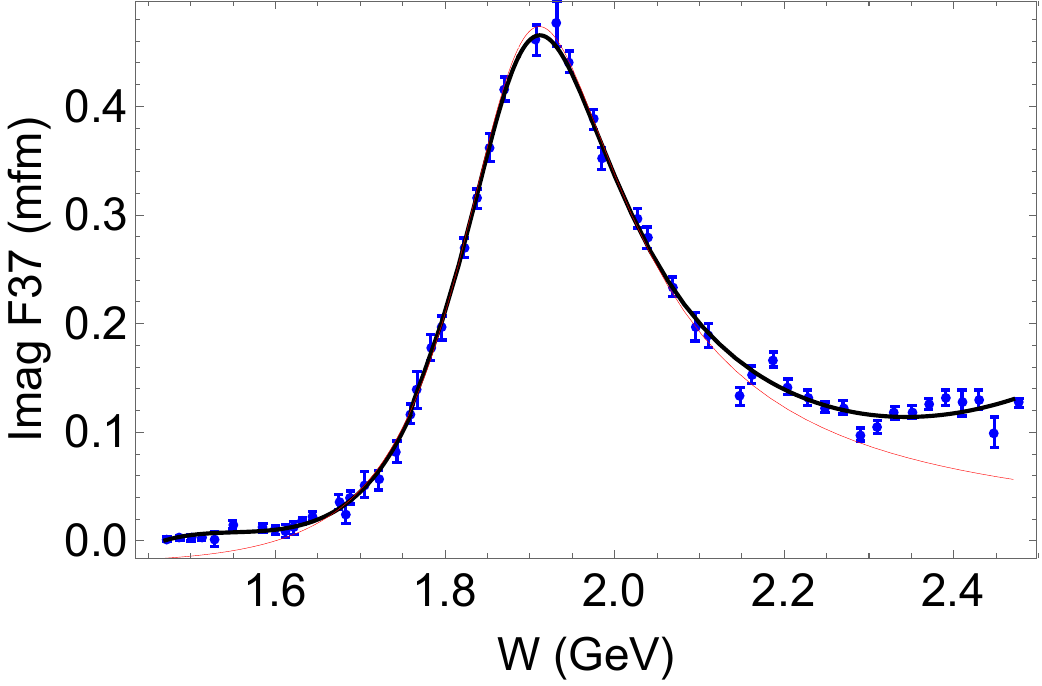}  \\
}
\centering
{
\includegraphics[width=0.37\textwidth]{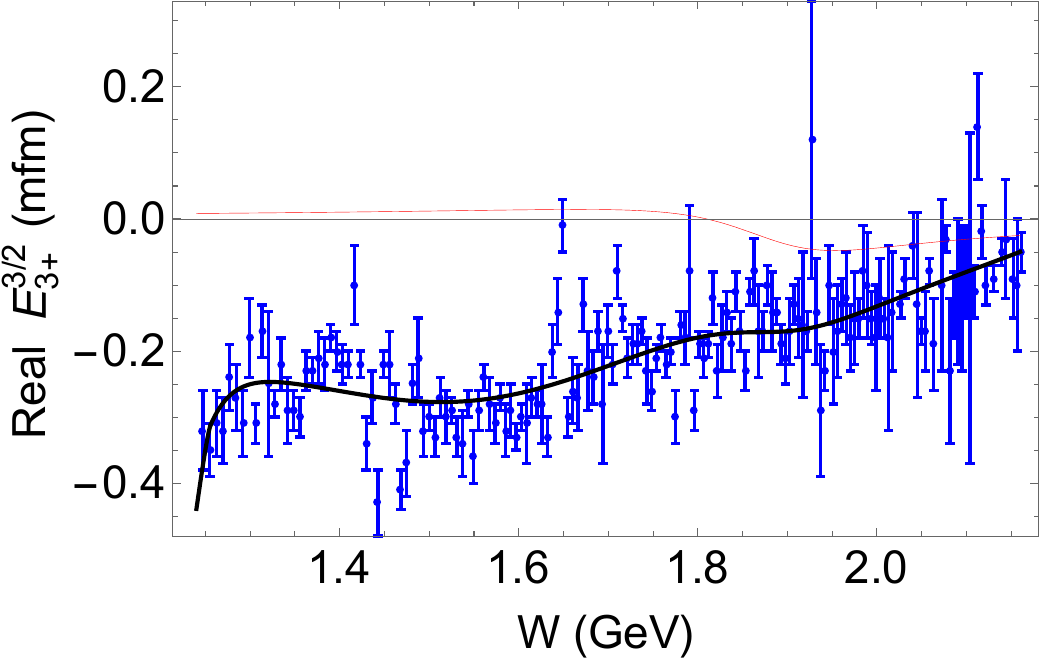} \hspace{0.5cm}
\includegraphics[width=0.37\textwidth]{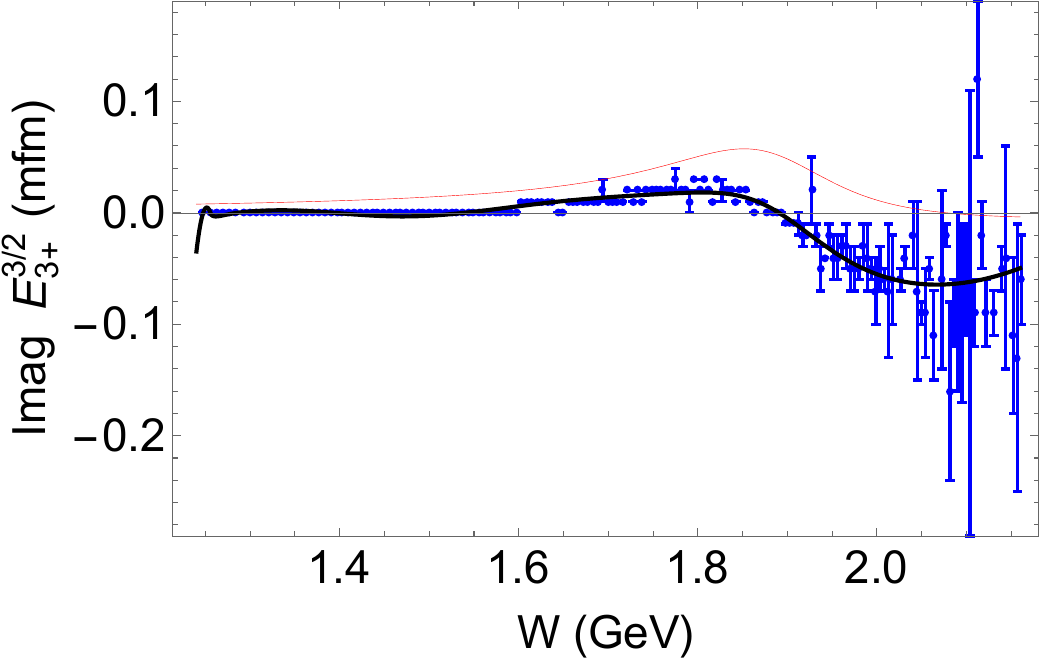}  \\
}
\centering
{
\includegraphics[width=0.37\textwidth]{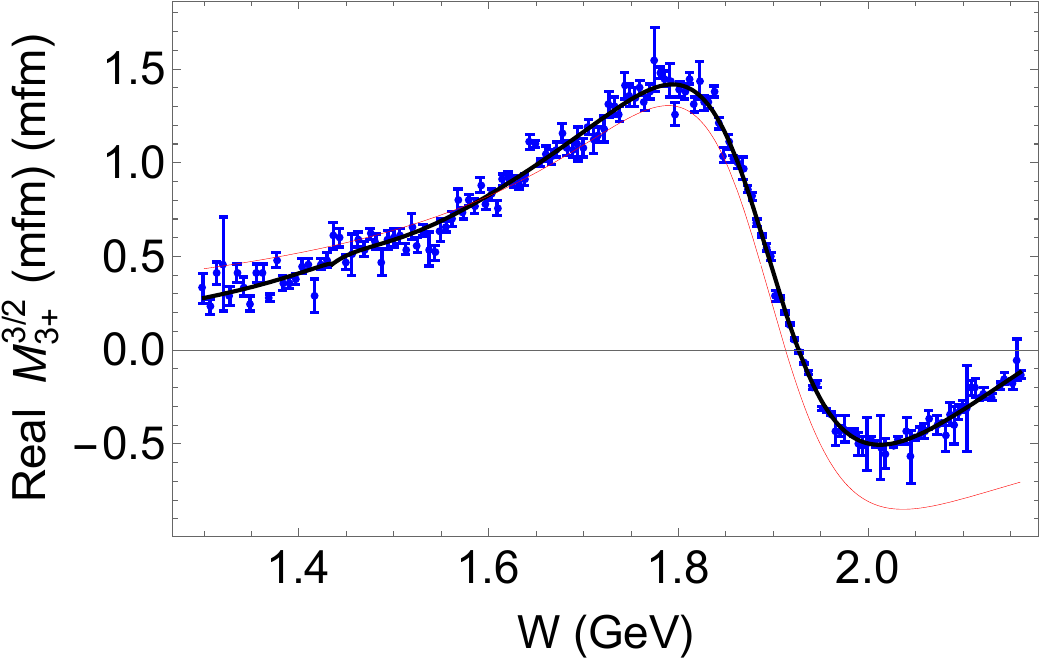} \hspace{0.5cm}
\includegraphics[width=0.37\textwidth]{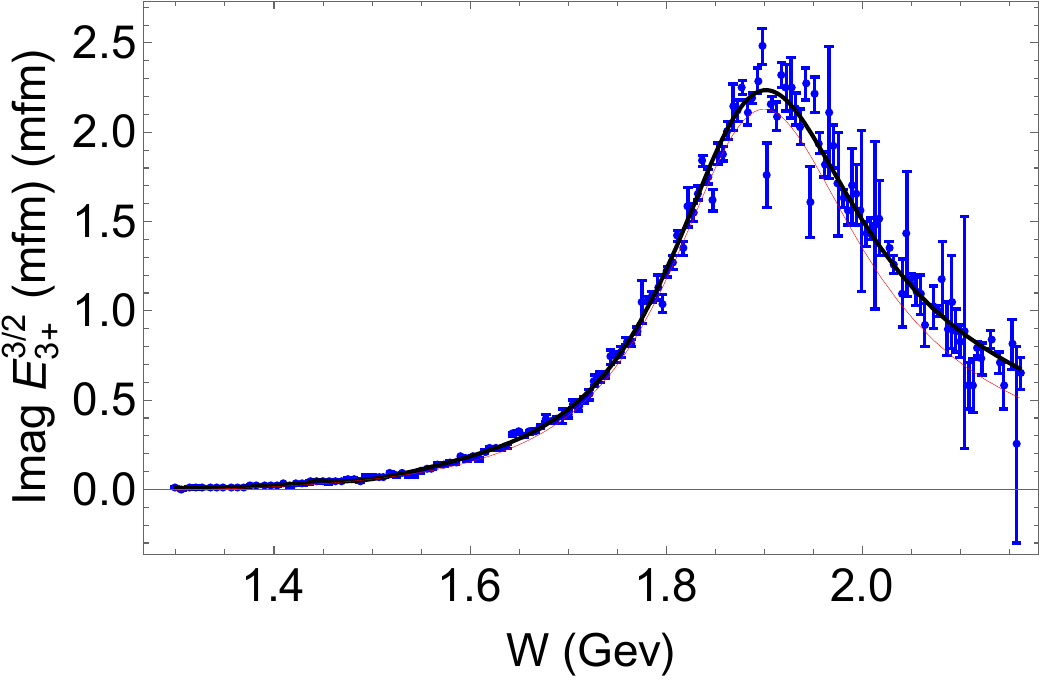}  \\
}

\caption{Samples of Laurent+Pietarinen (L+P) coupled fit of the $F_{37}$  $\pi N$ partial wave of the GWU-SAID WI08~\cite{Workman:2012hx} and SM22 SE4 GWU-SAID multipole solutions. Notation of the solutions is the same as in Fig.~\ref{fig:a1}.
%Blue symbols are the GWU-SAID solutions, solid black curves are the L+P coupled-multipole fit, and thin red curves are the resonant contribution in the L+P coupled-multipole fit.
}
\label{fig:a4}
\end{figure*}

%------------------------------------------------------------
%\clearpage
\section{Results and Conclusions}
\label{Sec:sum}

The present results update the SAID fit (CM12) which first utilized a 
Chew-Mandelstam K-matrix approach (as opposed to the Heitler K-matrix formalism used in the
original SAID analyses). The L+P method for pole parameter extraction has been extended to
simultaneously incorporate all connected $\pi N$ elastic and photoproduction amplitudes. 

The amplitude tables give pole positions and helicity amplitudes at the pole where available. Values for the $n\gamma$ amplitudes were not extracted in the 2014 SAID analysis; comparisons can now be made to multi-channel determinations. Complex amplitudes are given in terms of modulus and phase. In cases where a large phase is found, close to 180~degrees, a minus sign is commonly extracted to ease comparison with the real amplitudes found in older Breit-Wigner fits. The ``modulus'' then has a sign and a phase closer to zero. Here, however, the modulus remains positive.

In cases where the fitted multipoles have a clear canonical resonance variation, with a relatively small non-resonance contribution, comparison to the Bonn-Gatchina multi-channel analysis generally shows good agreement (to the 10$\%$ level). This includes the $\Delta (1232) 3/2^+$, $N(1520) 3/2^-$, $N(1680) 5/2^+$, and $\Delta (1905) 5/2^+$ and applies to both the $p\gamma$, and $n\gamma$ helicity amplitudes.

Comparisons are more complicated for states associated with the low-angular momentum states
$E_{0+}^{1/2}$ and $M_{1-}^{1/2}$. The $N(1535) 1/2^-$ and $N(1650) 1/2^-$ have some overlap and are close to the $\eta N$ threshold cusp. The $N(1440)$ is complicated by the close proximity of its pole position to the $\pi \Delta$ threshold. We note that differences in $N(1535) 1/2^-$ $p\gamma$  amplitudes disappear if one compares instead with the recent J\"ulich-Bonn analysis~\cite{Ronchen:2022hqk}. For the $n\gamma$ amplitudes, the agreement is qualitative and no J\"ulich-Bonn values are available. Qualitative agreement is also seen for the $N(1650) 1/2^-$. 

Agreement for the $\Delta (1700)3/2^-$ is good for the moduli and at least qualitative for the phases. For the $N(1720) 3/2^+$, within fairly large uncertainties, there is qualitative agreement of the helicity amplitude moduli, with less agreement at the level of phases.
Hunt and Manley~\cite{Hunt:2018wqz} note that the $N(1675) 5/2^-$ decays to $p \gamma$ violate the Moorhouse selection rule~~\cite{Moorhouse:1966jn}. We see the moduli of $p\gamma$ photo-decay amplitudes to be small but non-zero. 

In Figs.~\ref{fig:a1} - \ref{fig:a4}, we display L+P fits for the $D_{13}$ partial-wave and multipole amplitudes, where resonance behavior is clear and the dominant feature, and the $S_{11}$ amplitudes, where resonance overlap and a nearby $\eta N$ cusp complicate
this process.

%----------------------------------------------------------------------------------
\clearpage
\section*{Acknowledgments}

%\textcolor{red}{The authors express their gratitude Viktor Kashevarov...}
This work was supported in part by the U.~S.~Department of Energy, Office of Science, Office of Nuclear Physics, under Awards No.~DE--SC0016583 and No.~DE--SC0016582, and in part by the U.S. Department of Energy, Office of Science, Office of Nuclear Physics under contract DE-AC05-06OR23177.

%---------------------- REFERENCES ----------------------

%-----------------------------------------------
\end{document}